\documentclass[11pt,preprintnumbers,nofootinbib,prd]{revtex4}
\usepackage{graphicx}
\usepackage{amsmath}
\usepackage{epsfig}
\usepackage{mathrsfs}

\newcommand{\be}{\begin{equation}}
\newcommand{\ee}{\end{equation}}
\newcommand{\ba}{\begin{eqnarray}}
\newcommand{\ea}{\end{eqnarray}}

\begin{document}
\input{epsf}
\begin{flushright}
IC/2006/131\\
NSF-KITP-06-117\\
MCTP-06-34\\
\end{flushright}

\title{Explaining the Electroweak Scale and Stabilizing Moduli in $M$ theory}
\author{Bobby S. Acharya}
\affiliation{Abdus Salam International Centre for Theoretical
Physics, Strada Costiera 11, Trieste, Italy\\and\\INFN, Sezione di Trieste}
\author{Konstantin Bobkov}
\author{Gordon L. Kane}
\author{Piyush Kumar}
\author{Jing Shao}
\affiliation{Michigan Center for Theoretical Physics, University
of Michigan, Ann Arbor, MI 48109, USA}

\vspace{0.5cm}

\date{\today}

\vspace{0.3cm}

\begin{abstract}

In a recent paper \cite{Acharya:2006ia} it was shown that in
fluxless $M$ theory vacua with at least two hidden sectors
undergoing strong gauge dynamics and a particular form of the
K\"{a}hler potential, all moduli are stabilized by the effective
potential and a stable hierarchy is generated, consistent with
standard gauge unification. This paper explains the results of
\cite{Acharya:2006ia} in more detail and generalizes them, finding
an essentially unique de Sitter (dS) vacuum under reasonable
conditions. One of the main phenomenological consequences is a
prediction which emerges from this entire class of vacua: namely
gaugino masses are significantly suppressed relative to the
gravitino mass. We also present evidence that, for those vacua in
which the vacuum energy is small, the gravitino mass, which sets
all the superpartner masses, is automatically in the TeV - 100 TeV
range.

\end{abstract}
\maketitle
\newpage
\vspace{-1.2cm} \tableofcontents

\section{Introduction and Summary}

There are many good reasons why we study string theory as a theory of
particle physics. One of these, discovered some twenty or so years
ago \cite{Candelas:1985en}, is that a simple question, ``what properties do
four-dimensional heterotic string vacua generically have?''
has an extremely compelling answer: non-Abelian gauge symmetry,
chiral fermions, hierarchical Yukawa couplings and dynamical
supersymmetry breaking. One can also add other important properties
such as gauge coupling unification and doublet-triplet splitting.
Furthermore, after the dust of
the string duality revolution settled, a similar picture was
discovered in other perturbative corners of the landscape, eg Type
IIA, Type IIB and $M$ theory. The above four properties are the
most important properties of the Standard Model. The fifth one -
gauge coupling unification - is an important feature of the MSSM and
low energy supersymmetry.

There is another crucial feature of the Standard Model: namely its overall mass scale,
which is of order $M_W$ as opposed to some other scale such as $M_P$ or $M_{GUT}$.
This property of the Standard Model is much less well understood in string theory
or otherwise, and will be the focus of this paper.

In string/$M$ theory, masses and coupling constants, including
$M_W$ are all functions of the moduli field vevs. Thus the
hierarchy problem in string theory is double edged: one has to
both stabilize all the moduli {\it and} generate the hierarchy
simultaneously. In recent years there has been progress in moduli
stabilization via fluxes and other effects (for a recent review
see \cite{Douglas:2006es}), and with the imminent arrival of the
LHC it is appropriate to address the hierarchy problem in this
context. After all, any given string/$M$ theory vacuum either will
or will not be consistent with the LHC signal, but one cannot even
begin to address this in a meaningful way if the hierarchy is not
understood.

In Type IIB vacua, moduli are stabilized via a combination of
fluxes and quantum corrections
\cite{Giddings:2001yu,Kachru:2003aw}. In these vacua, hierarchies
arise in three ways: warp factors
\cite{Giddings:2001yu,Verlinde:1999fy} as in
\cite{Randall:1999ee}, the presence of non-perturbative effects
\cite{Kachru:2003aw} or by fine tuning the large number of fluxes.
However, flux vacua in Type IIA \cite{DeWolfe:2005uu}, $M$ theory \cite{Acharya:2002kv}
and heterotic string theory \cite{Gukov:2003cy} have the property that the (currently known and understood)
fluxes roughly are equal in number to the number of moduli. This
leads to a large value of the superpotential, and consequently, if
the volume of the extra dimensions is not huge, to a large
gravitino mass. This tends to give a large mass to all scalars via
the effective 4d supergravity potential, which leads to either
$M_W = 0$ or some other large value such as $M_P$ or $M_{GUT}$.
Therefore, for these vacua we require a good idea for generating
and stabilizing the hierarchy.

Thus far, there has been essentially one good idea proposed to
explain the relatively small value of the weak scale. This is that
the weak scale might be identified with, or related to, the strong
coupling scale of an asymptotically free theory which becomes
strongly coupled at low energies and exhibits a mass gap at that
strong coupling scale.
Holographically dual
to this is the idea of warped extra dimensions \cite{Randall:1999ee}.
Strong dynamics
(or its dual) can certainly generate a small scale in a natural
manner, but can it also be compatible with the stabilization of
all the moduli fields?

One context for this question, which we will see is particularly
natural, is $M$ theory compactification
on manifolds $X$ of $G_2$-holonomy without fluxes. In these vacua,
the only moduli one has are zero modes of the metric on $X$, whose
bosonic superpartners are axions. Thus each moduli supermultiplet
has a Peccei-Quinn shift symmetry (which originates from 3-form
gauge transformations in the bulk 11d supergravity). Since such
symmetries can only be broken by non-perturbative effects, the
entire moduli superpotential $W$ is non-perturbative. In general $W$
can depend on all the moduli. Therefore, in addition to the small
scale generated by the strong dynamics we might expect that all
the moduli are actually stabilized.
This paper will demonstrate in detail that this is
indeed the case.

Having established that the basic idea works well, the next
question we address is ``what are the phenomenological
implications?'' Since string/$M$ theory has many vacua, it would
be extremely useful if we could obtain a general prediction from
{\it all} vacua or at least some well-defined subset of vacua.
Remarkably, we are able to give such a prediction for all fluxless
$M$ theory vacua  within the supergravity
approximation\footnote{to which we are restricted for
calculability} with at least two hidden sectors undergoing strong
gauge dynamics and a particular form of the K\"{a}hler potential
as in (\ref{kahler}): {\it gaugino masses are generically
suppressed relative to the gravitino mass}.

A slightly more detailed elucidation of this result is that in all
de Sitter vacua within this class, gaugino masses are always
suppressed. In AdS vacua - which are obviously less interesting phenomenologically -
the gaugino masses are suppressed in `most' of the vacua. This will be
explained in more detail later.

The reason why we are able to draw such a generic conclusion is
the following: any given non-perturbative contribution to the
superpotential depends on various constants which are determined
by a specific choice of $G_2$-manifold $X$. {\it These constants
determine entirely the moduli potential}. They are given by the
constants $b_k$ which are related to the one-loop beta-function
coefficients, the normalization of each term $A_k$ and the
constants $a_i$ (see (\ref{vol})) which characterize the
K\"{a}hler potential for the moduli. Finally, there is a
dependence on the gauge kinetic function, and in $M$ theory this
is determined by a set of integers $N_i$ which specify the
homology class of the 3-cycle on which the non-Abelian gauge group
is localized. Rather than study a particular $X$ which fixes a
particular choice for these constants we have studied the
effective potential as a function of the $(A_k, b_k , a_i, N_i )$
. The result of gaugino mass suppression holds essentially for
arbitrary values of the $(A_k, b_k , a_i , N_i )$, at least in the
supergravity regime where we have been able to calculate. Thus,
any $G_2$-manifold which has hidden sectors with strong gauge
dynamics will lead to suppressed gaugino masses.

At a deeper level, however, the reason that this works is that the
idea of strong gauge dynamics to solve the hierarchy problem is a
good and simple idea which guides us to the answers directly. If
one's theory does not provide a simple mechanism for how the hierarchy is
generated, then it is difficult to see how one could obtain a
reliable prediction for, say, the spectrum of beyond the Standard
Model particles. In a particular subset of Type IIB compactifications,
Conlon and Quevedo have
also discovered some general results \cite{Conlon:2006us}. In fact,
they remarkably also find that gaugino masses are suppressed at tree level,
though the nature of the suppression is not quite the same. Some heterotic
compactifications also exhibit a suppression of tree-level gaugino masses \cite{Binetruy:2000md}.

The suppression of gaugino masses relative to $m_{3/2}$ applies
for all vacua in the supergravity regime arising out of these
compactifications, independent of the value of $m_{3/2}$. However,
in a generic vacuum the cosmological constant is too large. If we
therefore consider only those vacua in which the cosmological
constant is acceptable at leading order, this constrains the scale
of $m_{3/2}$ further. Remarkably, we find evidence that for such
vacua, $m_{3/2}$ is of order $1-100$ TeV. This result certainly
deserves much further investigation.

The fact that such general results emerge from these studies makes the
task of predicting implications for various collider observables
as well as distinguishing among different vacua with data from
the LHC (or any other experiment) easier. A more detailed study of
the collider physics and other phenomenology will appear in the future \cite{ToAppear}.
However, as we will see in section \ref{pheno}, it could be quite easy
to distinguish Type IIB and $M$ theory vacua using the forthcoming LHC data.

This paper is the somewhat longer companion paper to
\cite{Acharya:2006ia}. Given its length we thought that it would
be worthwhile to end this introduction with a guide to its
contents. Much of the bulk of the paper is devoted to analyzing
and explaining the details of why the potential generated by
strong dynamics in the hidden sector has vacua in which all moduli
are stabilized. At first we begin with the simplest non-trivial
example, two hidden sectors, without any charged matter: only
gauge bosons and gauginos. Section \ref{secII} calculates the
moduli potential in this case. Section \ref{susy} analyzes its
supersymmetric vacua: these are all isolated with a negative
vacuum energy. Section \ref{exampleG2} describes explicit examples
realizing the vacuum structure of Sections \ref{susy} and
\ref{nonsusyadsvac}. Section \ref{nonsusyadsvac} describes the
vacua which spontaneously break supersymmetry. These also have
negative vacuum energy and all moduli stabilized. Section
\ref{chargedmattervac} goes on to consider more complicated hidden
sectors, where it is argued that metastable de Sitter vacua can
also occur under very reasonable conditions, and that the
metastable de Sitter vacuum obtained for a given $G_2$-manifold is
essentially {\it unique}. In section \ref{distribution} we study
the distribution of $m_{3/2}$. In particular for the de Sitter
vacua it is shown that requiring the absence of a large
cosmological constant fixes the gravitino mass to be of
$\mathcal{O}$(1-100) TeV. Section \ref{pheno} discusses
phenomenology. In particular, we explain the suppression of
gaugino masses and discuss the other soft SUSY breaking couplings.
We conclude in section \ref{summary}, followed by an Appendix
which discusses the K\"{a}hler metric for visible charged matter
fields in $M$ theory.

\section{The Moduli Potential} \label{secII}

In this section we quickly summarize the basic relevant features
of $G_2$-compactifications, setup the notation and calculate the
potential for the moduli generated by strong hidden sector gauge
dynamics.

In $M$ theory compactifications on a manifold $X$ of
$G_2$-holonomy the moduli are in correspondence with the harmonic
3-forms. Since there are $N \equiv {b_3}(X)$ such independent 3-forms
there are $N$ moduli $z_i=t_i+is_i$. The real parts of these
moduli $t_i$ are axion fields which originate from the 3-form
field $C$ in $M$ theory and the imaginary parts $s_i$ are zero
modes of the metric on $X$ and characterize the size and shape of
$X$. Roughly speaking, one can think of the $s_i$'s as measuring
the volumes of a basis of the $N$ independent three dimensional
cycles in $X$.

Non-Abelian gauge fields are localized on three dimensional
submanifolds $Q$ of $X$ along which there is an orbifold
singularity \cite{Acharya:1998pm} while chiral fermions are
localized at point-like conical singularities
\cite{Atiyah:2001qf,Acharya:2004qe,Acharya:2001gy}. Thus these
provide $M$ theory realizations of theories with localized matter.
A particle localized at a point $p$ will be charged under a gauge
field supported on $Q$ if $p$ $\in Q$. Since generically, two
three dimensional submanifolds do not intersect in a seven
dimensional space, there will be no light matter fields charged
under both the standard model gauge group and any hidden sector
gauge group.  {\it Supersymmetry breaking is therefore gravity
mediated in these vacua.}\footnote{This is an example of the sort
of general result one is aiming for in string/$M$ theory. We can
contrast this result with Type IIA vacua. Here the non-Abelian
gauge fields are again localized on 3-cycles, but since
generically a pair of three cycles intersect at points in six
extra dimensions, {\it In Type IIA supersymmetry breaking will generically be
gauge mediated.}}

In general the K\"{a}hler potentials for the moduli are difficult
to determine in these vacua. However a set of K\"{a}hler
potentials, consistent with $G_2$-holonomy and known to describe
accurately some explicit examples of $G_2$ moduli dynamics were
given in \cite{Acharya:2005ez}. These models are given by

\be \label{kahler} K = -3 \ln(4\pi^{1/3}\,V_X) \ee where the
volume in 11-dimensional units as a function of $s_i$ is \be
\label{vol} V_X = \prod_{i=1}^{N} s_i^{a_i},\;\; \mathrm{with}
\;\; \sum_{i=1}^{N} a_i = 7/3 \ee

We will assume that this $N$-parameter family of K\"{a}hler
potentials represents well the moduli dynamics. More general
K\"{a}hler potentials outside this class have the volume
functional multiplied by a function invariant under rescaling of
the metric. It would be extremely interesting to investigate the extension
of our results to these cases.

As motivated in the introduction, we are interested in studying
moduli stabilization induced via strong gauge dynamics. We will
begin by considering hidden sector gauge groups with no chiral
matter. Later sections will describe the cases with hidden sector
chiral matter.

In this `no matter' case, a superpotential (in units of $m_p^3$)
of the following form is generated \be \label{generalsuper} W =
\sum_{k=1}^{M}A_ke^{ib_k f_{k}}\, \ee \noindent where $M$ is the
number of hidden sectors undergoing gaugino condensation,
$b_k=\frac {2\pi}{c_k}$ with $c_k$ being the dual coxeter numbers
of the hidden sector gauge groups, and $A_k$ are numerical
constants. The $A_k$ are RG-scheme dependent and also depend upon
the threshold corrections to the gauge couplings; the work of
\cite{Friedmann:2002ty} shows that their ratios (which should be
scheme independent) can in fact take a reasonably wide range of
values in the space of $M$ theory vacua. We will only consider the
ratios to vary from $\mathcal{O}$(0.1-10) in what follows.

The gauge coupling functions $f_{k}$ for these singularities are
integer linear combinations of the $z_i$, because a 3-cycle $Q$
along which a given non-Abelian gauge field is localized is a
supersymmetric cycle, whose volume is linear in the moduli. \be
\label{generalgaugecoupl} f_k=\sum_{i=1}^{N} N^k_i\,z_i\,. \ee
Notice that, given a particular $G_2$-manifold $X$ for the extra
dimensions, the constants $( a_i, b_k , A_k , N^k_i )$ are
determined. Then, the K\"{a}hler potential and superpotential for
that particular $X$ are {\it completely determined} by the
constants $( a_i, b_k , A_k , N^k_i )$. This is as it should be,
since $M$ theory has no free dimensionless parameters.

We are ultimately aiming for an answer to the question, ``do $M$
theory vacua in general make a prediction for the beyond the
standard model spectrum?''. For this reason, since a fluxless $M$
theory vacuum is completely specified by the constants $( a_i, b_k
, A_k , N^k_i )$ we will try as much as possible ${\it not}$ to
pick a particular value for the constants and try to first
evaluate whether or not there is a prediction for {\it general
values of the constants}. Our results will show that
at least within the supergravity approximation there is
indeed a general prediction: the suppression of gaugino masses
relative to the gravitino mass.

At this point the simplest possibility would be to consider a
single hidden sector gauge group. Whilst this does in fact
stabilize all the moduli, it is a) non-generic and b) fixes the
moduli in a place which is strictly beyond the supergravity
approximation. Therefore we will begin, for simplicity, by
considering two such hidden sectors, which is more representative
of a typical $G_2$ compactification as well as being tractable
enough to analyze. The superpotential therefore has the following
form \be \label{super} W^{np} = A_1e^{ib_1 f_1}+ A_2e^{ib_2
f_2}\,. \ee The metric corresponding to the K\"{a}hler potential
(\ref{kahler}) is given by \be \label{metric}
K_{i\,\bar{j}}=\frac{3a_i}{4s_i^2}\delta_{i\,\bar{j}}\,. \ee The
${\cal N}=1$ supergravity scalar potential given by \be V=e^K
(K^{i\,\bar{j}} F_i\bar F_{\bar{j}} -3|W|^2)\,, \ee where \be
F_i=\partial_iW+(\partial_iK)W \,, \ee can now be computed. The
full expression for the scalar potential is given by \ba
\label{fullpotential}
V&=&\frac{1}{48\pi V_X^3}\,[\sum_{k=1}^{2}\sum_{i=1}^{N}a_i{\nu_i^k}\left({\nu_i^k}b_k+3\right)b_kA_k^2e^{-2b_k\vec\nu^{\,k}\cdot\,\vec a}+3\sum_{k=1}^{2}A_k^2e^{-2b_k\vec\nu^{\,k}\cdot\,\vec a}\nonumber\\
& &+2{\rm cos}[(b_1\vec N^1-b_2\vec N^2)\cdot \vec t\,\,]\sum_{i=1}^{N}a_i\prod_{k=1}^2\nu_i^kb_kA_ke^{-b_k\vec\nu^{\,k}\cdot\,\vec a}\\
& &+3{\rm cos}[(b_1\vec N^1-b_2\vec N^2)\cdot \vec
t\,\,]\left(2+\sum_{k=1}^{2}b_k\,\vec\nu^{\,k}\cdot\vec a
\right)\prod_{j=1}^2A_je^{-b_j\vec\nu^{\,j}\cdot\,\vec
a}]\,\nonumber \ea where we introduced a variable \be \label{nu}
\nu_i^k\equiv\frac{N_i^k s_i}{a_i} \;\;(\mathrm{no\;\,sum}) \ee
such that \be {\rm Im}f_k=\vec\nu^{\,k}\cdot\,\vec a\,. \ee By
extremizing (\ref{fullpotential}) with respect to the axions $t_i$
we obtain an equation \be {\rm sin}[(b_1\vec N^1-b_2\vec N^2)\cdot
\vec t\,\,]=0\,, \ee which fixes only one linear combination of
the axions. In this case \be {\rm cos}[(b_1\vec N^1-b_2\vec
N^2)\cdot \vec t\,\,]=\pm 1\,. \ee It turns out that in order for
the potential (\ref{fullpotential}) to have minima, the axions
must take on the values such that ${\rm cos}[(b_1\vec N^1-b_2\vec
N^2)\cdot \vec t\,\,]=-1$ for $A_1, A_2 > 0$. Otherwise the
potential has a runaway behavior. After choosing the minus sign,
the potential takes the form \ba \label{potential}
V=\frac{1}{48\pi V_X^3}\,[\sum_{k=1}^{2}\sum_{i=1}^{N}a_i{\nu_i^k}\left({\nu_i^k}b_k+3\right)b_kA_k^2e^{-2b_k\vec\nu^{\,k}\cdot\,\vec a}+3\sum_{k=1}^{2}A_k^2e^{-2b_k\vec\nu^{\,k}\cdot\,\vec a}&&\\
-2\sum_{i=1}^{N}a_i\prod_{k=1}^2\nu_i^kb_kA_ke^{-b_k\vec\nu^{\,k}\cdot\,\vec a}
-3\left(2+\sum_{k=1}^{2}b_k\,\vec\nu^{\,k}\cdot\vec a \right)\prod_{j=1}^2A_je^{-b_j\vec\nu^{\,j}\cdot\,\vec a}]&&\,\nonumber
\ea

In the next section we will go on to analyze the vacua of this
potential with unbroken supersymmetry. The vacua in which
supersymmetry is spontaneously broken are described in sections
V and VI.
\section{Supersymmetric Vacua} \label{susy}

In this section we will discuss the existence and properties of
the supersymmetric vacua in our theory. This is comparatively easy
to do since such vacua can be obtained by imposing the
supersymmetry conditions instead of extremizing the full scalar
potential (\ref{potential}). Therefore, we will study this case
with the most detail. Experience has also taught us that
potentials possessing rigid isolated supersymmetric vacua, also
typically have other non-supersymmetric vacua with many
qualitatively similar features.

The conditions for a supersymmetric vacuum are: \be \label{Fterm}
F_i = 0\,, \nonumber \ee which implies \ba
(b_1N^1_i+\frac{3a_i}{2s_i})A_1+(b_2N^2_i+\frac{3a_i}{2s_i})A_2\,
[\cos[(b_1\vec N^1-b_2\vec N^2)\cdot \vec t\,]\,
e^{(b_1\vec {N}^1-b_2\vec {N}^2)\cdot \vec s}\nonumber \\
+\; i\sin[(b_1\vec {N}^1-b_2\vec {N}^2)\cdot \vec t\,]\,
e^{(b_1\vec {N}^1-b_2\vec {N}^2)\cdot \vec s}]\,\, = \,\, 0
\ea
Equating the imaginary part of (\ref{Fterm}) to zero, one
finds that
\ba
{\sin}[(b_1\vec N^1-b_2\vec N^2)\cdot \vec t\,\,] &=& 0\,, \nonumber\\
\mathrm{which} \;\; \mathrm{implies}\;\;
\cos[(b_1\vec N^1-b_2\vec N^2)\cdot \vec t\,\,] &=& \pm 1\,.
\ea

For $A_1,A_2>0$, a solution with positive values for the moduli
($s_i$) exists when the axions take on the values such that
$\cos[(b_1\vec N^1-b_2\vec N^2)\cdot \vec t\,]=-1\,$. Now,
equating the real part of (\ref{Fterm}) to zero, one obtains \ba
\label{systemeqns}
(b_1{N}^1_i+\frac{3a_i}{2s_i})A_1+(b_2{N}^2_i+\frac{3a_i}{2s_i})A_2
\,e^{(b_1\vec{N}^1-b_2\vec {N}^2)\cdot \vec s} = 0\,. \ea This is
a system of $N$ transcendental equations with $N$ unknowns. As
such, it can only be solved numerically, in which case, it is
harder to get a good understanding of the nature of solutions
obtained. Rather than doing a brute force numerical analysis of
the system (\ref{systemeqns}) it is very convenient to introduce a
new auxiliary variable $\alpha$ to recast (\ref{systemeqns}) into
a system of \emph{linear} equations with $N$ unknowns coupled to a
single transcendental constraint as follows: \ba
&&\alpha(b_1N^1_i+\frac{3a_i}{2s_i})-(b_2N^2_i+\frac{3a_i}{2s_i}) = 0 \label{linear}\\
&&\frac {A_2} {A_1}\, = \frac 1 \alpha e^{-(b_1\vec {N}^1-b_2\vec {N}^2)\cdot \vec s}\,.\label{constraint}
\ea
The system of linear equations (\ref{linear}) can then be formally
solved for $s_i$ in terms of $\{b_1,b_2,N^1_i,N^2_i,a_i\}$ and $\alpha$:
\be \label{si}
 s_i = -\frac{3\,a_i\,(\alpha-1)}{2\,(b_1N^1_i\alpha-b_2N^2_i)};\;\;\;\;i=1,2,..,N\,.
\ee One can then substitute the solutions for $s_i$ into the
constraint (\ref{constraint}) and self-consistently solve for the
parameter $\alpha$ in terms of the input quantities
$\{A_1,A_2,b_1,b_2,N^1_i,N^2_i,a_i\}$. This, of course, has to be
done numerically, but we have indeed verified that solutions
exist. Thus, we have shown explicitly that the moduli can be
stabilized. We now go on to discuss the solutions, in particular
those which lie within the supergravity approximation.

\subsection{Solutions and the Supergravity Approximation}

Not all choices of the constants
$\{A_1,A_2,b_1,b_2,N^1_i,N^2_i,a_i\}$ lead to solutions consistent
with the approximation that in the bulk of spacetime, eleven
dimensional supergravity is valid. Although this is not a
precisely (in the numerical sense) defined approximation, a
reasonable requirement would seem to be that the values of the
stabilized moduli ($s_i$) obtained from (\ref{si}) are greater
than 1. It is an interesting question, certainly worthy of further
study, whether or not this is the correct criterion. In any case,
this is the criterion that we will use and discuss further.

\noindent From (\ref{si}) and (\ref{constraint}), and requiring
the $s_i$ to be greater than 1, we get the following two branches
of conditions on parameter $\alpha$ : \ba a) \;\;\; \frac{A_2}{A_1} > 1;\;\;\;
min\,\{\frac{b_2N^2_i}{b_1N^1_i}\,;i=\overline{1,N}\,\} > \alpha > \;
max\,\{\frac{b_2N^2_i+3a_i/2}{b_1N^1_i+3a_i/2}\,;i=\overline{1,N}\,\}
\nonumber \\ b) \;\;\; \frac{A_2}{A_1} < 1;\;\;\;
max\,\{\frac{b_2N^2_i}{b_1N^1_i}\,;i=\overline{1,N}\,\} < \alpha < \;
min\,\{\frac{b_2N^2_i+3a_i/2}{b_1N^1_i+3a_i/2}\,;i=\overline{1,N}\,\}
 \label{inequality}\ea

\begin{figure}[h!]
\begin{center}
\leavevmode \epsfxsize 12 cm \epsfbox{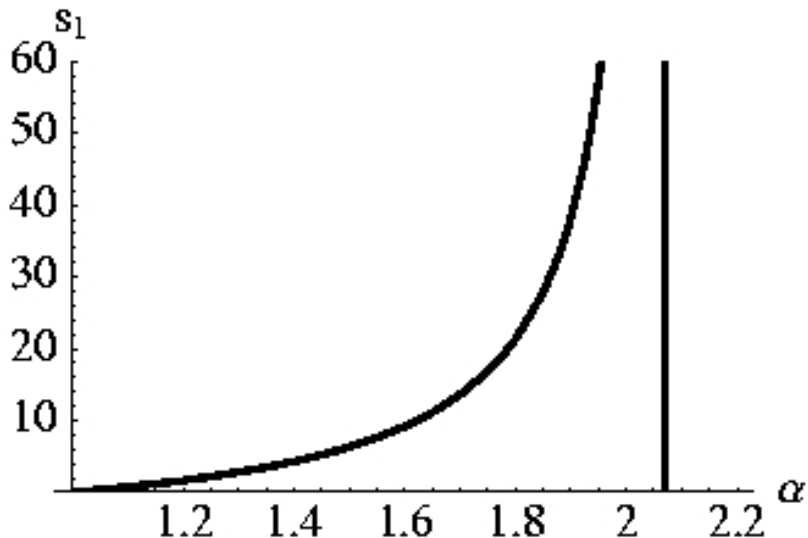}
\end{center}
\caption{ \footnotesize Positive values of $s_1$ plotted as
a function of $\alpha$ for a case with two condensates and three bulk moduli
for the following choice of constants
$b_1=\frac{2\pi}{30},b_2=\frac{2\pi}{29},N^1_i=\{1,2,2\},N^2_i=\{2,3,5\},a_i=\{1,1/7,25/21\}$. The
qualitative feature of this plot remains the same for different
choices of constants as well as for different $i$. The
vertical line is the locus for $\alpha=\frac{b_2N^2_i}{b_1N^1_i}$, where
the denominator of (\ref{si}) vanishes.} \label{sA2tilde}
\end{figure}

Notice that the solution for
$s_i$ (\ref{si}) has a singularity
at $\alpha=\frac{b_2N^2_i}{b_1N^1_i}$. This can be seen clearly from Figure (\ref{sA2tilde}).
We see that the modulus $s_1 \,(> 0) $ falls very rapidly as one moves
away from the vertical asymptote representing the singularity and can become smaller than
one very quickly, where the supergravity approximation fails to be
valid.
The relative location of the singularities for different moduli
will turn out to be very important as we will see
shortly.  From
 (\ref{inequality}), we know that there are two branches for
 allowed values of $\alpha$. Here we consider branch a) for
 concreteness, branch b) can be analyzed similarly.
\begin{figure}[h!]
    \begin{tabular}{cc}
      \leavevmode \epsfxsize 9 cm \epsfbox{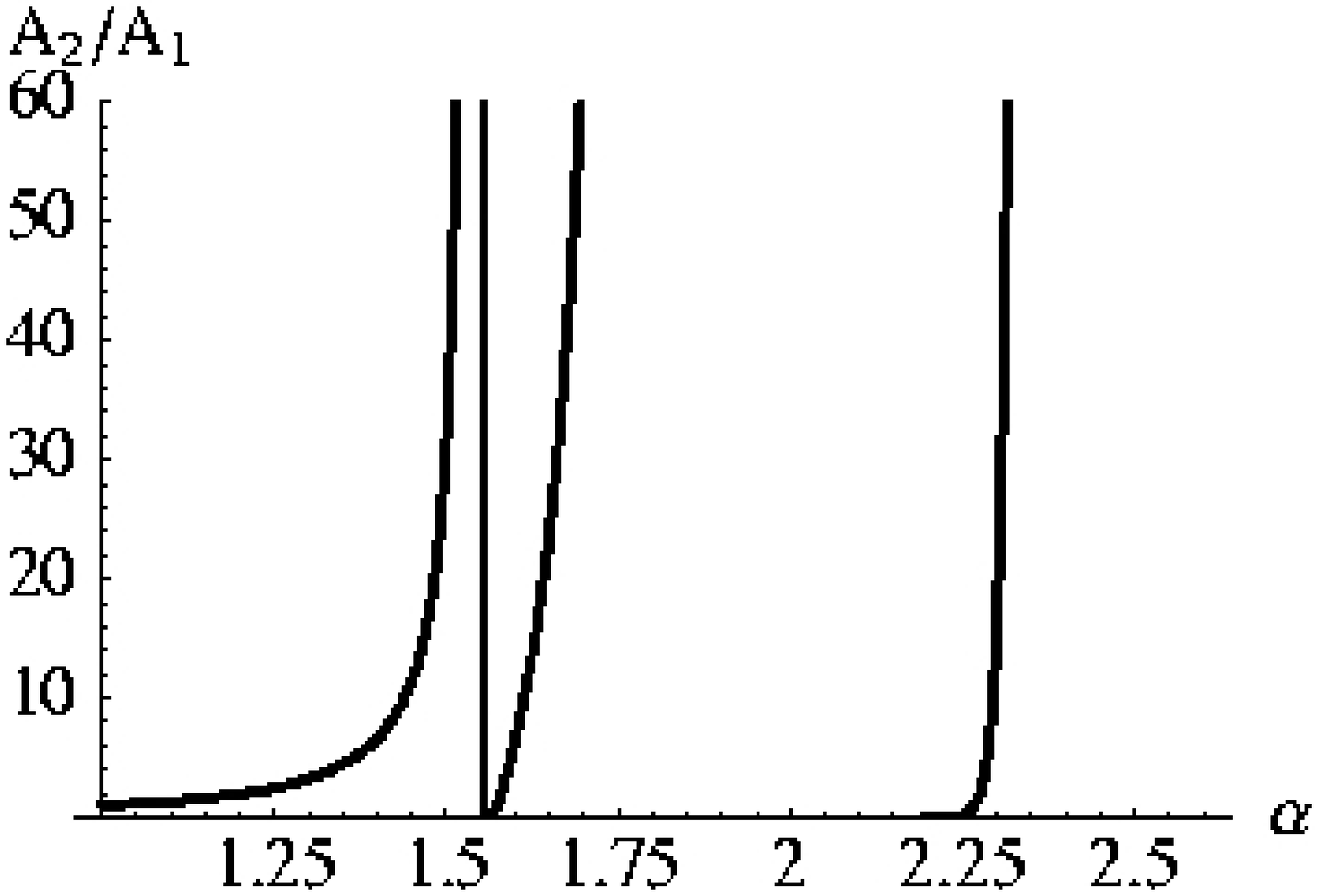}&
      \leavevmode \epsfxsize 9 cm \epsfbox{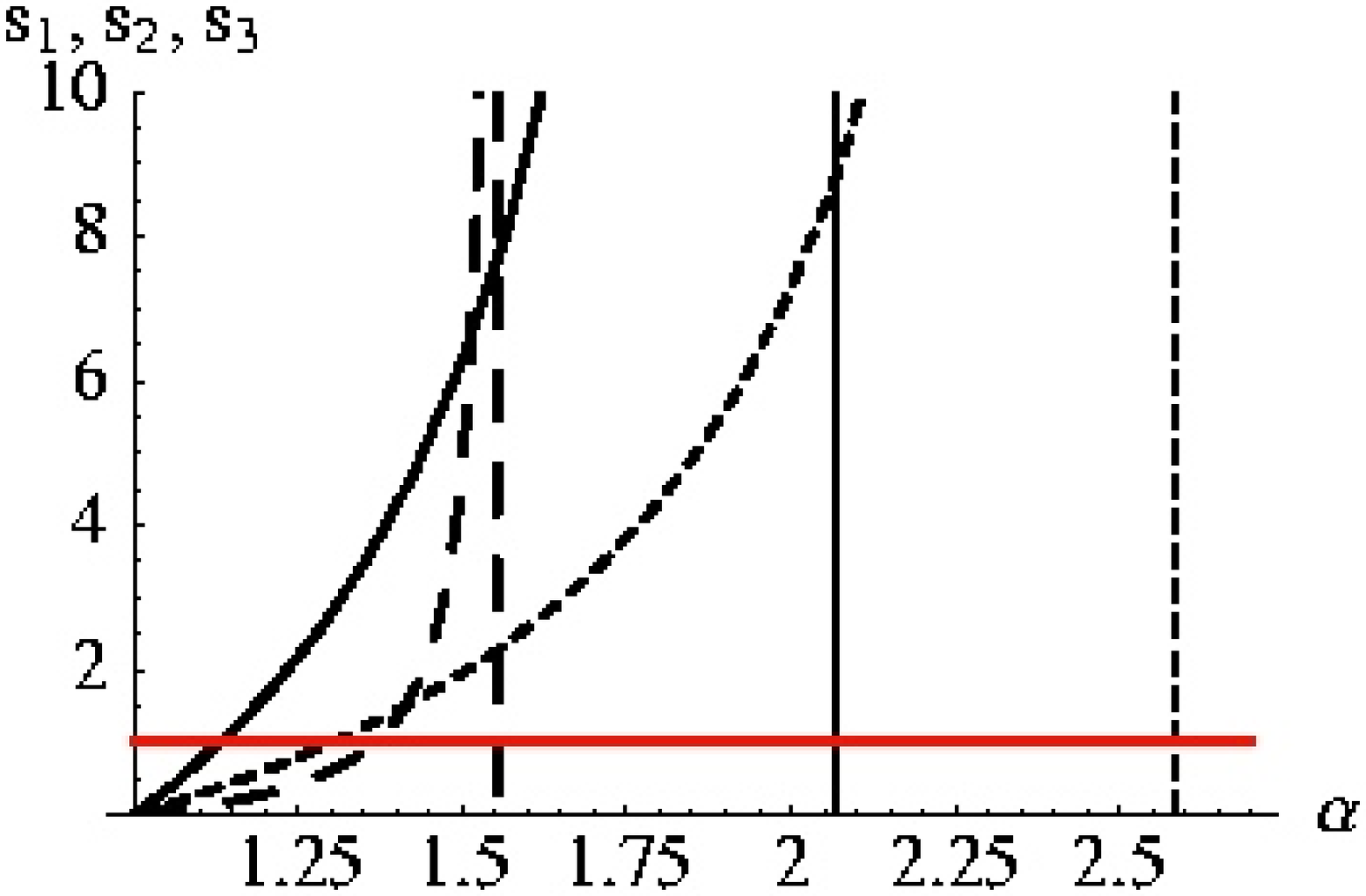} \\
    \end{tabular}
    \caption{\footnotesize Left - $A_2/A_1$ plotted as a function of $\alpha$ for a case with two
condensates and three bulk
    moduli. The function diverges as it approaches the loci of singularities of
    (\ref{si}), \textit{viz.} $\alpha=\frac{b_2N^2_i}{b_1N^1_i}$.
    \newline
    Right - Positive $s_i,\,i=1,2,3$ for the same case plotted as functions
    of $\alpha$. $s_1$ is represented by the solid curve, $s_2$ by the long dashed curve and $s_3$ by the short dashed
    curve.The vertical lines again represent the loci of singularities of (\ref{si}) which the respective moduli $s_i$ asymptote
    to. The horizontal solid (red) line shows the value unity for the
    moduli, below which the supergravity approximation is not
    valid.
    \newline
    Both plots are for $b_1=\frac{2\pi}{30},b_2=\frac{2\pi}{29},N^1_i=\{1,2,2\},N^2_i=\{2,3,5\},a_i=\{1,1/7,25/21\}$.}
    \label{Plots2-3}
\end{figure}
Figure (\ref{Plots2-3}) shows plots for $A_2/A_1$ and $s_i$ as
functions of $\alpha$ for a case with two condensates and
three bulk moduli. The plots are for a given choice of the constants $\{b_1,b_2,N^1_i,N^2_i,a_i,\,i=1,2,3\}$.
The qualitative feature of the plots remains the same
even if one has a different value for the constants.

Since the $s_i$ fall very rapidly as one goes to the left of the
vertical asymptotes, there is a small region of $\alpha$ between the origin and the
leftmost vertical asymptote which yields allowed values for all $s_i>1$.
Thus, for a solution in the supergravity regime
all (three) vertical lines representing the loci of
singularities of the (three) moduli $s_i$ should be (sufficiently)
close to each other. This means that the positions of the
vertical line for the $i$th modulus
($\alpha=\frac{b_2N^2_i}{b_1N^1_i}$) and the $j$th modulus
($\alpha=\frac{b_2N^2_j}{b_1N^1_j}$) can not be too far
apart. This in turn implies that the ratio of integer coefficients
$(N^1_i/N^2_i)$ and $(N^1_j/N^2_j)$ for the $i$th and $j$th modulus cannot be too
different from each other in order to remain within the approximation. Effectively, this means that
the integer combinations in the gauge kinetic functions
(\ref{generalgaugecoupl}) of the two hidden sector gauge groups in
(\ref{super}) can not be too linearly independent.
We will give explicit examples of $G_2$ manifolds in which $(N^1_i/N^2_i)$ and $(N^1_j/N^2_j)$
are the same for all $i$ and $j$, so the constraint of being within the supergravity
approximation is satisfied.
\begin{figure}[h!]
    \begin{tabular}{cc}
      \leavevmode \epsfxsize 9 cm \epsfbox{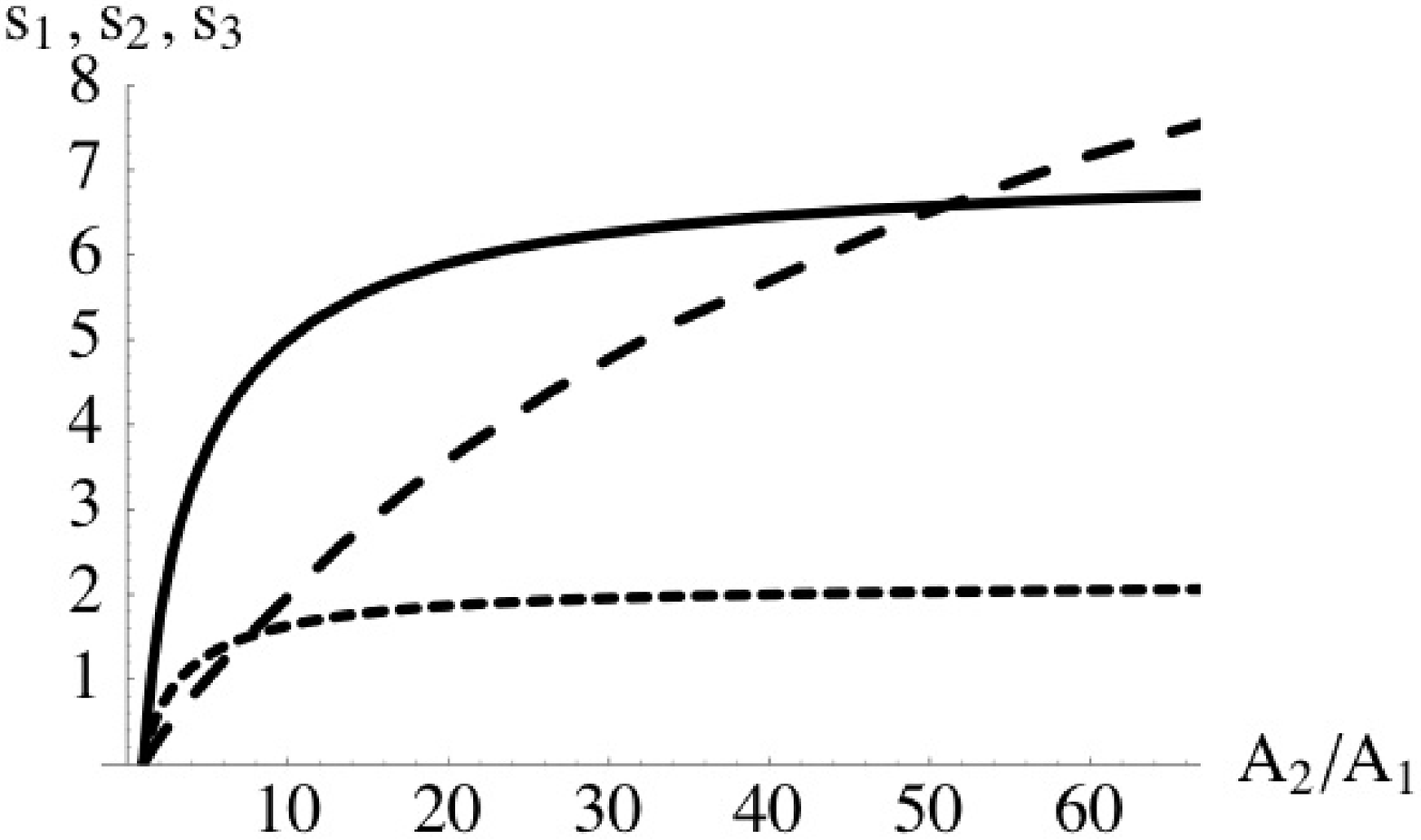}&
      \leavevmode \epsfxsize 9 cm \epsfbox{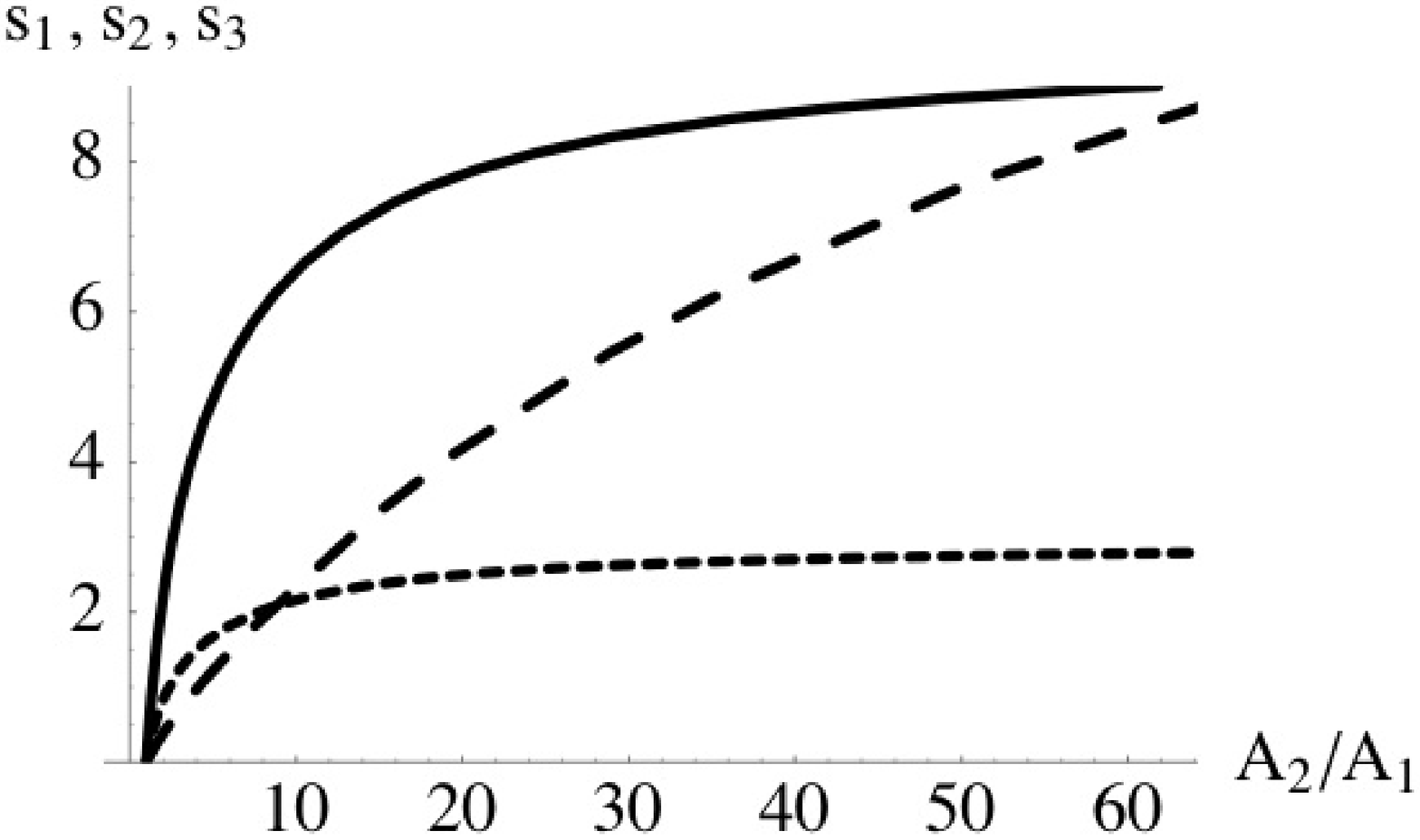} \\
      \leavevmode \epsfxsize 9 cm \epsfbox{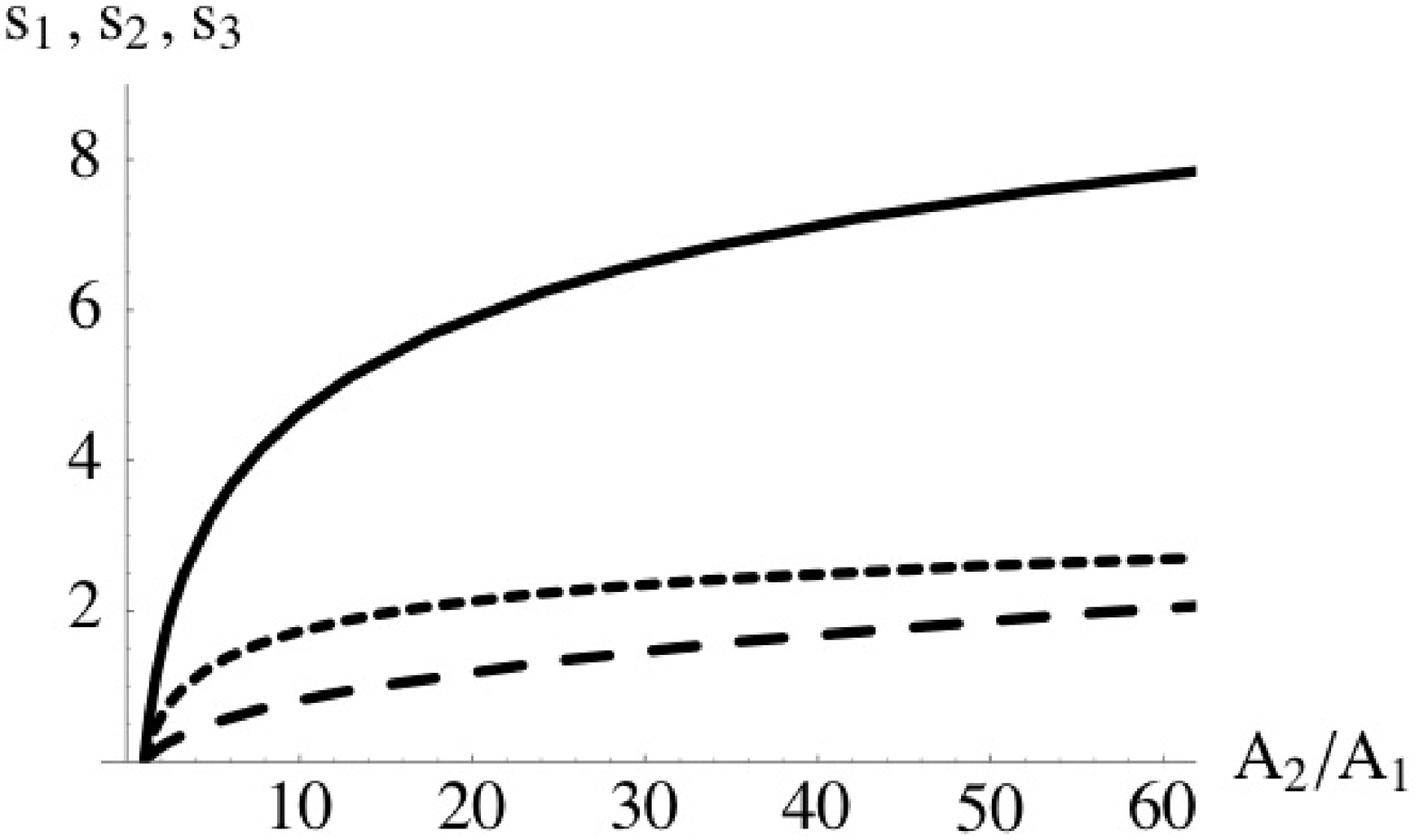}&
      \leavevmode \epsfxsize 9 cm \epsfbox{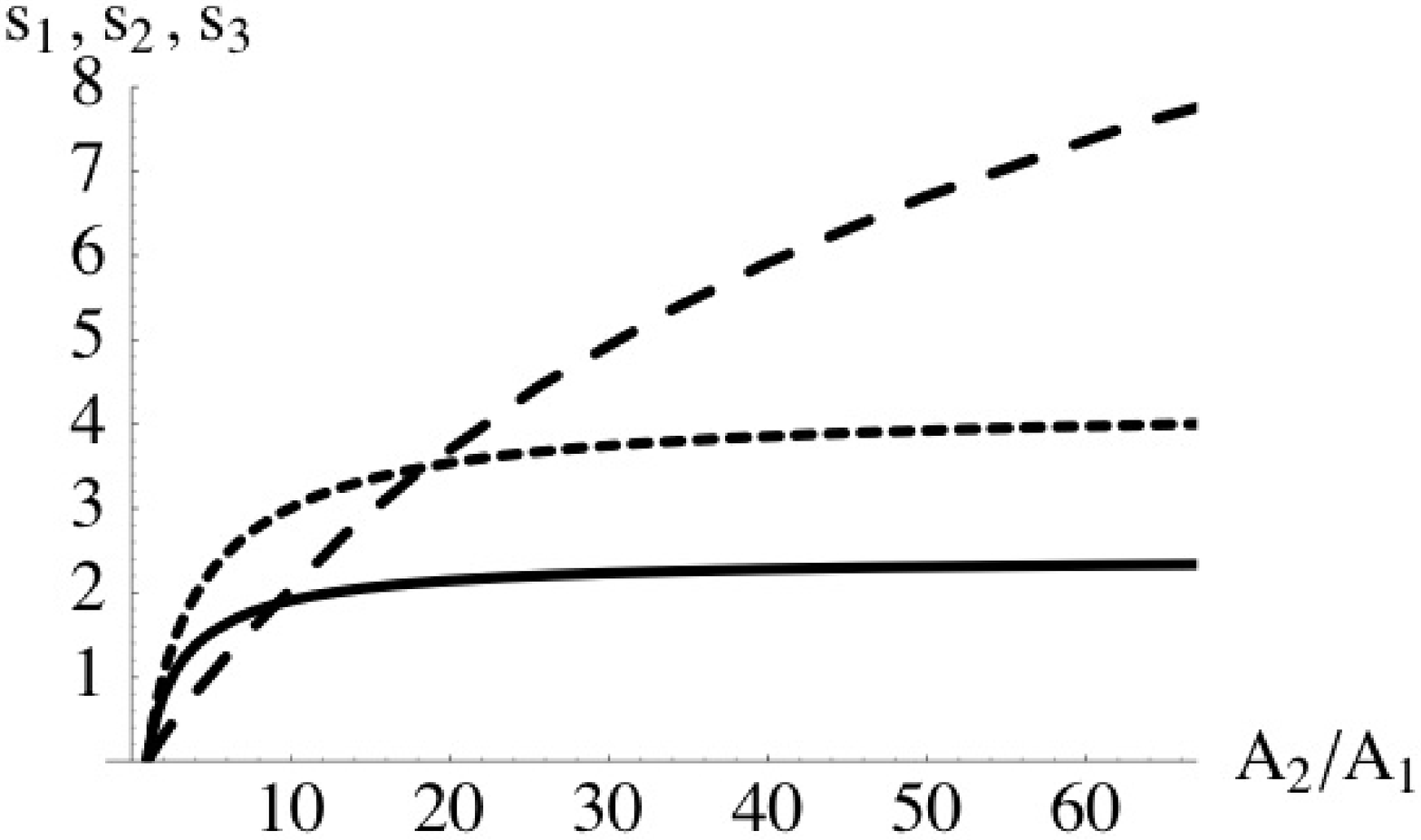} \\
    \end{tabular}
    \caption{\footnotesize Plots of positive $s_i,i=1,2,3$ as functions of $A_2/A_1$.\newline
    Top Left: Same choice of constants as in Figure(\ref{Plots2-3}), i.e.
    $b_1=\frac{2\pi}{30},\,b_2=\frac{2\pi}{29},\,N^1_i=\{1,2,2\},\,N^2_i=\{2,3,5\},\,a_i=\{1,1/7,25/21\}.$\newline
    Top Right: We increase the ranks of the gauge groups but keep them close (keeping everything else same) -
    $b_1=\frac{2\pi}{40},\,b_2=\frac{2\pi}{38}$.\newline
    Bottom Left: We introduce a large difference in the ranks of
    the gauge groups (with everything else same) - $b_1=\frac{2\pi}{40},\,b_2=\frac{2\pi}{30}$.\newline
    Bottom Right: We keep the ranks of the gauge groups as in Top
    Left but change the integer coefficients to
    $N^1_i=\{1,2,2\},\,N^1_i=\{3,3,4\}$.}
    \label{Plots4-7}
\end{figure}

We now turn to the effect of the other constants on the nature of
solutions obtained. From the top right plot in Figure
(\ref{Plots4-7}), we see that increasing the ranks of the gauge
groups while keeping them close to each other (with all other
constants fixed) increases the size of the moduli in general. On
the other hand, from the bottom left plot we see that introducing
a large difference in the ranks leads to a decrease in the size of
the moduli in general. Hence, typically it is easier to find
solutions with comparatively large rank gauge groups which are
close to each other. The bottom right plot shows the sizes of the
moduli as functions of $A_2/A_1$ while keeping the ranks of the
gauge groups same as in the top left plot but changing the integer
coefficients. We typically find that if the integer coefficients
are such that the two gauge kinetic functions are almost
dependent, then it is easier to find solutions with values of
moduli in the supergravity regime.

The above analysis performed for three moduli can be easily
extended to include many more moduli. Typically, as the number of
moduli grows, the values of $a_i$ in (\ref{si}) decrease because
of (\ref{vol}). Therefore the ranks of the gauge groups should be
increased in order to remain in the supergravity regime as one can
see from the structure of (\ref{si}). At the same time, for
reasons described above, the integer combinations for the two
gauge kinetic functions should not be too linearly independent. In
addition, the integers $N^k_i$ should not be too large as they
also decrease the moduli sizes in (\ref{si}).

\begin{figure}[h!]
    \begin{tabular}{cc}
      \leavevmode \epsfxsize 9 cm \epsfbox{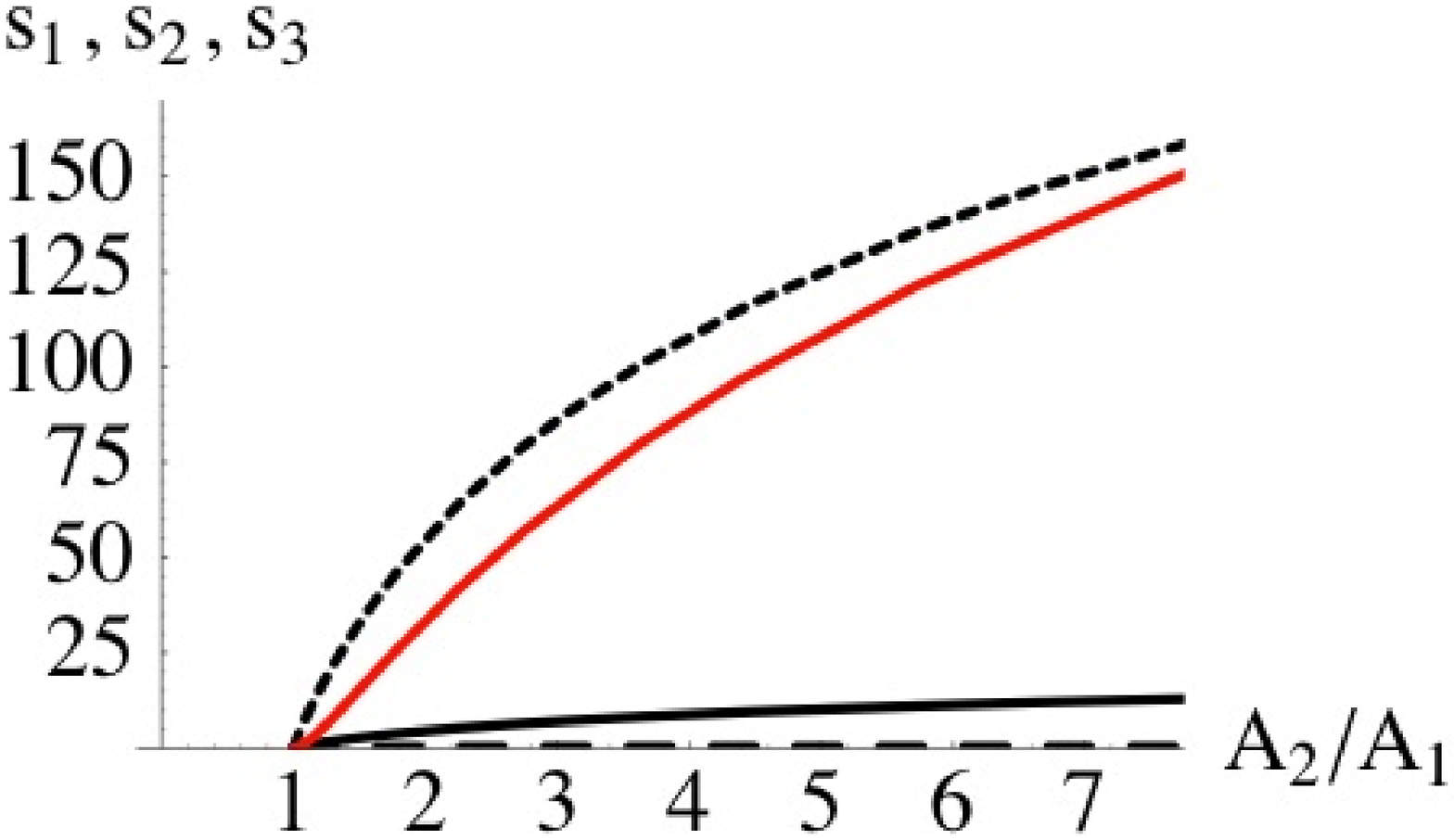}&
      \leavevmode \epsfxsize 9 cm \epsfbox{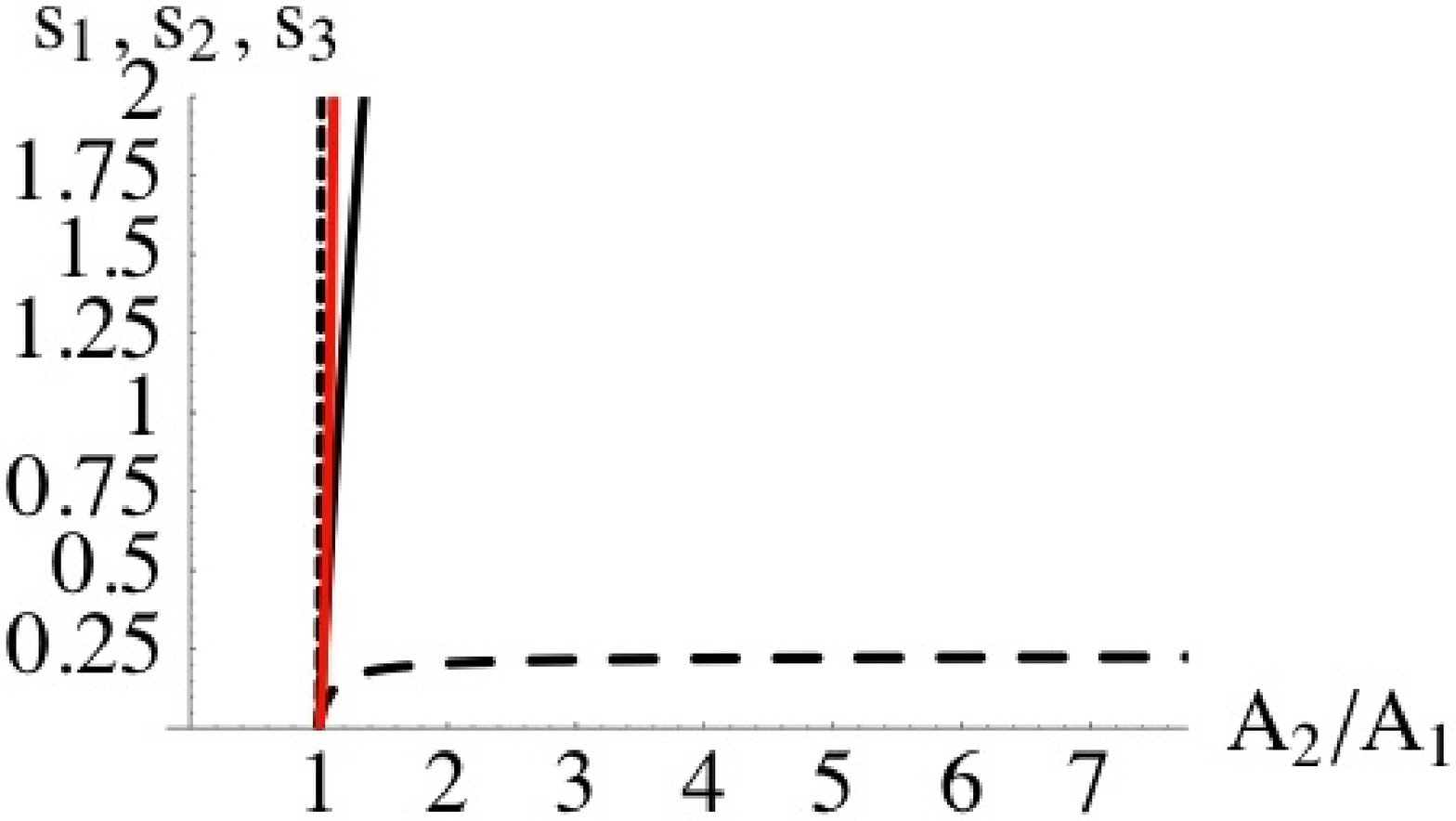} \\
    \end{tabular}
    \caption{\footnotesize Plots of positive $s_i,i=1,2,3$ as functions of $A_2/A_1$.
    The constants are $b_1=\frac{2\pi}{30},b_2=\frac{2\pi}{29},N^1_i=\{1,0,1\},N^2_i=\{1,1,1\},a_i=\{1/10,1,37/30\}$. $s_1$
    is represented by the solid curve, $s_2$ by the long dashed curve and $s_3$ by the short dashed curve. The
    red curve represents the volume of the internal manifold as a function of $A_2/A_1$.
    \newline
    Right - the same plot with the vertical plot range decreased.}
    \label{Plot15}
\end{figure}

What happens if some of the integers $N_i^1$ or $N_i^2$ are zero.
Figure \ref{Plot15} corresponds to this type of a situation when
the integer combinations  are given by
$N^1_i=\{1,0,1\},\,N^1_i=\{1,1,1\}$.
As we can see from the plots, all the moduli can still be
stabilized although one of the moduli, namely $s_2$ is stabilized
at values less than one in 11-dim Planck units. This gets us back
to the previous discussion as to when the supergravity
approximation can be valid. We will not have too much to say about
this point, except to note that a) the volume of $X$ can still be
large ((\ref{vol}) is large, greater than one in 11-dim Plank
units), b) the volumes of the associative three-cycles $Q_k$ which
appear in the gauge kinetic function (\ref{generalgaugecoupl}),
i.e. $Vol(Q_k)=\sum_{i=1}^nN_i^k\,s_i$  can also be large and c)
that the top Yukawa in these models comes from a small modulus vev
\cite{Atiyah:2001qf}. From Figure \ref{Plot15} we see that
although the modulus $s_2$ is always much smaller than one, the
overall volume of the manifold $V_X$ represented by the solid red
curve is much greater than one. Likewise, the volumes of the
associative three cycles $Vol(Q_1)=s_1+s_3$ and
$Vol(Q_2)=s_1+s_2+s_3$ are also large. Therefore if one interprets
the SUGRA approximation in this way, it seems possible to have
zero entries in the gauge kinetic functions for some of the moduli
and still stabilize all the moduli, as demonstrated by the
explicit example given above. In general, however, there is no
reason why any of the integers should vanish in the basis in which
the K\"{a}hler metric is given by (\ref{metric}).

\subsection{Special Case} \label{spl}
A very interesting special case arises when the gauge kinetic
functions $f_1$ and $f_2$ in (\ref{super}) are equal. Since in
this case $N^1_i = N^2_i$, the moduli vevs are larger in the
supersymmetric vacuum; hence this case is representative of the
vacua to be found within the supergravity approximation. Even
though this is a special case, in section \ref{exampleG2}, we will
describe explicit examples of $G_2$ manifolds in which $N^1_i =
N^2_i$.

\begin{figure}[h!]
\begin{center}
\leavevmode \epsfxsize 11 cm \epsfbox{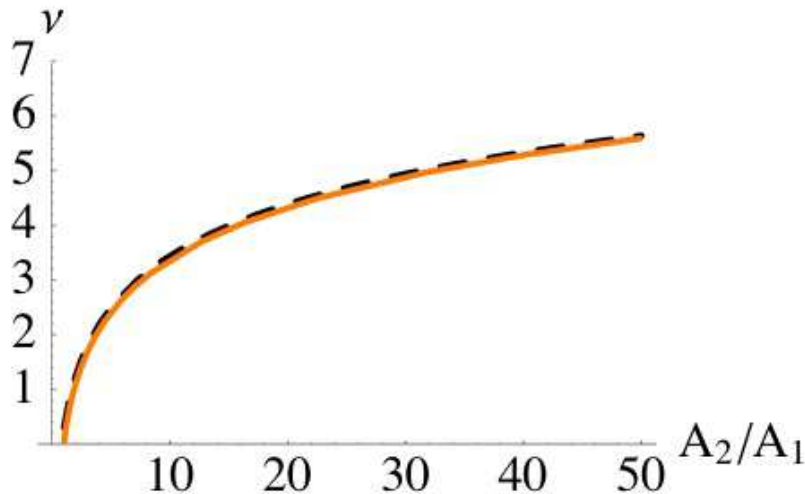}
\end{center}
\caption{\footnotesize Plot of $\nu$ as a function of $A_2/A_1$
for the choice $b_1=\frac{2\pi}{5},\,b_2=\frac{2\pi}{4}$. The
solid (red) curve represents the exact numerical solution whereas
the black dashed curve is the leading order approximation given by
(\ref{nu10}).}
    \label{Plot8}
\end{figure}

In the special case, we have \be N^1_i=N^2_i \equiv N_i\,, \ee and
therefore \be \label{nu1} \nu_i^1=\nu_i^2=\nu_i\equiv\frac{N_i
s_i}{a_i}\, \ee For this special case, the system of equations
(\ref{systemeqns}) can be simplified even further. We have
\be\label{constraint2} (b_1{\nu}_i+\frac{3}{2})A_1 -
(b_2{\nu}_i+\frac{3}{2})A_2e^{(b_1-b_2)\vec {\nu} \cdot \vec a}= 0
\ee with $\nu_i$ actually \emph{independent} of $i$. Thus, we are
left with just \emph{one} simple algebraic equation and one
transcendental constraint. The solution for $\nu_i$ is given by :
\ba\label{nu5} \nu_i \equiv \nu = -\frac{3(\alpha-1)}{2(\alpha
b_1-b_2)}\,, \ea with \ba\label{con}
 \frac{A_2}{A_1}\, =\frac 1 \alpha e^{-\frac 7 3(b_1-b_2)\nu}\,.
\ea Since $\nu_i$ is independent of $i$, it is also independent of
the number of moduli $N$.
In Figure
(\ref{Plot8}) we plotted $\nu$ as a function of $A_2/A_1$ when the
hidden sector gauge groups are $SU(5)$ and $SU(4)$. Notice
that here the ranks of the gauge groups don't have to be large for
the moduli to be greater than one. This is in contrast with the
linearly independent cases plotted in Figure (\ref{Plots4-7}).
Once $\nu$ is determined in terms of $A_2/A_1$, the moduli are
given by: \be s_i=\frac {a_i\nu}{N_i}\,. \ee Therefore, the
hierarchy between the moduli sizes is completely determined by the
ratios $a_i/N_i$ for different values of $i$.
In addition, from Figure \ref{Plot8} it can be seen that $\nu$
keeps increasing indefinitely if we keep increasing $A_2/A_1$
(though theoretically there may be a reasonable upper limit for
$A_2/A_1$), which is not possible for the general case as there
are $N$ $\nu_i$'s. This implies that it is possible to have a wide
range of the constants which yield a solution in the supergravity
regime.

Although the numerical solutions to the system (\ref{nu5}-\ref{con})
described above are easy to generate, having an explicit analytic solution,
even an approximate one, which could capture the dependence of $\nu$
on the constants $A_2/A_1$, $b_1$ and $b_2$ would be very useful.

Fortunately there exists a good approximation, namely a large $\nu$
limit, which allows us to find an analytical solution for $\nu$ in
a straightforward way. Expressing $\alpha$ from (\ref{nu5}), in the leading order approximation when $\nu$ is large we obtain
\be\label{alpha}
\alpha^{(0)}=\frac{b_2}{b_1}\,.
\ee

After substituting (\ref{alpha}) into (\ref{con}) we obtain the
approximate solution for $\nu$ in the leading order:
\be\label{nu10} \nu^{(0)}=\frac 3 7 \frac
1{b_2-b_1}\ln\left(\frac{A_2\,b_2}{A_1\,b_1}\right)=\frac 3
{14\pi} \frac {P\,Q}{P-Q}
\ln\left(\frac{A_2\,P}{A_1\,Q}\right)\,, \ee where the last
expression corresponds to $SU(P)$ and $SU(Q)$ hidden sector
gauge groups. For the moduli to be positive either
of the two following conditions have to be satisfied
\ba\label{branches}
&&a) \;\;\; A_1 Q < A_2 P;\;\;\;P>Q \nonumber \\
&&b) \;\;\; A_1 Q > A_2 P;\;\;\;P<Q\,. \ea From the plots in
Figure \ref{Plot8} we notice that the above approximation is
fairly accurate even when $\nu$ is $O(1)$. This is very helpful
and can be seen once we compute the first subleading contribution.
By substituting (\ref{nu10}) back into (\ref{nu5}) and solving for
$\alpha$ we now have up to the first subleading order:
\be\label{alpha4} \alpha=\alpha^{(0)}+\alpha^{(1)}=\frac P Q-\frac
7
{2\ln\left(\frac{A_2\,P}{A_1\,Q}\right)}\left(\frac{P-Q}{Q}\right)^2\,.
\ee It is then straightforward to compute $\nu$ which includes the
first subleading order contribution \be\label{nu40}
\nu=\nu^{(0)}+\nu^{(1)}=\frac 3 {14\pi} \frac
{P\,Q}{P-Q}\ln\left(\frac{A_2\,P}{A_1\,Q}\right)-\frac 3
{4\pi}\frac{P-Q} {\ln\left(\frac{A_2\,P}{A_1\,Q}\right)}\,. \ee We
can now examine the accuracy of the leading order approximation
when $\nu$ is $O(1)$ by considering the region where the ratio
$A_2/A_1$ is small. A quick check for the $SU(5)$ and $SU(4)$
hidden sector gauge groups chosen in the case presented in Figure
\ref {Plot8} yields for $A_2/A_1=4$: \ba
&&\alpha=\alpha^{(0)}+\alpha^{(1)}=1.25-0.136\,,
\\
&&\nu=\nu^{(0)}+\nu^{(1)}=2.195-0.148\,. \ea which results in a
12\% and 7\% error for $\alpha^{(0)}$ and $\nu^{(0)}$
respectively. The errors get highly suppressed when $\nu$ becomes
$O(10)$ and larger. Also, when the ranks of the gauge groups
$SU(P)$ and $SU(Q)$ are $O(10)$ and $P-Q$ is small, the ratio
$A_2/A_1$ can be $O(1)$ and still yield a large $\nu$. The
dependence of $\nu$ on the constants in (\ref{nu10}) is very
similar to the moduli dependence obtained for SUSY Minkowski vacua
in the Type IIB racetrack models \cite{Krefl:2006vu}.

We have demonstrated that there exist isolated supersymmetric
vacua in $M$ theory compactifications on $G_2$-manifolds with two
strongly coupled hidden sectors which give non-perturbative
contributions to the superpotential. Given the existence of
supersymmetric vacua, it is very likely that the potential also
contains non-supersymmetric critical points. Previous examples
have certainly illustrated this \cite{Acharya:2005ez}. Before
analyzing the non-supersymmetric critical points, however, we will
now present some examples of vacua which give rise to two strongly
coupled hidden sectors.

\section{Examples of $G_2$ Manifolds}\label{exampleG2}

Having shown that the potential stabilizes all the moduli, it is
of interest to construct explicit examples of $G_2$-manifolds
realizing these vacua. To demonstrate the existence of a
$G_2$-holonomy metric on a compact 7-manifold is a difficult
problem in solving non-linear equations \cite{Kovalev:2001zr}.
There is no analogue of Yau's theorem for Calabi-Yau manifolds
which allows an ``algebraic'' construction. However, Joyce and
Kovalev have successfully constructed many smooth examples
\cite{Kovalev:2001zr}. Furthermore, dualities with heterotic and
Type IIA string vacua also imply the existence of many singular
examples. The vacua of interest to us here are those with two or
more hidden sector gauge groups These correspond to
$G_2$-manifolds which have two three dimensional submanifolds
$Q_1$ and $Q_2$ along which there are orbifold singularities. In
order to describe such examples we will a) outline an extension of
Kovalev's construction to include orbifold singularities and b)
use duality with the heterotic string.

Kovalev constructs $G_2$ manifolds which can be described as the
total space of a fibration. The fibres are four dimensional $K3$
surfaces, which vary over a three dimensional sphere. Kovalev
considers the case in which the $K3$ fibers are generically
smooth, but it is reasonably straightforward to also consider
cases in which the (generic) $K3$ fiber has orbifold
singularities. This gives $G_2$-manifolds which also have orbifold
singularities along the sphere and give rise to Yang-Mills fields
in $M$ theory. For example if the generic fibre has both an
$SU(4)$ and an $SU(5)$ singularity, then the $G_2$ manifold will
have two such singularities, both parameterized by disjoint copies
of the sphere. In this case $N^1_i$ and $N^2_i$ are equal because
$Q_1$ and $Q_2$ are in the same homology class, which is precisely
the special case that we consider both above and below.

We arrive at a very similar picture by considering the $M$ theory dual of the heterotic string on
a Calabi-Yau manifold at large complex structure. In this limit, the Calabi-Yau is $T^3$ fibered
and the $M$ theory dual is $K3$-fibered, again over a three-sphere (or a discrete quotient thereof).
Then, if the hidden sector $E_8$ is broken by the background gauge field to, say, $SU(5) \times SU(2)$
the $K3$-fibers of the $G_2$-manifold generically have $SU(5)$ and $SU(2)$ singularities, again with
$N^1_i$ = $N^2_i$. More generally, in $K3$ fibered examples, the homology class of $Q_1$ could be
$k$ times that of $Q_2$ and in this case $N^1_i = k N^2_i$. As a particularly interesting example, the
$M$ theory dual of the heterotic vacua described in \cite{Braun:2005nv} include a $G_2$ manifold
whose singularities are such that they give rise to an observable sector with precisely the matter
content of the MSSM whilst the hidden sector has gauge group $G = E_8$.

Finally, we also note that Joyce's examples typically can have several sets of orbifold singularities which
often fall into the special class \cite{Kovalev:2001zr}. We now go on to describe the vacua in which supersymmetry is
spontaneously broken.

\section{Vacua with spontaneously broken Supersymmetry}\label{nonsusyadsvac}

The potential (\ref{fullpotential}) also possesses vacua in which
supersymmetry is spontaneously broken. Again these are isolated,
so the moduli are all fixed. These all turn out to have negative
cosmological constant. We will see in section
\ref{chargedmattervac} that adding matter in the hidden sector can
give a potential with de Sitter vacua.

Since the scalar potential (\ref{fullpotential}) is extremely
complicated, finding solutions is quite a non-trivial task. As for
the supersymmetric solution, it is possible to simplify the system
of $N$ transcendental equations obtained. However, unlike the
supersymmetric solution, we have only been able to do this so far
for the special case as in \ref{spl}. Therefore, for simplicity we
analyze the special case in detail. As we described above, there
are examples of vacua which fall into this special class.
Moreover, as explained previously, we expect that typically vacua
not in the special class are beyond the supergravity
approximation.

\noindent By extremizing (\ref{potential}) with respect to $s_k$
we obtain the following system of equations \ba \label{e1}
&&2\nu^2_k(b_1\alpha-b_2)^2-\nu_k[2(b_1\alpha-b_2)(b_1^2\alpha-b_2^2)\sum_{i=1}^{N}a_i\nu^2_i+3(\alpha-1)(b_1^2\alpha-b_2^2)
\vec\nu\cdot\vec a \nonumber\\
&&+3(b_1\alpha-b_2)^2\vec\nu\cdot\vec a+3(\alpha-1)(b_1\alpha-b_2)]-3[(b_1\alpha-b_2)^2\sum_{i=1}^{N}a_i\nu^2_i\nonumber\\
&&+3(\alpha-1)(b_1\alpha-b_2)\vec\nu\cdot\vec
a+3(\alpha-1)^2] \;=\; 0\,, \ea

\noindent where we have again introduced an auxiliary variable $\alpha$
defined by \be
\label{aux} \frac {A_2}{A_1}\equiv \frac 1 {\alpha}\,
e^{-(b_1-b_2)\vec\nu\cdot\vec a}\,.
\ee

\noindent similar to that in section \ref{spl}. The definition
(\ref{aux}) together with the system of polynomial equations
(\ref{e1}) can be regarded as a coupled system of equations for
$\alpha$ and $\nu_k$. We introduce the following notation:
\be\label{xyzw}
x\equiv(\alpha-1)\,,\,\,y\equiv(b_1\alpha-b_2)\,,\,\,z\equiv(b_1^2\alpha-b_2^2)\,,\,\,w\equiv
\frac{xz}{y^2}\,. \ee In this notation, from (\ref{e1}) (divided
by $x^2$) we obtain the following system of coupled equations \be
\label{u2} 2\frac {y^2} {x^2} \nu_k^2-\left(2\,\frac {y^2}
{x^2}w\sum_{i=1}^{N}a_i\nu^2_i+3\frac y x(w+1)\vec\nu\cdot\vec
a+3\right)\frac y x\nu_k-3\left(\frac {y^2}
{x^2}\sum_{i=1}^{N}a_i\nu_i^2+3\frac y x\vec\nu\cdot\vec
a+3\right)=0\,. \ee It is convenient to recast this system of $N$
{\it cubic} equations into a system of $N$ {\it quadratic}
equations plus a constraint. Namely, by introducing a new variable
$T$ as\be \label{T} 4\,T\equiv2\frac {y^2}
{x^2}w\sum_{i=1}^{N}a_i\nu^2_i+3\frac y x(w+1)\vec\nu\cdot\vec
a+3\, , \ee where the factor of four has been introduced for
future convenience, the system in (\ref{u2}) can be expressed as
\be \label{e2} 2\frac {y^2} {x^2} \nu_k^2-4\,T\frac y
x\nu_k-3\left(\frac {y^2} {x^2}\sum_{i=1}^{N}a_i\nu_i^2+3\frac y
x\vec\nu\cdot\vec a+3\right)=0\,. \ee

An important property of the system (\ref{e2}) is that {\it all of
its equations are the same  independent of the index k}. However,
since the combination in the round brackets in (\ref{e2}) is not a
constant with respect to $\vec\nu$ this system of quadratic
equations does not decouple. Nevertheless, because both the first
and the second monomials in (\ref{e2}) with respect to $\nu_k$ are
{\it independent of $\vec\nu$}, the standard solution of a
quadratic equation dictates that the solutions for $\nu_k$ of
(\ref{e2}) have the form \be \label{sol14} \nu_k=\frac x y
\left({T}+m_k H\right)\,,\,\, {\rm with}\,\, m_k=\pm
1\,,\,\,k=\overline{1,N}\,, \ee where we introduced another
variable $H$ and pulled out the factor of $x/y$ for future
convenience.

We have now reduced the task of determining $\nu_k$ for each
$k=\overline{1,N}$ to finding {\it only two} quantities - $T$ and
$H$. By substituting (\ref{sol14}) into equations
(\ref{T}-\ref{e2}) and using (\ref{vol}), we obtain a system of
two coupled quadratic equations
\ba \label{e3} \frac{14w}3\left({T_{A}^2}+2 A{T_{A}} H_{A}+H_{A}^2\right)+7(w+1) \left({T_{A}}+A H_{A}\right)+3-4T_{A}=0\,\,\,&& \\
9\left({T_{A}^2}+2 A {T_{A}}
H_{A}+H_{A}^2\right)-4H_{A}\left(H_{A}+A{T_{A}}
\right)+21\left({T_{A}}+AH_{A}\right)+9=0\,,&& \nonumber \ea where
parameter $A$ defined by \be \label{parA} A\equiv\frac 3 7\vec
m\cdot\vec a\,, \ee is now labelling each solution. Note that by
factoring out $x/y$ in (\ref{sol14}), the system obtained in
(\ref{e3}) is independent of either $x$ or $y$. However, it does
couple to the constraint (\ref{aux}) via $w$. In subsection
\ref{approxsol} we will see that there exists a natural limit when
the system (\ref{e3}) completely decouples from the constraint
(\ref{aux}). Since both $T_A$ and $H_A$ both depend on the
parameter A, the solution in (\ref{sol14}) is now written as \be
\label{sol1} \nu_k^{A}=\frac x y \left(T_{A}+m_k H_{A}\right)\,.
\ee Since $k=\overline{1,N}$ and $m_k=\pm 1$, vector $\vec m$
represents one of $2^{N}$ possible combinations. Thus, parameter
$\vec m\cdot\vec a$ can take on $2^{N}$ possible rational values
within the range: \be -\frac 7 3\leq \vec m\cdot\vec a\leq\frac 7
3\,, \ee so that parameter $A$ defined in (\ref{parA}) labelling
each solution can take on $2^{N}$ rational values in the range:
\be -1\leq A \leq 1\,. \ee
 For example, when $N=2$, there are four possible
combinations for $\vec m=(m_1,m_2)$, namely
\be
(m_1,m_2)=\{(-1,-1),\,(1,-1),\,(-1,1),\,(1,1)\}\,.
\ee
These combinations result in the following four possible values for $A$:
\be
A=\{-1,\,\frac3 7(a_1-a_2)\,,\frac3 7(-a_1+a_2)\,,1\}\,,
\ee
where we used (\ref{vol}) for the first and last combinations.

In general, for an arbitrary value of $A$, system
(\ref{e3}) has four solutions. However, with the exception of the case
when $A=1$, out of the four solutions only two are actually real,
as we will see later in subsection \ref{approxsol}.
The way to find those solutions is the following:

Having found $\nu_k^{A}$ analytically in terms of $\alpha$ and the
other constants, we can substitute it into the transcendental
constraint (\ref{aux}) to determine $\alpha$ numerically for
particular values of  $\{A_1, \,A_2,\,b_1,\,b_2,\,N_k,\,a_k\}$.
Again, in general there will be more than one solution for
$\alpha$. We can then substitute those values back into the
analytical solution for $\nu_k^{A}$ to find the corresponding
extrema, having chosen only those $\alpha$, obtained numerically
from (\ref{aux}), which result in real values of $\nu_k^{A}$. We
thus have $2^{N +1}$ real extrema. However, after a closer look at
the system of equations (\ref{e3}) we notice that when
$A\rightarrow -A$, equations remain invariant if
$H_{-A}\rightarrow - H_{A}$, and $T_{-A}\rightarrow T_{A}$, thus
simply exchanging the solutions $\nu_{(k,+)}^{A}$ where $m_k=1$
with $\nu_{(k,-)}^{A}$ where $m_k=-1$, i.e. \be\label{symmet}
\nu_{(k,+)}^{-A}=\nu_{(k,-)}^{A}\,, \ee which implies that the
scalar potential (\ref{potential}) in general has a total number
of $2^{N}$ real independent extrema. However, as we will see later
in section \ref{approxsol}, many of those vacua will be
incompatible with the supergravity approximation.

For general values of $A$, equations (\ref{e3})
have analytical solutions that are too complicated to be presented
here. In addition to restricting to the situation with the
same gauge kinetic function $f$ in both hidden sectors,
we now further restrict to special situations where $A$ takes special
values, so that the expressions are simple. However, it is
important to understand that they still capture the main features
of the general solution. In the following, we provide explicit
solutions (in the restricted situation as mentioned above) for $M$
theory compactifcations on $G_2$ manifolds with one and two moduli
respectively. In subsection \ref{approxsol} we will generalize our
results to the case with many moduli and give a complete classification
of all possible solutions. We will then consider the limit when
the volume of the associative cycle $Vol(Q)=\vec\nu\cdot\vec a$
is large and obtain explicit analytic solutions for the moduli.
\subsection{One modulus case}
The first, and the simplest case is to consider a manifold with
only one modulus, i.e. $N=1$, $a=\frac 7 3$. In this case, $A=\pm
1$. From the previous discussion we only need to consider the case
$A=1$. It turns out that this is a special case for which the
system (\ref{e3}) degenerates to yield three solutions instead of
four. All three are real, however, only two of them result in
positive values of the modulus: \be \label{NH1}
T_{1}^{(1)}=-\frac{15} 8\,,\,\,\,\,H_{1}^{(1)}=\frac 3 8 \ee and
\ba \label{NH2} T_{1}^{(2)}&=&\frac{3}{28 \left(243-441 w+196
w^2\right)}
(-13419+\frac{3645}{w}+15288 w-5488 w^2\nonumber\\
&&-329 \sqrt{729-1701 w+1323 w^2-343 w^3}\\
&&+\frac{135} w \sqrt{729-1701 w+1323 w^2-343 w^3}\nonumber\\
&&+196 w \sqrt{729-1701 w+1323 w^2-343 w^3})\nonumber\\
H_{1}^{(2)}&=&\frac{3} {28 w}\left(-27+28 w-\sqrt{729-1701 w+1323 w^2-343 w^3}\right)\,,\nonumber
\ea
which give the following two values for the modulus
\ba
\label{sol2}
&&s^{(1)}=\frac a{N_1}\frac x y\left(T_{1}^{(1)}+H_{1}^{(1)}\right)=-\frac {7x}{2Ny}\\
&&s^{(2)}=\frac a{N_1}\frac x
y\left(T_{1}^{(2)}+H_{1}^{(2)}\right)=-\frac
{x}{Ny}\left(\frac{3+\sqrt{9-7w}}w\right)\,.\nonumber \ea In
addition, each solution in (\ref{sol2}) is a function of the
auxiliary variable $\alpha$ defined in (\ref{aux}). By
substituting (\ref{sol2}) into (\ref{aux}) we obtain two equations
for $\alpha^{(1)}$ and $\alpha^{(2)}$ \be \label{e4} \frac
{A_2}{A_1}= \frac 1
{\alpha^{(1)}}\,e^{-(b_1-b_2)s^{(1)}N_1}\,,\,\,\,\, \frac
{A_2}{A_1}= \frac 1 {\alpha^{(2)}}\,e^{-(b_1-b_2)s^{(2)}N_1} \,.
\ee The transcendental equations (\ref{e4}) can only be solved
numerically. Here we will choose the following values for this
simple toy model \be \label{choice}
A_1=0.12\,,\,\,\,A_2=2\,,\,\,\,b_1=\frac
{2\pi}{8}\,,\,\,\,b_2=\frac {2\pi}{7}\,,\,\,\,N_1=1\,. \ee By
solving (\ref{e4}) numerically and keeping only those solutions
that result in real positive values for the modulus $s$ in
(\ref{sol2}) we get \be \label{sol3}
s^{(1)}=26.101\,,\,\,\,s^{(2)}=27.185\,. \ee with \be
\alpha^{(1)}=1.122\,,\,\,\,\alpha^{(2)}=1.267\,. \ee
\begin{figure}[ht]
\begin{center}
\leavevmode
\epsfxsize 10 cm
\epsfbox{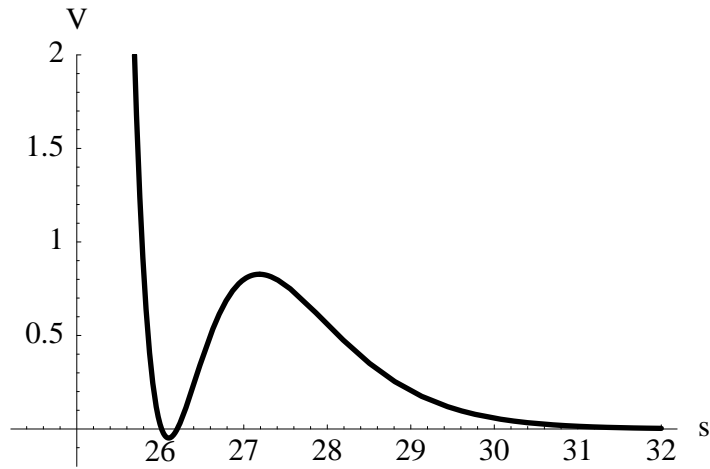}
\end{center}
\caption{ \footnotesize Potential multiplied by $10^{32}$ plotted
as a function of one modulus $s$. For our particular choice of
constants in (\ref{choice}), the modulus is stabilized at the
supersymmetric AdS minimum $s^{(1)}=26.101$. The maximum is de
Sitter, given by $s^{(2)}=27.185$.} \label{onemodulusplot}
\end{figure}
In figure (\ref{onemodulusplot}) we see that the two solutions in
(\ref{sol3}) correspond to an AdS minimum and a de Sitter maximum.
In fact, the AdS minimum at $s^{(1)}$ is supersymmetric. The
general solution for $s^{(1)}$ given in (\ref{sol2}) can also be
obtained by methods of section \ref{susy}, imposing the SUSY
condition on the corresponding $F$-term by setting it to zero,
while introducing the same auxiliary constraint as in (\ref{aux}).
\subsection{Two moduli case}
While the previous example with one modulus is interesting, it
does not capture some very important properties of the vacua which
arise when two or more moduli are considered. In particular, in
this subsection we will see that the supersymmetric AdS minimum,
obtained in the one-dimensional case, actually turns into a saddle
point whereas the stable minima are AdS with spontaneously broken
supersymmetry. Let us now consider a particularly simple example
with two moduli. Here we will choose both moduli to appear on an
equal footing in the K\"{a}hler potential (\ref{kahler}) by
choosing \be a_1=\frac 7 6\,,\,\,\,a_2=\frac 7 6\,. \ee We now
have four possible combinations for $\vec m=(m_1,m_2)$: \be
(1,1)\,,\,\,\,(1,-1)\,,\,\,\,(-1,1)\,,\,\,\,(-1,-1)\,, \ee
corresponding to the following possible values of $A$: \be
1\,,\,\,\,0\,,\,\,\,0\,,\,\,\,-1\,, \ee where only two of the four
actually produce independent solutions. The case when $A=1$ has
been solved in the previous subsection with $T_{1}^{(1)}$,
$H_{1}^{(1)}$ and  $T_{1}^{(2)}$, $H_{1}^{(2)}$ given by
(\ref{NH1}-\ref{NH2}) with the moduli taking on the following
values for the supersymmetric AdS extremum \ba \label{sol5}
&&s^{(1)}_1=\frac {a_1x}{N_1y}\left(-\frac 3 2\right)=-\frac {7x}{4N_1y}\,,\\
&&s^{(1)}_2=\frac {a_2x}{N_2y}\left(-\frac 3 2\right)=-\frac
{7x}{4N_2y}\,,\nonumber \ea and the de Sitter extremum \ba \label{sol6}
&&s^{(2)}_1=\frac {a_1x}{N_1y}\left(-\frac 3{7w}(3+\sqrt{9-7w})\right)=-\frac {x}{2N_1y}{\left(\frac{3+\sqrt{9-7w}}w\right)}\,,\\
&&s^{(2)}_2=\frac {a_2x}{N_2y}\left(-\frac
3{7w}(3+\sqrt{9-7w})\right)=-\frac {x}{2N_2y}{\left(\frac{3+\sqrt{9-7w}}w\right)}\,.\nonumber \ea As mentioned earlier,
the supersymmetric solution can also be obtained by the methods of
section \ref{susy}. Now, we also have a new case when $A=0$. The corresponding two real solutions for $T_{0}$ and $H_{0}$ are
\ba \label{NH3}
&&T_{0}^{(1)}=\frac 3{112w}(15 - 63w - D)\,,\\
&&H_{0}^{(1)}=\frac 1{4\sqrt{5}}\sqrt{-\frac{585}8 - \frac{18225}{392w^2} + \frac{3915}{28w} +
\frac{1215}{392w^2}D- \frac{225}{56 w}D}\,,\nonumber
\ea
and
\ba
\label{NH4}
&&T_{0}^{(2)}=\frac 3{112w}(15 - 63w - D)\,,\\
&&H_{0}^{(2)}=-\frac 1{4\sqrt{5}}\sqrt{-\frac{585}8 - \frac{18225}{392w^2} + \frac{3915}{28w} +
\frac{1215}{392w^2}D- \frac{225}{56 w}D}\,,\nonumber
\ea
where we defined
\be
D\equiv\sqrt{225 - 770w + 833w^2}\,.
\ee
The moduli are then extremized at the values given by
\ba
&&s^{(3)}_1=\frac {a_1x}{N_1y}\left({T_{0}^{(1)}}+H_{0}^{(1)}\right)\,,\\
&&s^{(3)}_2=\frac {a_2x}{N_2y}\left({T_{0}^{(1)}}-H_{0}^{(1)}\right)\,,
\nonumber
\ea
and
\ba
&&s^{(4)}_1=\frac {a_1x}{N_1y}\left({T_{0}^{(2)}}+H_{0}^{(2)}\right)\,,\\
&&s^{(4)}_2=\frac
{a_2x}{N_2y}\left({T_{0}^{(2)}}-H_{0}^{(2)}\right)\,.\nonumber \ea
To completely determine the extrema we again need to substitute
the solutions given above into the constraint equation (\ref{aux})
and choose a particular set of values for $A_1$, $A_2$, $b_1$, and
$b_2$ to find numerical solutions that result in real positive
values for the moduli $s_1$ and $s_2$. Here we again use the same
values as we chose in the previous case given by \be
\label{choice1} A_1=0.12\,,\,\,\,A_2=2\,,\,\,\,b_1=\frac
{2\pi}{8}\,,\,\,\,b_2=\frac
{2\pi}{7}\,,\,\,\,N_1=1\,,\,\,\,N_2=1\,. \ee For the SUSY extremum
we have \be \label{sol7}
s^{(1)}_1=13.05\,,\,\,\,s^{(1)}_2=13.05\,. \ee The de Sitter
extremum is given by \be \label{sol8}
s^{(2)}_2=13.59\,,\,\,\,s^{(2)}_2=13.59\,. \ee The other two
extrema are at the values \be \label{sol9}
s^{(3)}_1=2.61\,,\,\,\,s^{(3)}_2=23.55\,,\,{\rm
and}\,\,s^{(4)}_1=23.55\,,\,\,\,s^{(4)}_2=2.61 \ee
\begin{figure}[ht]
\begin{center}
\leavevmode
\epsfxsize 12 cm
\epsfbox{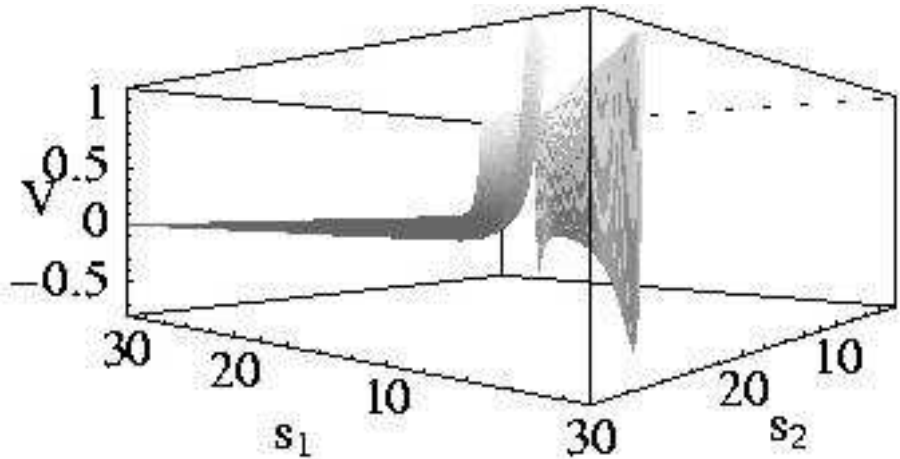}
\end{center}
\caption{ \footnotesize
Potential multiplied by $10^{32}$ plotted as a function of two moduli $s_1$ and $s_2$ for the
values in (\ref{choice1}). The SUSY AdS extremum given by (\ref{sol7})
is a saddle point, located between the non-supersymmetric AdS minima given by (\ref{sol9}).}
\label{twomoduliplot1}
\end{figure}

It is interesting to note that the supersymmetric extremum in
(\ref{sol7}) is no longer a stable minimum but instead, a saddle
point. The two symmetrically located stable minima seen in figure
(\ref{twomoduliplot1}) are non-supersymmetric. Thus we have an
explicit illustration of a potential where spontaneous breaking of
supersymmetry can be realized. The stable minima appear
symmetrically since both moduli were chosen to be on an equal
footing in the scalar potential. With a slight deviation where
$a_1\not=a_2$ and/or  $N_1\not =N_2$ one of the minima will be
deeper that the other. It is important to note that at both
minima, the volume given by (\ref{vol}) is stabilized at the value
$V_X=122.28$ which is large enough for the supergravity analysis
presented here to be valid.
\subsection{Generalization to many moduli}\label{approxsol}

\begin{figure}[h!]
    \begin{tabular}{cc}
      \leavevmode \epsfxsize 8.5 cm \epsfbox{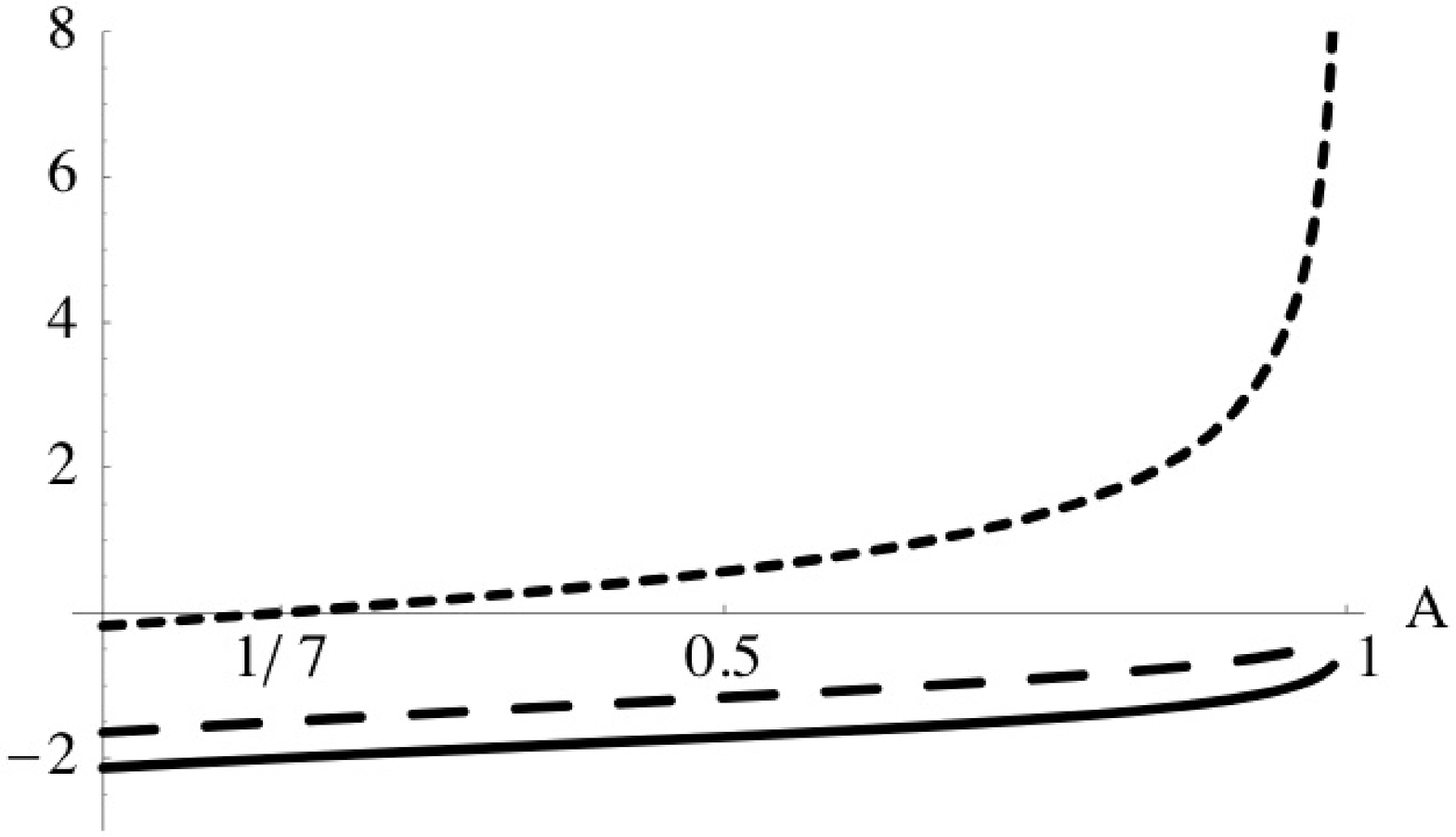}&
      \leavevmode \epsfxsize 8.5 cm \epsfbox{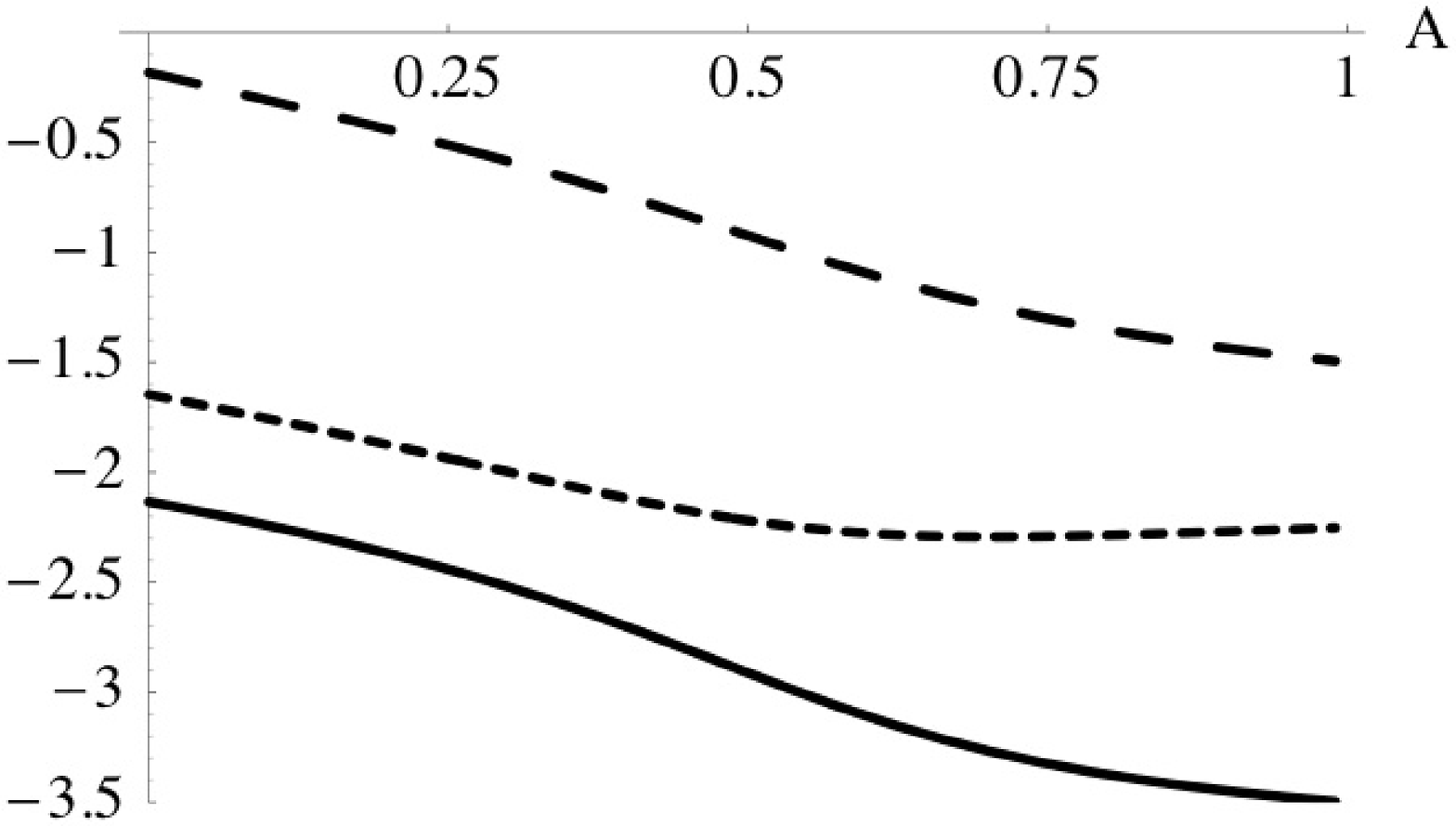} \\
    \end{tabular}
    \caption{\footnotesize Plots of $L_{A,\,+}^{(c)}$,  $L_{A,\,-}^{(c)}$ and ${B_A^{(c)}}$, where $c=\overline{1,2}$,
corresponding to the two real solutions of the system (\ref{e20}) as functions of parameter $A$ in the range $0\leq A<1$.
Both left and right graphs have $L_{A,\,+}^{(c)}$ - long dashed line,
$L_{A,\,-}^{(c)}$ -short dashed line, ${B_A^{(c)}}$ - solid line.
\newline
Left: Plots of $L_{A,\,+}^{(1)}$,  $L_{A,\,-}^{(1)}$ and
${B_A^{(1)}}$ corresponding to the first real solution at each
$A$. There is a critical value $A=1/7$ where $L_{A,\,-}^{(1)}=0$
and becomes positive for $A>1/7$.
    \newline
    Right: Plots of $L_{A,\,+}^{(2)}$,  $L_{A,\,-}^{(2)}$ and ${B_A^{(2)}}$ corresponding to the second real solution at each $A$. }
    \label{Plot24}
\end{figure}

In the previous section we demonstrated the existence of stable
vacua with broken SUSY for the special case with two moduli. Here
we will extend the analysis to include cases with an arbitrary
number of moduli for any value of the parameter $A$. It was
demonstrated in section \ref{spl} that the SUSY extremum has an
approximate analytical solution given by (\ref{nu10}). Therefore,
it would be highly desirable to obtain approximate analytical
solutions for the other extrema in a similar way. We will start
with the observation that for the SUSY extremum (\ref{sol5})
obtained for the special case when $A=1$, both $T_1^{(1)}$ and
$H_1^{(1)}$ given by (\ref{NH1}) are independent of $w$. On the
other hand, if in the leading order parameter $\alpha$ is given by
(\ref{alpha}), from the definitions in (\ref{xyzw}) it follows
that in this case \be\label{lim} y\rightarrow 0\,,\,\,{\rm
and}\,\,\,w\rightarrow -\infty\,. \ee Thus, if we consider the
system (\ref{e3}) in the limit when $w\rightarrow -\infty$, we
should be able to still obtain the SUSY extremum exactly. In
addition, one might also expect that the solutions for the vacua
with broken SUSY may also be located near the loci where
$y\rightarrow 0$. With this in mind we will take the limit
(\ref{lim}) which results in the following somewhat simplified
system of equations for $T_A$ and $H_A$:
\ba \label{e20} &&2\left({T_{A}^2}+2 A{T_{A}} H_{A}+H_{A}^2\right)+ 3\left({T_{A}}+A H_{A}\right)=0 \\
&&9\left({T_{A}^2}+2 A {T_{A}}
H_{A}+H_{A}^2\right)-4H_{A}\left(H_{A}+A{T_{A}}
\right)+21\left({T_{A}}+AH_{A}\right)+9=0\,. \nonumber \ea
Because system (\ref{e20}) is completely decoupled from the constraint
(\ref{aux}) and hence the microscopic constants, we can perform a completely
general analysis of the vacua valid for arbitrary values of the microscopic constants,
at least when the limit (\ref{lim}) is a good approximation.
It is straightforward to see that (\ref{NH1}) is an exact solution to
the above system when $A=1$. Moreover, unlike the general case
when $w$ is finite, where the system had three real solutions two
of which resulted in positive moduli, system (\ref{e20}) above
completely degenerates when $A=1$ yielding only one solution
corresponding to the SUSY extremum. On the other hand, for an
arbitrary $0\leq A < 1$ the system has four solutions. One can
check that at every point $A$ in the range $0\leq A < 1$ exactly
two out of these four solutions are real. The corresponding plots
are presented in Figure \ref{Plot24}. Before we discuss the plots
we would like to introduce some new notation: \be\label{lk}
L_{A,\,k}^{(c)}=T_A^{(c)}+m_kH_A^{(c)}\,, \ee where $c=\overline
{1,2}$ corresponding to the two real solutions. In this notation
(\ref{sol1}) can be reexpressed as: \be\label{sol32}
\nu_{A,\,k}^{(c)}=\frac x yL_{A,\,k}^{(c)}\,. \ee The volume of
the associative three cycle $Q$ for these vacua is then:
\be\label{deft} {\cal T}^{\,(c)}_A\equiv Vol(Q)^{(c)}_A=
Im(f^{(c)}_A)=\sum_{i=1}^{N}N_i\,
s^{(c)}_{A,\,i}=\sum_{i=1}^{N}a_i\,\nu^{(c)}_{A,\,i}=\frac x
y\,\vec a\cdot\vec L_{A}^{\,(c)}\,. \ee For future convenience we
will also introduce \be\label{BAc} B_A^{(c)}\equiv\vec a\cdot\vec
L_{A}^{\,(c)}=\frac 7 3\left(T_A^{(c)}+AH_A^{(c)}\right)\,. \ee
Constraint (\ref{aux}) is then given by: \be
\label{aux2}{\alpha^{(c\,)}_A}= \frac {A_1}{A_2}\,
e^{-(b_1-b_2){\cal T}^{\,(c)}_A}\,, \ee which is coupled to
\be\label{eq32} \frac x
y=\frac{{\alpha^{(c\,)}_A}-1}{b_1\,{\alpha^{(c\,)}_A}-b_2}=\frac
{{\cal T}^{\,(c)}_A}{B_A^{(c)}}\,, \ee where definitions
(\ref{xyzw}) were used to substitute for $x$ and $y$. Both
$L_{A,\,k}^{(c)}$ and $B_A^{(c)}$ are completely determined by the
system (\ref{e20}), whereas ${\cal T}^{\,(c)}_A$ is determined
from (\ref{aux2}-\ref{eq32}). Then solution (\ref{sol32}) can be
conveniently expressed as \be\label{sol34} \nu_{A,\,k}^{(c)}=\frac
{{\cal T}^{\,(c)}_A}{B_A^{(c)}}\,L_{A,\,k}^{(c)}\,. \ee
 Recall that $m_k=\pm 1$. Thus the only two
possibilities for $L_{A,\,k}^{(c)}$ for any $k=\overline{1,N}$ are
\be\label{lpm} L_{A,\,\pm}^{(c)}=T_A^{(c)}\pm H_A^{(c)}\,. \ee As
we vary parameter $A$ over the range $0\leq A<1$ point by point,
system (\ref{e20}) always has exactly two real solutions. In
Figure \ref{Plot24} we present plots of $L_{A,\,+}^{(c)}$,
$L_{A,\,-}^{(c)}$ and ${B_A^{(c)}}$, where $c=\overline{1,2}$ as
functions of $A$.
We only need to consider the positive range $0\leq A<1$ because of
the symmetry (\ref{symmet}).

What happens to these solutions when $A=1$? We already know from
the previous discussion that the system (\ref{e20}) obtained in
the limit $w\rightarrow -\infty$ degenerates for $A=1$ and one
obtains the solution that corresponds to the SUSY extremum
explicitly. The solutions plotted in Figure \ref{Plot24} were
obtained assuming $A\not =1$ and therefore have an apparent
singularity when $A=1$. Thus they cannot capture either the SUSY
or the de Sitter extrema that arise in this special case. To
explain what happens to the de Sitter extremum we need to examine
the exact solution in (\ref{NH2}), in the same limit. Indeed,
bearing in mind that $w$ is negative, from (\ref{NH2}) we have \be
L_{A,+}^{(2)}=T_1^{(2)}+H_1^{(2)}=-\frac 3
7\left(\frac{3+\sqrt{9-7w}}w\right)\,. \ee Here we see immediately
that in the limit $w\rightarrow -\infty$ for the solution above
$L_{A,+}^{(2)}\rightarrow 0$. Therefore we conclude that the de
Sitter extremum cannot be obtained from (\ref{e20}) which
correlates with the previous observation that for $A=1$
(\ref{e20}) has only one solution - the SUSY extremum.
Nevertheless, as we will see in the next subsection the real
solutions plotted in Figure \ref{Plot24} are a very good
approximation to the exact numerical solutions corresponding to
the AdS vacua with spontaneously broken supersymmetry.

Now we would like to classify which of these AdS vacua have all
the moduli stabilized at positive values. Indeed if some of the
moduli are fixed at negative values we can automatically exclude
such vacua from further consideration since the supergravity
approximation assumes that all the moduli are positive. Since the
volume ${{\cal T}^{\,(c)}_A}$ is always positive by definition,
from (\ref{sol34}) we see immediately that for all moduli to be
stabilized in the positive range, all three quantities
$L_{A,\,+}^{(c)}$,  $L_{A,\,-}^{(c)}$ and ${B_A^{(c)}}$ must have
the same sign. In Figure {\ref{Plot24}} the plots on the right
satisfy this requirement for the entire range $0\leq A<1$. On the
other hand, the short-dashed curve corresponding to
$L_{A,-}^{(1)}$ on the left plot is negative when $0\leq A<1/7$,
features a zero at $A=1/7$ and becomes positive for $1/7< A<1$.
Yet, both $L_{A,\,+}^{(1)}$ and ${B_A^{(1)}}$ remain negative
throughout the entire range. Moreover, it is easy to verify that
the solution with $T_{1/7}^{(1)}=-3/4$ and $H_{1/7}^{(1)}=-3/4$,
such that $L_{1/7,-}^{(1)}=0$ is also an exact solution for the
general case (\ref{e3}) when $w$ is finite. Therefore, all
solutions compatible with the SUGRA approximation can be
classified as follows:

\vspace{0.5cm}

Given a set of $\{a_i\}$ with $i=\overline{1,N}$, there are
$2^{N}$ possible values of $A$, including the negative ones. From
the symmetry in (\ref{symmet}), only half of those give
independent solutions. This narrows the possibilities to $2^{N-1}$
positive combinations that fall in the range $0\leq A\leq 1$. For
each $A$ in the range $0\leq A<1/7$ there exist exactly two
solutions describing AdS vacua with broken SUSY with all the
moduli fixed at positive values.

For each  $A$ in the range $1/7\leq A<1$ there exists exactly one
solution describing an AdS vacuum with broken SUSY with all the
moduli stabilized in the positive range of values. For $A=1$ there
are exactly two solutions with all the moduli stabilized in the
positive range - de Sitter extremum in (\ref{NH2}) and the SUSY
AdS extremum in (\ref{NH1}). These two solutions are always present
for any set of $\{a_i\}$.


\subsection{Explicit approximate solutions}\label{approxmod}
In this section we will complete our analysis of the AdS vacua and
obtain explicit analytic solutions for the moduli. We will take an
approach similar to the one we employed in section \ref{spl} when
we obtained an approximate formula (\ref{nu10}). Expressing
$\alpha^{(c\,)}$ from (\ref{eq32}) we obtain \be \label{eq37}
\alpha^{(c\,)}_A=\frac{b_2{{{\cal T}^{\,(c)}_A}}-{B_A^{(c)}}}
{b_1{{{\cal T}^{\,(c)}_A}}-{B_A^{(c)}}}\,. \ee There exists a
natural limit when the volume of the associative cycle ${\cal
T}^{\,(c)}_A$ is large. Just like in the approximate SUSY case in
(\ref{alpha}), the leading order solution to (\ref{eq37}) in this
limit is given by \be \label{eq38}
\alpha^{(c\,)}_A=\frac{b_2}{b_1}\,, \ee independent of $A$ and
$c$. Plugging this into (\ref{aux2}) and solving for ${\cal
T}^{\,(c)}_A$ we have in the leading order \be\label{eq39} {\cal
T}^{\,(c)}_A=\frac
1{b_2-b_1}\ln\left(\frac{A_2b_2}{A_1b_1}\right)=\frac
1{2\pi}\frac{P\,Q}{P-Q}\ln\left(\frac{A_2P}{A_1Q}\right)\,, \ee
where we again assumed the hidden sector gauge groups to be
$SU(P)$ and $SU(Q)$. Notice that this approximation
automatically results in the limit $w\rightarrow -\infty$ and
therefore, $L_{A,\,+}^{(c)}$,  $L_{A,\,-}^{(c)}$ and ${B_A^{(c)}}$
computed by solving (\ref{e20}) and plotted in Figure \ref{Plot24}
are consistent with this approximation. Thus, combining
(\ref{eq39}) with (\ref{sol34}) and (\ref{nu1}) we have the
following approximate analytic solution for the moduli in the
leading order: \be\label{eq43} s^{\,(c)}_{A,\,k}=\frac
1{2\pi}\left(\frac{a_k}{N_k}\right)\left(\frac
{L_{A,\,k}^{(c)}}{B_A^{(c)}}\right)\frac{P\,Q}{P-Q}\ln\left(\frac{A_2P}{A_1Q}\right)\,.
\ee To verify the approximation we can check it for the previously
considered special case with two moduli when $a_1=a_2=7/6$, i.e.
the case when $A=0$. By solving (\ref{e20}) we obtain: \ba
&&L_{0,\,+}^{(1)}=L_{0,\,-}^{(2)}=\frac
3{16}\left(-9+\sqrt{17}+\sqrt{-26+10\sqrt{17}}\right)\,,\\\nonumber
&&L_{0,\,-}^{(1)}=L_{0,\,+}^{(2)}=\frac
3{16}\left(-9+\sqrt{17}-\sqrt{-26+10\sqrt{17}}\right)\,,\\\nonumber
&&B_{0}^{(1)}=B_{0}^{(2)}=\frac 7{16}\left(-9+\sqrt{17}\right)\,.
\ea Thus, we have the following two solutions for the moduli for
the AdS vacua with broken SUSY: \ba\label{mod43}
&&s^{\,(1)}_{0,\,1}=\left(1-\frac{\sqrt{-26+10\sqrt{17}}}{9-\sqrt{17}}\right)\frac
1{4\pi{N_1}}\frac{P\,Q}{P-Q}\ln\left(\frac{A_2P}{A_1Q}\right)\sim
\frac
{0.016}{N_1}\frac{P\,Q}{P-Q}\ln\left(\frac{A_2\,P}{A_1\,Q}\right)\,,\\\nonumber
&&s^{\,(1)}_{0,\,2}=\left(1+\frac{\sqrt{-26+10\sqrt{17}}}{9-\sqrt{17}}\right)\frac
1{4\pi{N_2}}\frac{P\,Q}{P-Q}\ln\left(\frac{A_2P}{A_1Q}\right)\sim
\frac
{0.143}{N_2}\frac{P\,Q}{P-Q}\ln\left(\frac{A_2\,P}{A_1\,Q}\right)\,.\nonumber
\ea and \ba\label{mod44}
&&s^{\,(2)}_{0,\,1}=\left(1+\frac{\sqrt{-26+10\sqrt{17}}}{9-\sqrt{17}}\right)\frac
1{4\pi{N_1}}\frac{P\,Q}{P-Q}\ln\left(\frac{A_2P}{A_1Q}\right)\sim
\frac
{0.143}{N_2}\frac{P\,Q}{P-Q}\ln\left(\frac{A_2\,P}{A_1\,Q}\right)\,,\\\nonumber
&&s^{\,(2)}_{0,\,2}=\left(1-\frac{\sqrt{-26+10\sqrt{17}}}{9-\sqrt{17}}\right)\frac
1{4\pi{N_2}}\frac{P\,Q}{P-Q}\ln\left(\frac{A_2P}{A_1Q}\right)\sim
\frac
{0.016}{N_1}\frac{P\,Q}{P-Q}\ln\left(\frac{A_2\,P}{A_1\,Q}\right)\,.\nonumber
\ea The choice of the constants given in (\ref{choice1}) results
the following values: \be
s^{\,(1)}_{0,\,1}=s^{\,(2)}_{0,\,2}=2.62\,,\,\,\,\,\,s^{\,(1)}_{0,\,2}=s^{\,(2)}_{0,\,1}=23.64\,.
\ee A quick comparison with the exact values in (\ref{sol9})
obtained numerically leads us to believe that the approximate
analytical solutions presented here are highly accurate. This is
especially true when the volume of the associative cycle ${\cal
T}^{\,(c)}_A$ is large. For the particular choice above the
approximate value is: \be {\cal T}^{\,(1)}_0={\cal T}^{\,(2)}_0=
26.16\,, \ee which is indeed fairly large. To complete the
picture, we also would like to include the first subleading order
contributions to the approximate solutions presented here. After a
straightforward computation we have the following: \be
\label{eq45} \alpha^{(c\,)}_A=\frac P
Q+\frac{B_A^{(c)}}{\ln\left(\frac{A_2 P}{A_1 Q}\right)}
\left(\frac{P-Q}Q\right)^2\,, \ee and \be \label{eq46} {\cal
T}^{\,(c)}_A=\frac
1{2\pi}\frac{P\,Q}{P-Q}\ln\left(\frac{A_2P}{A_1Q}\right) +\frac
{B_A^{(c)}}{2\pi}\left(\frac{P-Q}{\log\left(\frac{A_2 P}{A_1
Q}\right)}\right)\,. \ee By combining (\ref{eq46}) with
(\ref{sol34}) and (\ref{nu1}) it is easy to obtain the
corresponding expressions for the moduli that include the first
subleading order correction: \be\label{eq47}
s^{\,(c)}_{A,\,k}=\frac
1{2\pi}\left(\frac{a_k}{N_k}\right)\left(\frac
{L_{A,\,k}^{(c)}}{B_A^{(c)}}\right)\frac{P\,Q}{P-Q}\ln\left(\frac{A_2P}{A_1Q}\right)
+\frac
{L_{A,\,k}^{(c)}}{2\pi}\left(\frac{a_k}{N_k}\right)\left(\frac{P-Q}{\ln\left(\frac{A_2
P}{A_1 Q}\right)}\right)\,. \ee


\section{Vacua with charged matter in the Hidden Sector}\label{chargedmattervac}
Thus far, we have studied in reasonable detail, the vacuum
structure in the cases when the hidden sector has two strongly
coupled gauge groups without any charged matter. It is of interest
to study how the addition of matter charged under the hidden
sector gauge group changes the conclusions. We argue that
the addition of charged matter can give rise to
Minkowski or metastable de Sitter (dS) vacua due to additional $F$-terms for the hidden sector matter fields.
Hence, dS vacua are obtained without adding any anti-branes which explicitly break supersymmetry. This possibility
was first studied in \cite{Lebedev:2006qq}. Moreover, we explain why it is reasonable
to expect that for a given choice of $G_2$-manifold, the dS vacuum obtained is \emph{unique}.

\subsection{Scalar Potential}

Generically we would expect that a hidden sector gauge theory can
possess a fairly rich particle spectrum which, like the visible
sector, may include chiral matter. For example, an $SU(N_c)$ gauge
theory apart from the ``pure glue'' may also include massless
quark states $Q$ and $\tilde Q$ transforming in $N_c$ and $\bar
N_c$ of $SU(N_c)$. When embedded into $M$ theory the effective
superpotential due to gaugino condensation for such a hidden
sector with $N_f$ ($N_f < N_c$) quark flavors has the following
form \cite{Seiberg:1994bz}: \be \label{mattersup}
W=A_1\,e^{i\frac{2\pi}{N_c-N_f}\sum_{i=1}^{N} N_i^{(1)} z_i}
\det(Q\tilde Q)^{-\frac{1}{N_c-N_f}}\,. \ee We can introduce an
effective meson field $\phi$ to replace the quark bilinear \be
\label{meson} \phi\equiv\left(2Q\tilde
Q\right)^{1/2}=\phi_0e^{i\theta}\,, \ee and for notational brevity
we define \be \label{b1anda}
b_1\equiv\frac{2\pi}{N_c-N_f}\,,\,\,\,\,\,\,\,a\equiv-\frac
2{N_c-N_f}\,. \ee Here we will consider the case when the hidden
sector gauge groups are $SU(N_c)$ and  $SU(Q)$ with $N_f$ flavors
of the quarks $Q$ ($\tilde Q$) transforming as $N_c$ ($\bar N_c$)
under $SU(N_c)$ and as singlets under $SU(Q)$. In this case, when
$N_f=1$, the effective nonperturbative superpotential has the
following form: \ba \label{mattersuptwosectors}
W=A_1{\phi}^a\,e^{ib_1\,f}+A_2e^{ib_2\,f}\,. \ea

One serious drawback of considering hidden sector matter is that we cannot
explicitly calculate the moduli dependence of the matter K\"{a}hler potential.
Therefore we will have to make some (albeit reasonable) assumptions, unlike the
cases studied in the previous sections.
In what follows
we will assume that we work in a particular region of the moduli
space where the K\"{a}hler metric for the matter fields in the
hidden sector is a very slowly varying function of the moduli,
essentially a constant. This assumption is based on the fact that
the chiral fermions are localized at point-like conical
singularities so that the bulk moduli $s_i$ should have very
little effect on the local physics. In general, a singularity supporting a chiral fermion
has no local moduli, since there are no flat directions constructed from a single chiral matter
representation. Our assumption is further justified by the $M$ theory lift of some calculable Type IIA
matter metrics as described in the appendix. It is an interesting and extremely important
problem to properly derive the matter K\"{a}hler potential in $M$ theory and test our assumptions.

Thus we will consider the case when the hidden sector chiral
fermions have ``modular weight zero'' and assume a canonically
normalized K\"{a}hler potential. The scalar potential is invariant
under $Q \leftrightarrow {\tilde Q}$ and $Q = {\tilde Q}$ along
the $D$-flat direction. For the sake of simplicity, we will first
study the case $N_f=1$, but later it will be shown that all the
results also hold true for $N_f>1$. The meson field $
\phi\equiv(2Q\tilde Q)^{1/2}$ along the $D$-flat direction is such
that the corresponding K\"{a}hler potential for $\phi$ is
canonical. The total K\"{a}hler potential, i.e. moduli plus matter
thus takes the form: \be \label{kahlerwithmatter} K = -3
\ln(4\pi^{1/3}\,V_X)+ Q^{\dag}Q + {\tilde Q}^{\dag} {\tilde Q} =
-3 \ln(4\pi^{1/3}\,V_X)+\phi\bar\phi\,. \ee The moduli $F$-terms
are then given by \ba\label{modf}
F_k&=&ie^{i b_2\vec N\cdot \vec t}[N_k(b_1A_1\phi_0^a e^{-b_1\vec N\cdot\vec s+i(b_1-b_2)\vec N\cdot \vec t+
ia\theta}+b_2A_2 e^{-b_2\vec N\cdot\vec s})\nonumber\\
&&+\frac{3a_k}{2s_k}(A_1\phi_0^a e^{-b_1\vec N\cdot\vec
s+i(b_1-b_2)\vec N\cdot \vec t+ia\theta}+A_2 e^{-b_2\vec
N\cdot\vec s})]\,. \ea

In addition, an $F$-term due to the meson field is also generated
\be\label{mesf} F_\phi=\phi_0e^{-i\theta+i b_2\vec N\cdot \vec
t}\left[\left(\frac a{\phi_0^2}+1\right)A_1\phi_0^a e^{-b_1\vec
N\cdot\vec s+i(b_1-b_2)\vec N\cdot \vec t+ia\theta}+A_2
e^{-b_2\vec N\cdot\vec s}\right]\,. \ee The supergravity scalar
potential is then given by: \ba \label{potential_with_matter}
V&=&\frac{e^{\phi_0^2}}{48\pi
V_X^3}\,[(b_1^2A_1^2\phi_0^{2a}e^{-2b_1\vec\nu\cdot\,\vec
a}+b_2^2A_2^2e^{-2b_2\vec\nu\cdot\,\vec a}
+2b_1b_2A_1A_2\phi_0^{a}e^{-(b_1+b_2)\vec\nu\cdot\,\vec a}{\rm cos}((b_1-b_2)\vec N\cdot \vec t+a\theta))\nonumber\\\nonumber\\
&&\times\sum_{i=1}^{N}a_i({\nu_i})^2+3(\vec\nu\cdot\,\vec a)(b_1A_1^2\phi_0^{2\alpha}e^{-2b_1\vec\nu\cdot\,\vec a}+b_2A_2^2e^{-2b_2\vec\nu\cdot\,\vec a}+(b_1+b_2)A_1A_2\phi_0^{a}e^{-(b_1+b_2)\vec\nu\cdot\,\vec a}\,\nonumber\\\nonumber\\
&&\times\,{\rm cos}((b_1-b_2)\vec N\cdot \vec t+a\theta))
+3(A_1^2\phi_0^{2a}e^{-2b_1\vec\nu\cdot\,\vec a}+A_2^2e^{-2b_2\vec\nu\cdot\,\vec a}+2A_1A_2\phi_0^{a}e^{-(b_1+b_2)\vec\nu\cdot\,\vec a}\,\\\nonumber\\
&&\times{\rm cos}((b_1-b_2)\vec N\cdot \vec t+a\theta))
+\frac 3 4{\phi_0^2}\,(A_1^2\phi_0^{2\alpha}\left(\frac a{\phi_0^2}+1\right)^2e^{-2b_1\vec\nu\cdot\,\vec a}+A_2^2e^{-2b_2\vec\nu\cdot\,\vec a}\,\nonumber\\\nonumber\\
&&+2A_1A_2\phi_0^{a}\left(\frac
a{\phi_0^{2}}+1\right)e^{-(b_1+b_2)\vec\nu\cdot\,\vec a} {\rm
cos}((b_1-b_2)\vec N\cdot \vec t+a\theta))]\,.\nonumber \ea
Minimizing this potential with respect to the axions and $\theta$
we obtain the following condition: \be {\rm sin}((b_1-b_2)\vec
N\cdot \vec t+a\theta)=0\,. \ee The potential has local minima
with respect to the moduli $s_i$ when \be\label{axcond} {\rm
cos}((b_1-b_2)\vec N\cdot \vec t+a\theta)=-1\,. \ee In this case
(\ref{potential_with_matter}) reduces to \ba
\label{potential_noaxions}
V&=&\frac{e^{\phi_0^2}}{48\pi V_X^3}\,[(b_1A_1\phi_0^{a}e^{-b_1\vec\nu\cdot\,\vec a}-b_2A_2e^{-b_2\vec\nu\cdot\,\vec a})^2\sum_{i=1}^{N}a_i({\nu_i})^2\nonumber\\\nonumber\\
&&+3(\vec\nu\cdot\,\vec a)(A_1\phi_0^{a}e^{-b_1\vec\nu\cdot\,\vec a}-A_2e^{-b_2\vec\nu\cdot\,\vec a})(b_1A_1\phi_0^{a}e^{-b_1\vec\nu\cdot\,\vec a}-b_2A_2e^{-b_2\vec\nu\cdot\,\vec a})\\\nonumber\\
&&+3(A_1\phi_0^{a}e^{-b_1\vec\nu\cdot\,\vec
a}-A_2e^{-b_2\vec\nu\cdot\,\vec a})^2+\frac 3
4(A_1\phi_0^{a}\left(\frac a\phi_0+\phi_0\right)e^{-b_1\vec\nu\cdot\,\vec
a}-A_2\phi_0e^{-b_2\vec\nu\cdot\,\vec a})^2]\,.\nonumber \ea

\subsection{Supersymmetric extrema}
Here we consider a case when the scalar potential
(\ref{potential_noaxions}) possess SUSY extrema and find
approximate solutions for the moduli and the meson field vevs.
Taking into account (\ref{axcond}) and setting the moduli
$F$-terms (\ref{modf}) to zero we obtain \be\label{nueq}
\nu_k=\nu=-\frac 3 2\frac{\tilde\alpha-1}{b_1\tilde\alpha-b_2}\,,
\ee together with the constraint \be\label{trcons}
\tilde\alpha\equiv\frac {A_1}{A_2}\phi_0^ae^{-\frac 7
3(b_1-b_2)\nu}\,. \ee At the same time, setting the matter
$F$-term (\ref{mesf}) to zero results in the following condition:
\be\label{mescon} \left(\frac
a{\phi_0^2}+1\right)\tilde\alpha-1=0\,. \ee Expressing
$\tilde\alpha$ from (\ref{nueq}) and substituting it into
(\ref{mescon}) we obtain the following solution for the meson vev
at the SUSY extremum: \be\label{phisol} \phi_0^2=a\frac{b_2+
3/(2\nu)}{b_1-b_2}\,. \ee Recall that in our analysis we are
considering the case when $P\equiv N_c-N_f>0$, which implies that
parameter $a$ defined in (\ref{b1anda}) is negative. Thus, since
the left hand side of (\ref{phisol}) is positive, for the SUSY
solution to exist, it is necessary to satisfy \be\label{con21}
b_2>b_1\,\,\,\,\,\,=>\,\,\,\,\,\,\,P>Q\,. \ee Recall that for the
moduli to be positive, the constants have to satisfy certain
conditions resulting in two possible branches (\ref{branches}).
Therefore, condition (\ref{con21}) implies that the SUSY AdS
extremum exists only for branch a) in (\ref{branches}). In the
limit, when $\nu$ is large, the approximate solution is given by:
\ba\label{app21} &&\tilde\alpha=\frac P Q\,,\\\nonumber
&&s_i=\frac{a_i\nu}{N_i}\,,\,\,\,\,\,\,{\rm with}\,\,\,\,\,\,\,
\nu=\frac
3{14\pi}\frac{P\,Q}{P-Q}\ln\left(\frac{A_2P}{A_1Q}\right)\,,\\\nonumber
&&\phi_0^2=\frac 2{P-Q}+\frac
7{P\,\ln\left(\frac{A_2P}{A_1Q}\right)}\,,\nonumber \ea where we
also assumed that $P\sim {\mathcal O}(10)$, such that
$\phi_0^a\approx 1$. For the case with two moduli where
$a_1=a_2=7/6$ and the choice \be \label{choice3}
A_1=4.1\,,\,\,\,A_2=30\,,\,\,\,b_1=\frac
{2\pi}{30}\,,\,\,\,b_2=\frac
{2\pi}{27}\,,\,\,\,N_1=1\,,\,\,\,N_2=1\,, \ee the numerical
solution for the SUSY extremum obtained by minimizing the scalar
potential (\ref{potential_noaxions}) gives \be \label{sol17}
s_1\approx44.5\,,\,\,\,s_2\approx44.5\,,\,\,\,
\phi_0\approx0.883\,, \ee whereas the approximate analytic
solution obtained in (\ref{app21}) yields \be \label{sol18}
s_1\approx45.0\,,\,\,\,s_2\approx45.0\,,\,\,\,
\phi_0\approx0.882\,. \ee This vacuum is very similar to the SUSY
AdS extremum obtained previously for the potential arising from
the ``pure glue'' Super Yang-Mills (SYM) hidden sector gauge
theory. Thus, we will not discuss it any further and instead move
to the more interesting case, for which condition (\ref{con21}) is
not satisfied.
\subsection{Metastable de Sitter (dS) minima}
Below we will use the same approach and notation we used in
section \ref{nonsusyadsvac}, to describe AdS vacua with broken
SUSY. Again, for brevity we denote \be\label{xyzw1} \tilde
x\equiv(\tilde\alpha-1)\,,\,\,\tilde
y\equiv(b_1\tilde\alpha-b_2)\,,\,\,\tilde
z\equiv(b_1^2\tilde\alpha-b_2^2)\,,\,\, \tilde w\equiv\frac{\tilde
x\tilde z}{{\tilde y}^2}\,. \ee Extremizing
(\ref{potential_noaxions}) with respect to the moduli $s_i$ and
dividing by ${\tilde x}^2$ we obtain the following system of
coupled equations \ba \label{u24} 2\frac {{\tilde y}^2} {{\tilde
x}^2} \nu_k^2&-&\left(2\,\frac {{\tilde y}^2} {{\tilde x}^2}\tilde
w\sum_{i=1}^{N}a_i\nu^2_i+3\frac {\tilde y} {\tilde x}\left(\tilde
w+1\right)\vec\nu\cdot\vec a+3+\frac 3 2
\phi_0^2\left(\frac{a\tilde\alpha}{\phi_0^2\,\tilde x}+1\right)
\left(\frac{a\tilde\alpha b_1}{\phi_0^2\,\tilde y}+1\right)\right)\frac {\tilde y}{\tilde x}\nu_k\,\\
&-&3\left(\frac {{\tilde y}^2} {{\tilde
x}^2}\sum_{i=1}^{N}a_i\nu_i^2+3\frac {\tilde y}{\tilde x}
\vec\nu\cdot\vec a+3+\frac 3 4
\phi_0^2\left(\frac{a\tilde\alpha}{\phi_0^2\,\tilde
x}+1\right)^2\right)=0\,,\nonumber \ea plus the constraint
(\ref{trcons}). Next, we extremize (\ref{potential_noaxions}) with
respect to $\phi_0$ and divide it by $2\phi_0{\tilde x}^2$ to
obtain: \ba \label{u25} \frac {{\tilde y}^2} {{\tilde
x}^2}\sum_{i=1}^{N}a_i\nu_i^2+\frac 32\frac {\tilde y}{\tilde
x}\vec\nu\cdot\vec a+\frac 34 \left(2\frac {\tilde y}{\tilde
x}\vec\nu\cdot\vec a+\frac{a\tilde\alpha}{\tilde
x}\left(\frac{a-1}{\phi_0^2}+2\right)+
5+\phi_0^2\right)\left(\frac{a\tilde\alpha}{\phi_0^2\,\tilde x}+1\right)&&\\
+\frac{a\tilde\alpha b_1}{\phi_0^2\,\tilde x}\left(\frac {{\tilde
y}} {{\tilde x}}\sum_{i=1}^{N}a_i\nu_i^2+ \frac 3
2\vec\nu\cdot\vec a\right)=0\,.&&\,\nonumber \ea To solve the
system of $N$ cubic equations (\ref{u24}), we introduce a
quadratic constraint \be \label{Nt} 4\,\tilde T\equiv2\,\frac
{{\tilde y}^2} {{\tilde x}^2}\tilde
w\sum_{i=1}^{N}a_i\nu^2_i+3\frac {\tilde y} {\tilde x}\left(\tilde
w+1\right)\vec\nu\cdot\vec a+3+\frac 3 2
\phi_0^2\left(\frac{a\tilde\alpha}{\phi_0^2\,\tilde
x}+1\right)\left(\frac{a\tilde\alpha b_1}{\phi_0^2\,\tilde
y}+1\right)\,, \ee such that (\ref{u24}) turns into a system of
$N$ coupled quadratic equations: \be \label{u26} 2\frac {{\tilde
y}^2} {{\tilde x}^2} \nu_k^2-4\tilde T\frac {\tilde y}{\tilde
x}\nu_k -3\left(\frac {{\tilde y}^2} {{\tilde
x}^2}\sum_{i=1}^{N}a_i\nu_i^2+3\frac {\tilde y}{\tilde
x}\vec\nu\cdot\vec a+3+\frac 3 4
\phi_0^2\left(\frac{a\tilde\alpha}{\phi_0^2\,\tilde
x}+1\right)^2\right)=0\,. \ee Again, the standard solution of a
quadratic equation dictates that the solutions for $\nu_k$ of
(\ref{u26}) have the form \be \label{sol24} \nu_k=\frac {\tilde x}
{\tilde y} \left({\tilde T}+m_k \tilde H\right)\,,\,\, {\rm
with}\,\, m_k=\pm 1\,,\,\,k=\overline{1,N}\,. \ee We have now
reduced the task of determining $\nu_k$ for each
$k=\overline{1,N}$ to finding {\it only two} quantities - $\tilde
T$ and $\tilde H$. By substituting (\ref{sol24}) into equations
(\ref{u25}-\ref{u26}) and using (\ref{vol}), we obtain a system of
three coupled equations \ba \label{e43}
\frac 7 3\left({\tilde T_{A}^2}+2 A{\tilde T_{A}} \tilde H_{A}+\tilde H_{A}^2\right)+\frac 72\left({\tilde T_{A}}+A \tilde H_{A}\right)+\frac 3 4\left(\frac{14}3\left({\tilde T_{A}}+A \tilde H_{A}\right)+\frac{a\tilde\alpha}x\left(\frac{a-1}{\phi_0^2}+2\right)+5+\phi_0^2\right)\,\,\,\,\,\,\,\,\,&&\,\\
\times\left(\frac{a\tilde\alpha}{\phi_0^2\,\tilde x}+1\right)+\frac{a\tilde\alpha b_1}{\phi_0^2}\frac {7} {3{\tilde y}}\left(\left({\tilde T_{A}^2}+2 A{\tilde T_{A}} \tilde H_{A}+\tilde H_{A}^2\right)+\frac 3 2\left({\tilde T_{A}}+A \tilde H_{A}\right)\right)=0&&\,\nonumber\\
\frac{14w}3\left({\tilde T_{A}^2}+2 A{\tilde T_{A}} \tilde H_{A}+\tilde H_{A}^2\right)+7(w+1) \left({\tilde T_{A}}+A \tilde H_{A}\right)+3+\frac 3 2 \phi_0^2\left(\frac{a\tilde\alpha}{\phi_0^2\,\tilde x}+1\right)\left(\frac{a\tilde\alpha b_1}{\phi_0^2\,\tilde y}+1\right)-4\tilde T_{A}=0\,\nonumber&& \\
9\left({\tilde T_{A}^2}+2 A {\tilde T_{A}} \tilde H_{A}+\tilde
H_{A}^2\right)-4\tilde H_{A}\left(\tilde H_{A}+A{\tilde T_{A}}
\right)+21\left({\tilde T_{A}}+A\tilde H_{A}\right)+9+\frac 9 4
\phi_0^2\left(\frac{a\tilde\alpha}{\phi_0^2\,\tilde
x}+1\right)^2=0\,,&& \nonumber \ea plus the constraint
(\ref{trcons}). Note that each solution is again labelled by
parameter $A$ so that (\ref{sol24}) becomes \be \label{sol36}
\nu_k^{A}=\frac {\tilde x} {\tilde y} \left(\tilde T_{A}+m_k
\tilde H_{A}\right)\,. \ee Let us consider the case when $A=1$. In
this case, the solution is given by \be \label{sol37}
\nu_k^{1}=\nu=\frac {\tilde x} {\tilde y} \left(\tilde
T_{1}+\tilde H_{1}\right)=\frac {\tilde x} {\tilde y}\tilde
L_{1,+}\,. \ee and (\ref{e43}) is reduced to \ba \label{e46}
\frac 7 3\left(\tilde T_1+\tilde H_1\right)^2+\frac 72\left({\tilde T_1}+\tilde H_1\right)+\frac 3 4\left(\frac {14}3\left(\tilde T_1+\tilde H_1\right)+\frac{a\tilde\alpha}x\left(\frac{a-1}{\phi_0^2}+2\right)+\phi_0^2+5\right)\left(\frac{a\tilde\alpha}{\phi_0^2\,\tilde x}+1\right)&&\,\\
+\frac{a\tilde\alpha b_1}{\phi_0^2}\frac {7} {3{\tilde y}}\left(\left(\tilde T_1+\tilde H_1\right)^2+\frac 3 2\left({\tilde T_1}+\tilde H_1\right)\right)=0&&\,\nonumber\\
\frac{14w}3\left(\tilde T_1+\tilde H_1\right)^2+7(w+1) \left({\tilde T_1}+\tilde H_1\right)+3+\frac 3 2 \phi_0^2\left(\frac{a\tilde\alpha}{\phi_0^2\,\tilde x}+1\right)\left(\frac{a\tilde\alpha b_1}{\phi_0^2\,\tilde y}+1\right)-4\tilde T_1=0\,\nonumber&& \\
9\left(\tilde T_1+\tilde H_1\right)^2-4\tilde H_1\left(\tilde
H_1+{\tilde T_1} \right)+21\left({\tilde T_1}+\tilde
H_1\right)+9+\frac 9 4
\phi_0^2\left(\frac{a\tilde\alpha}{\phi_0^2\,\tilde
x}+1\right)^2=0\,.&& \nonumber \ea In the notation introduced in
(\ref{xyzw1}), the SUSY condition (\ref{mescon}) can be written as
\be\label{sucon} \frac{a\tilde\alpha}{\phi_0^2}+\tilde x=0\,. \ee
It is then straightforward to check that in the SUSY case, the
system (\ref{e46}) yields \be \tilde T_1=-\frac {15}
8\,,\,\,\,\,\,\,\tilde H_1=\frac 3 8\,,\,\,\,\,\,\,\tilde
L_{1,+}=-\frac 3 2\,, \ee as expected. We will now consider branch
b) in (\ref{branches}) for which (\ref{sucon}) is not satisfied.
Moreover, in order to obtain analytical solutions for the moduli
and the meson vev $\phi_0$ we will again consider the large three
cycle volume approximation. Recall that in this case we take
$\tilde y \rightarrow 0$ and $\tilde w\rightarrow -\infty$ limit
to obtain the following reduced system of equations when $A=1$ for
$\tilde L_{1,+}$ and $\phi_0$: \ba \label{e47}
\frac 7 3\left(\tilde L_{1,+}\right)^2+\frac 72\tilde L_{1,+}+\frac 3 4\left(\frac{14}3\tilde L_{1,+}+\frac{a\tilde\alpha}x\left(\frac{a-1}{\phi_0^2}+2\right)+\phi_0^2+5\right)\left(\frac{a\tilde\alpha}{\phi_0^2\,\tilde x}+1\right)&&\,\\
+\frac{a\tilde\alpha b_1}{\phi_0^2}\frac 7 {3{\tilde y}}
\left(\left(\tilde L_{1,+}\right)^2+\frac 3 2\tilde L_{1,+}\right)=0&&\,\nonumber\\
\frac 23\left(\tilde L_{1,+}\right)^2+\tilde L_{1,+}+\frac
{3{a\tilde\alpha b_1}\tilde y} {14 \tilde x\tilde z}
\left(\frac{a\tilde\alpha} {\phi_0^2\,\tilde x}+1\right)=0\,,&&
\nonumber \ea Note that in (\ref{e47}), we have dropped the third
equation since for $A=1$ we only need to know $\tilde{L}_{1,+}$
and the third equation in (\ref{e46}) determines $\tilde{H}_{1,+}$
in terms of $\tilde{L}_{1,+}$. We also kept the first subleading
term in the second equation. Note that the term in the second line
of the first equation proportional to $\sim 1/{\tilde y}$ appears
to blow up as $\tilde y \rightarrow 0$. However, from the second
equation one can see that the combination $\left(\tilde
L_{1,+}\right)^2+\frac 3 2\tilde L_{1,+}$ is proportional to
$\tilde y$ which makes the corresponding term finite. By keeping
the subleading term in the second equation, we can express
\be\label{e51} \left(\tilde L_{1,+}\right)^2=-\frac 3 2\tilde
L_{1,+}-\frac {9{a\tilde\alpha b_1}\tilde y} {28 \tilde x\tilde z}
\left(\frac{a\tilde\alpha}{\phi_0^2\,\tilde x}+1\right) \ee from
the second equation to substitute into the first equation to
obtain in the leading order \be \label{e49} \left(\frac {14}
3\tilde
L_{1,+}+5+\phi_0^2+\frac{a\tilde\alpha}x\left(\frac{a-1}{\phi_0^2}+2\right)-\frac
1 {\tilde z\tilde x}\left(\frac{a\tilde\alpha
b_1}{\phi_0}\right)^2\right)\left(\frac{a\tilde\alpha}{\phi_0^2\,\tilde
x}+1\right)=0\,. \ee Since we are now considering branch b) in
(\ref{branches}), the second factor in (\ref{e49}) is
automatically non-zero. Therefore, the first factor in (\ref{e49})
must be zero. Thus, after substituting \be\label{e61} \tilde
L_{1,+}\approx-\frac 3 2+\frac {3{a\tilde\alpha b_1}\tilde y} {14
\tilde x\tilde z} \left(\frac{a\tilde\alpha}{\phi_0^2\,\tilde
x}+1\right)\,, \ee obtained from (\ref{e51}), we have the
following equation for $\phi_0$ \be \label{e50} \phi_0^2-2+\frac
{{a\tilde\alpha b_1}\tilde y} {\tilde x\tilde z}
\left(\frac{a\tilde\alpha}{\phi_0^2\,\tilde x}+1\right)+
\frac{a\tilde\alpha}x\left(\frac{a-1}{\phi_0^2}+2\right)-\frac 1
{\tilde z\tilde x}\left(\frac{a\tilde\alpha
b_1}{\phi_0}\right)^2=0\,. \ee

Also, since in the leading order $\tilde L_{1,+}=-3/2$, using the
definitions in (\ref{xyzw1}) we can express $\tilde\alpha$ from
(\ref{sol37}) in the limit when $\nu$ is large, including the
first subleading term \be\label{e54}
\tilde\alpha\approx\frac{b_2}{b_1}+\frac{3(b_1-b_2)}{2b_1^2\,\nu}\,.
\ee By combining (\ref{trcons}) with the leading term in
(\ref{e54}) and taking into account that $\phi_0^a\sim 1$ we again
obtain \be\label{app24} s_i=\frac{a_i\nu}{N_i}\,,\,\,\,\,\,\,{\rm
with}\,\,\,\,\,\,\, \nu\approx\frac
3{14\pi}\frac{P\,Q}{Q-P}\ln\left(\frac{A_1Q}{A_2P}\right)\,. \ee
Thus, from (\ref{e54}) we have \be\label{e57}
\tilde\alpha\approx\frac{P}{Q}+\frac{7\,(Q-P)^2}{2\,Q^2\,\ln\left(\frac
{A_1Q}{A_2P}\right)}\,. \ee Finally, using (\ref{e57}) along with
the definitions of $\tilde x$, $\tilde y$ and $\tilde z$ in
(\ref{xyzw1}) in terms of $\tilde\alpha$ we can solve for
$\phi_0^2$ from (\ref{e50}) and assuming that $Q-P\sim {\mathcal
O}(1)$, in the limit when $P$ is large we obtain \be\label{e58}
\phi_0^2\approx1-\frac 2{Q-P}+\sqrt{1-\frac 2{Q-P}}-\frac
7{P\,\ln\left(\frac{A_1\,Q}{A_2 P}\right)}\left(\frac 3
2+\sqrt{1-\frac 2{Q-P}}\right)\,. \ee We notice immediately that
since $\phi_0^2$ is real and positive it is necessary that
\be\label{e74} Q-P>2\,. \ee We will show shortly that the extremum
we found above corresponds to a metastable minimum. Also, for a
simple case with two moduli, via an explicit numerical check we
have confirmed that if $Q-P\leq 2$ the local minimum is completely
destabilized yielding a runaway potential. Also note that for
(\ref{e58}) to be accurate, it is not only $P$ which has to be
large but also the product $P\,\ln\left(\frac{A_1\,Q}{A_2
P}\right)$ has to stay large to keep the subleading terms
suppressed. To check the accuracy of the solution we again
consider a manifold with two moduli where the values of the
microscopic constants are:
\begin{equation} \label{setdS} a_1=a_2=7/6,\; P=20,\; Q=23,\; A_1=27,\; A_2=2,\;
N_1=N_2=1. \end{equation} The exact values obtained numerically
are:
\begin{equation}\label{exact23}
s_1\approx 33.470\,,\,\,\,s_2\approx 33.470\,,\,\,\phi_0\approx
0.810\,.
\end{equation}
The approximate equations above yield the following values:
\begin{equation}\label{approxim23}
s_1\approx 33.463\,,\,\,\,s_2\approx 33.463\,,\,\,\phi_0\approx
0.803\,.
\end{equation}
Note the high accuracy of the leading order approximation for the
moduli $s_i$.

It is now straightforward to compute the vacuum energy using the
approximate solution obtained above. First, we compute
\be\label{e63} K^{i\,\bar j}F_i{\bar F_{\bar
j}}-3|W|^2=4\left(A_2\tilde x\right)^2\left(\frac 7
9\left(L_{1,+}\right)^2+\frac 7 3L_{1,+}+1\right) \left(\frac
{A_1Q}{A_2P}\right)^{-\frac{2P}{Q-P}} \ee and \be\label{e62}
K^{\phi\bar{\phi}}F_{\phi}{\bar F_{\bar{\phi}}}=\left(A_2\tilde
x\phi_0\right)^2\left(\frac{a\tilde\alpha}{\phi_0^2\,\tilde
x}+1\right)^2 \left(\frac {A_1Q}{A_2P}\right)^{-\frac{2P}{Q-P}}
\ee

\begin{figure}[h!]
      \leavevmode \epsfxsize 12 cm \epsfbox{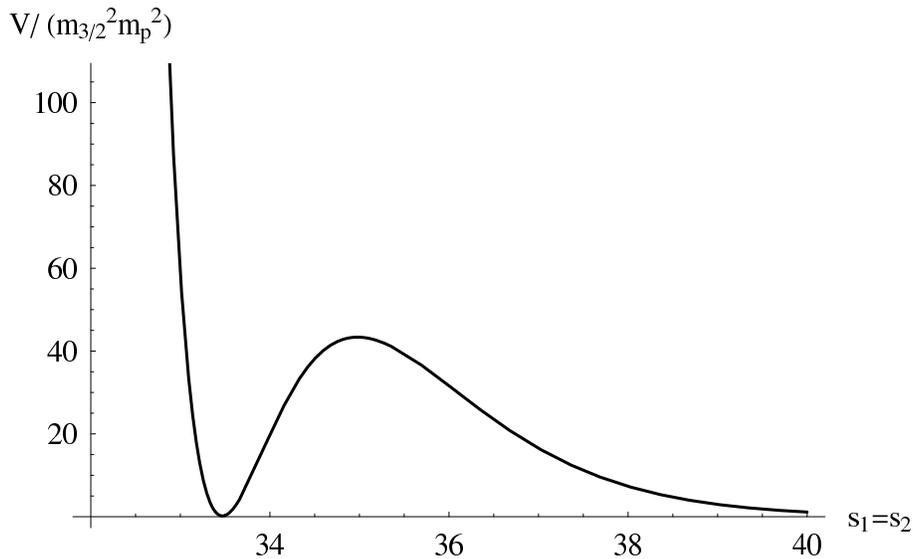}
    \caption{\footnotesize Potential in units of $m_{3/2}^2m_{p}^2$ along the slice $s_1=s_2$ for a manifold with two moduli with the
    meson field equal to its value at the minimum of the potential (\ref{e58}). The microscopic constants are as in (\ref{setdS}). Although
    hard to see from the graph, the value of the potential at the minimum (i.e. the cosmological constant) is
$0.194\,m_{3/2}^2m_p^2$.  }
    \label{dS-plot1}
\end{figure}

Using (\ref{e61}), (\ref{e63}) and (\ref{e62}) we obtain the
following expression for the potential at the extremum with
respect to the moduli $s_i$ as a function of $\phi_0$
\be\label{po45} V_0=\frac {(A_2\tilde x)^2}{64\pi
V_X^3}\left[\phi_0^4+ \left(\frac {2\,a\tilde\alpha}{\tilde
x}-3\right)\phi_0^2+\left(\frac {a\tilde\alpha}{\tilde
x}\right)^2\right]\frac {e^{\phi_0^2}}{\phi_0^2}\left(\frac
{A_1Q}{A_2P}\right)^{-\frac{2P}{Q-P}}\,, \ee where the terms
linear in $\tilde y$ cancelled and the quadratic terms were
dropped. A quick look at the structure of the potential
(\ref{po45}) as a function of $\phi_0^2$, where $\phi_0^2>0$, is
enough to see that there is a single extremum with respect to
$\phi_0^2$ which is indeed, a minimum. The polynomial in the
square brackets is quadratic with respect to $\phi_0^2$. Moreover,
the coefficient of the $\phi_0^4$ monomial is equal to unity and
therefore is always positive. This implies that for the minimum of
such a biquadratic polynomial to be positive, it is necessary for
the corresponding discriminant to be negative, which results in
the following condition:
\begin{equation}\label{e70}
3-4\frac {a\tilde\alpha}{\tilde x}<0\,.
\end{equation}
Again, since in the leading order $\tilde L_{1,+}=-3/2$, using the
definitions in (\ref{app24}), we can express $\tilde\alpha$ from
(\ref{sol37}) in terms of $\nu$ to get
\begin{equation}\label{eqab}
\frac{\tilde\alpha}{\tilde
x}=\frac{\tilde\alpha}{\tilde\alpha-1}=\frac{P}{P-Q}+\frac{3PQ}{4\pi\nu(P-Q)}\,.
\end{equation}
We then substitute $\nu$ from (\ref{app24}) into (\ref{eqab}) and
use it together with $a=-2/P$ we obtain from (\ref{e70}) the
following condition
\begin{equation}\label{e80}
3-\frac 8{Q-P}-\frac{28}{P\ln\left(\frac{A_1Q}{A_2P}\right)}<0\,.
\end{equation}
The above equation is the leading order requirement for the energy
density at the minimum to be positive. It is also clear that the
minimum is {\it metastable}, as in the decompactification limit
($V_X \rightarrow \infty$), the scalar potential vanishes from
above, leading to an absolute Minkowski minimum. Figure
\ref{dS-plot1} shows the scalar potential for a manifold with two
moduli along the slice $s_1=s_2$ with the meson field $\phi$ equal
to its value at the minimum of the potential (\ref{e58}). The
microscopic constants are the same as in (\ref{setdS}).


\subsection{The uniqueness of the dS vacuum}\label{unique}
In the previous subsection we found a particular solution of the
system in (\ref{e43}) corresponding to $A=1$. Here we would like
to investigate if solutions for $0\leq A<1$ are possible when the
vacuum for $A=1$ is de Sitter. Just like for the pure Super
Yang-Mills (SYM) case, we can recast (\ref{sol36}) as
\be\label{eq786} \nu^A_{k}=\frac{{\cal T}_A}{\tilde B_A}\,\tilde
L_{A,\,k}\,, \ee where the volume of the associative three cycle
$Q$ is again \be\label{def456} {\cal T}_A\equiv Vol(Q)_A=\vec
a\cdot\vec\nu^{A}=\frac {\tilde x}{\tilde y}\,\tilde B_{A}\,, \ee
and we have introduced \be\label{BAcbt} \tilde B_A\equiv\vec
a\cdot\vec{\tilde L}_{A}=\frac 7 3\left(\tilde T_A+A\tilde
H_A\right)\,. \ee Just like we did in equation (\ref{eq37}) for
the pure SYM case, we can also express $\tilde\alpha_A$ as \be
\label{eq370} \tilde\alpha_A=\frac{b_2{{{\cal T}_A}}-{\tilde
B_A}}{b_1{{{\cal T}_A}}-{\tilde B_A}}\,. \ee If we again consider
the large associative cycle volume limit and take $\tilde y
\rightarrow 0$ and $\tilde w\rightarrow -\infty$, the second and
third equations in (\ref{e43}) in the leading order reduce to
\ba \label{e202} &&2\left({{\tilde T}_{A}^2}+2 A{\tilde T_{A}} \tilde H_{A}+\tilde H_{A}^2\right)+ 3\left({\tilde T_{A}}+A \tilde H_{A}\right)=0 \\
&&9\left({\tilde T_{A}^2}+2 A {\tilde T_{A}} \tilde H_{A}+\tilde
H_{A}^2\right)-4\tilde H_{A}\left(\tilde H_{A}+A{\tilde T_{A}}
\right)+21\left({\tilde T_{A}}+A\tilde H_{A}\right)+9+\frac 9 4
\phi_0^2\left(\frac{a\tilde\alpha}{\phi_0^2\,\tilde
x}+1\right)^2=0\,. \nonumber \ea Note that the only difference
between (\ref{e20}) and (\ref{e202}) is the presence of the term
\be\label{exter} \delta\equiv\frac 9 4
\phi_0^2\left(\frac{a\tilde\alpha}{\phi_0^2\,\tilde
x}+1\right)^2\,, \ee which couples the system (\ref{e202}) to the
first equation in (\ref{e43}) which determines $\phi_0$. Instead
of solving the full system to determine $\tilde T_A$, $\tilde H_A$
and $\phi_0$ and analyzing the solutions we choose a quicker
strategy for our further analysis. Namely, we can solve the system
of two equations in (\ref{e202}) and regard $\delta$ as a
continuous deformation parameter. One may object to this
proposition because $\tilde\alpha$ and therefore $\tilde
x=\tilde\alpha-1$ are not independent of parameter $A$. However,
in the limit when ${\cal T}_A$ is large, we notice from
(\ref{eq370}) that in the leading order, $\tilde\alpha_A$ is
indeed {\em independent} of $A$.

Recall that in the pure SYM case the system (\ref{e20})
corresponding to the case when $\delta=0$ has two {\em real}
solutions for all $0\leq A\leq 1$. Thus, one may expect that as we
continuously dial $\delta$, the system may still yield real
solutions for $A<1$. Let us first determine the range of possible
values of parameter $\delta$. A quick calculation yields that the
combination in (\ref{exter}) is the smallest with respect to
$\phi_0$ when $\phi_0^2=\frac{a\tilde\alpha}{\tilde x}$. In this
case \be\label{e97} \delta=9 \frac{a\tilde\alpha}{\tilde x}\,. \ee
Now, recall from the previous subsection that for the solution
corresponding to $A=1$ to have a positive vacuum energy, condition
(\ref{e70}) must hold. Since $\tilde\alpha$ and $\tilde x$ are
independent of $A$ in the leading order, condition (\ref{e70})
implies that \be\label{e970} \delta>\frac {27}4. \ee Again, since
the volume ${\cal T}_A$ is always positive, from (\ref{eq786}) we
see that for all moduli to be stabilized in the positive range,
all three quantities $\tilde L_{A,\,+}$, $\tilde L_{A,\,-}$ and
$\tilde B_A$ must have the same sign. For $\delta=27/4$, the
system (\ref{e202}) has two real solutions when $0.877781<A<1$.
\begin{figure}[h!]
    \begin{tabular}{cc}
      \leavevmode \epsfxsize 8 cm \epsfbox{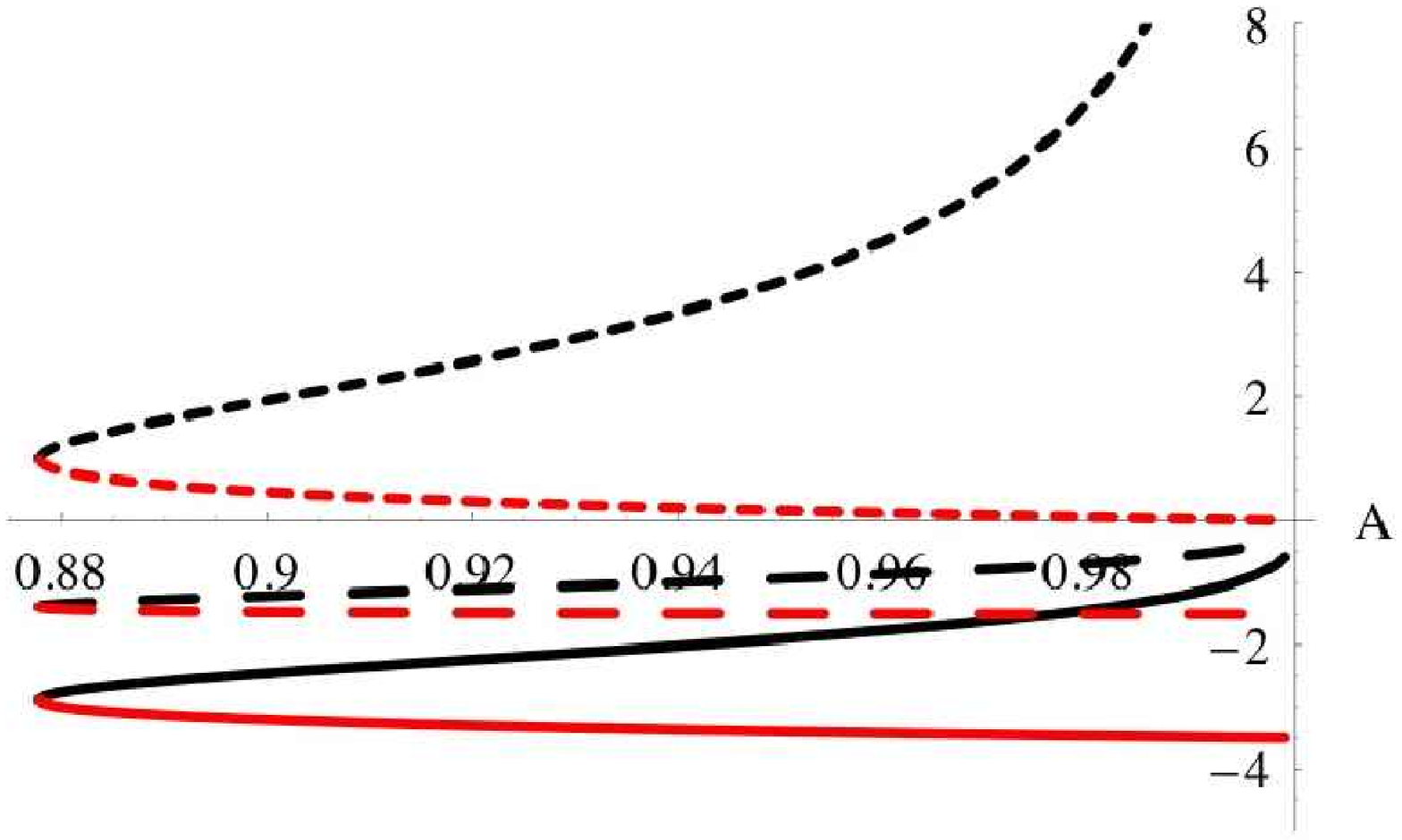}&
      \leavevmode \epsfxsize 8 cm \epsfbox{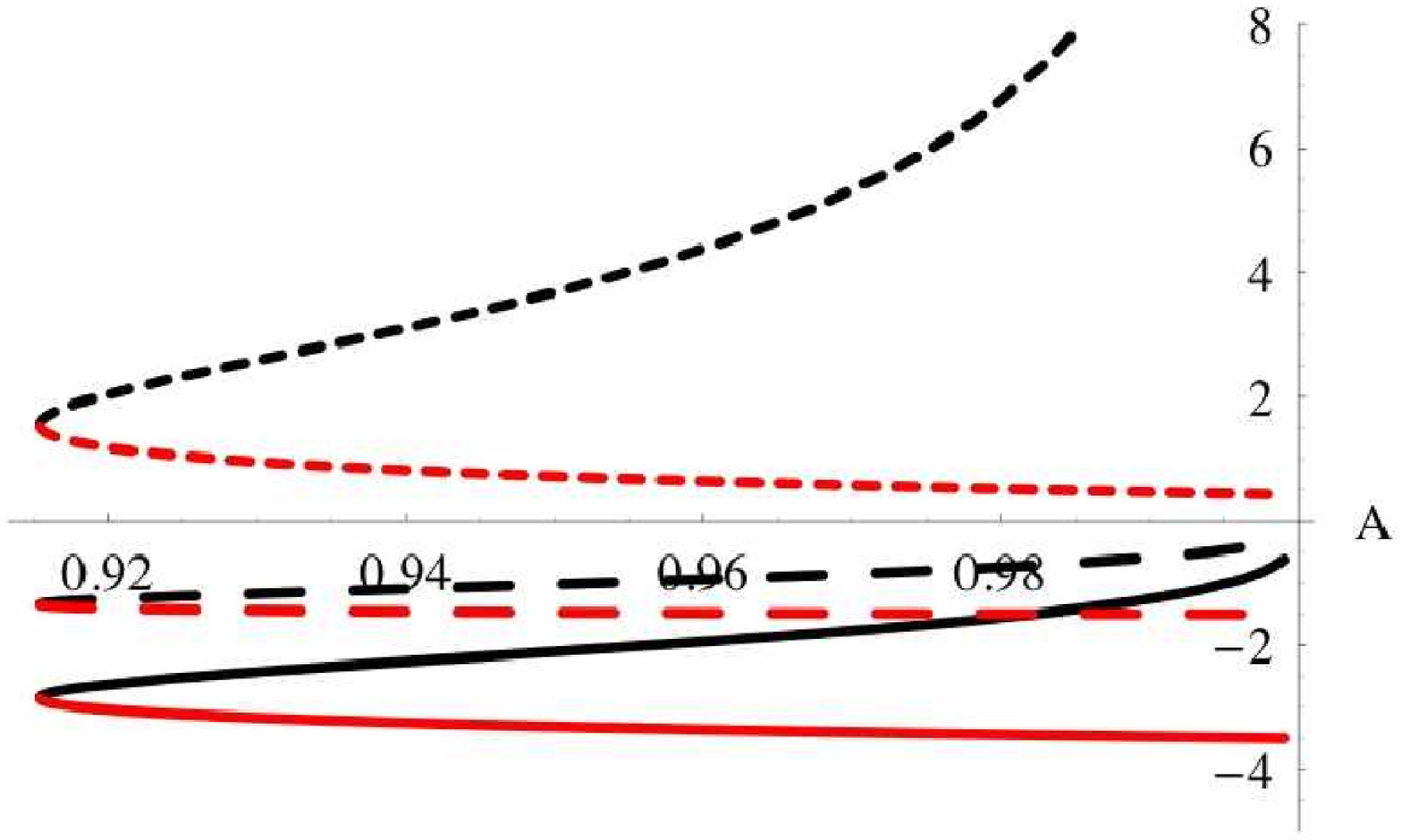}\\
    \end{tabular}
    \caption{\footnotesize Plots of $\tilde L_{A,\,+}^{(c)}$,  $\tilde L_{A,\,-}^{(c)}$ and ${\tilde B_A^{(c)}}$, where $c=\overline{1,2}$, corresponding to the two real solutions of the system (\ref{e202})  as functions of parameter $A$. $\tilde L_{A,\,+
}^{(c)}$ - long dashed curve, $\tilde L_{A,\,-}^{(c)}$ -short
dashed curve, ${\tilde B_A^{(c)}}$ - solid curve. Black color:
$\tilde L_{A,\,+}^{(1)}$,  $\tilde L_{A,\,-}^{(1)}$ and ${\tilde
B_A^{(1)}}$ corresponding to the first real solution. Red color:
$\tilde L_{A,\,+}^{(2)}$,  $\tilde L_{A,\,-}^{(2)}$ and ${\tilde
B_A^{(2)}}$ corresponding to the second real solution. Left plot:
when $\delta=27/4$ the real solutions exist in the range
$0.877781< A<1$. Right plot: when $\delta=8$ the real solutions
exist st in the range $0.915342< A<1$.}
    \label{Plot240}
\end{figure}
However, from the left plot in Figure {\ref{Plot240}}
corresponding to the minimum value $\delta=27/4$ we see that
neither of the two solutions satisfy the above requirement since
both short-dashed curves corresponding to $\tilde L_{A,-}$ for the
two solutions are {\em always} positive for the entire range
$0.877781<A<1$, whereas both $\tilde L_{A,\,+}$ and $\tilde B_A$
remain negative. Therefore for $\delta=27/4$ and $A<1$ there are
no solutions for which all the moduli are stabilized at positive
values. Moreover, as parameter $\delta$ is further increased, the
range of possible values of $A$ for which the system has two real
solutions gets smaller and more importantly, the values of $\tilde
L_{A,-}$ remain positive and only increase while both $\tilde
L_{A,\,+}$ and $\tilde B_A$ remain negative, which can be seen
from the right plot in Figure {\ref{Plot240}}, where $\delta=8$.
This trend continues as we increase $\delta$.

Thus, we can make the following general claim: If the solution for
$A=1$ has a positive vacuum energy, condition (\ref{e970}) must
hold. When  this condition is satisfied the system (\ref{e202})
has no solutions in the range $0\leq A<1$ for which all the moduli
are stabilized at positive values. Therefore, if the vacuum found
for $A=1$ is de Sitter it is the only possible vacuum where all
the moduli are stabilized at positive values. Although the above
analysis was done in the limit when ${\cal T}_A$ is large, we have
run a number of explicit numerical checks for a manifold with two
moduli and various values of the constants confirming the
above claim. In addition, although we have not proved it, it seems
plausible from many numerical checks we carried out that it is
also not possible to have a metastable dS minimum for values of
$A$ different from unity, even if the dS condition on the $A=1$
vacuum is {\it not} imposed.

Finally, it should be noted that the situation with a ``unique" dS
vacuum is in sharp contrast to that when one obtains anti-de
Sitter vacua, where there are between $2^{N-1}$ and $2^N$
solutions for $N$ moduli depending on the value of $A$ (see
section \ref{nonsusyadsvac}). Let us explain this in a bit more
detail. Since the dS solution found for $A=1$ is located right in
the vicinity of the ``would be AdS SUSY extremum''\footnote{This
can be seen by comparing the leading order expression for the
moduli vevs in the dS case (\ref{app24}) with the corresponding
formula for the SUSY AdS extremum (\ref{alpha4}).} where the
moduli $F$-terms are nearly zero, it is the large contribution
from the matter $F$-term (\ref{e62}) which cancels the $-3|W|^2$
term in the scalar potential resulting in a positive vacuum
energy. Recall that in the leading order all the AdS vacua with
the moduli vevs $s_{A,i}^{(c)}$ are located within the hyperplane
\footnote{c=1,2 labels the two real solutions of the system
(\ref{e20}).} \be\label{hyp} \sum_{i=1}^{N}s_{A,i}^{(c)}N_i =\frac
1{2\pi}\frac{P\,Q}{P-Q}{\ln}\left(\frac{A_2P}{A_1Q}\right)=constant\,.
\ee The matter $F$-term contribution to the scalar potential
$K^{\phi\bar{\phi}}F_{\phi}{\bar F_{\bar{\phi}}}$ evaluated  at
the same $s_{A,i}^{(c)}$ but arbitrary $\phi_0$ is therefore also
{\em constant} along the hyperplane (\ref{hyp}). Thus, while the
matter $F$-term contribution stays constant, as we move along the
hyperplane (\ref{hyp}) away from the dS minimum, where the moduli
$F$-terms are the smallest, the moduli $F$-term contributions can
only get larger so that the scalar potential becomes even more
positive. This implies that the AdS minima with broken SUSY found
in Section \ref{nonsusyadsvac} completely disappear, as the AdS
SUSY extremum becomes a dS minimum.


\section{Relevant Scales}\label{distribution}

We have demonstrated above that in fluxless $M$ theory vacua,
strong gauge dynamics can generate a potential which stabilizes
all the moduli. Since the entire potential is generated by this
dynamics, and the strong coupling scale is below the Planck scale,
we also have a hierarchy of scales. In this section we calculate
some of the basic scales in detail. In particular, the gravitino
mass, which typically controls the scale of supersymmetry breaking
is calculated. By uniformly scanning over the constants $( N, P, Q
, A_k)$ with $N_i$ order one, we demonstrate in
\ref{samplingsection} that a reasonable fraction of choices of
constants have a TeV scale gravitino mass. We do not know if the
space of $G_2$-manifolds uniformly scans the $( P, Q , A_k , N_i
)$ or not, and more importantly, the scale of variation of the
$A_k$'s in the space of manifolds is not clear. The variation of
the $A_k$'s is the most important issue here, since one can
certainly vary $P$ and $Q$ over an order of magnitude. We begin
with a discussion of the basic scales in the problem. We will
begin with the AdS vacua, then go on to discuss the de Sitter
case. In particular, in the dS case, requiring a small vacuum
energy seems to lead to superpartners at around the TeV scale. It
will also be shown that including more than one flavor of quarks
in the hidden sector or including matter in both hidden sectors
does not change this result. The section will end with an
estimation of the height of the potential barrier in these vacua.

\subsection{Scales: AdS Vacua}

As an example, we consider one of the non-SUSY minima in our toy
model given by (\ref{sol9}) and compute some of the quantities
relevant for phenomenology.
Namely, the vacuum energy
\be
\label{lambda} \Lambda_0=-(5.1\times 10^{10}\,\mathrm{GeV})^4\,,
\ee
the gravitino mass
\be \label{mgrav}
M_{3/2}=m_pe^{K/2}|W| \approx2.081\,\mathrm{TeV}\,,
\ee
the 11-dimensional Planck scale
\be \label{m11} M_{11}=\frac{\sqrt{\pi}m_p}
{V_X^{1/2}}\approx3.9\times 10^{17}\, \mathrm{GeV}\,,
\ee
the scale of gaugino condensation in the
hidden sectors
\ba
\Lambda^{(1)}_{g} &=& m_p\,e^{-\frac{b_1}{3}\Sigma_iN_is^i}\approx2.6\times 10^{15}\, \mathrm{GeV}\\
\Lambda^{(2)}_{g} &\approx& 9.7\times10^{14}\,\mathrm{GeV}
\ea

where $m_p=(8\pi G_N)^{-1/2}=2.43\times 10^{18}$ GeV is the reduced
four-dimensional Planck mass.

\noindent From (\ref{mgrav}) and (\ref{m11}), we see that it is
possible to have a TeV scale gravitino mass together with $M_{11}
\geq M_{unif} (2\times 10^{16}$GeV). This feature survives in more
general cases as well, {\it implying that standard gauge
unification is compatible with low scale SUSY in these vacua.}

\subsection{Gravitino mass}
By definition, the gravitino mass is given by: \be\label{gr1}
m_{3/2}=m_p\,e^{K/2}|W|\,. \ee For the particular $M$ theory
vacua with K\"{a}hler potential given by (\ref{kahler}) and
the non-perturbative superpotential as in (\ref{super}) with
$SU(P)$ and $SU(Q)$ hidden sector gauge groups we have:
\be\label{gr2} m_{3/2}=\,\frac
{m_p}{8\sqrt{\pi}{V_X}^{3/2}}\left|A_1e^{-{\frac{2\pi}{P}{\rm
Im}f}}-A_2e^{-{\frac{2\pi}{Q}{\rm Im}f}}\right|\,, \ee where the
relative minus sign inside the superpotential is due to the
axions. Before we get to the gravitino mass we first compute the
volume of the compactified manifold $V_X$ for the AdS vacua with
broken SUSY. By plugging the approximate leading order solution
for the moduli (\ref{eq43}) into the definition (\ref{vol}) of
$V_X$ we obtain:
\be\label{volume} (V_X)^{(c)}_A=\left[\frac
1{2\pi}\frac{PQ}{P-Q} {\rm
ln}\left(\frac{A_2P}{A_1Q}\right)\right]^{7/3}\prod_{i=1}^{N}
\left(\frac{a_i\,L_{A,\,i}^{(c)}}{N_i\,B_A^{(c)}}\right)^{a_i}\,.
\ee
Recalling the definition (\ref{deft}) of ${\cal T}^{\,(c)}_A$
and using (\ref{eq39}) together with (\ref{volume}) to plug into
(\ref{gr2}) the gravitino mass for these vacua in the leading
order approximation is given by:
\be\label{gr3}
(m_{3/2})^{(c)}_A=\sqrt{2}\pi^3\,A_2P
\left|\frac{P-Q}{PQ}\right|\left[\frac{PQ}{P-Q}{\rm
ln}\left(\frac{A_2P}{A_1Q}\right) \right]^{-\frac
72}\left[\frac{A_2P}{A_1Q}\right]^{-\frac{P}{P-Q}}\prod_{i=1}^{N}
\left(\frac{N_i\,B_A^{(c)}}{a_i\,L_{A,\,i}^{(c)}}\right)^{\frac{3a_i}2}\,.
\ee For the special case with two moduli when $a_1=a_2=7/6$,
considered in the previous sections we obtain the following:
\begin{eqnarray}\label{gr4}
(m_{3/2})^{(1,2)}_0&=&m_p\,2^{1/2}\pi^3\left(7+\sqrt{17}\right)^{\frac 7 4}\left(N_1\,N_2\right)^{\frac 7 4}\,A_2\,P\left|\frac{P-Q}{P\,Q}\right|\left(\frac{A_2\,P}{A_1\,Q}\right)^{-\frac P{P-Q}}\left(\frac{PQ}{P-Q}{\rm ln}\frac{A_2\,P}{A_1\,Q}\right)^{-\frac 7 2}\nonumber\\
&\sim&m_p\,2.97\times 10^3\left(N_1\,N_2\right)^{\frac 7 4}\,A_2\,P\left|\frac{P-Q}{P\,Q}\right|\left(\frac{A_2\,P}{A_1\,Q}\right)^{-\frac P{P-Q}}\left(\frac{PQ}{P-Q}{\rm ln}\frac{A_2\,P}{A_1\,Q}\right)^{-\frac 7 2}
\end{eqnarray}
For the choice of constants as in (\ref{choice1}) the leading order approximation (\ref{gr4}) yields:
\be
(m_{3/2})^{(1,2)}_0=2061 {\rm GeV}\,,
\ee
whereas the exact value computed numerically for the same choice of constants is:
\be
m_{3/2}=2081 {\rm GeV}\,.
\ee
Again, we see a good agreement between the leading order approximation and the
exact values.


\subsection{Scanning the Gravitino mass}\label{samplingsection}

In previous sections we found explicit solutions describing vacua
with spontaneously broken supersymmetry. Moreover, we also
demonstrated that for a particular set of the constants these
solutions can result in $m_{3/2}\sim O(1)\,{\rm TeV}$. It
would be extremely interesting and worthwhile to estimate (even
roughly) the fraction of all possible solutions which exhibit
spontaneously broken SUSY at the scales of $O(1)$-$O(10)$ TeV. We would first like to
do this for generic AdS/dS vacua with a large magnitude of the cosmological constant ($\sim
m^2_{3/2}m^2_{p}$). The analysis for the AdS vacua is given below but as we will see,
the results obtained for the fraction of vacua are quite similar
for the dS case as well. In subsection \ref{CClowsusy}, we impose the requirement of a
small cosmological constant as a constraint and try to understand its
repercussions for the gravitino mass.

We do not yet know the range that the constants $(N, P , Q , A_1, A_2 )$ take in the space
of all $G_2$ manifolds. Nevertheless, we do have a
rough idea about some of them. For example, we expect that the
quantity given by the ratio \be\label{rat}
\rho\equiv\frac{A_2\,P}{A_1\,Q}\,, \ee which appears in several
equations, does deviate from unity. One reason for this may be due
the threshold corrections {\cite{Friedmann:2002ty}} which in turn depend
on the properties of a particular $G_2$-holonomy manifold. For concreteness, we
take an upper bound $\rho \leq 10$. Also, based on the duality
with the Heterotic String we can get some idea on the possible
range of integers $P$ and $Q$ corresponding to the dual coxeter
numbers of the hidden sector gauge groups. Namely, since for both
$SO(32)$ and $E_8$ gauge groups appearing in the Heterotic String
theories the dual coxeter numbers are $h^v=30$, we can
tentatively assume that both $P$ and $Q$ can be at least as
large as $30$. Of course, we do not rule out any values higher
than $30$ but in this section we will assume an upper bound
$P,Q\leq 30$.

We now turn our attention to equation (\ref{gr3}) which will
be used to estimate the gravitino mass scale. It is clear from
the structure of the formula that $m_{3/2}$ is extremely sensitive
to $P$, $Q$ as well as the ratio $\rho$, given by (\ref{rat}).
On the other hand it is less sensitive to the other constants
appearing in the equation such as $N_i$, $a_i$ and the ratios
$B^{(c)}_A/L^{(c)}_{A,i}$. This is because the powers $3a_i/2$
for each term under the product get much less than one as the number
of moduli increases because of the constraint
on $a_i$ in (\ref{vol}). This will smooth any differences between the
contributions coming from the individual factors inside the product.
Since for $0\leq A\leq 1$ ($A$ is defined in (\ref{sol1})), the ratios $B^{(c)}_A/L^{(c)}_{A,i}$
vary only in the range $O(1)$-$O(10)$, for our purposes it will
be sufficient to simply consider (\ref{gr3}) for the case
when $A=1$ corresponding to the SUSY extremum
so that ${B^{(1)}_1}/L^{(1)}_{1,i}=7/3$ for all $i$. This is certainly
good enough for the order of magnitude estimates we are interested in.
It also seems reasonable to assume that the integers $N_i$ are all
of $O(1)$. Yet, even if some $N_i$ are unnaturally large, their
individual contributions are generically washed out since they are
raised to the powers that are much less than one. Thus, for simplicity
we will take $N_i=1$ for all $i=\overline{1,N}$.
Finally, from field theory computations \cite{Finnell:1995dr}, $A_2 = Q$ (in
a particular RG scheme) up to threshold corrections. We therefore take
$A_2 \sim Q$ for simplicity, allowing $A_1$ to vary.

Thus, the gravitino mass in our analysis is given by
\be\label{gr7}
m_{3/2}\sim\sqrt{2}\pi^3\,PQ
\left(\frac{P-Q}{PQ}\right)^{\frac 92}\left[{\rm ln}\rho
\right]^{-\frac 72}(\rho)^{-\frac{P}{P-Q}}\prod_{i=1}^{N}
\left(\frac 7{3 a_i}\right)^{\frac{3a_i}2}\,.
\ee
Finally, with regard to the constants $a_i$ which are a subject to
the constraint
\be\label{con2}
\sum_{i=1}^{N}a_i=\frac 7 3\,,
\ee
we will narrow our analysis to two opposite cases. For the first case
we make the following choice
\be
{\rm 1)}\,\,\,\,\,\,\,\,\,a_1=2\,,\,\,{\rm and}\,\,\,
a_i=\frac 1 {3(N-1)},\,\,{\rm for}\,\,\,i=\overline{2,N}\,,
\ee
such that one modulus is generically large and all the other moduli
are much smaller. This is a highly anisotropic $G_2$-manifold.
The second case is
\be
{\rm 2)}\,\,\,\,\,\,\,\,a_i=\frac7{3\,N}\,,\,\,\,{\rm for\,all}\,\,\,i=\overline{1,N}\,,\,\,\,\,\,\,\,\,\,\,\,\,\,\,\,\,\,\,\,\,\,\,\,\,\,\,\,\,\,\,\,\,\,\,\,\,\,\,\,\,\,\,
\ee
with all the moduli being on an equal footing.
Therefore, by considering these opposite cases we expect
that most other possible sets of $a_i$ will give similar
results that are somewhere in between.
For each set of $a_i$ above, equation (\ref{gr7}) gives
\be\label{gr8}
{\rm 1)}\,\,\,\,\,\,\,\,\,m_{3/2}^{(1)}\sim\frac{343\sqrt{14}\pi^3\,}{216}\,PQ
\left(\frac{P-Q}{PQ}\right)^{\frac 92}\left[{\rm ln}\rho
\right]^{-\frac 72}(\rho)^{-\frac{P}{P-Q}}
\left(N-1\right)^{\frac 12}\,,
\ee
\be\label{gr9}
{\rm 2)}\,\,\,\,\,\,\,\,\,m_{3/2}^{(2)}\sim\sqrt{2}\pi^3\,PQ
\left(\frac{P-Q}{PQ}\right)^{\frac 92}\left[{\rm ln}\rho
\right]^{-\frac 72}(\rho)^{-\frac{P}{P-Q}}
\left(N\right)^{\frac 72}\,.\,\,\,\,\,\,\,\,\,\,\,\,\,\,\,\,\,\,\,\,\,\,\,\,\,
\ee
For a typical compactification we expect $N\sim O(100)$, therefore
the variation of $m_{3/2}$ due to an $O(1)$ change in the number
of moduli for the first case is $O(1)$ whereas in the
second case it can be as large as $O(10)$. Thus, if we choose
$N=100$, we expect that our order of magnitude analysis will be fairly
robust for case $1)$. For case $2)$, however, we will perform
the same analysis for $N=100$ and $N=50$ to see how different
the results will be.
Before we proceed further we need to impose a restriction on the
possible solutions to remain within the SUGRA framework. Using
(\ref{volume}), condition that $V_X$ must
remain greater than one for the two cases under consideration
translates into the following two conditions:
\be\label{vol8}
{\rm 1)}\,\,\,\,\,\,\,\,\,\frac 37\left(\frac {64}{3(N-1)}\right)^{\frac 1 7}\frac 1{2\pi}\frac{P\,Q}{P-Q}{\rm ln}\,\rho\,>\,1\,,
\ee
\be\label{vol9}
{\rm 2)}\,\,\,\,\,\,\,\,\,\frac 1{2\pi\,N}\frac{P\,Q}{P-Q}{\rm ln}\,\rho\,>\,1\,.\,\,\,\,\,\,\,\,\,\,\,\,\,\,\,\,\,\,\,\,\,\,\,\,\,\,\,\,\,\,\,\,\,\,\,\,
\ee
Then, as long as conditions (\ref{vol8}-\ref{vol9}) hold,
volume of the associative cycle $\cal T$ is greater than one is satisfied
automatically - a necessary condition for the validity of supergravity.
This is obvious from comparing the right hand side of (\ref{eq39}) with
each condition above. From (\ref{vol8}-\ref{vol9}) we can find a critical value of
$\rho=\rho_{crit}$ for both cases at which $V_X=1$:
\be\label{rho8}
{\rm 1)}\,\,\,\,\,\,\,\,\,\rho_{crit}^{(1)}\equiv{\rm Exp}\left[\frac {14\pi}3\left(\frac {3(N-1)}{64}\right)^{\frac 1 7}\frac{P-Q}{P\,Q}\right]\,,
\ee
\be\label{rho9}
{\rm 2)}\,\,\,\,\,\,\,\,\,\rho_{crit}^{(2)}\equiv{\rm Exp}\left[{2\pi N}\left(\frac{P-Q}{P\,Q}\right)\right]\,.\,\,\,\,\,\,\,\,\,\,\,\,\,\,\,\,\,\,\,\,\,\,\,\,\,\,\,\,
\ee
By substituting (\ref{rho8}-\ref{rho9}) into (\ref{gr8}-\ref{gr9}) we can find the corresponding
upper limits on $m_{3/2}$ as functions of $P$ and $Q$, below which our
solutions are going to be consistent with the SUGRA approximation.

\begin{figure}[h!]
\begin{tabular}{cc}
      \leavevmode \epsfxsize 9 cm \epsfbox{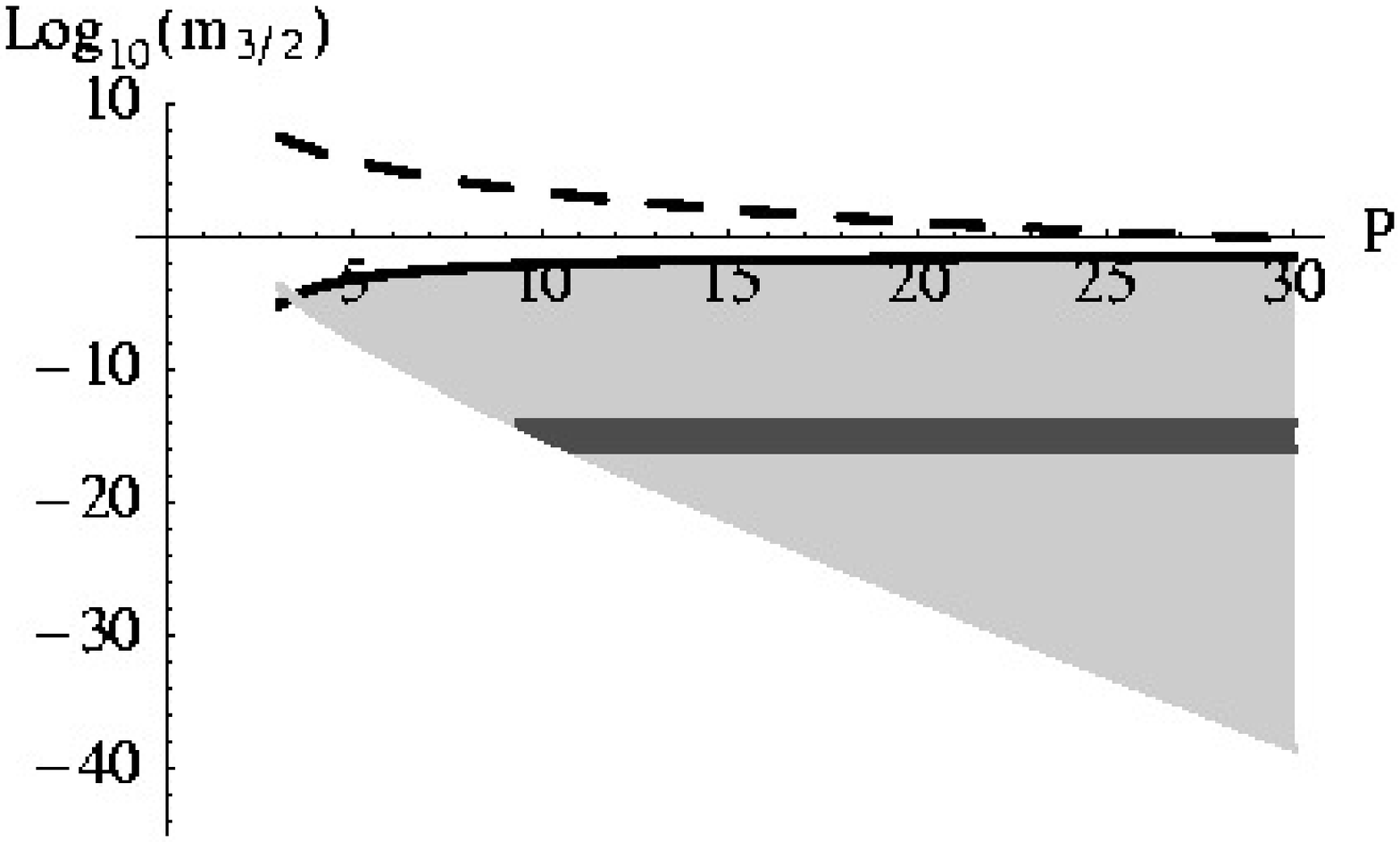}&
      \leavevmode \epsfxsize 9 cm \epsfbox{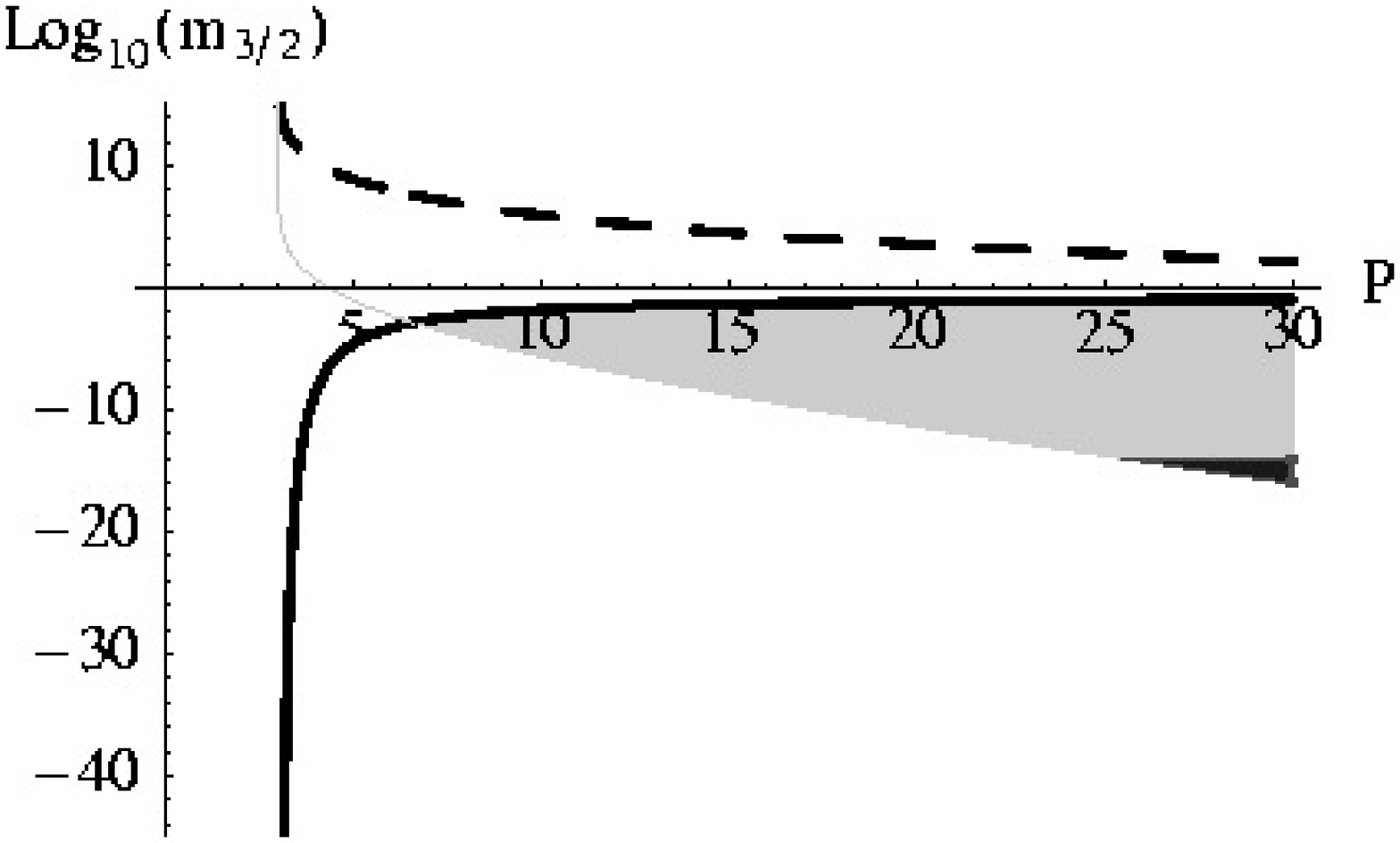} \\
      \leavevmode \epsfxsize 9 cm \epsfbox{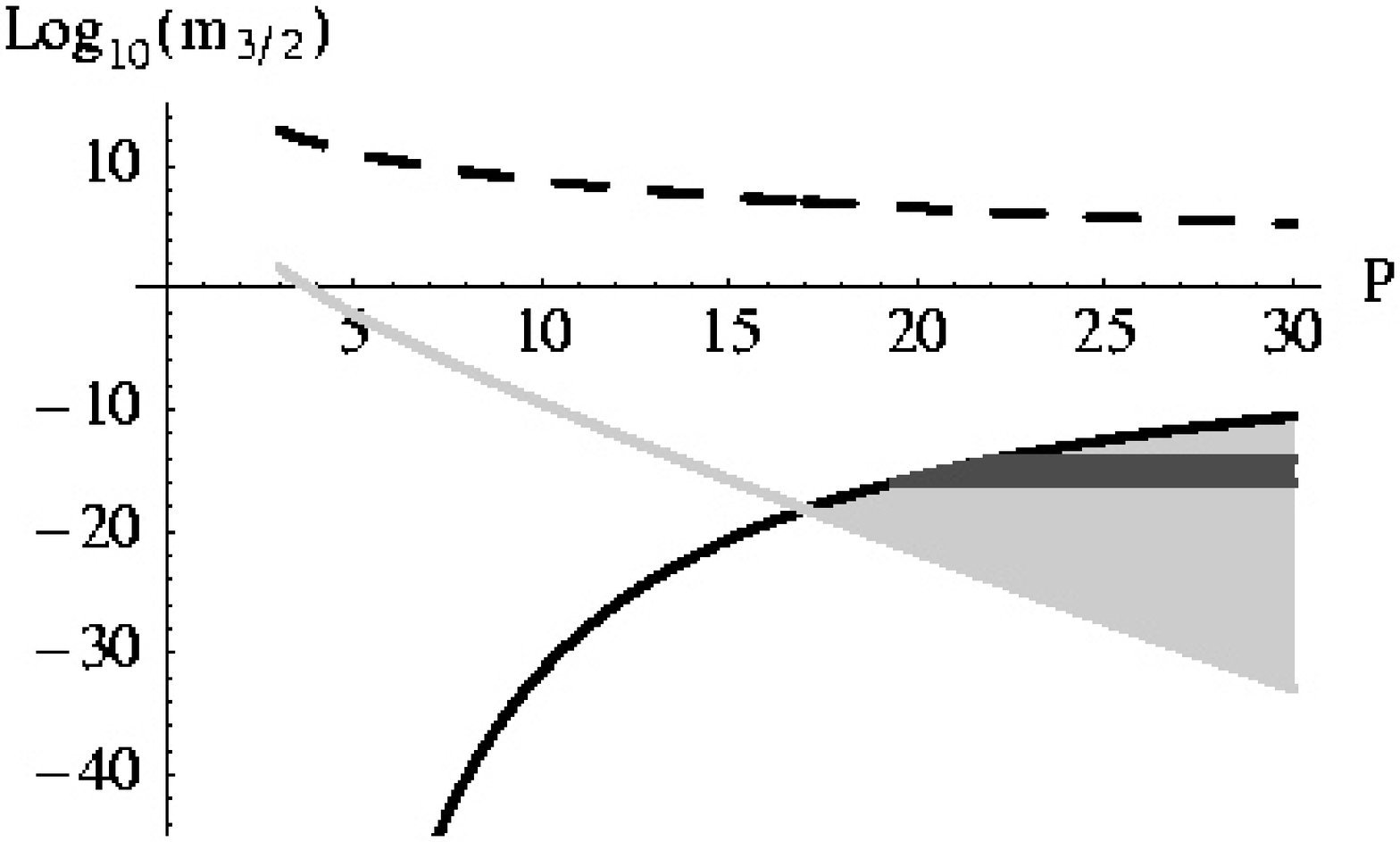}&
      \leavevmode \epsfxsize 9 cm \epsfbox{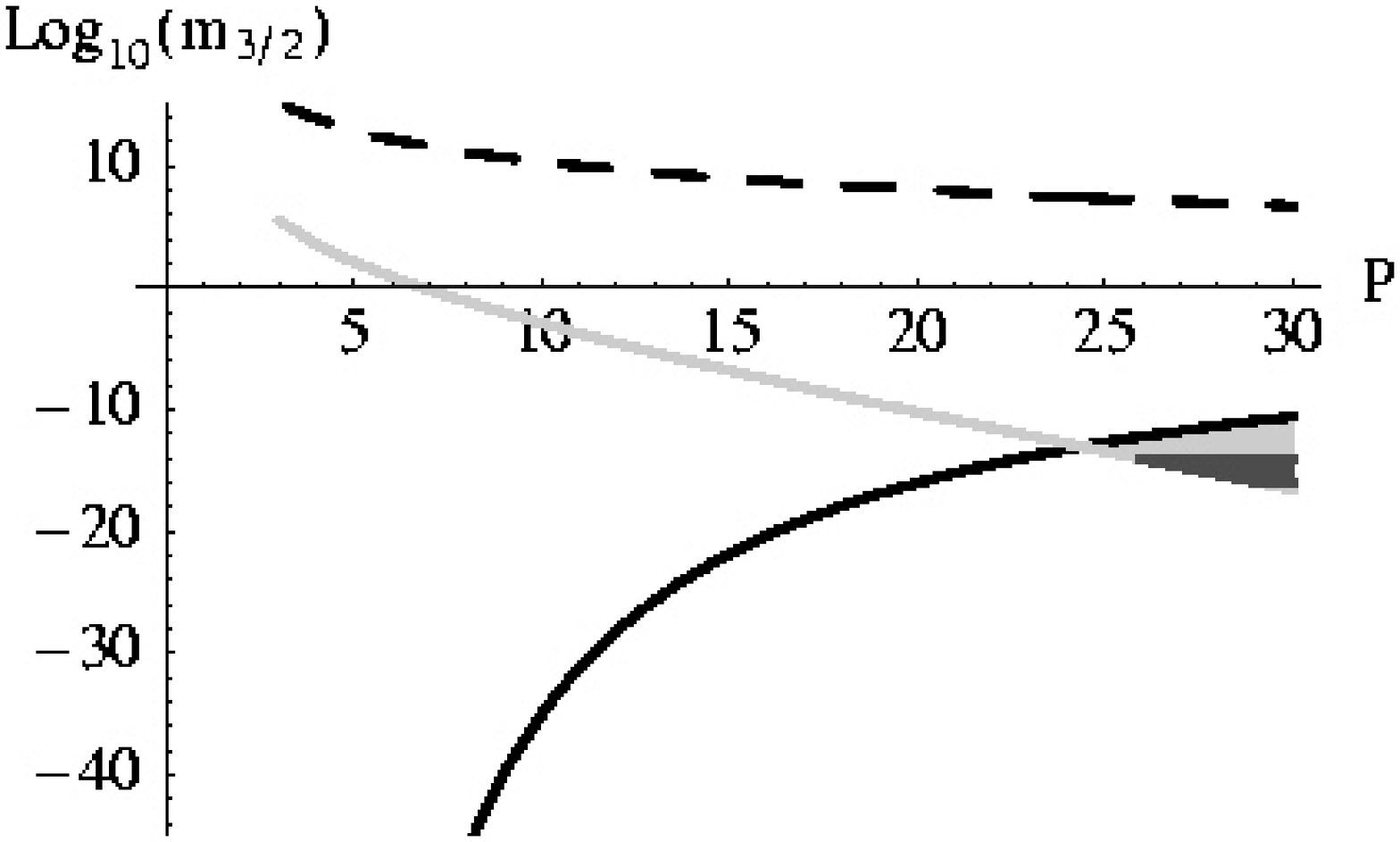} \\
    \end{tabular}
\caption{\footnotesize ${\rm log}_{10}(m_{3/2})$ as a function of $P$ for case $1)$ - top plots and
case $2)$ - bottom plots.
The light grey area represents possible values of ${\rm log}_{10}(m_{3/2})$
in the range where $\rho_{crit}\leq\rho\leq 10$ consistent
with the SUGRA approximation. The dark area indicates the
region of interest where $-16\leq {\rm log}_{10}(m_{3/2})\leq -14$ such that
$240\,{\rm GeV}\leq m_{3/2}\leq 24\,{\rm TeV}$.
The dark solid curve corresponds to ${\rm log}_{10}(m_{3/2})$ when $\rho=\rho_{crit}$.
The lower boundary of the light grey area represents the ${\rm log}_{10}(m_{3/2})$ curve when $\rho=10$.
The dashed curve corresponds to $\rho=1.01$. Top left: Case $1)$ when $P-Q=1$. Top right: Case $1)$ when $P-Q=3$. Bottom left:
Case $2)$ when $P-Q=1$. Bottom right: Case $2)$ when $P-Q=2$.}
\label{scans1}
\end{figure}

In Figure \ref{scans1} we present plots of ${\rm
log}_{10}(m_{3/2})$ for both cases as a function of $P$ in the
range where ${\rho}_{crit}\leq \rho\leq 10$ for different values
of $P-Q$.
On all the plots the light grey area represents possible values of
${\rm log}_{10}(m_{3/2})$ consistent with the supergravity
framework. For the sake of completeness we have also
included the formal plot of ${\rm log}_{10}(m_{3/2})$ corresponding
to  $\rho=1.01$ represented by the dashed curve.
>From the plots it is clear that as the difference $P-Q$ is
increased from $1$ to $3$ - top and from $1$ to $2$ - bottom, both
the light grey area representing all possible values of ${\rm
log}_{10}(m_{3/2})$ consistent with the SUGRA approximation and
the dark area corresponding to $-16\leq {\rm
log}_{10}(m_{3/2})\leq -14$ get significantly smaller. If we
further increase $P-Q$, the light grey region shrinks even more
for case $1)$, and does not exist for case $2)$, while the dark
region completely disappears in both cases. Therefore, for case
$2)$ the plots on the bottom of Figure \ref{scans1} are the only
possibilities where solutions for $P\leq30$ and $N=100$ consistent
with the SUGRA approximation are possible, implying un upper bound
$(P-Q)_{max}=2$. It turns out that for case $1)$ the upper bound
on $(P-Q)$ where such solutions are possible is much higher
$(P-Q)_{max}=23$.

Assuming that all values of the constants such as $P$, $Q$ and
$\rho$ are equally likely to occur in the ranges chosen above we
can perform a crude estimate of the number of solutions with
$-16\leq {\rm log}_{10}(m_{3/2})\leq -14$ relative to the total
number of possible solutions consistent with the SUGRA approximation.
In doing so we will use the following
approach. For each value of $(P-Q)$ in the range
$1\leq (P-Q)\leq (P-Q)_{max}$ we
compute the area of the grey region for each plot and then add
all of them to find the total volume corresponding to all possible values of
${\rm log}_{10}(m_{3/2})$ consistent with the supergravity approximation.
\be
\Omega_{\,tot}=\sum_{\,\,\,\,\,\,\,\,(P-Q)\,=1}^{\,\,\,\,\,\,\,\,\,\,\,\,(P-Q)_{max}}\int_{P_{min}}^{30}dP\,{\rm log}_{10}(m_{3/2})_{|_{\{\rho_{crit}\leq\rho\leq10\}}}\,.
\ee
Likewise, we add all the dark areas for each plot to find the volume corresponding to the region where $-16\leq {\rm log}_{10}(m_{3/2})\leq -14$
\be
\Omega_{\,0}=\sum_{\,\,\,\,\,\,\,\,(P-Q)\,=1}^{\,\,\,\,\,\,\,\,\,\,\,\,(P-Q)^{*}_{max}}\int\,dP\,{\rm log}_{10}(m_{3/2})_{|_{{\left\{{\rho_{crit}\leq\rho\leq10}\atop{-16\leq {\rm log}_{10}(m_{3/2})\leq -14}\right\}}}}\,,
\ee
where $(P-Q)^{*}_{max}$ is un upper bound before the dark region
completely disappears. From the previous discussion, for case $1)$: $(P-Q)^{*}_{max}=3$; for case $2)$: $(P-Q)^{*}_{max}=2$ .
Then, the fraction of the volume where $240\,{\rm GeV}\leq m_{3/2}\leq 24\,{\rm TeV}$ is given by the ratio
\be
\Delta=\frac {\Omega_{\,0}}{\Omega_{\,tot}}\,.
\ee
Numerical computations yield the following values for the two cases when
$N=100$:
\be\label{delta1}
{\rm 1)}\,\,\,\,\,\,\,\,\,\Delta_1=3.5\%\,,
\ee
\be\label{delta2}
{\rm 2)}\,\,\,\,\,\,\,\,\,\Delta_2=13.6\%\,.
\ee
Because of the significant difference in the dependence of
$\rho_{crit}$ on the number of moduli $N$ in (\ref{rho8})
versus (\ref{rho9}), the number of solutions consistent
with the SUGRA approximation is cut down dramatically in
case $2)$ compared to case $1)$.
This also occurs because of the different dependence of $m_{3/2}$
in (\ref{gr8}) and (\ref{gr9}) on $N$.
Namely, for $N\sim O(100)$, the values of $m_{3/2}$ for case
$2)$ in (\ref{gr9}) are $\sim O(10^6)$ greater than those for case $1)$.
Furthermore, for the same reasons, it turns out that if we keep increasing
the number of moduli $N$, for case $2)$ there is un upper bound $N\leq 157$ for the solutions with
$240\,{\rm GeV}\leq m_{3/2}\leq 24\,{\rm TeV}$ compatible with the SUGRA approximation to exist at all.
Of course, this upper bound can be higher if we allow $P$ to
be greater than $30$.

By performing the same analysis for $N=50$ we get the following
estimates \be\label{delta3} {\rm
1)}\,\,\,\,\,\,\,\,\,\Delta_1=3.4\%\,, \ee \be\label{delta4}
\,\,\,{\rm 2)}\,\,\,\,\,\,\,\,\,\Delta_2=10.7\%\,. \ee As
expected, decreasing the number of moduli by a half has produced
little effect on $\Delta_1$ while decreasing $\Delta_2$ by a few
present. These numbers coming from our somewhat crude analysis
already demonstrate that a comparatively large fraction of vacua
in $M$ theory generate the desired hierarchy between the Planck
and the electroweak scale physics. Also, one can easily check that
all the solutions consistent with the SUGRA framework for which
$240\,{\rm GeV}\leq m_{3/2}\leq 24\,{\rm TeV}$ for any number of
moduli $N$ satisfy the following bound on the eleven dimensional
scale \be 3.6\times 10^{16}\,{\rm GeV}\leq m_{11}\leq 4.3\times
10^{18}\,{\rm GeV}\,, \ee which makes them compatible with the
standard unification at $M_{GUT}\sim 2\times 10^{16}$ GeV. This is
also a nice feature. Of course, apart from determining the upper
and lower bounds on the constants, it would be desirable to know
their distribution for all possible manifolds of $G_2$ holonomy.
In this case instead of using the flat statistical measure as we
did here, each solution would be assigned a certain weight making
the sampling analysis more accurate. However, this is an extremely
challenging task which goes beyond the scope of this work.

The simple analysis presented in this section clearly points to a very
restrictive nature of the solutions. Namely, the requirement of
consistency with the supergravity regime results in very strict
bounds on the properties of the compactification manifold.
Further requirements coming from the SUSY breaking scale to be in
the range required for supersymmetry to solve the hierarchy problem
narrows down the class of possible $G_2$ holonomy
manifolds even more. It would be extremely interesting to know to what extent these results
extend
into the small volume, "stringy" regime, about which we have nothing to
say here.

\subsubsection{Results for dS Vacua}
We will now show that the results obtained in the previous subsection also hold true for dS vacua
with K\"{a}hler potential given by (\ref{kahlerwithmatter}) and
the non-perturbative superpotential as in
(\ref{mattersuptwosectors}) with $SU(N_c)$ and $SU(Q)$ hidden
sector gauge groups. For this case, we have : \be\label{gr22} m_{3/2}=m_p\,\frac
{e^{\phi_0^2/2}}{8\sqrt{\pi}{V_X}^{3/2}}\left|A_1\phi_0^a\,e^{-{\frac{2\pi}{P}{\rm
Im}f}}-A_2e^{-{\frac{2\pi}{Q}{\rm Im}f}}\right|\,, \ee where the
relative minus sign inside the superpotential is due to the axions
and $P\equiv N_c-1$. Before we get to the gravitino mass we first
compute the volume of the compactified manifold $V_X$ for the
metastable dS vacuum with broken SUSY. By substituting the
approximate leading order solution for the moduli (\ref{app24})
into the definition (\ref{vol}) of $V_X$ we obtain:
\be\label{volume2} V_X=\left(\frac
1{2\pi}\right)^{7/3}\left[\frac{PQ}{Q-P}\ln\left(\frac{A_1Q}{A_2P}\right)\right]^{7/3}\prod_{i=1}^{N}
\left(\frac{3a_i}{7N_i}\right)^{a_i}\,. \ee Recalling the
definition of Im($f$) in terms of $\nu$ and using the solution for
$\nu$ (eqn.(\ref{app24})) together with (\ref{volume2}), the
gravitino mass for the dS vacuum in the leading order
approximation is given by: \be\label{gravitino3}
m_{3/2}=m_p\sqrt{2}\pi^3\,A_2 \left|\frac P Q \,\phi_0^{-\frac
2P}-1\right|\left(\frac {P\,Q}{Q-P}\ln\left(\frac
{A_1Q}{A_2P}\right) \right)^{-\frac 72}\left(\frac
{A_1Q}{A_2P}\right)^{-\frac{P}{Q-P}}\prod_{i=1}^{N} \left(\frac {7
N_i}{3 a_i}\right)^{\frac{3a_i}2}e^{\phi_0^2/2}\,, \ee where
$\phi_0^2$ is given by (\ref{e58}). Since $\phi_0^{-2/P} \sim 1$ from section
\ref{chargedmattervac} and $A_2 \sim Q$, we see that the expression for the gravitino mass
for dS vacua is almost the same as that for the AdS vacua (eqn. \ref{gr7}) provided we replace
$\rho$ in (\ref{gr7}) by $\tilde{\rho}=A_1Q/A_2P$ in (\ref{gravitino3}) and $P-Q$ in (\ref{gr7})
by $Q-P$ in (\ref{gravitino3}).

\begin{figure}[h!]
\begin{tabular}{cc}
      \leavevmode \epsfxsize 9 cm \epsfbox{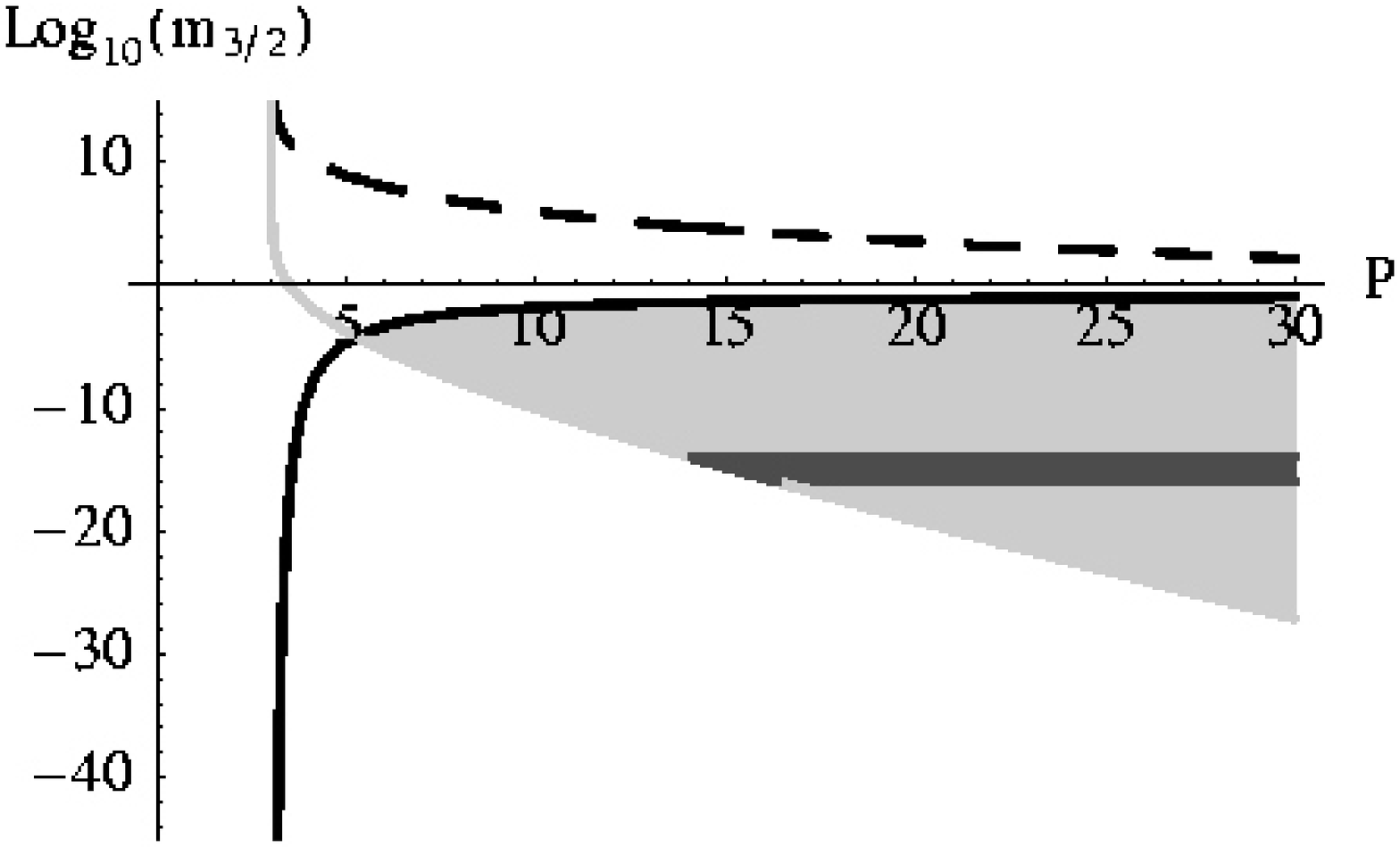}&
      \leavevmode \epsfxsize 9 cm \epsfbox{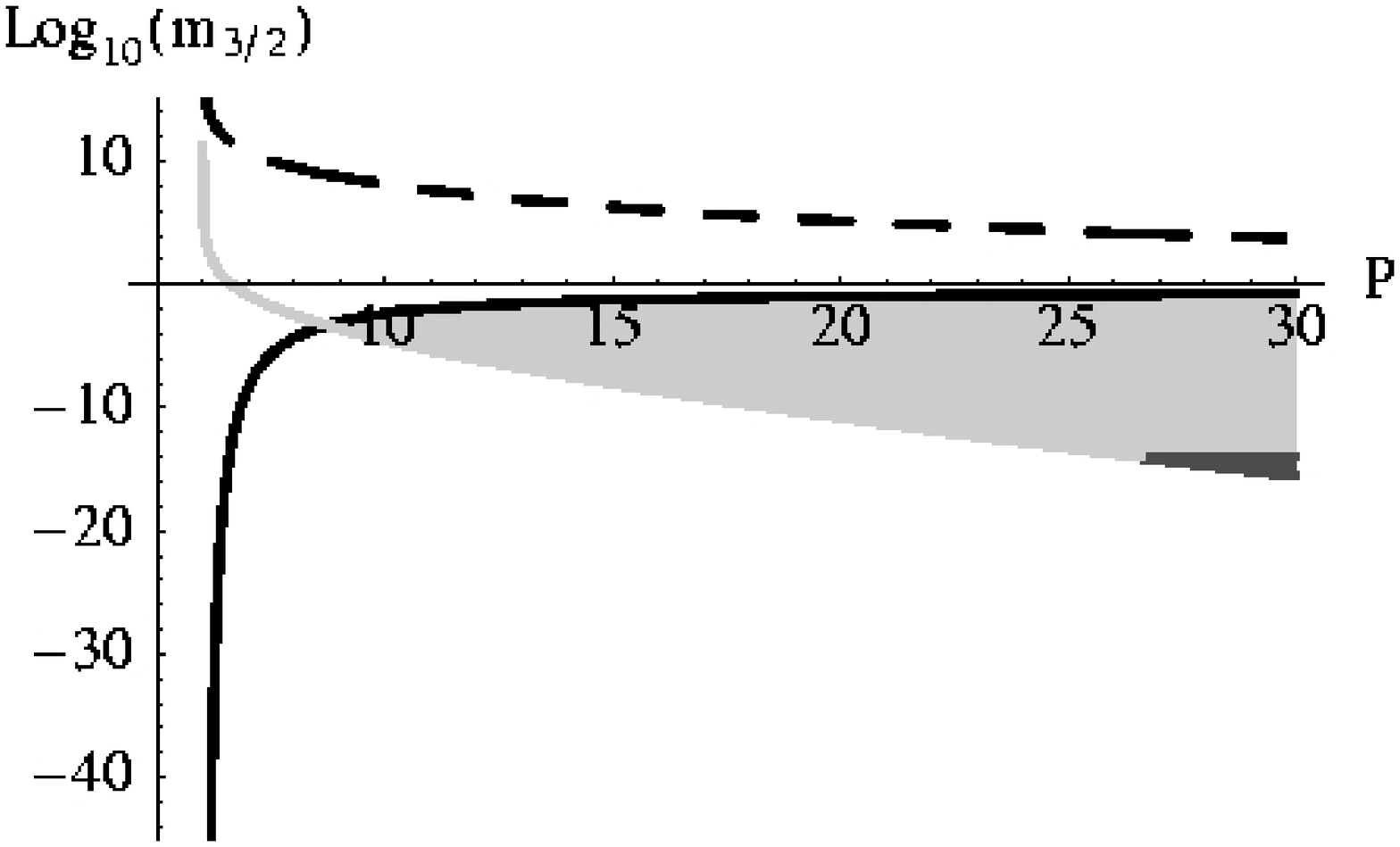} \\
      \leavevmode \epsfxsize 9 cm \epsfbox{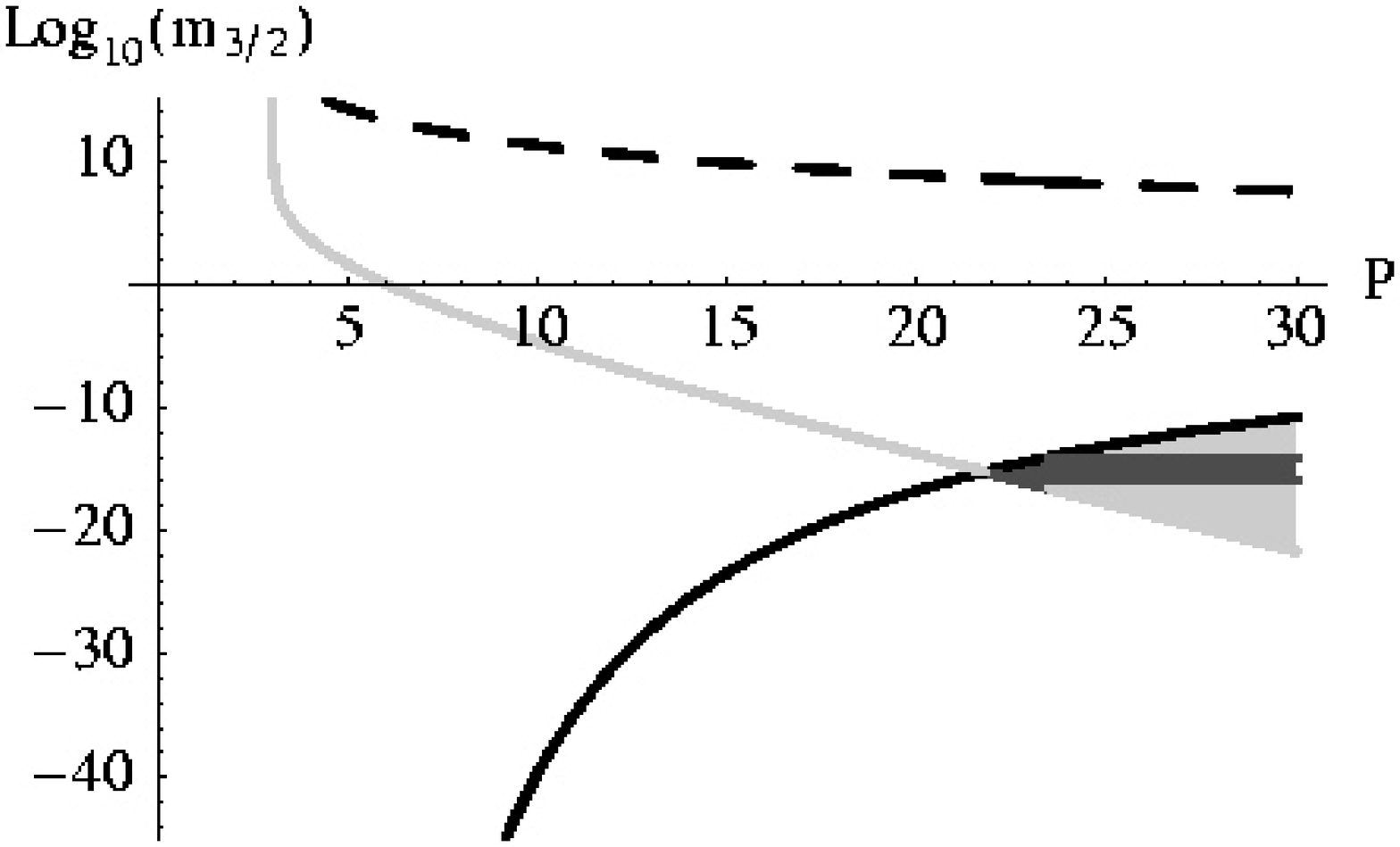}&
      \leavevmode \epsfxsize 9 cm \epsfbox{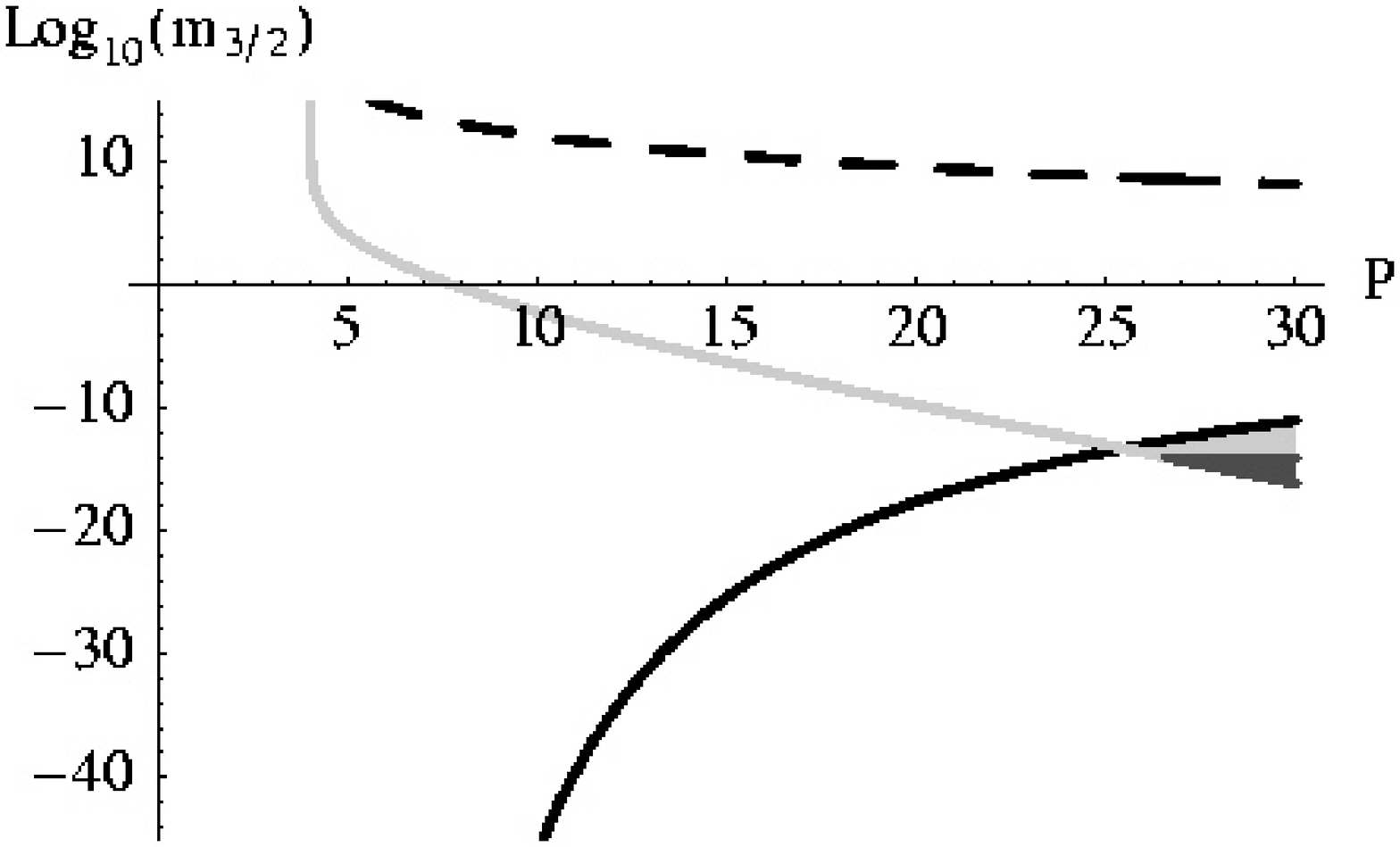} \\
    \end{tabular}
\caption{\footnotesize ${\rm log}_{10}(m_{3/2})$ as a function of
$P$ for case $1)$ - top plots and case $2)$ - bottom plots. The
light grey area represents possible values of ${\rm
log}_{10}(m_{3/2})$ in the range where
$\tilde{\rho}_{crit}\leq\tilde{\rho}\leq 100$ consistent with the
SUGRA approximation. The dark area indicates the region of
interest where $-16\leq {\rm log}_{10}(m_{3/2})\leq -14$ such that
$240\,{\rm GeV}\leq m_{3/2}\leq 24\,{\rm TeV}$. The dark solid
curve corresponds to ${\rm log}_{10}(m_{3/2})$ when
$\tilde{\rho}=\tilde{\rho}_{crit}$. The lower boundary of the
light grey area represents the ${\rm log}_{10}(m_{3/2})$ curve
when $\tilde{\rho}=100$. The dashed curve corresponds to
$\tilde{\rho}=1.01$. Top left: Case $1)$ when $Q-P=3$. Top right:
Case $1)$ when $Q-P=6$. Bottom left: Case $2)$ when $Q-P=3$.
Bottom right: Case $2)$ when $Q-P=4$.} \label{scans17}
\end{figure}

For the de Sitter vacua, we use a less restrictive upper bound
$\tilde{\rho} \equiv \frac{A_1\,Q}{A_2\,P}\leq100$. For a manifold
with $N=100$ moduli we obtain: \be\label{delta8} {\rm
1)}\,\,\,\,\,\,\,\,\,\Delta_1=3\%\,, \ee \be\label{delta9}
\,\,{\rm 2)}\,\,\,\,\,\,\,\,\,\Delta_2=31\%\,. \ee In Figure
\ref{scans17} we present plots of ${\rm log}_{10}(m_{3/2})$ for
both cases as a function of $P$ in the range where
$\tilde{\rho}_{crit}\leq\tilde{\rho}\leq 100$ for different values
of $Q-P$.
Recall that the smallest possible value of $Q-P$ for de Sitter vacua
is $(Q-P)_{min}=3$. In this case the region where $240\,{\rm GeV}\leq m_{3/2}\leq 24\,{\rm TeV}$,
exists for $3\leq Q-P\leq 6$ for the anisotropic case in (\ref{delta8}),
and for $3\leq Q-P\leq 4$ for the ``democratic'' case in (\ref{delta9}).


\subsection{Small Cosmological Constant implies Low-scale Supersymmetry in dS Vacua}\label{CClowsusy}

In this subsection we will study the distribution of SUSY breaking scales in the de Sitter vacua
which as we showed earlier, can arise when the hidden sector has chiral matter. In particular
we will see that the requirement
of a small cosmological constant leads to a scale of SUSY breaking of $\mathcal{O}(1-100)$ TeV.

In section VI, we saw that the minimum obtained is de
Sitter if the discriminant of the quadratic polynomial with
respect to $\phi_0^2$ in eqn. (\ref{po45}) is negative, while it is
anti-de Sitter if the discriminant in (\ref{po45}) is positive.
For $m_{3/2}\sim {\mathcal O}(1-10\,{\rm TeV})$ the magnitude of
the vacuum energy in both cases can be estimated to be \be |V_0|
\sim m_{p}^2\,m_{3/2}^2\sim (10^{10} {\rm GeV})^4 - (10^{11} {\rm
GeV})^4\,. \ee

On the other hand, if the discriminant in (\ref{po45}) vanishes,
one obtains a vanishing cosmological constant (to leading order in
the approximation). At present it is not known if there is a
physical principle which imposes this condition. However, one can
still use it as an observational constraint since the observed
value of the cosmological constant is known to be extremely small.
For instance, it could happen that the space of $G_2$ manifolds
scans the constants $(A_i, P, Q, N)$ finely enough such that there
exist vacua for which the vacuum energy is acceptably low. In
particular, the constants $A_i, i=1,2$ which are determined by the
threshold corrections have been shown to depend on
integers\footnote{This is because the threshold corrections can be
related to certain topological invariants of the associative
three-cycle.} \cite{Friedmann:2002ty}. In this work, we will
assume that to be the case. A detailed computation to show this in
a convincing manner is currently being attempted and will be
reported in the future. It should also be kept in mind that a
different mechanism for solving the cosmological constant,
completely decoupled with particle physics, could exist. Such a
mechanism, if present, would not affect any predictions for low
energy particle physics.

By setting the left hand side in (\ref{e80}) to zero, we can then
express
\begin{equation}\label{er765}
{P\ln\left(\frac{A_1Q}{A_2P}\right)}=\frac{28(Q-P)}{3(Q-P)-8}\,,
\end{equation}

Of course, since the above constraint was
obtained in the leading order, the vacuum energy is only zero in
the leading order in our analytic expansion.
The subleading contributions we neglected in
(225) although smaller than the leading contributions, are still
much larger than the observed value of the cosmological constant.
However, one can {\em in principle} take into account all the
subleading corrections and tune the ratio $A_1Q/A_2P$ inside the
logarithm to set the vacuum energy to a very small value
compatible with the observations. As will be seen later, since
the expression in the R.H.S. of (\ref{er765}) turns out to be
large, the subleading corrections which affect the value of the
cosmological constant will have little effect on the
phenomenological quantities calculated by imposing the constraint
to leading order.

We would now like to analyze in detail the phenomenological
implications of the solutions obtained by imposing (\ref{er765})
as a constraint. The most important phenomenological quantity in
this regard is the gravitino mass as it sets the scale of
all soft supersymmetry breaking parameters. We
focus on the gravitino mass in this section. The soft
supersymmetry breaking parameters will be discussed
in section \ref{pheno}.

\subsubsection{Gravitino mass with a small positive cosmological constant}

By substituting the constraint (\ref{er765}) into the gravitino mass formula (\ref{gravitino3}),
we obtain:\begin{equation}\label{gravitino76} m_{3/2}=m_p\sqrt{2}\pi^3\,A_2
\left|\frac P {Q} \,\phi_0^{-\frac 2P}-1\right|\left(\frac
{28Q}{3(Q-P)-8} \right)^{-\frac
72}e^{-\frac{28}{3(Q-P)-8}}\prod_{i=1}^{N} \left(\frac {7 N_i}{3
a_i}\right)^{\frac{3a_i}2}e^{\phi_0^2/2}\,,
\end{equation} where the meson vev is now given by: \begin{equation}\label{mesonvevconstr}
\phi_0^2\approx-\frac 18+\frac 1{Q-P}+\frac 1 4\sqrt{1-\frac
2{Q-P}}+\frac 2{Q-P}\sqrt{1-\frac 2{Q-P}}\,.
\end{equation} In this case, the moduli vevs are given by
\begin{equation}\label{app247}
s_i=\frac{a_i\nu}{N_i}\,,\,\,\,\,\,\,{\rm with}\,\,\,\,\,\,\,
\nu\approx\frac{6\,Q}{{\pi}(3(Q-P)-8)}\,.
\end{equation}

From (\ref{gravitino76}), one notes that the gravitino mass (when
the cosmological constant is made tiny) is completely determined
by the dual coxeter numbers of the hidden sector gauge groups
$N_c\,{\rm and}\,Q$, the rational numbers $a_i$ (see (\ref{vol}))
characterizing the volume of the $G_2$ manifold and the integers
$N_i$.\footnote{From field theory computations
\cite{Finnell:1995dr}, $A_1 = N_c-N_f=P$ and $A_2=Q$ (in a
particular RG scheme), up to threshold corrections. We can
therefore express $A_1$, $A_2$ as $A_1=P\,C_1$ and $A_2=Q\,C_2$,
where coefficients $C_1$ and $C_2$ depend {\em only} on the
threshold corrections and are constant with respect to the moduli
\cite{Friedmann:2002ty}. In this case, the quantity
$\ln(A_1Q/A_2P)=\ln(C_1/C_2)$ is fixed by imposing (\ref{er765}).}

The rationals $a_i$ are subject to the constraint $\sum_{i=1}^N a_i = 7/3$.
It is reasonable to consider a ``democratic'' choice for $a_i$,
$a_i=7/(3N)$ for all $i=\overline{1,N}$ and also to take for
simplicity all the integers $N_i=1$. The integers $N_i$ will
generically be of ${\mathcal O}(1)$; even if some of the $N_i$ are
unnaturally large, their individual contributions will be
typically washed out as they are raised to powers that are much
less than unity (see (\ref{gravitino3}) and the expression for
$a_i$ for the democratic choice). In this case, after setting $A_2=Q\,C_2$,
the gravitino mass formula is given by
\begin{equation}\label{gravitino44}
m_{3/2}=m_p\sqrt{2}\pi^3 C_2\left|P \,\phi_0^{-\frac
2P}-Q\right|\left(\frac{N(3(Q-P)-8)} {28Q} \right)^{\frac
72}e^{-\frac{28}{3(Q-P)-8}}e^{\phi_0^2/2}\,,
\end{equation}and the moduli vevs are
\begin{equation}\label{app243}
s_i=\frac{14\,Q}{\pi\,N(3(Q-P)-8)}\,.
\end{equation} From (\ref{gravitino44}), the gravitino mass depends on just four
constants - $C_2$, $P$, $Q$ and $N$ (the total number of moduli), determined
by the topology of the manifold. It should be kept in mind that
for the solution to exist, it is necessary that $Q-P>2$ (see
(\ref{e74})). For the smallest possible value $(Q-P)_{min}=3$, the
expression for $m_{3/2}$ simplifies even
further\begin{equation}\label{gravitino49}
m_{3/2}=m_p\sqrt{2}\pi^3 C_2\left|P(\phi_0^{-\frac
2P}-1)-3\right|\left(\frac N{28(P+3)} \right)^{\frac
72}e^{-{28}}e^{\phi_0^2/2}\approx m_p\,3\sqrt{2}\pi^3 C_2\left(\frac N{28(P+3)} \right)^{\frac
72}e^{-{28}}e^{\phi_0^2/2}\,.
\end{equation}together with\begin{equation}\label{e772}
\phi_0^2\approx\frac 1{72}\left(15+22\sqrt{3}\right)\approx
0.7376\,,\,\,\,\,\,\,s_i=\frac{14\,(P+3)}{{\pi\,N}}\,.\nonumber
\end{equation}
Note that the dependence on $N$ and $P$ in (\ref{gravitino49}) is due solely
to the volume $V_X$ dependence on those parameters.
The expression for the gravitino mass has a more transparent form if we don't
substitute the expression for the volume (\ref{volume2}) into (\ref{gr22}).
For $Q-P=3$ we obtain
\be\label{gr780}
m_{3/2}\approx m_p\,\frac
{3\,e^{\phi_0^2/2}}{8\sqrt{\pi}{V_X}^{3/2}}e^{-{28}}C_2
\approx 514\,{\rm TeV}\frac{C_2}{V_X^{3/2}}\,,
\ee
where the detailed dependence on $a_i$, $N_i$, $P$ and the number of moduli
$N$ is completely encoded inside the seven-dimensional volume $V_X$ which
appears to be the more relevant physical quantity. Furthermore, in the supergravity
approximation the volume $V_X>1$, which translates into an upper bound on the
gravitino mass when $Q-P=3$
\be
m_{3/2}< {\cal O}(100\,{\rm TeV})\,.
\ee

\subsubsection{The Gravitino mass Distribution and its consequences}

The gravitino mass (\ref{gravitino44}) depends on three integers: the two gauge group
dual coxeter numbers $N_c$ , $Q$ and the number of moduli $N$.\footnote{We can set $C_2=1$
for the order of magnitude estimates we are doing here.}
This will give us an idea about the distribution of
the gravitino mass (which sets the superpartner masses)
obtained after imposing the constraint (\ref{er765}) that the vacuum
energy is acceptable.
Asides from only considering the vacua within the
supergravity approximation (ie $s_i >1$) we expect an upper bound on the dual
coxeter numbers of the hidden sector gauge groups $P$ and $Q$.
Based on duality with the heterotic string, it
seems reasonable to assume that they can be at least as large as 30 -
the dual coxeter number of $E_8$. Of course, values of $P,Q$
larger than 30 cannot be ruled out, and here
we assume an upper bound $P \leq 100$.
Notice that from the constraint in (\ref{er765}), the ratio $(A_1Q)/(A_2P)$
can get exponentially large when $Q-P=3$ and the values of $P$ are small.
Here we are going to completely relax the requirement on the upper
bound of the ratio because as we will see, for generic manifolds with a
large numer of moduli, $P$ has to be large for the supergravity
approximation to hold.
The distribution can be constructed as follows. The three
integers: $P$, $Q-P$ and $N$ are varied subject to (\ref{e74})
and the supergravity constraint $s_i>1$.
For each point in the resulting two dimensional subspace,
$\log_{10}(m_{3/2})$ can be computed and rounded off to the
closest integer value. One can then count how many times each
integer value is encountered in the entire scan and plot the
corresponding distribution.
\begin{figure}[h!]
    \begin{tabular}{cc}
      \leavevmode \epsfxsize 5.5 cm \epsfbox{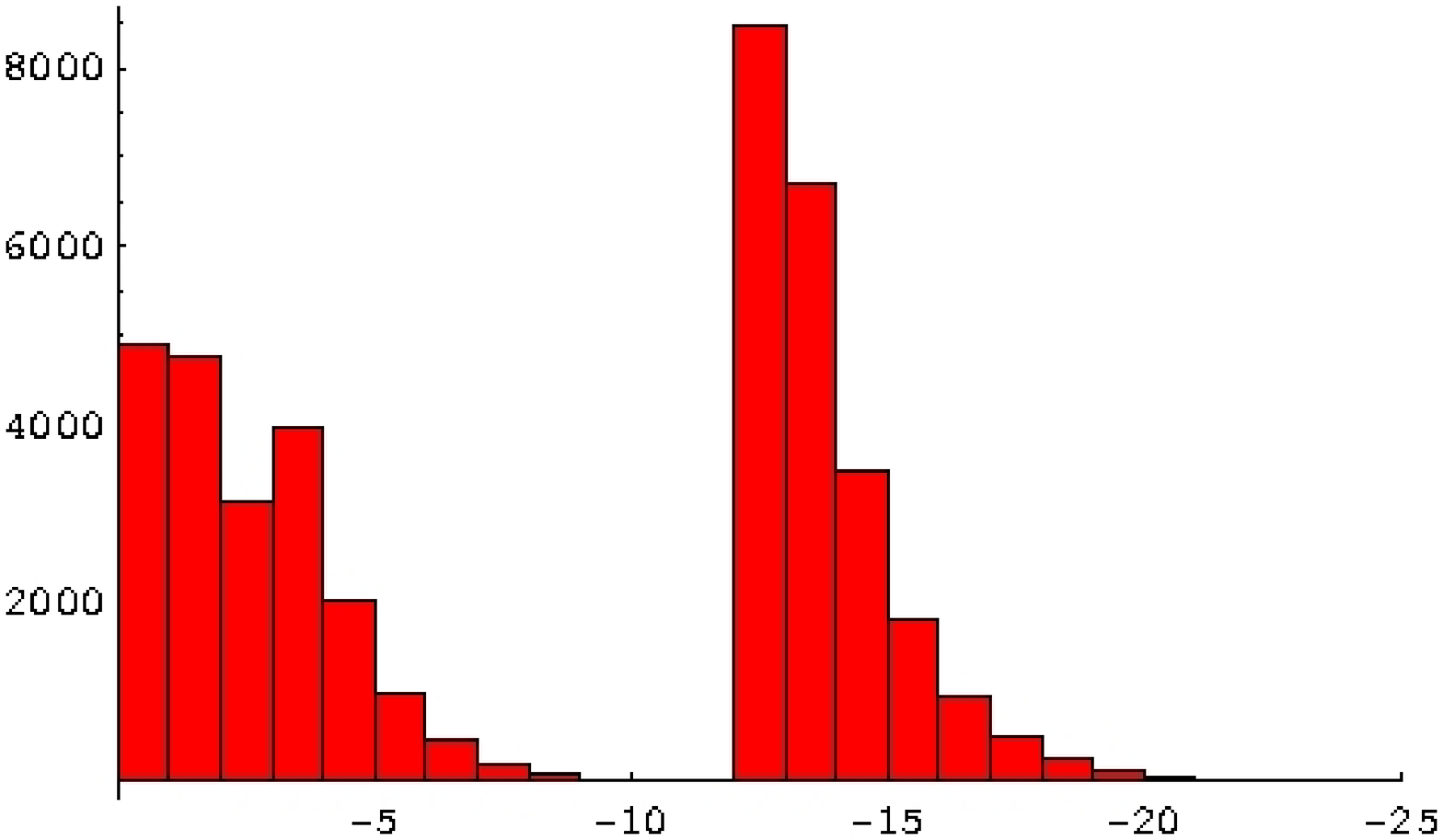}&
      \leavevmode \epsfxsize 5.5 cm \epsfbox{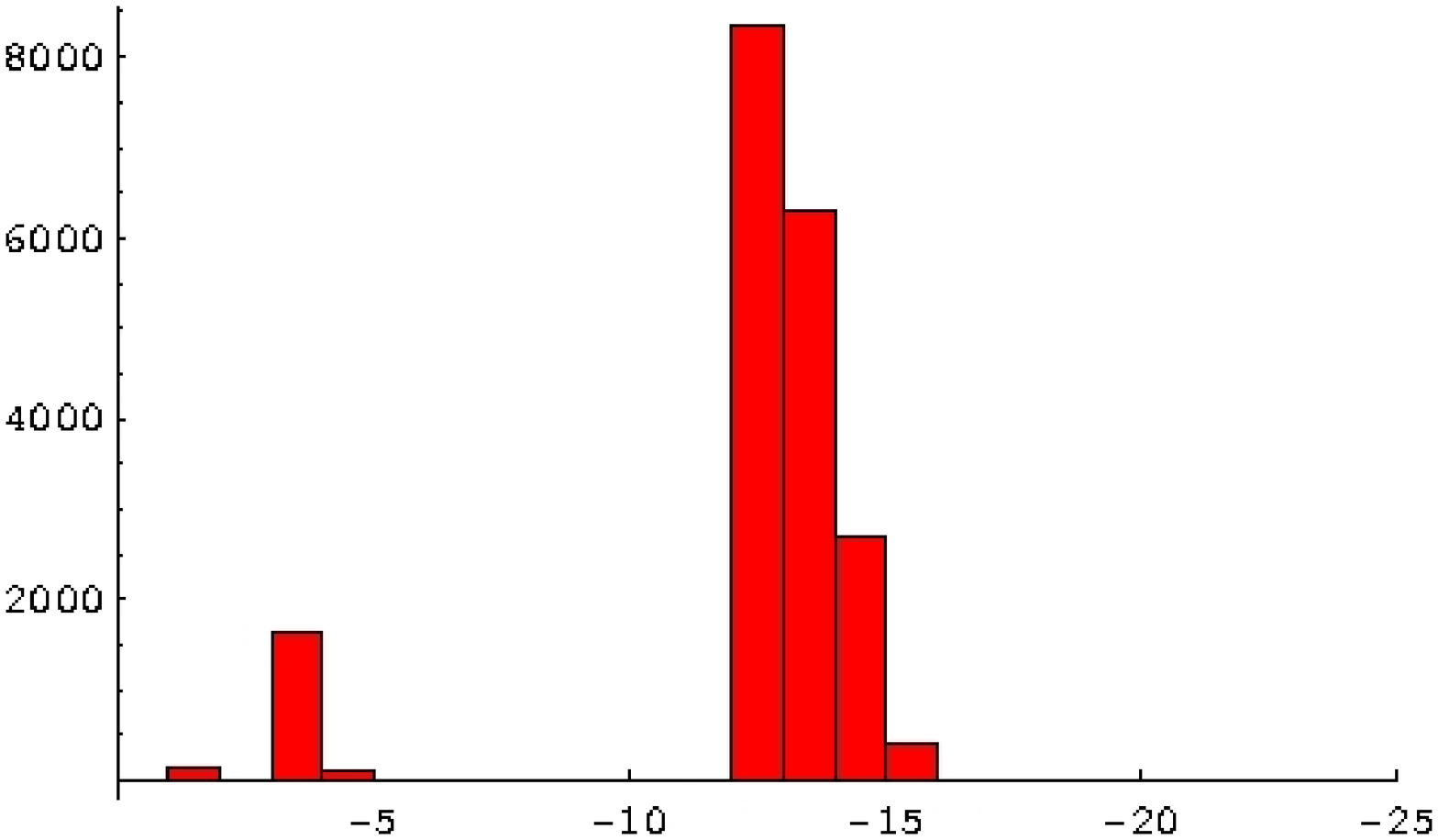}
      \leavevmode \epsfxsize 5.5 cm \epsfbox{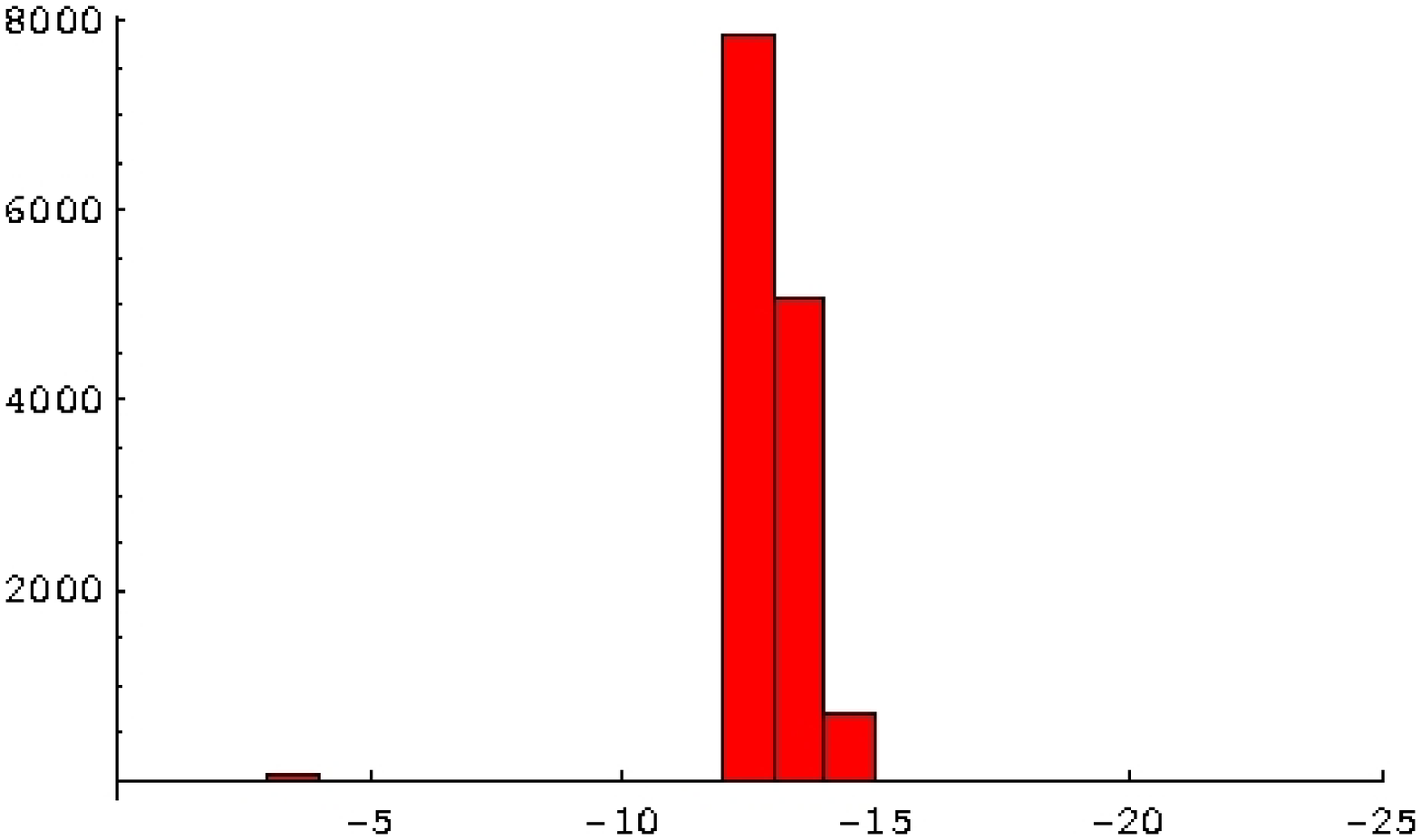} \\
    \end{tabular}
    \caption{\footnotesize The gravitino mass distribution with the x-axis
    denoting the logarithm of the gravitino mass (to base 10). Left: Distribution corresponding to scan one in (\ref{scan1}).
Middle: Distribution corresponding to scan two in (\ref{scan2})
for which manifolds with the number of moduli $N<50$ were excluded
from the scan. Right: Distribution corresponding to scan three in
(\ref{scan3}) for which manifolds with the number of moduli
$N<100$ were excluded from the scan.}
    \label{Plot79}
\end{figure}

In the first three scans scan we cover a broad range of values by
choosing $P_{max}=100$. Taking into account the SUGRA constraint
($s_i>1$), we have the following ranges of integers for
the first scan: \begin{equation}\label{scan1} 3\leq P\leq
100\,;\,\,\,\,\,\,\,\, 3\leq(Q-P)\leq 100-P\,;\,\,\,\,\,\,\,\,
2\leq N<\frac{14\,(P+(Q-P))}{\pi\,(3(Q-P)-8)}\,.
\end{equation} In the second scan we have excluded the small N region
and considered only the manifolds with $N\geq50$. Thus we have the
following ranges of constants for the second scan:
\begin{equation}\label{scan2} 3\leq P\leq 100\,;\,\,\,\,\,\,\,\,
3\leq(Q-P)\leq 100-P\,;\,\,\,\,\,\,\,\, 50\leq
N<\frac{14\,(P+(Q-P))}{\pi\,(3(Q-P)-8)}\,.
\end{equation} In the third scan we have
only considered manifolds with $N\geq100$. Thus we have the
following ranges of integers for the second scan:
\begin{equation}\label{scan3} 3\leq P\leq 100\,;\,\,\,\,\,\,\,\,
3\leq(Q-P)\leq 100-P\,;\,\,\,\,\,\,\,\, 100\leq
N<\frac{14\,(P+(Q-P))}{\pi\,(3(Q-P)-8)}\,.
\end{equation} The first two distributions in Figure \ref{Plot79}
clearly have several prominent peaks. Amazingly, in all three plots one of
the peaks landed right in the $m_{3/2}\sim {\mathcal O}(1-100)\,{\rm
TeV}$ range! The high scale peaks on the left plot appear to be around
$m_{3/2}\sim 10^{14}\,{\rm GeV}$ and the GUT scales.
However, for the middle plot the GUT scale peak almost disappears.
Recall that the middle plot corresponds to scan two in
(\ref{scan2}) where we excluded all the manifolds for which the
number of moduli $N$ is less than $50$. Therefore, the high scale
peaks are largely dominated by contributions from the $G_2$
manifolds with a small number of moduli $N<50$. As seen from the
right plot, when $G_2$ manifolds with $N<100$ are excluded from
the scan, the peak at the $m_{3/2}\sim 10^{14}\,{\rm GeV}$ scale has all but disappeared,
whereas the peak at $m_{3/2}\sim {\mathcal O}(1-100)\,{\rm TeV}$
remains virtually unchanged.

\begin{figure}[h!]
    \begin{tabular}{cc}
      \leavevmode \epsfxsize 5.5 cm \epsfbox{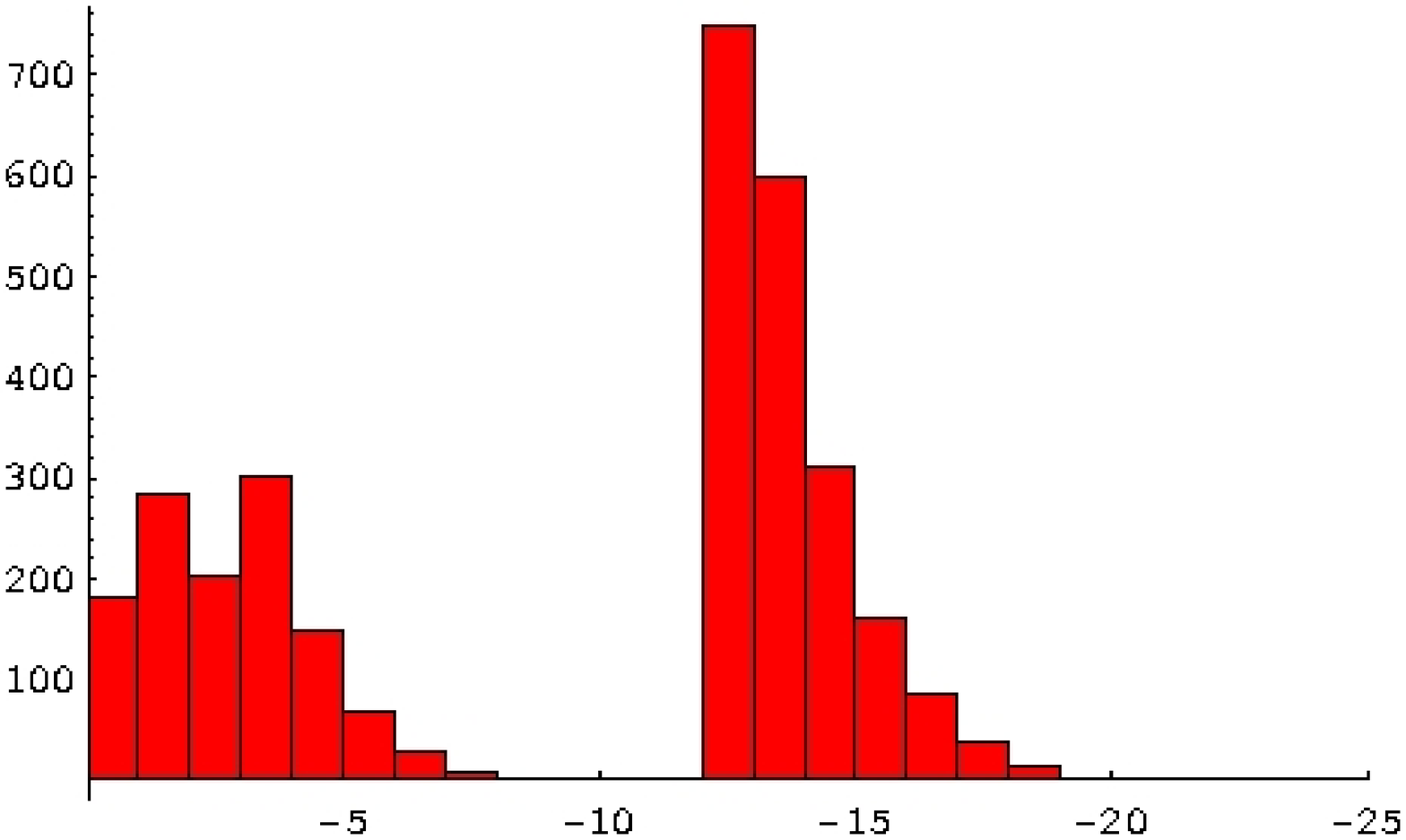}&
      \leavevmode \epsfxsize 5.5 cm \epsfbox{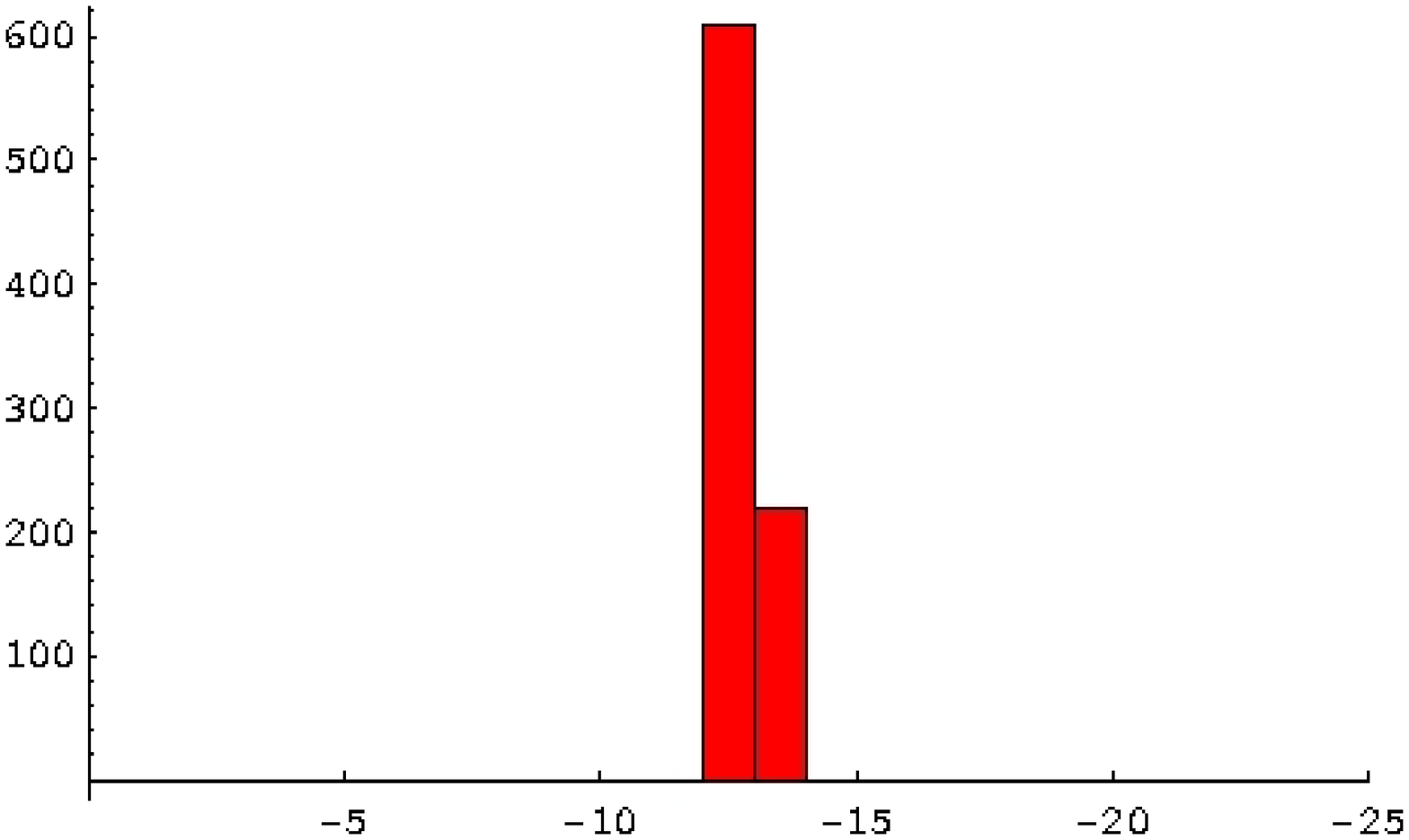}
      \leavevmode \epsfxsize 5.5 cm \epsfbox{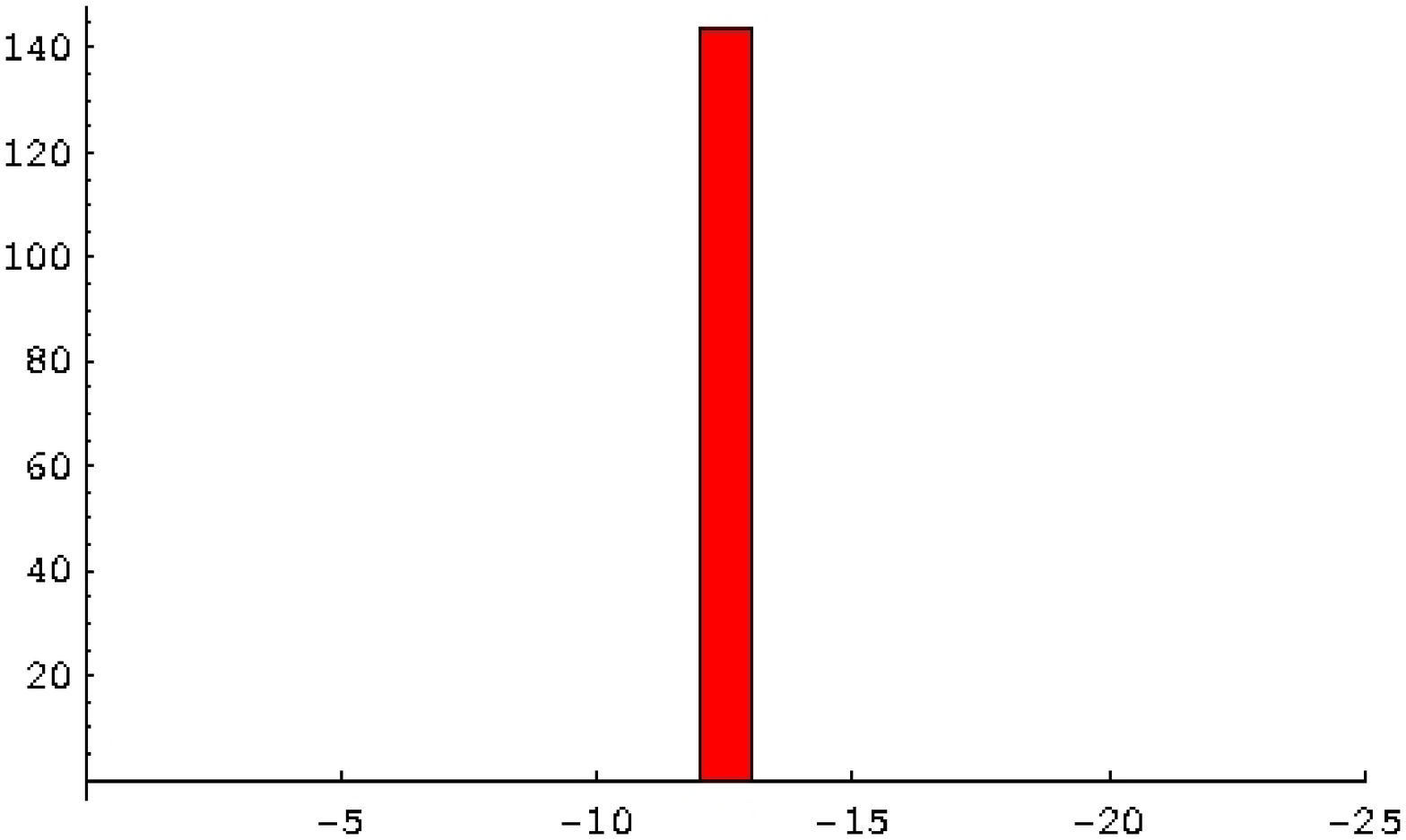} \\
    \end{tabular}
    \caption{\footnotesize The gravitino mass distribution with the x-axis
    denoting the logarithm of the gravitino mass (to base 10). Left: Distribution corresponding to scan four in (\ref{scan4}).
Middle: Distribution corresponding to scan five in (\ref{scan5})
for which manifolds with the number of moduli $N<50$ were excluded
from the scan. Right: Distribution corresponding to scan six in
(\ref{scan6}) for which manifolds with the number of moduli
$N<100$ were excluded from the scan.} \label{Plot80}
\end{figure}

\begin{figure}[h!]
    \begin{tabular}{cc}
      \leavevmode \epsfxsize 5.5 cm \epsfbox{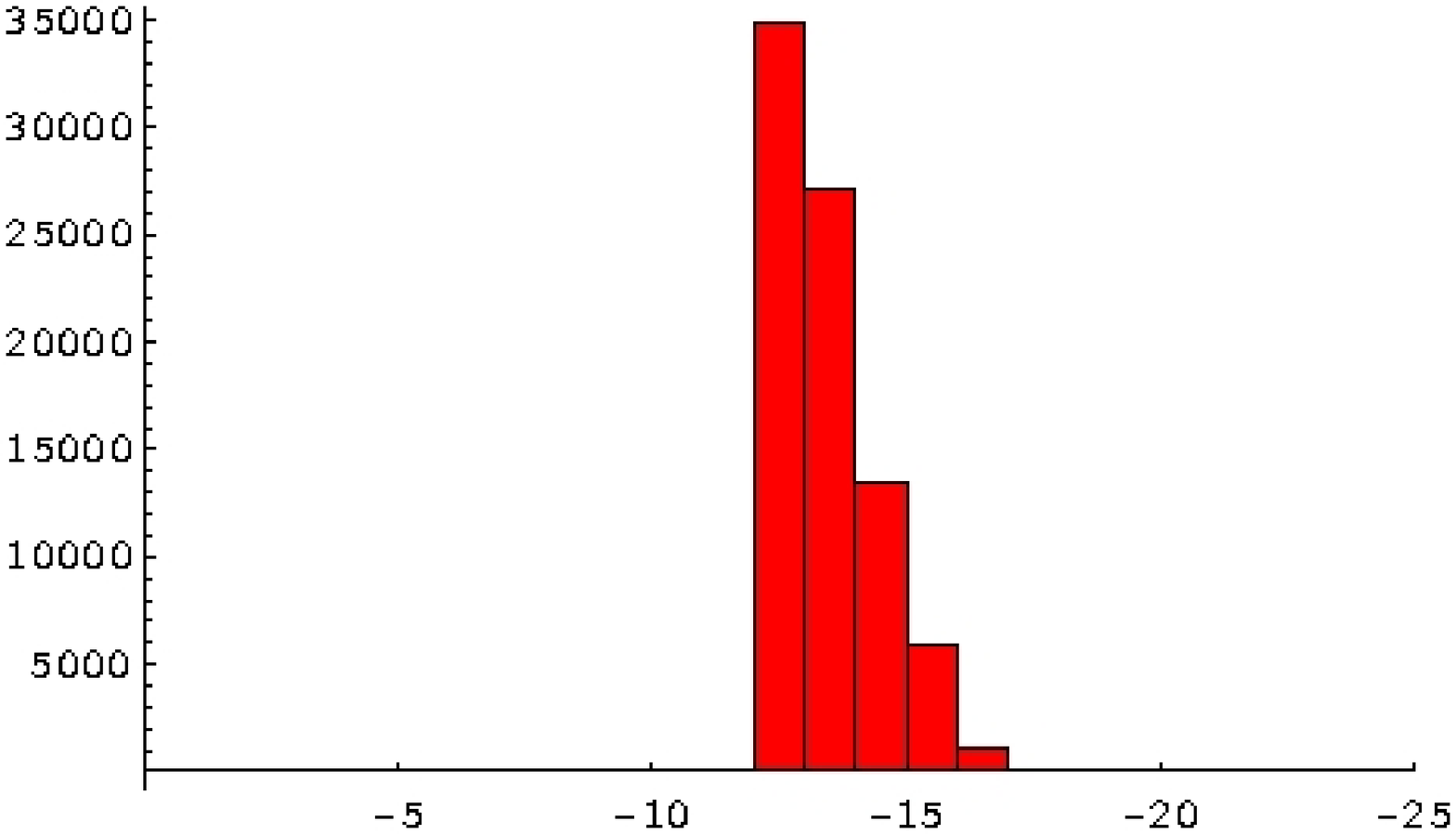}&
      \leavevmode \epsfxsize 5.5 cm \epsfbox{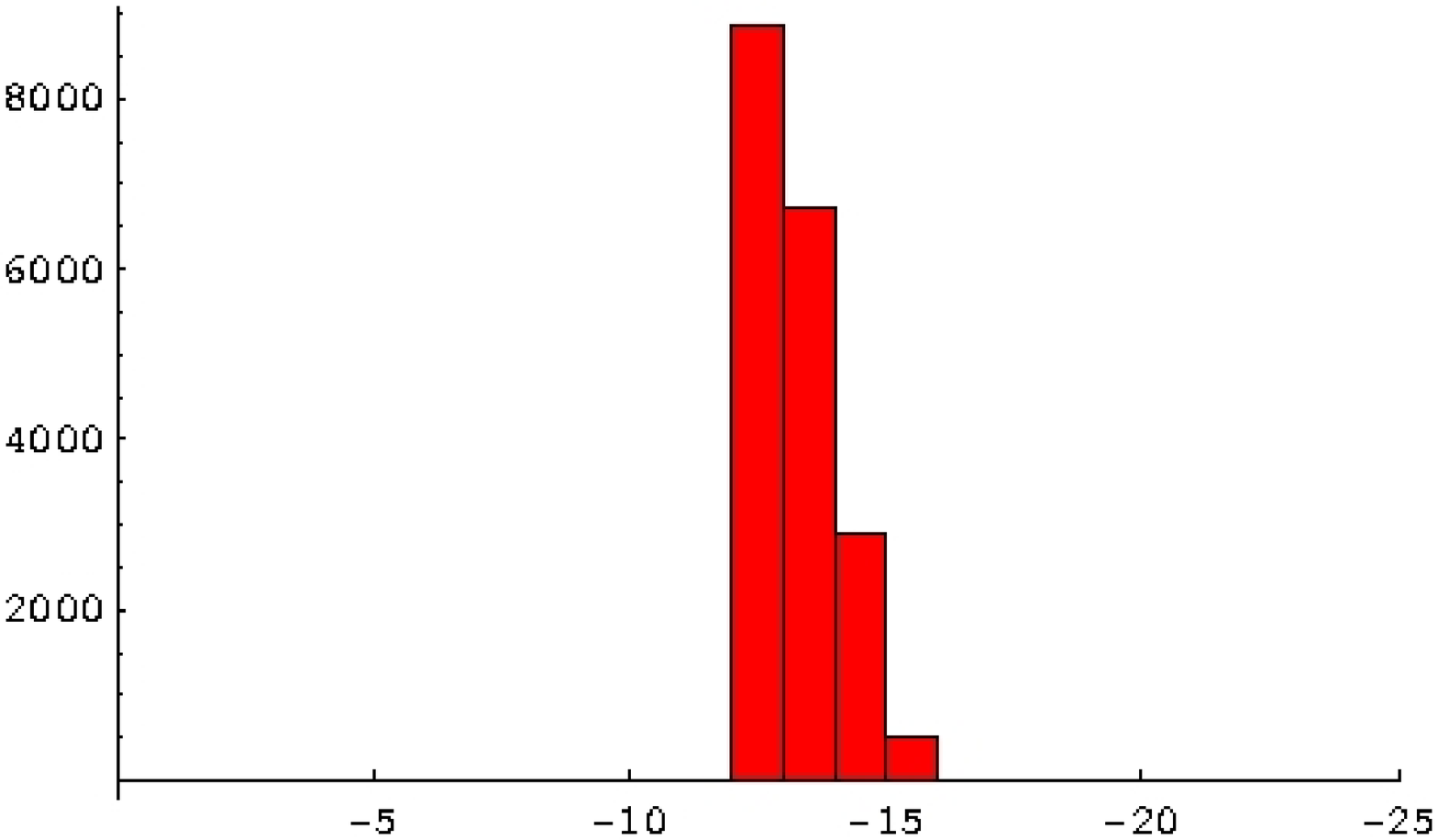}
      \leavevmode \epsfxsize 5.5 cm \epsfbox{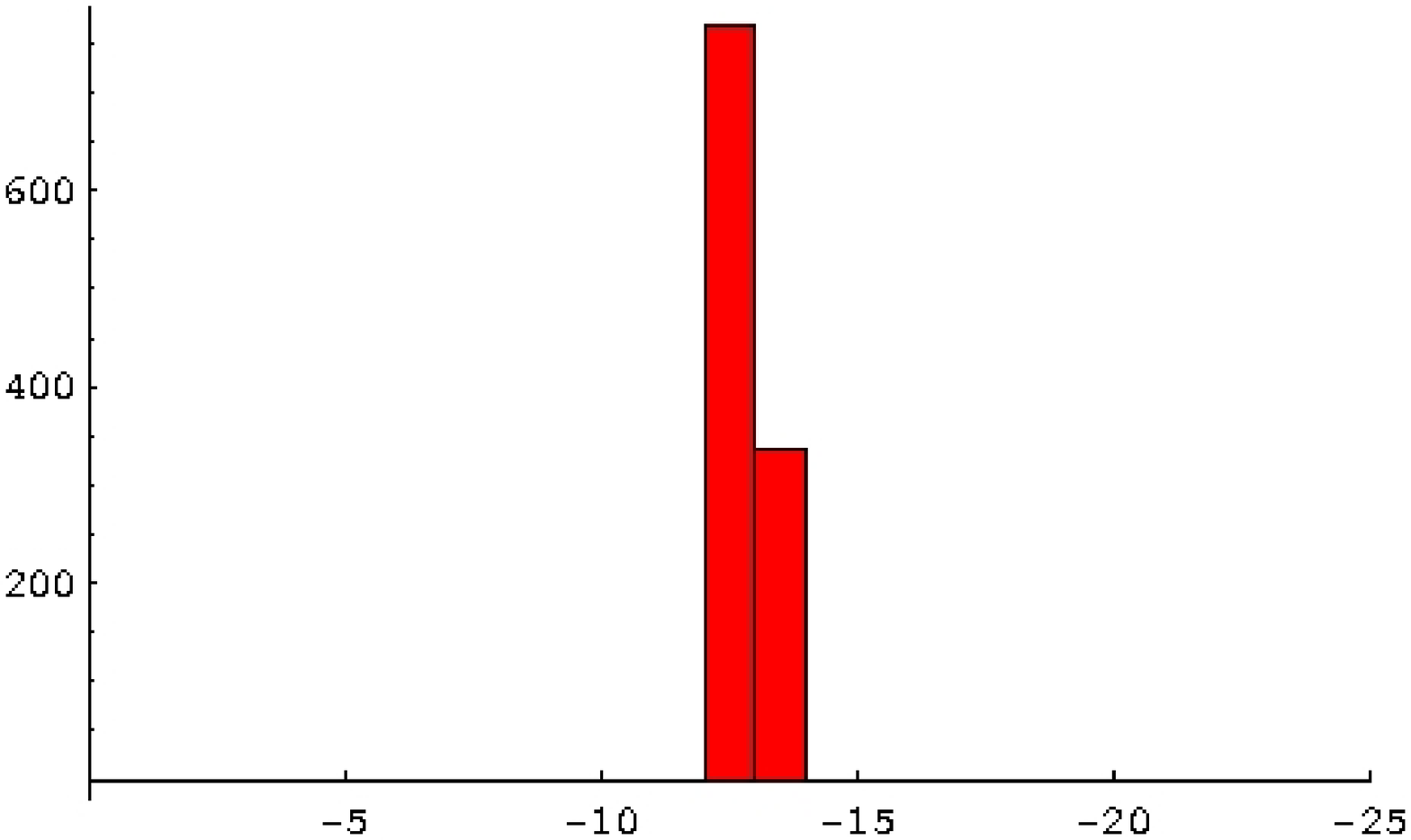}
    \end{tabular}
    \caption{\footnotesize The gravitino mass distribution with the x-axis
    denoting the logarithm of the gravitino mass (to base 10). Scans for the smallest possible choice $(Q-P)_{min}=3$.
Left: Distribution corresponding to the scan with $P_{max}=200$.
Middle: Distribution corresponding to the scan with $P_{max}=100$.
Right: Distribution corresponding to the scan with $P_{max}=30$.}
    \label{Plot81}
\end{figure}

In Figure \ref{Plot80} we included three more scans for which the
upper bound on $P$ was reduced to $P_{max}=30$. The
fourth scan has the following ranges:\begin{equation}\label{scan4} 3\leq P\leq
30\,;\,\,\,\,\,\,\,\, 3\leq(Q-P)\leq 30-P\,;\,\,\,\,\,\,\,\, 2\leq
N<\frac{14\,(P+(Q-P))}{\pi\,(3(Q-P)-8)}\,.
\end{equation} In the fifth scan we again excluded the small N region an
considered only the manifolds with $N\geq50$ and considered
$P_{max}=30$:
\begin{equation}\label{scan5}
3\leq P\leq 30\,;\,\,\,\,\,\,\,\, 3\leq(Q-P)\leq
30-P\,;\,\,\,\,\,\,\,\, 50\leq
N<\frac{14\,(P+(Q-P))}{\pi\,(3(Q-P)-8)}\,.
\end{equation} In the sixth scan we considered only the manifolds with $N\geq
100$ and $P_{max}=30$: \begin{equation}\label{scan6} 3\leq P\leq
30\,;\,\,\,\,\,\,\,\, 3\leq(Q-P)\leq 30-P\,;\,\,\,\,\,\,\,\,
100\leq N<\frac{14\,(P+(Q-P))}{\pi\,(3(Q-P)-8)}\,.
\end{equation}
Again, in Figure \ref{Plot80} we notice that the
${\mathcal O}(1-100)\,{\rm TeV}$ peak narrows around $m_{3/2}\sim{\mathcal O}(100)\,{\rm TeV}$,
as we exclude manifolds with small number of moduli. As the same time,
the peaks at the high scale completely disappear for $G_2$ manifolds with $N>50$.
Finally, in Figure \ref{Plot81} we chose the smallest possible
value $Q-P=3$
and scanned integers $P$ and $N$ in the following ranges:
\begin{eqnarray}\label{scan7}
&&3\leq P\leq 200\,;\,\,\,\,\,50\leq N<\frac{14\,(P+3)}{\pi}\,,\\
&&3\leq P\leq 100\,;\,\,\,\,\,50\leq N<\frac{14\,(P+3)}{\pi}\,,\nonumber\\
&&3\leq P\leq 30\,;\,\,\,\,\,50\leq
N<\frac{14\,(P+3)}{\pi}\,.\nonumber
\end{eqnarray}
In all three plots in Figure \ref{Plot81} we see the same peak at
$m_{3/2}\sim {\mathcal O}(1-100)\,{\rm TeV}$, which narrows around
$m_{3/2}\sim{\mathcal O}(100)\,{\rm TeV}$ as $P_{max}$ is decreased.

Therefore, from the above distributions we conclude that the peak
corresponding to $m_{3/2}\sim(1-100)\,{\rm TeV}$ is entirely due to
the smallest possible value $(Q-P)_{min}=3$. This can be explained
if we examine the gravitino mass formula in (\ref{gravitino49}).
In particular the constant factor $e^{-28}\sim 10^{-12}$ is most
crucial in lowering the gravitino mass to the TeV scale. It is
easy to trace the origin of this factor to the constraint
(\ref{er765}), imposed by the requirement to have a zero
cosmological constant (to leading order). When (\ref{er765}) is
used along with the requirement $Q-P=3$ we simply get
\begin{equation}\label{ior} P\ln\left(\frac{A_1Q}{A_2P}\right)=84\,.
\end{equation}
When this is substituted into the gravitino mass
(\ref{gravitino3}), the corresponding suppression factor turns
into the constant
\begin{equation}\label{crucial}
\left(\frac {A_1Q}{A_2P}\right)^{-\frac{P}{Q-P}}=e^{-28}\,.
\end{equation}
Physically, this suppression factor corresponds to the hidden sector
gaugino condensation scale (cubed).
Recall that for an $SU(Q)$ hidden sector gauge group, the scale of gaugino
condensation is given by
\be\label{gaugcond}
\Lambda_g=m_p\,e^{-\frac{8\pi^2}{3Q\,g^2}}=m_p\,e^{-\frac{2\pi}{3Q}{\rm Im}f}\,.
\ee
The moduli vevs in (\ref{app247}) completely determine the gauge kinetic function.
Taking $Q-P=3$ we obtain
\be\label{gaugkinf}
{\rm Im}f=\sum_{i=1}^N N_i s_i=\frac{14\,Q}{\pi}.
\ee
Substituting (\ref{gaugkinf}) into (\ref{gaugcond}) we obtain the following
scale of gaugino condensation
\be
\Lambda_g=m_p\,e^{-28/3}\approx 2.15\times 10^{14}\,{\rm GeV}\,.
\ee
It is important to note that the expression in the R.H.S. of
(\ref{er765}) is quite large ($=84$, when $Q-P=3$) in the leading
order, and the quantity ($A_1Q/A_2P$) which is fixed by imposing
the vacuum energy constraint is inside a logarithm. Therefore,
even when one incorporates all the higher order corrections and
tunes the ratio $A_1Q/A_2P$ inside the logarithm to set the
cosmological constant equal to the observed value, the constant on
the R.H.S. (=84), crucial in obtaining the ${\mathcal O}(100)\,{\rm TeV}$
scale peak, is hardly affected.

\begin{figure}[h!]
    \begin{tabular}{cc}
      \leavevmode \epsfxsize 11 cm \epsfbox{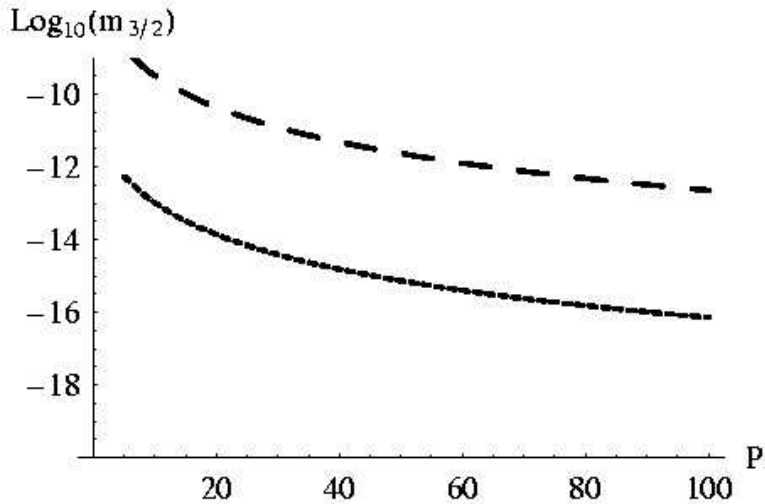}
    \end{tabular}
    \caption{\footnotesize Plot of $\log_{10}(m_{3/2})$ as a function of $P$ for $Q-P=3$.
Short-dashed curve corresponds to $N=50$. Long-dashed curve
corresponds to $N=500$.}
    \label{Plot82}
\end{figure}

The dominance of the ${\mathcal O}(1-100)\,{\rm TeV}$ range also becomes clear from Figure
\ref{Plot82}, where $\log_{10}(m_{3/2})$ as a function of $P$ for
$Q-P=3$ is plotted for a manifold with $N=50$ moduli -
short-dashed curve, and a manifold with $N=500$ moduli -
long-dashed curve.
Indeed, even when we do not impose the SUGRA constraint, from the
above plot we can see that the ${\mathcal O}(1-100)\,{\rm TeV}$
range is covered by a large
swath of the vacuum space and it is not so surprising that the
corresponding distribution peaks at that scale.
This essentially follows from the formula
for the gravitino mass in (\ref{gravitino49}).

An important point which should be emphasized is that for $Q-P=3$,
the gravitino mass dependence on $P$ and $N$ appears {\em only}
through the volume $V_X$, as can be seen from (\ref{gr780}). Thus,
the distribution in Figure \ref{Plot81} directly correlates with
the corresponding distribution of the stabilized volume of the
seven-dimensional manifold $V_X$ as a function of $P$ and $N$.
Therefore, it is the dominance of the vacua with a relatively
small volume, which results in the peak at ${\mathcal
O}(100)\,{\rm TeV}$.

Also note that in the above analysis we simply set the constant
coefficient due to the threshold corrections $C_2$ to unity. It would be
interesting to get a handle on this quantity and include
its variation into the gravitino mass distribution study.

One could argue that even though $Q-P=(Q-P)_{min}=3$ gives a peak
for the gravitino mass distribution at around ${\mathcal
O}(100)\,{\rm TeV}$ scale, it seems plausible from a theoretical
point of view to have many examples of gauge singularities in
$G_2$ manifolds such that $Q-P>3$. However, by imposing the
supergravity constraint that all moduli $s_i$ are larger than
unity (which is the regime in which the entire analysis is valid),
one sees that having $Q-P>3$ drastically reduces the upper bound
on $N$ compared to that for $Q-P=3$ (see eqn(\ref{app247})).
Therefore, the peaks in the gravitino mass distribution obtained
for $-2\leq \log_{10}(m_{3/2})\leq-5$ in Figures \ref{Plot79} and
\ref{Plot80} come from vacua with a \emph{small} number of moduli
as well as $Q-P>3$, compatible with the analysis in the
supergravity regime. Since it is presumably true that the number
of $G_2$ manifolds with the required gauge singularities which
have a large number of moduli is much larger than those with a
small number of moduli, it seems reasonable to expect that the
peak of the gravitino mass distribution obtained at around
${\mathcal O}(100)\,{\rm TeV}$ scale is quite robust and is
representative of the most generic class of $G_2$ manifolds with
the appropriate gauge singularities.
Notice also that in the case of manifolds with a large number of moduli $N\geq100$,
because of the constraint from the supergravity approximation, the actual minimal value
of $P$ in the scans (\ref{scan3}) and (\ref{scan6}) is quite large, i.e. $P_{min}=20$.
Hence, if $P\geq20$, from (\ref{er765}) the value of the ratio when $Q-P=3$ is bounded from above
\begin{equation}\label{er7658}
\frac{A_1Q}{A_2P}\lesssim 67.7\,
\end{equation}
On the other hand, for less generic manifolds with a small number of moduli
the supergravity constraint allows $P$ to be small. In this case in order
to satisfy (\ref{er765}), $(A_1Q)/(A_2P)$ has to be exponentially large.
Whether this is possible remains to be seen. However, because for the more generic case
when $N\geq100$ the upper bound in (\ref{er7658}) is quite reasonable, our analysis
remains robust.

One could also contrast the results obtained above with those obtained for the Type IIB flux vacua.
In Type IIB flux vacua, one has to {\em independently} tune both the gravitino mass to a TeV scale (if one
requires low scale supersymmetry to solve the hierarchy problem) as well as the cosmological constant to
its observed value. This is quite different to what we are finding here. Finally, we should emphasize that
imposing the supergravity approximation was crucial in obtaining low scale SUSY breaking.
Plausibly, the vacua which exist in the $M$ theoretic, small
volume regime will have a much higher SUSY breaking scale.
However, such vacua presumably also have the incorrect electroweak
scale (either zero or $M_{11}$).

\subsection{Including more than one flavor of quarks in the hidden sector}

In the previous analysis we assumed a single flavor for the quarks
in the hidden sector, i.e $N_f=1$. In order to check that the way
we obtain a dS metastable minimum is robust and not dependent on a
particular choice of chiral hidden sector matter spectrum, we
would like to extend our analysis to include more than one flavor
(but still with $N_f < N_c$) so that the meson fields are given
by: \be \phi^{\,\sigma}_{\,\bar \sigma}\equiv
\left(2\,Q^{\sigma}\tilde Q_{\bar \sigma}\right)^{1/2}\,, \ee
where $\sigma,\bar \sigma=\overline{1,N_f}$. In the absence of a
perturbative superpotential (which is guaranteed in the absence of
fluxes), one has along the $D$-flat direction: \begin{eqnarray} Q
= {\tilde Q} = \frac{1}{\sqrt{2}}\,\left( \begin{array}{cccc}
{\phi}^{1}_{1} &  &  &\\
  & {\phi}^{2}_{2} &  &\\
 &  & \ddots  & \\
 &  &   & {\phi}^{N_f}_{N_f}
\end{array} \right)
\end{eqnarray}
Thus, the determinant appearing in (\ref{mattersup}) becomes \be
{\rm det}(\phi^{\,\sigma}_{\,\bar
\sigma})=\prod_{\sigma=1}^{N_f}{\phi_\sigma}\,,\,\,\,\,\,\,{\rm
where} \,\,\,\, {\phi_\sigma}\equiv\phi^{\,\sigma}_{\,\sigma}\,.
\ee The nonperturbative superpotential and the K\"{a}hler
potential are then given by \ba \label{mattersuptwosectors45}
&&W=A_1\prod_{\sigma=1}^{N_f}{\phi_\sigma}^a\,e^{ib_1\,f}+A_2e^{ib_2\,f}\,\\
&&K = -3\ln(4\pi^{1/3}\,V_X)+\sum_{\sigma=1}^{N_f}\phi_{\sigma}\bar\phi_{\sigma}\,, \nonumber
\ea
where we again denoted $b_1\equiv 2\pi/P$, $b_2\equiv 2\pi/Q$, $P\equiv N_c-N_f$ and $a\equiv -2/P$.
After minimizing with respect to the axions, the scalar potential is given by
\ba
\label{potential_noaxions56}
V&=&\frac{e^{\sum_{\sigma=1}^{N_f}{(\phi_0)_{\sigma}^2}}}{48\pi V_X^3}\,[(b_1A_1\prod_{\sigma=1}^{N_f}(\phi_0)_{\sigma}^ae^{-b_1\vec\nu\cdot\,\vec a}-b_2A_2e^{-b_2\vec\nu\cdot\,\vec a})^2\sum_{i=1}^{N}a_i({\nu_i})^2+3(A_1\prod_{\sigma=1}^{N_f}(\phi_0)_{\sigma}^ae^{-b_1\vec\nu\cdot\,\vec a}-A_2e^{-b_2\vec\nu\cdot\,\vec a})^2\nonumber\\\nonumber\\
&&+3(\vec\nu\cdot\,\vec a)(A_1\prod_{\sigma=1}^{N_f}(\phi_0)_{\sigma}^ae^{-b_1\vec\nu\cdot\,\vec a}-A_2e^{-b_2\vec\nu\cdot\,\vec a})(b_1A_1\prod_{\sigma=1}^{N_f}(\phi_0)_{\sigma}^ae^{-b_1\vec\nu\cdot\,\vec a}-b_2A_2e^{-b_2\vec\nu\cdot\,\vec a})\\\nonumber\\
&&+\sum_{\gamma=1}^{N_f}\frac 3
4{(\phi_0)_{\gamma}^2}((\frac{a}{(\phi_0)_{\gamma}^2}+1)A_1\prod_{\sigma=1}^{N_f}(\phi_0)_{\sigma}^a
e^{-b_1\vec\nu\cdot\,\vec a}-A_2e^{-b_2\vec\nu\cdot\,\vec
a})^2]\,.\nonumber \ea Instead of presenting a full analysis of
this more general case we would simply like to check that we have
a metastable dS vacuum, and that the main feature of the dS
vacuum, namely the emergence of the TeV scale when the tree-level
cosmological constant is set to zero, survives when $N_f>1$. For
this purpose we need to compute the scalar potential at the
minimum with respect to the moduli $s_i$ as a function of the
meson fields $\phi_{\sigma}$.

The generalization of the equations minimizing the scalar
potential is fairly straightforward. In particular, in the limit
when the size of the associative cycle ${\rm
Im}f=\vec\nu\cdot\,\vec a$ is large, for $A=1$ the generalization
of the second equation in (\ref{e47}), which determines $\tilde
L_{1,+}$ takes on the following form \be \label{e479} \frac
23\left(\tilde L_{1,+}\right)^2+\tilde
L_{1,+}+\sum_{\sigma=1}^{N_f}\frac {3{a\beta b_1}\hat y} {14 \hat
x\hat z} \left(\frac{a\beta} {(\phi_0)_{\sigma}^2\,\hat
x}+1\right)=0\, \nonumber \ee where we again defined $\beta\equiv
\frac{A_1}{A_2}\prod_{\sigma=1}^{N_f}(\phi_0)_{\sigma}^ae^{-(b_1-b_2)\vec\nu\cdot\,\vec
a}$, $\hat x\equiv \beta -1$, $\hat y\equiv b_1\beta -b_2$ and
$\hat z\equiv b_1^2\beta -b_2^2$. Thus, in the large three-cycle
limit we again have $\beta\approx b_2/b_1=P/Q$ so that $\hat
y\rightarrow 0$ and the leading order solution for
$\tilde{L}_{1,+}$ is again given by \be\label{lplus}
\tilde{L}_{1,+}\approx -\frac 3 2\,. \ee In this case the moduli
are stabilized at the same values given by (\ref{app24}). Since
both the superpotential and the K\"{a}hler potential are
completely symmetric with respect to the meson fields, it seems
reasonable to expect that there is a vacuum where all
$\phi_{\sigma}$ are stabilized at the same value, i.e.
$(\phi_0)_{\sigma}=\tilde\phi_0$ for all
$\sigma=\overline{1,N_f}$. Using the solution for the moduli vevs
(\ref{app24}) and the above assumption we obtain the following
expression for the potential at the extremum with respect to the
moduli $s_i$ as a function of $\tilde\phi_0$ \be\label{po4590}
V_0=N_f\frac {(A_2\hat x)^2}{64\pi V_X^3}\left[\tilde\phi_0^4+
\left(\frac {2\,a\beta}{\hat x}-\frac
3{N_f}\right)\tilde\phi_0^2+\left(\frac {a\beta}{\hat
x}\right)^2\right]\frac
{e^{N_f\tilde\phi_0^2}}{\tilde\phi_0^2}\left(\frac
{A_1Q}{A_2P}\right)^{-\frac{2P}{Q-P}}\,. \ee

By setting the discriminant of the biquadratic polynomial in the
square brackets to zero we again obtain the leading order
condition on the tree-level cosmological constant to vanish:
\begin{equation}\label{e800}
\frac 3{N_f}-\frac 8{Q-P}-\frac{28}{P\ln\left(\frac{A_1Q}{A_2P}\right)}=0\,.
\end{equation}
Since the solutions for the moduli in the dS case correspond to branch b) where $Q>P$
and ${A_1Q}>{A_2P}$, zero vacuum energy condition (\ref{e800}) can be satisfied only when
\be
\frac 3{N_f}>\frac 8{Q-P}\,\,\,\,\Rightarrow\,\,\,\, (Q-P)>\frac 8 3N_f\,.
\ee
Therefore, a vanishing tree-level cosmological constant in the leading order
results in the following set of conditions:
\be\label{fr765}
{P\ln\left(\frac{A_1Q}{A_2P}\right)}=\frac{28(Q-P)N_f}{3(Q-P)-8N_f}\,,\,\,\,\,{\rm and}\,\,\,\,\,(Q-P)>\frac 8 3N_f\,.
\ee
Recall that the key to obtaining the TeV scale gravitino mass was the exponential
suppression factor $e^{-28}$ when $Q-P=(Q-P)_{min}=3$, related to the scale of gaugino condensation.
In the present case, up a factor of order one, we have
\be
m_{3/2}\sim \frac {m_p}{V_X^{3/2}}\left(\frac{A_1Q}{A_2P}\right)^{-\frac P{Q-P}}=
\frac {m_p}{V_X^{3/2}}e^{-\frac{28 N_f}{3(Q-P)-8N_f}}
\ee
Consider a few examples where $N_f>1$. From (\ref{fr765}) we have the following set
\ba\label{exfl}
&&N_f=2,\,\,\,\,\,(Q-P)_{min}=6,\,\,\,\,\,{P\ln\left(\frac{A_1Q}{A_2P}\right)}=
168,\,\,\,\,\,\,\,\,m_{3/2}\sim\frac {m_p}{V_X^{3/2}}e^{-\frac{28 N_f}{3(Q-P)-8N_f}}=
\frac {m_p}{V_X^{3/2}}e^{-{28}} \\
\nonumber\\
&&N_f=3,\,\,\,\,\,(Q-P)_{min}=9,\,\,\,\,\,{P\ln\left(\frac{A_1Q}{A_2P}\right)}=252,
\,\,\,\,\,\,\,\,m_{3/2}\sim\frac {m_p}{V_X^{3/2}}e^{-\frac{28 N_f}{3(Q-P)-8N_f}}=
\frac {m_p}{V_X^{3/2}}e^{-{28}}\nonumber\\
\nonumber\\
&&N_f=4,\,\,\,\,\,(Q-P)_{min}=11,\,\,\,\,\,{P\ln\left(\frac{A_1Q}{A_2P}\right)}=1232,
\,\,\,\,\,\,\,\,m_{3/2}\sim\frac {m_p}{V_X^{3/2}}e^{-\frac{28 N_f}{3(Q-P)-8N_f}}=
\frac {m_p}{V_X^{3/2}}e^{-{112}}\nonumber\\
\nonumber\\
&&N_f=4,\,\,\,\,\,\,\,\,\,\,\,\,Q-P=12,\,\,\,\,\,\,\,\,\,\,\,\,
{P\ln\left(\frac{A_1Q}{A_2P}\right)}=336,\,\,\,\,\,\,\,\,m_{3/2}
\sim\frac {m_p}{V_X^{3/2}}e^{-\frac{28 N_f}{3(Q-P)-8N_f}}=\frac {m_p}{V_X^{3/2}}e^{-{28}}\nonumber
\ea

Remarkably, in all but one cases listed above we obtain {\em the same} suppression factor
$e^{-28}\approx 7\times 10^{-13}$ which was the reason for the peak at
$m_{3/2}\sim{\cal O}(1-100) {\rm TeV}$! Note that the only example which did not fall into this range
was the third case for which the condition on the cosmological constant to vanish
was ${P\ln\left(\frac{A_1Q}{A_2P}\right)}=1232$, which is too unrealistic anyway, as
it requires either extremely large dual coxeter numbers for the gauge groups $N_c,Q\sim{\cal O}(1000)$
or an exponentially large ratio inside the logarithm.
On a similar note, as can be seen from the third entry in each line in (\ref{exfl}),
increasing the number of flavors $N_f$ even further would again require either $P,Q\,>300$ or an extremely large
ratio $\left(\frac{A_1Q}{A_2P}\right)$, which appears inside the logarithm. Therefore, limiting
our analysis to the cases with $N_f<5$ seems quite reasonable.

Recall that for $N_f=1$ the TeV scale appeared for the minimum value $(Q-P)_{min}=3$ whereas the vacua
corresponding to the higher values of $Q-P$ generally failed to satisfy the SUGRA constraint
for more generic $G_2$ manifolds with a large number of  moduli. For this reason, considering
larger values of $Q-P$ for the examples listed above is probably unnecessary. Hence, for more than
one flavor of quarks, we only need to take
\be
Q-P=3N_f\,.
\ee
Thus, given that the assumptions we made in the beginning of this subsection are reasonable,
it appears that the connection of the TeV scale SUSY breaking to the requirement that
the tree-level vacuum energy is very small is a fairly robust feature of these vacua,
independent of the number of flavors.


\subsection{Including matter in both hidden sectors}
In the previous analysis we tried to be minimalistic and included chiral matter in only
one of the hidden sectors. Due to this asymmetry, we obtained two types of solutions -
a supersymmetric AdS extremum when $P>Q$ corresponding to branch a)
and a dS minimum when $Q>P$ (when condition (\ref{e80} holds), corresponding to branch b).
Using this result it is then fairly straightforward to figure out what happens when both hidden
sectors produce F-terms due to chiral matter. For the sake of simplicity, we will again
consider the case when $N_f=1$ in both hidden sectors.
In this case, the K\"{a}hler potential is given by
\be
K = -3\ln(4\pi^{1/3}\,V_X)+\phi\bar\phi+\psi\bar\psi\,.
\ee
After minimizing with respect to the axions, the non-perturbative superpotential (up to a phase)
is given by
\ba \label{mattersuptwosectors093}
W=-A_1{\phi}^{a_1}\,e^{-\frac{2\pi}P\,{\rm Im}f}+A_2{\psi}^{a_2}\,e^{-\frac{2\pi}Q\,{\rm Im}f}\,,
\ea
where $a_1\equiv-2/P$ and $a_2\equiv-2/Q$.
We will now check to see if it is still possible to obtain SUSY extrema when both hidden sectors
have chiral matter. Setting the moduli $F$-terms
to zero we obtain \be\label{nueq02} \nu_k=\nu=-\frac {3\,PQ}{4\pi}\frac{\tilde\beta-1}{Q\tilde\beta-P}\,, \ee
where $\tilde\beta\equiv \frac{A_1{\phi}^{a_1}}{A_2{\psi}^{a_2}}
\,e^{-(\frac{2\pi}{P}-\frac{2\pi}Q)\,{\rm Im}f}$.
At the same time, setting the matter $F$-terms to zero results in
the following conditions:
\be\label{mescon01}
\left(\frac{a_1}{\phi_0^2}+1\right)\tilde\beta-1=0\,.
\ee
\be\label{mescon02}
-\frac{a_2}{\psi_0^2}+\tilde\beta-1=0\,.
\ee
Expressing $\tilde\beta$ from (\ref{nueq02}) and substituting it into
(\ref{mescon01}-\ref{mescon02}) and using the definitions for $a_1$ and $a_2$,
we obtain the following expressions for the meson field vevs:
\ba\label{phisol01}
&&\phi_0^2=\frac{2+3Q/(2\pi\nu)}{P-Q}\,,\,\\
&&\psi_0^2=\frac{2+3P/(2\pi\nu)}{Q-P}\,.
\ea
Since $\nu$ as well as both $\phi_0^2$ and $\psi_0^2$ are positive definite, we have
the following two possibilities:
\ba
&&a)\,\,\,\,\,P>Q:\,\,\,\,\,\,\Rightarrow\,\,\,\,\,\,F_{\phi}=0\,\,\,{\rm and}\,\,\,\,F_{\psi}\not=0,\,\\
&&b)\,\,\,\,\,P<Q:\,\,\,\,\,\,\Rightarrow\,\,\,\,\,\,F_{\phi}\not=0\,\,\,{\rm and}\,\,\,\,F_{\psi}=0.\,\nonumber
\ea
Thus, when both hidden sectors have chiral matter, supersymmetric extrema are absent. Instead,
when condition (\ref{e80}) holds (for branch a) we simply swap $P$ and $Q$, $A_1$ and $A_2$
in (\ref{e80})), for each branch we obtain a dS vacuum where only one of the matter $F$-terms is non-zero.
Keep in mind that although in the above analysis we used condition (\ref{nueq02})
obtained by setting the moduli $F$-terms to zero, even in the dS case when the moduli $F$-terms are non-zero,
one of the two mesons will be stabilized at a value such that the corresponding matter $F$-term is zero.
The zero $F$-term has no effect on the analysis of the dS solution and apart from replacing $\tilde\alpha$
with $\tilde\beta$ defined above, the same solution obtained previously for the dS vacuum applies.
In this case, the only difference will be in the meson field vevs:
\ba
a)\,\phi_0^2\approx\frac 2{P-Q}+\frac 7{P\,\ln\left(\frac{A_2P}{A_1Q}\right)},\,
\psi_0^2\approx1-\frac 2{P-Q}+\sqrt{1-\frac 2{P-Q}}-\frac
7{Q\,\ln\left(\frac{A_2\,P}{A_1 Q}\right)}\left(\frac 32+\sqrt{1-\frac 2{P-Q}}\right)\,,&&\nonumber\\
\\
b)\,\psi_0^2\approx\frac 2{Q-P}+\frac 7{Q\,\ln\left(\frac{A_1Q}{A_2P}\right)},
\,\phi_0^2\approx1-\frac 2{Q-P}+\sqrt{1-\frac 2{Q-P}}-\frac
7{P\,\ln\left(\frac{A_1\,Q}{A_2 P}\right)}\left(\frac 3
2+\sqrt{1-\frac 2{Q-P}}\right)\,.&&\nonumber
\ea
Therefore, the dS solution obtained for the minimal case when
only one of the hidden sectors has chiral matter does not change
even when we include chiral matter in both hidden sectors.

\subsection{Height of the Potential barrier}

For simplicity, we first compute the height of the potential
barrier for the case with a pure SYM hidden sector. The two
solutions for $A=1$ in eq.(50) and eq.(51) exist for any number of
moduli and therefore the analysis below extends to the general
case with an arbitrary number of moduli. The solution in eq.(51)
corresponds to the dS maximum which determines the height of the
barrier. Using eq.(44), from eq.(51) we have
\begin{equation}\label{t1}
\nu^{1}_{k}=\nu=-\frac 3 7\frac x
y\left(\frac{3+\sqrt{9-7w}}w\right)\,.
\end{equation}
Therefore, we can express the volume of the associative cycle
$Vol(Q)\equiv\vec\nu\cdot\vec a$ as
\begin{equation}\label{t2}
\vec\nu\cdot\vec a=-\frac x y\left(\frac{3+\sqrt{9-7w}}w\right)\,,
\end{equation}
where we used the fact that $\nu^1_k=\nu$ in (\ref{t1}) is
independent of $k$ and $\sum_{i=1}^N a_i=\frac 7 3$. Using the
definitions of $x$, $y$, $z$, and $w$ in eq.(37) in terms of
$\alpha$, from (\ref{t2}) we can solve for $\alpha$
\begin{equation}\label{t3}
\alpha=\frac{P^2\left(7Q^2+12\pi Q\vec\nu\cdot\vec
a+4\pi^2\left(\vec\nu\cdot\vec
a\right)^2\right)}{Q^2\left(7P^2+12\pi P\vec\nu\cdot\vec
a+4\pi^2\left(\vec\nu\cdot\vec a\right)^2\right)}\,.
\end{equation}
From eq.(36) we can express $\vec\nu\cdot\vec a$ as
\begin{equation}\label{t4}
\vec\nu\cdot\vec a=\frac
1{2\pi}\frac{PQ}{P-Q}\ln\left(\frac{A_2}{A_1}\alpha\right)\,.
\end{equation}
In the limit when the volume of the associative cycle
$\vec\nu\cdot\vec a$ is large we can solve (\ref{t3}) and
(\ref{t4}) order by order to obtain
\begin{equation}\label{t5}
\alpha\approx\frac{P^2}{Q^2}\left(1+\frac{6(P-Q)^2}{PQ\ln\left(\frac{A_2P^2}{A_1Q^2}\right)}
+\frac{(29P-7Q)(P-Q)^3}{P^2Q^2\ln^2\left(\frac{A_2P^2}{A_1Q^2}\right)}\right)
\end{equation}
together with
\begin{equation}\label{t6}
\vec\nu\cdot\vec
a\approx\left(\frac{PQ\ln\left(\frac{A_2P^2}{A_1Q^2}\right)}{2\pi(P-Q)}
-\frac{3(P-Q)}{\pi\ln\left(\frac{A_2P^2}{A_1Q^2}\right)}+\frac{11(P-Q)^2(P+Q)}{2\pi
PQ\ln^2\left(\frac{A_2P^2}{A_1Q^2}\right)}\right)\,.
\end{equation}
The moduli vevs at the barrier are then given by
\begin{equation}\label{t7}
s_i\approx\frac{3a_i}{7N_i}\left(\frac{PQ\ln\left(\frac{A_2P^2}{A_1Q^2}\right)}{2\pi(P-Q)}
-\frac{3(P-Q)}{\pi\ln\left(\frac{A_2P^2}{A_1Q^2}\right)}+\frac{11(P-Q)^2(P+Q)}{2\pi
PQ\ln^2\left(\frac{A_2P^2}{A_1Q^2}\right)}\right)\,.
\end{equation}
In the leading order
\begin{equation}\label{t8}
s_i\approx\frac{3a_i}{7N_i}\left(\frac{PQ\ln\left(\frac{A_2P^2}{A_1Q^2}\right)}{2\pi(P-Q)}\right)\,
\end{equation}
and
\begin{equation}\label{t9}
\alpha\approx\frac{P^2}{Q^2}\,.
\end{equation}
Using (\ref{t8}) and (\ref{t9}), the value of the potential at the
barrier in the leading order is given by
\begin{eqnarray}\label{t10}
V_{b}&\approx&m_p^4\frac{8A_2^2\pi^6}{7Q^4}\left(7(P^2-Q^2)^2+PQ\ln\left(\frac{A_2P^2}{A_1Q^2}\right)\left(
7(P^2-Q^2)+PQ\ln\left(\frac{A_2P^2}{A_1Q^2}\right)\right)\right)\nonumber\\
&\times&\left(\frac{PQ}{P-Q}\ln\left(\frac{A_2P^2}{A_1Q^2}\right)\right)^{-7}\left(\frac{A_2P^2}{A_1Q^2}\right)^{-\frac{2P}{P-Q}}\prod_{i=1}^N\left(\frac{7N_i}{3a_i}\right)^{3a_i}\,.
\end{eqnarray}
Recall that the value of the gravitino mass at the SUSY AdS
extremum in the leading order is given by
\begin{equation}\label{t11}
m_{3/2}=m_p\sqrt{2}\pi^3A_2P\left|\frac{P-Q}{PQ}\right|
\left(\frac{PQ}{P-Q}\ln\left(\frac{A_2P}{A_1Q}\right)\right)^{-7/2}\left(\frac{A_2P}{A_1Q}\right)^{-\frac{P}{P-Q}}\prod_{i=1}^N\left(\frac{7N_i}{3a_i}\right)^{\frac{3a_i}2}\,.
\end{equation}
Therefore, we can express the value of the potential at the
barrier in the leading order as
\begin{eqnarray}\label{t12}
V_b&\approx&m_p^2m_{3/2}^2\left(7(P^2-Q^2)^2+PQ\ln\left(\frac{A_2P^2}{A_1Q^2}\right)\left(
7(P^2-Q^2)+PQ\ln\left(\frac{A_2P^2}{A_1Q^2}\right)\right)\right)\nonumber\\
&\times&\frac
{4}{7Q^2(P-Q)^2}\left(\frac{P}{Q}\right)^{-\frac{2P}{P-Q}}\left(\frac{\ln\left(\frac{A_2P}{A_1Q}\right)}{\ln\left(\frac{A_2P^2}{A_1Q^2}\right)}\right)^7\,.
\end{eqnarray}
Note that the above expression is independent of the number of
moduli and parameters $a_i$ and $N_i$.
\begin{figure}[h!]
    \begin{tabular}{cc}
      \leavevmode \epsfxsize 7 cm \epsfbox{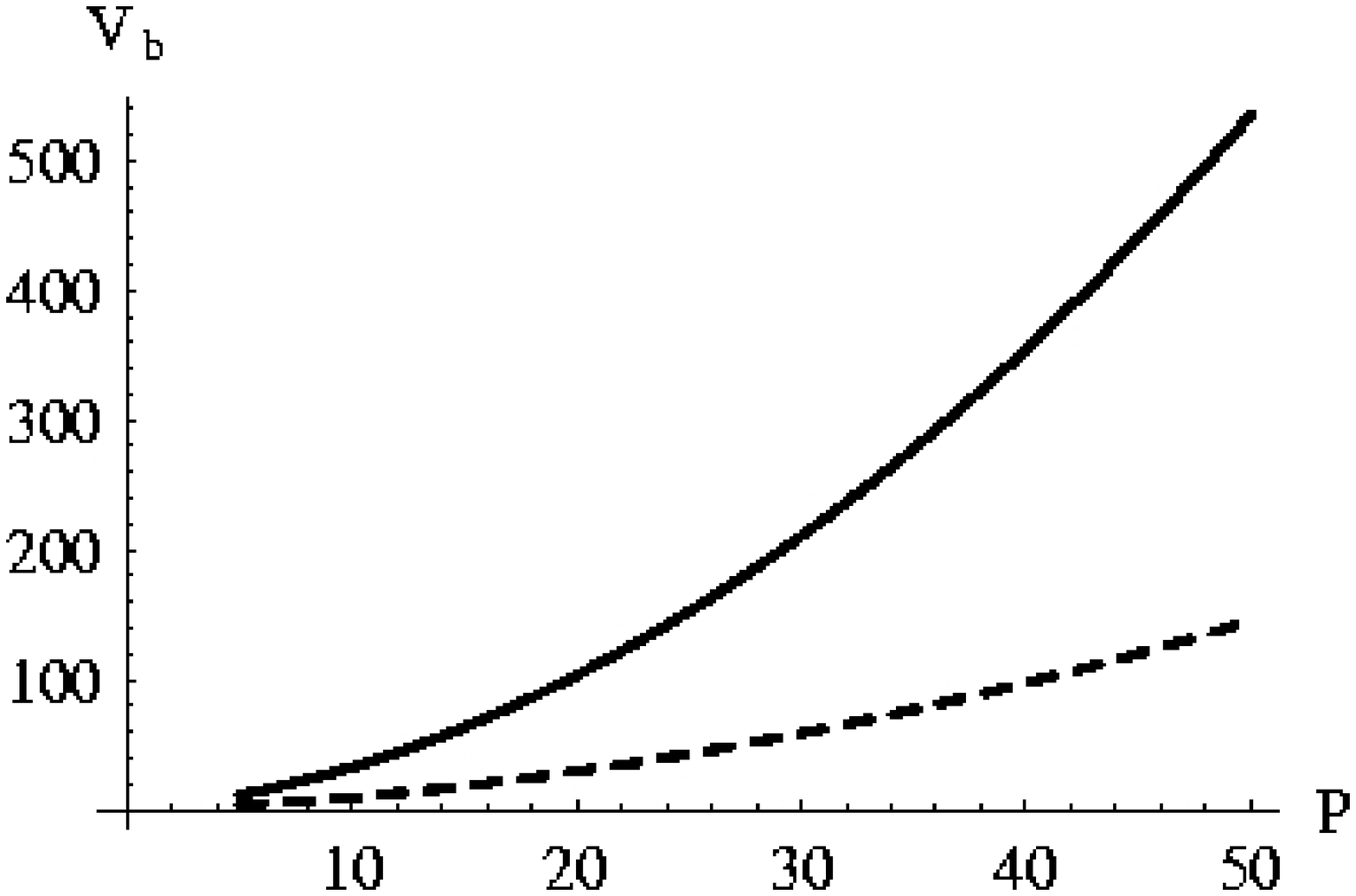}&
      \leavevmode \epsfxsize 7 cm \epsfbox{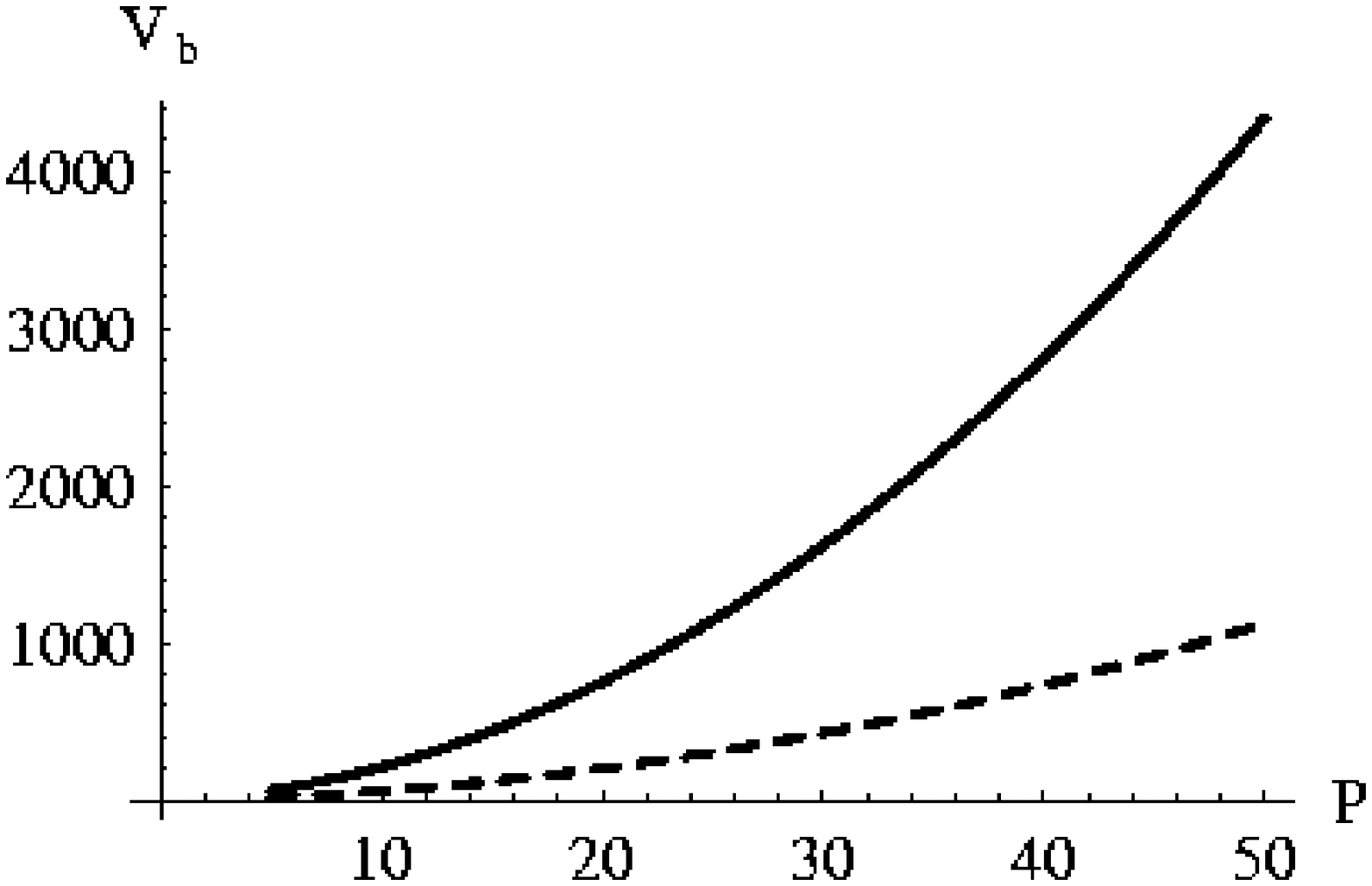}\\
    \end{tabular}
    \caption{\footnotesize Plots of the potential at the barrier in units of $m_p^2m_{3/2}^2$
as a function of P. Solid line corresponds to
$\frac{A_2P}{A_1Q}=100$ while the dashed line corresponds to
$\frac{A_2P}{A_1Q}=10$. Left: Plot for $P=Q+3$. Right:  Plot for
$P=Q+1$.}
    \label{Plot810}
\end{figure}
This formula is very accurate when compared to the exact numerical
values. For example, for the values of the parameters in eq.(54),
the numerically obtained result is
\begin{equation}\label{t13}
V_b\approx 51.55\times m_p^2m_{3/2}^2\,,
\end{equation}
whereas from the leading order expression in (\ref{t12}) we get
\begin{equation}\label{t13}
V_b\approx 49.92\times m_p^2m_{3/2}^2\,.
\end{equation}
\begin{figure}[h!]
      \leavevmode \epsfxsize 13 cm \epsfbox{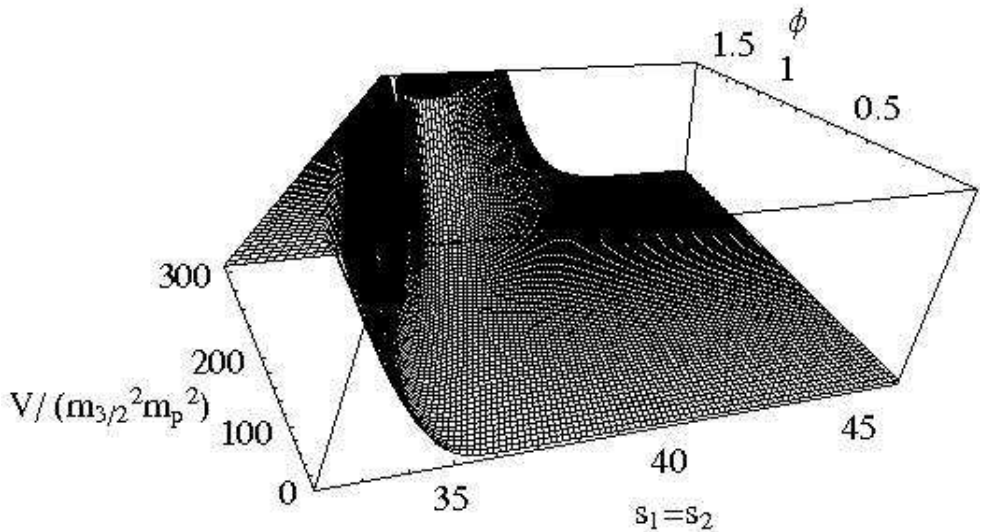}
    \caption{\footnotesize Potential in units of $m_{3/2}^2m_{p}^2$ plotted as a function of the meson field and the moduli along the slice
    $s_1=s_2$, for a manifold with two moduli and the microscopic constants as in (\ref{setdS}).} \label{ds-plot2}
\end{figure}
\begin{figure}[h!]
    \begin{tabular}{cc}
      \leavevmode \epsfxsize 8 cm \epsfbox{potential-dS_gr1.eps}&
      \leavevmode \epsfxsize 8 cm \epsfbox{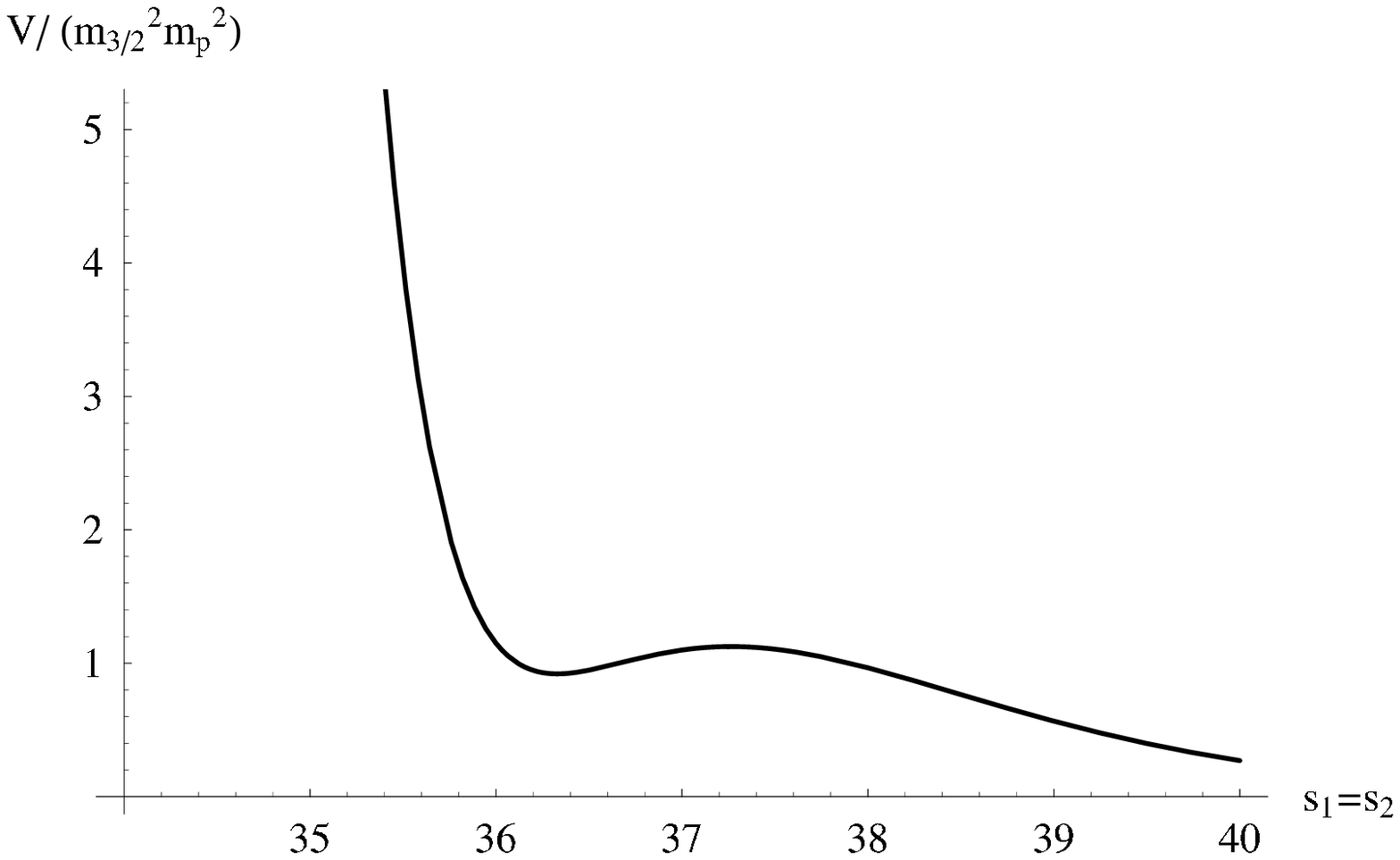}
    \end{tabular}
    \caption{\footnotesize Left: Potential in units of $m_{3/2}^2m_{p}^2$ along the slice $s_1=s_2$ for a manifold with two moduli with the
    meson field equal to its value at the minimum of the potential (as in (\ref{exact23})). Right: Potential in units of $m_{3/2}^2m_{p}^2$
    along the slice $s_1=s_2$ for a manifold with two moduli with the meson field $\phi=0.102$ (this is such that the height of the potential barrier
    is at its minimum). The microscopic constants for both cases are the same as in (\ref{setdS}).} \label{ds-plot3}
\end{figure}
In Figure \ref{Plot810} below we plotted the value of the scalar
potential at the barrier in units of $m_p^2m_{3/2}^2$ as a
function of P for $P-Q=3$ and $P-Q=1$ and two different values of
parameter $\rho\equiv\frac{A_2P}{A_1Q}$.

For the case of vacua with charged matter in the hidden sector,
the situation is more complicated as the potential depends on an
additional field (meson). The height of the potential barrier
changes as one moves along the meson and the moduli directions. To
illustrate this with an example, we have done a numerical analysis
for a manifold with two moduli and one meson field with the
microscopic constants as in (\ref{setdS}). Figure \ref{ds-plot2}
shows a three-dimensional plot of the potential as a function of
the moduli along the slice $s_1=s_2$, and the meson field $\phi$.
In figure \ref{ds-plot3}, in the left plot the potential is
plotted along the slice $s_1=s_2$ with the meson field equal to
its value at the minimum (\ref{exact23}), while in the right plot
the potential is plotted along the slice $s_1=s_2$ with the meson
field equal to a value such that the height of the potential
barrier is at its minimum.


\section{Phenomenology}\label{pheno}

In this section, we will begin the analysis of more detailed particle physics
features of the vacua, with emphasis on the soft supersymmetry breaking parameters, since we
are particularly interested in predicting collider physics
observables that will be measured at the LHC.

The low energy physics observables are determined by
the k\"{a}hler potential, superpotential and the gauge kinetic
function of the effective $\mathcal{N}=1,d=4$ supergravity. The
gauge kinetic function ($f$) has already been discussed. The
K\"{a}hler potential and the superpotential can be written in
general as follows: \ba \label{eq:KW} K &=& \hat{K}(s_i,{\phi}_h,\bar{\phi}_h) +
\tilde{K}_{\bar{\alpha}\beta}(s_i) \,\bar{\Phi}^{\bar{\alpha}}
\Phi^{\beta} + Z_{\alpha\beta}(s_i,{\phi}_h)\,{\Phi}^{\alpha}{\Phi}^{\beta} + ... \\
 W &=& \hat{W}(z_i) +
{\mu}'{\Phi}^{\alpha}{\Phi}^{\beta} + Y'_{\alpha\beta\gamma}\,
{\Phi}^{\alpha}{\Phi}^{\beta}{\Phi}^{\gamma} + ...\nonumber \ea

\noindent where ${\Phi}^{\alpha}$ are the visible sector chiral
mater fields, $\tilde{K}_{\bar{\alpha}\beta}$ is their K\"{a}hler
metric and $Y'_{\alpha\beta\gamma}$ are their \emph{unnormalized}
Yukawa couplings. ${\phi}^h$ denote the hidden sector matter
fields. The first terms in $K$ and $W$ depend only on the bulk
moduli and have been already studied earlier. In general there can
be a mass term (${\mu}'$) in the superpotential, but as explained
in \cite{decon}, natural discrete symmetries can exist which
forbid it, in order to solve the doublet-triplet splitting
problem. The quantity $Z_{\alpha\beta}$ in the K\"{a}hler
potential will be important for generating an effective $\mu$ term
as we will see later.

Since the vacua have low scale supersymmetry, the effective
lagrangian must be equivalent to the MSSM plus couplings involving
possibly additional fields beyond the MSSM. For simplicity in this
section we will assume an observable sector which is precisely the
MSSM, although it would also be interesting to consider natural
$M$ theoretic extensions. The MSSM lagrangian is characterized by
the Yukawa and gauge couplings of the standard model and the soft
supersymmetry breaking couplings. These are the scalar squared
masses $m_i^2$, the trilinear couplings $A_{ijk}$, the $\mu$ and
$B\mu$ mass parameters and the gaugino masses. In $M$ theory all
of these couplings become functions of the various constants
$(A_i, N, P, Q, N_k)$ which are determined by the {\it particular}
$G_2$-manifold $X$. In addition, because we are now discussing the
observable sector, we have to explain the origin of observable
sector gauge, Yukawa and other couplings in $M$ theory. As we have
already explained, all gauge couplings are integer linear
combinations of the $N$ moduli, the $N$ integers determining the
homology class of the three dimensional subspace of $X$ which
supports that particular gauge group. Furthermore, the entire
superpotential is generated by membrane instantons, as we have
already discussed. Therefore mass terms and Yukawa couplings in
the superpotential are also determined by integer linear
combinations of the moduli fields. Hence, in addition to the
constants $(A_i, N, P, Q, N_k)$ which determine the moduli
potential, additional integers enter in determining the observable
sector superpotential. Generically, though, we do not expect these
integers to be large in the basis that the moduli K\"{a}hler
metric is given by $(6)$.

We determine the values of the soft SUSY breaking couplings at
$M_{unif}$ in the standard way : The moduli fields, hidden sector
matter fields as well as their auxiliary fields are replaced by
their {\it vevs} in the $\mathcal{N}=1,d=4$ SUGRA lagrangian. One
then takes the flat limit $M_p \rightarrow \infty$ with $m_{3/2}$
fixed. This gives a global SUSY lagrangian with soft SUSY breaking
terms \cite{Nilles:1983ge}. Unfortunately, in $M$ theory the
matter K\"{a}hler potential is difficult to compute. This leads to
theoretical uncertainties in the calculation of the scalar masses
and $A$, $B$ and $\mu$ parameters. Fortunately though we are able
to calculate the gaugino masses. Our main phenomenological result
is that the tree level gaugino masses are suppressed relative to
the gravitino mass. After explaining this, we will go on to
discuss the other soft terms in a certain, calculable limit.

\subsection{Suppression of Gaugino masses}\label{treegaugino}
Grand Unification is particularly natural in $G_2$ vacua of $M$ theory \cite{Friedmann:2002ty}.
This implies that the gaugino masses at tree level (at the unification scale)
are \emph{universal}, i.e. the gauginos of the three SM gauge groups
have the same mass. In order to compute the SM sector gaugino mass
scale at tree-level we need
the Standard Model gauge kinetic function, $f_{sm}$. In general this will be an integer
linear combination of the moduli, with integers $N^{sm}_i$,
which is linearly
independent of the hidden sector gauge kinetic function in general. The expression for
the tree-level MSSM gaugino masses in general $\mathcal{N}=1,d=4$ SUGRA is
given by: \be\label{ga1}
M_{1/2}=m_p\frac{e^{\hat K/2}\hat{K}^{n\bar{m}}F_{\bar{m}}\partial_n\,f_{sm}}{2{i\,\rm
Im}f_{sm}}\,, \ee
Note that the gauge kinetic
function is independent of the hidden sector matter fields. Therefore, the large hidden sector
matter $F$ term responsible for the dS minimum does not contribute to the gaugino masses at tree
level. We will now proceed to evaluate this expression explicitly both for the
AdS and dS vacua. We will find that generically, the gaugino masses are suppressed
relative to the gravitino mass.

\subsubsection{Gaugino masses in AdS Vacua}

Choosing the hidden sector to be pure SYM with gauge
groups $SU(P)$ and $SU(Q)$, the normalized gaugino mass in these
compactifications can be expressed as \be M_{1/2}=-\frac
{m_p\,e^{-i\gamma_W}}{8\sqrt{\pi}V_X^{3/2}}\left[\frac {4\pi}
3\left(\frac{A_1}Pe^{-{\frac{2\pi}{P}{\rm
Im}f}}-\frac{A_2}Qe^{-{\frac{2\pi}{Q}{\rm
Im}f}}\right)\frac{\sum_{i=1}^{N}{N^{sm}_is_i\nu_i}}{\sum_{i=1}^{N}{N^{sm}_is_i}}+A_1e^{-{\frac{2\pi}{P}{\rm
Im}f}}-A_2e^{-{\frac{2\pi}{Q}{\rm Im}f}}\right] \ee where
$\gamma_W$ is the phase of the superpotential $W$. In the leading
order, the last two terms in the brackets can be combined as
\be\label{sup} A_1e^{-{\frac{2\pi}{P}{\rm
Im}f}}-A_2e^{-{\frac{2\pi}{Q}{\rm
Im}f}}=A_2\left[\frac{P-Q}Q\right]
\left[\frac{A_2P}{A_1Q}\right]^{-\frac P{P-Q}}\,. \ee On the other
hand, the two terms in the round brackets coming from the partial
derivative of the superpotential cancel in the leading order.
Therefore we need to take into account the first subleading order
contribution (\ref{eq46}). In this order, we obtain \footnote{Recall that
$\rm{Im}(f)_A^{(c)}\equiv{\cal T}^{\,(c)}_A$ (see (\ref{deft})).}:
\be\label{part} \left(\frac{A_1}Pe^{-{\frac{2\pi}{P}{\rm
Im}f}}-\frac{A_2}Qe^{-{\frac{2\pi}{Q}{\rm Im}f}}\right) =\frac
1{2\pi}A_2\left[\frac{P-Q}Q\right]
\left[\frac{A_2P}{A_1Q}\right]^{-\frac
P{P-Q}}\frac{B_A^{(c)}}{{\cal T}^{\,(c)}_A}\,. \ee From
(\ref{sup}) and (\ref{part}) we notice that the absolute value of
gaugino mass can now be conveniently expressed in terms of the
gravitino mass (for a given value of $A$ and $c$, as discussed in
previous sections) as \be\label{ga2} |M_{1/2}|^{(c)}_A=\frac2
3\frac{\sum_{i=1}^{N}a_i\,L_{A,\,k}^{(c)}
\left(L_{A,\,k}^{(c)}+3/
2\right)({N^{sm}_i}/{N_i})}{\sum_{i=1}^{N}a_i\,L_{A,\,k}^{(c)}
({N^{sm}_i}/{N_i})}\times (m_{3/2})^{(c)}_A\,, \ee where we also
used (\ref{sol34}) and (\ref{eq43}). Finally, using (\ref{lk}) and
the first equation in (\ref{e20}), after some algebra we arrive at
the following expression for the gaugino mass:
\be\label{gauginomass} |M_{1/2}|^{(c)}_A=\left(\frac 4
3{T}^{(c)}_A+1\right)\frac{q-A}{q+\frac{{T}^{(c)}_A}{H^{(c)}_A}}\times
(m_{3/2})^{(c)}_A\,, \ee where we have introduced a new quantity
\be\label{q}
q=\frac{\sum_{i=1}^{N}m_ia_i({N^{sm}_i}/{N_i})}{\sum_{i=1}^{N}a_i({N^{sm}_i}/{N_i})}\,,
\ee such that the range of possible values for $q$ is \be -1\leq
q\leq 1\,. \ee

Note that the general formula
(\ref{gauginomass}) which relates the gravitino and gaugino masses
is completely independent of the number of moduli.

\begin{figure}[h!]
 \begin{tabular}{cc}
      \leavevmode \epsfxsize 9 cm \epsfbox{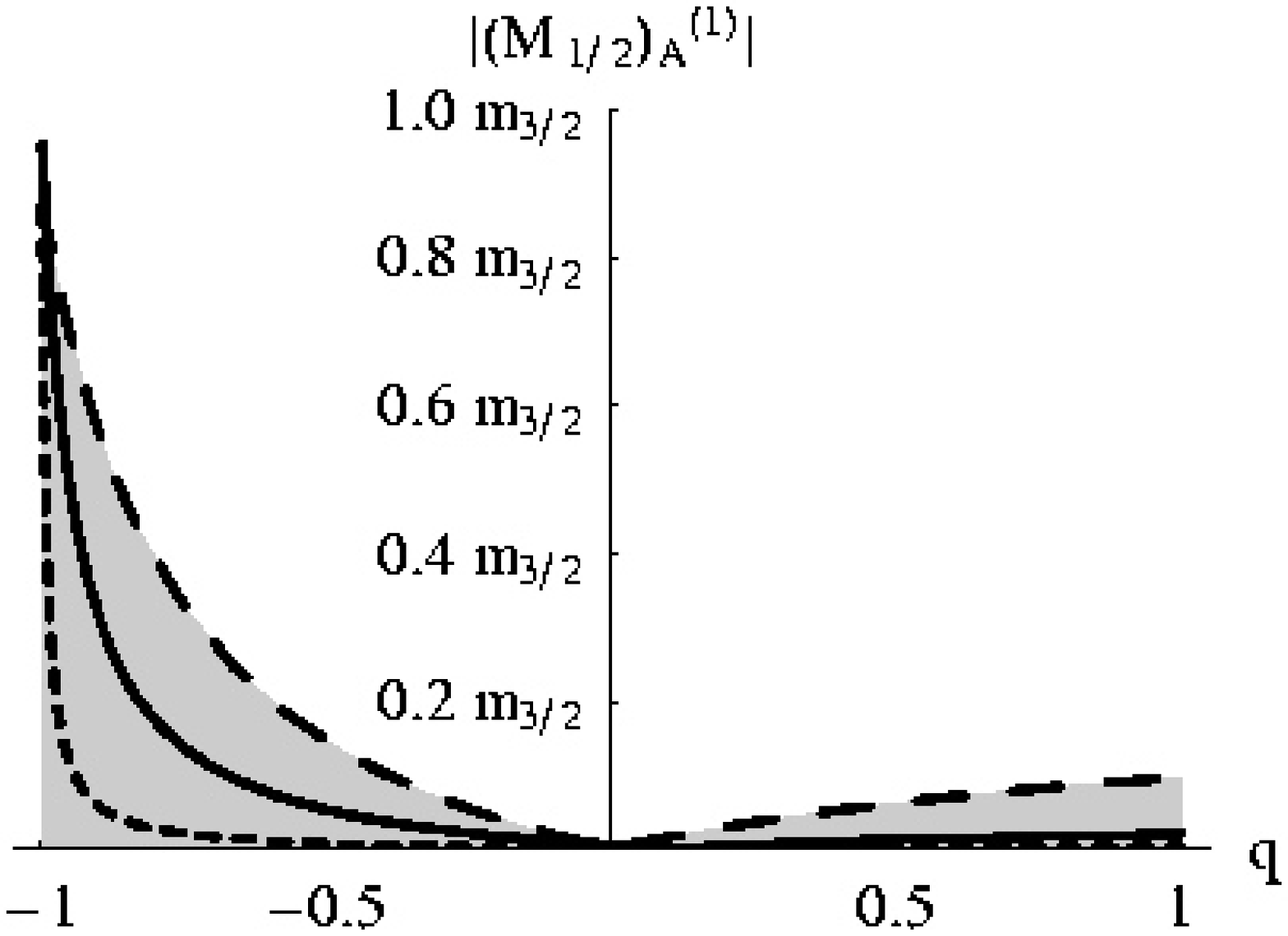}&
      \leavevmode \epsfxsize 9 cm \epsfbox{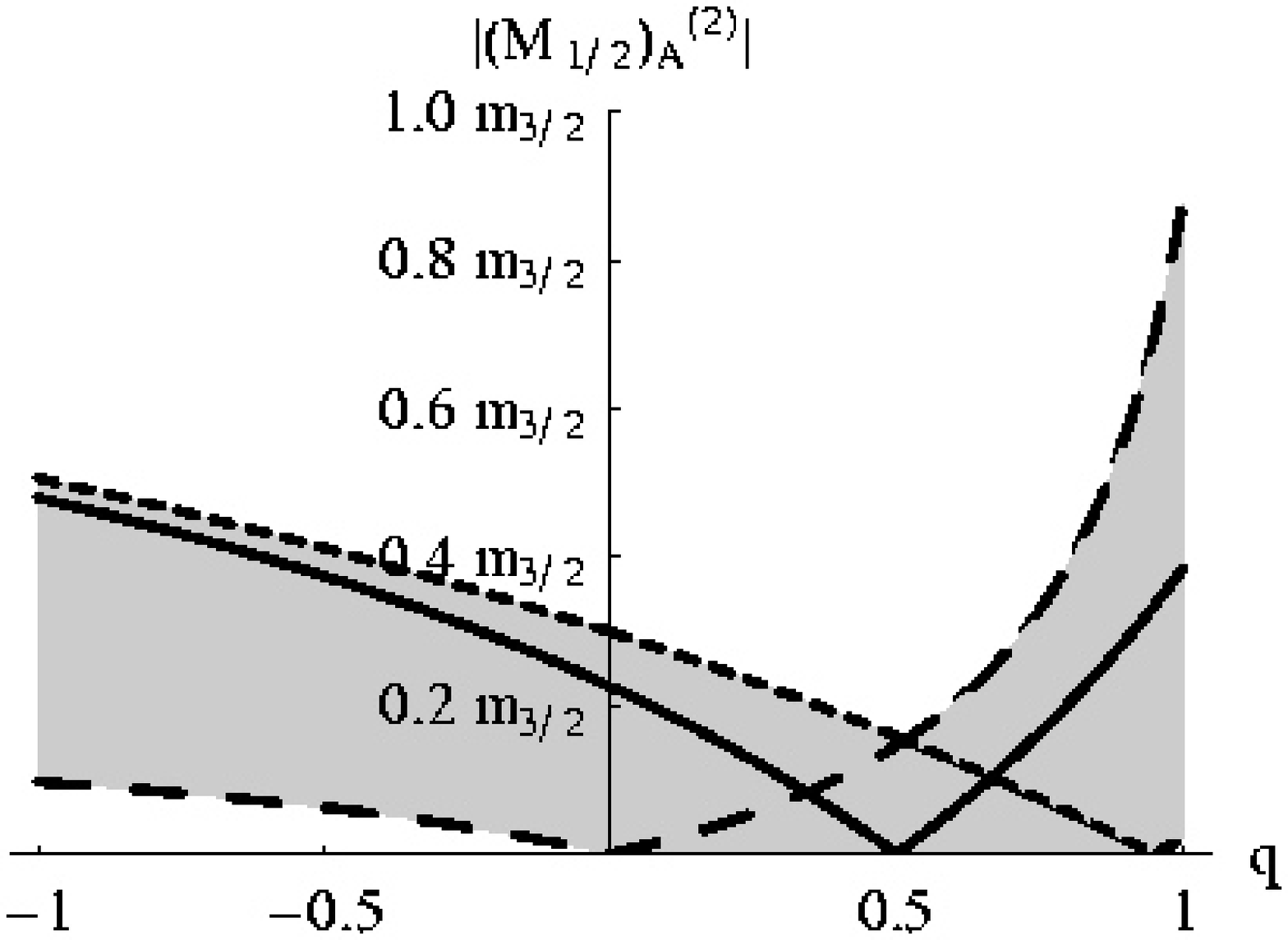} \\
    \end{tabular}
\caption{ \footnotesize Absolute values of
$(M_{1/2})^{(1)}_A$-left and $(M_{1/2})^{(2)}_A$ -right in units
of gravitino mass as functions of $q$. As parameter $A$ varies
over $0\leq A <1/7$ - on the
left and $0\leq A\leq 1$ - on the right and  the whole light grey region is covered.
\newline
Left plot: $A=0$- long dashed line, $A=1/9$ - solid line,
$A=5/36$ - short dashed line.
\newline
Right plot: $A=0$- long dashed line, $A=0.5$ - solid line, $A=0.95$
- short dashed line.} \label{twomoduliplot20}
\end{figure}

When all $m_k$ have the same sign the gaugino mass in
(\ref{gauginomass}) automatically vanishes. This is expected since
the solution when $A=\pm 1$ is the SUSY extremum. In Figure
{\ref{twomoduliplot20}} we have plotted absolute values of
$(M_{1/2})^{(1)}_A$ and $(M_{1/2})^{(2)}_A$ as functions of $q$.

For a significant fraction of the space in both plots we have
$(M_{1/2})^{(1,2)}_A\leq 0.2\,(m_{3/2})^{(1,2)}_A$, so
the gaugino masses are typically suppressed compared to the gravitino mass for these
AdS vacua. Note also that the suppression factor in (\ref{gauginomass}) is independent of
the gravitino mass. This result is different from the small
hierarchy between $M_{1/2}$ and $m_{3/2}$ in the Type IIB flux
vacua \cite{Conlon:2006us}, where the gaugino mass is generically
suppressed by ${{\rm ln}\left(m_{3/2}\right)}$.

For the special case $A=0$, system (\ref{e20}) yields two
solutions with positive moduli. Therefore, there will be two
different values for the gaugino mass corresponding to these
solutions. After some algebra we obtain
\begin{eqnarray}\label{specialgaugino}
|M_{1/2}|^{(1,2)}_0&&=\left(\frac{5-{\sqrt{17}}}{4}\right)\left|\frac
q{q\pm\frac{-9+\sqrt{17}}{\sqrt{-26+10\sqrt{17}}}}\right|
\times (m_{3/2})^{(1,2)}_0 \nonumber\\
&&\sim 0.22\,\left|\frac q{q\mp 1.25}\right|\times
(m_{3/2})^{(1,2)}_0\,.
\end{eqnarray}
Again, this relation is valid for any AdS vacuum with broken SUSY
with $A=0$ with an arbitrary number of moduli.

To check the accuracy of the approximate gaugino mass formula we
again try the special case with two moduli $a_1=a_2=7/6$ with the
same choice of the constants as in (\ref{choice1}) and the
integer combination for the Standard Model gauge kinetic function
$\{N^{sm}_1=2,N^{sm}_2=1\}$. In this case equation
(\ref{specialgaugino}) for the absolute value of $M_{1/2}$ yields:
\be (M_{1/2})^{(1)}_0=164.4 \,{\rm
GeV},\,\,\,\,\,(M_{1/2})^{(2)}_0=95 \,{\rm GeV}\,, \ee whereas the
exact values computed numerically for the same choice of constants
are: \be (M_{1/2})^{(1)}_0=165.4\, {\rm
GeV},\,\,\,\,\,(M_{1/2})^{(2)}_0=97 \,{\rm GeV}\,. \ee This
demonstrates a high degree of accuracy of our approximation,
similar to that for the gravitino mass.


\subsubsection{Gaugino masses in dS Vacua}

From the formula for the gaugino mass in (\ref{ga1}), the gaugino
mass for the dS vacua in general can be expressed as \ba
\label{e59} M_{1/2} &=&
\frac{e^{-i{\gamma}_W}m_p\,e^{\phi_0^2/2}}{8\sqrt{\pi}V_X^{3/2}}\left[\frac
{2} 3\,\tilde
y\,\frac{\sum_{i=1}^{N}{N^{sm}_is_i\nu_i}}{\sum_{i=1}^{N}{N^{sm}_is_i}}+\tilde
x\right]A_2{e^{-b_2\vec\nu\cdot\vec a}} \nonumber \\
\implies M_{1/2} &=& -e^{-i{\gamma}_W}\left(\frac 2 3L_{1,+}
+1\right)\,m_{3/2}\,, \ea where in the second equality we used
(\ref{sol37}) and the fact that for these vacua $\nu_i=\nu$ for
all $i=\overline{1,N}$, independent of $i$. Also, by including the
minus sign we took into account that $m_{3/2}=e^{K/2}\left|\tilde
x\right|A_2{e^{-b_2\vec\nu\cdot\vec a}}$  but $\tilde x < 0$,
since $Q-P\geq3$. From (\ref{e51}) we can find $\tilde L_{1,+}$
including the first subleading contribution \be\label{e60} \tilde
L_{1,+}\approx -\frac 3 2+\frac {3{a\tilde\alpha b_1}\tilde y} {14
\tilde x\tilde z}\left(\frac{a\tilde\alpha}{\phi_0^2\,\tilde
x}+1\right)\,. \ee For $\tilde x$, $\tilde y$ and $\tilde z$ in
(\ref{e60}) we use the definitions (\ref{xyzw1}) and substitute
the approximate result (\ref{e57}) for $\tilde\alpha$. Then after
substituting(\ref{e60}) into (\ref{e59}) and assuming that
$Q-P\sim {\mathcal O}(1)$, in the limit when $P$ is large the
approximate tree level MSSM gaugino mass is given by
\be\label{gaugino33} M_{1/2} \approx -\frac {e^{-i{\gamma}_W}}
{P\,\ln\left(\frac {A_1Q}{A_2P}\right)}\left(1+\frac
2{\phi_0^2\,(Q-P)}+\frac 7{\phi_0^2\,P\,\ln\left(\frac
{A_1Q}{A_2P}\right)}\right)\times {m_{3/2}}\,, \ee where we use
(\ref{e58}) to substitute for $\phi_0^2$. It is important to note
some features of the above equation. First, equation
(\ref{gaugino33}) is completely independent of the choice of
integers $N_i^{sm}$ for the Standard Model gauge kinetic function
as well as the integers $N_i$ for the hidden sector. Second, it is
independent of the number of moduli $N$ and moreover, it is also
independent of the particular details of the internal manifold
described by the rational numbers $a_i$ appearing in the
K\"{a}hler potential (\ref{vol}). These properties imply that
relation (\ref{gaugino33}) is  universal for all $G_2$ holonomy
compactifications consistent with our aproximations, independent
of many internal details of the manifold. Furthermore, the
denominator $P\ln(A_1Q/A_2P)$ turns out to be always greater than
unity for choices of microscopic parameters consistent with all
constraints such as the supergravity regime constraint and the dS
minimum constraint. In fact for reasonable choices of parameters,
it is typically of $\mathcal{O}(10-100)$. Since the expression in
the round brackets in (\ref{gaugino33}) is slowly varying and for
the range under consideration is of $\mathcal{O}(1)$, we see that
{\it gaugino masses are always suppressed relative to the
gravitino for these dS vacua}.

After one imposes the constraint equation (\ref{er765})
((\ref{ior}) when $Q-P=3$) to make the cosmological constant very
small, one can get rid of one of the constants in
(\ref{gaugino33}), and further simplify the expression for the
universal tree level gaugino mass parameters for the dS vacuum
with a very small cosmological
constant:\begin{equation}\label{gaugino36} M_{1/2}\approx-\frac
{e^{-i{\gamma}_W}} {84}\left(1+\frac 2{3\phi_0^2}+\frac
7{84\phi_0^2}\right)\times
{m_{3/2}}=-e^{-i{\gamma}_W}\frac{139+396\sqrt{3}}{34356}\times
{m_{3/2}}\approx -e^{-i{\gamma}_W}0.024\times {m_{3/2}}\,.
\end{equation} As in the more general case
(eqn.(\ref{gaugino33})), the tree level gaugino mass is suppressed
compared to the gravitino mass, and the suppression factor can
also be predicted.

One would like to understand the physical origin of the
suppression of the gaugino masses at tree level, especially for
the dS vacua which are phenomenologically relevant. As mentioned
earlier, since the matter $F$ term does not contribute to the
gaugino masses, the gaugino masses can only get contributions from
the moduli $F$ terms, which, as explained in the last paragraph in
section \ref{unique}, vanish in the leading order of our
approximation. The first subleading contribution is suppressed by
the inverse power of the volume of the associative three-cycle,
causing the gaugino masses to be suppressed relative to the
gravitino. Since the inverse volume of this three-cycle is
essentially $\alpha_{hidden}$ - the hidden sector gauge coupling
in the UV - the suppression is due to the fact that the hidden
sector is asymptotically free. In large volume type IIB
compactifications, the moduli $F$ terms also vanish in the leading
order, leading to suppressed gaugino masses as well
\cite{Conlon:2006us}. However, in contrast to our case, there the
subleading contribution is suppressed by the inverse power of the
volume of the compactification manifold. Note that a large
associative cycle on a $G_2$ manifold does \emph{not} translate
into a large volume compactification manifold. Thus, unlike large
volume type IIB compactifications, these $M$ theory vacua have a
much higher compactification scale and hence are consistent with
standard gauge coupling unification.

\subsection{Other parameters and Flavor issues}\label{others}

The trilinears, scalars, anomaly mediated contributions to gaugino
masses and the $B\mu$ parameter depend more on the microscopic
details of the theory -- the Yukawa couplings, the ${\mu}'$
parameter and the K\"{a}hler metric for visible sector matter
fields. The flavor structure of the Yukawa matrices as well as
that of the K\"{a}hler metric for matter fields is crucial for
estimating flavor changing effects. We will comment on these at
appropriate places.

The (un-normalized) Yukawa couplings in these vacua
arise from membrane instantons which connect singularities where
chiral superfields are supported (if some singularities coincide,
there could also be order one  contributions). They are given by:
\be \label{yukawas1} Y'_{\alpha\beta\gamma} =
C_{\alpha\beta\gamma}\,e^{i2\pi\sum_i l^{\alpha\beta\gamma}_i
z^i}\ee \noindent where $C_{\alpha\beta\gamma}$ is an ${\mathcal
O}(1)$ constant and $l^{\alpha\beta\gamma}_i$ are integers.

The moduli dependence of the matter K\"{a}hler metric is
notoriously difficult to compute in generic string and $M$ theory
vacua, and the vacua under study here are no exception. The best
we can do here is to consider the Type IIA limit of these vacua.
The matter K\"{a}hler metric has been computed in type IIA
intersecting $D$6-brane vacua on toroidal orientifolds
\cite{Bertolini:2005qh} building on earlier work
\cite{Cvetic:2003ch}. Since chiral fermions living at
intersections of D6-branes lift to chiral fermions supported at
conical singularities in $M$ theory \cite{Atiyah:2001qf,
Cvetic:2001kk}, we will simply uplift the IIA calculation to $M$
theory. Thus the results of this section are strictly only valid
in the Type IIA limit.

Lifting the Type IIA result
to $M$ theory, one gets (see Appendix
for details): \ba \label{kmetric} \tilde{K}_{\bar{\alpha}\beta}
&=& {\delta}_{\bar{\alpha}\beta}\,\prod_{i=1}^n
\left(\frac{\Gamma(1-\theta^{\alpha}_i)}
{\Gamma(\theta^{\alpha}_i)} \right)^{1/2}\nonumber \\
\tan(\pi \theta^{\alpha}_i)&=&
c^{\alpha}_i\,(s_i)^l;\;\;c^{\alpha}_i = {\mathrm
constant};\;l={\mathrm rational\;number\;of\;{\mathcal O}(1)}.\ea

In the type IIA toroidal orientifolds, the underlying
symmetries always allow us to have a diagonal K\"{a}hler metric
\cite{Bertolini:2005qh}. We have assumed for simplicity the K\"{a}hler metric to
be diagonal in the analysis below. Now, we will write down the
general expressions for the physical Yukawa couplings and the soft
parameters - the trilinears and the scalars and then estimate
these in $M$ theory compactifications. The $\mu$ and $B\mu$
parameters will be discussed in section \ref{muBmu}.

The K\"{a}hler potential for the chiral matter fields is
non-canonical for any compactification in general. In determining
physical implications however, it is much simpler to work in a
basis with a canonical K\"{a}hler potential.  So, to canonically
normalize the matter field K\"{a}hler potential, we introduce the
normalization matrix $\mathcal{Q}$ : \ba \label{norm} \Phi
\rightarrow \mathcal{Q} \cdot \Phi, \;\;\; s.t. \;\;\;
\mathcal{Q}^{\dag}\tilde{K}\mathcal{Q} = 1. \ea The $\mathcal{Q}s$
are themselves only defined up to a unitary transformation, i.e.
$\mathcal{Q}'=\mathcal{Q}\cdot\mathcal{U}$ is also an allowed
normalization matrix if $\mathcal{U}$ is unitary. If the
K\"{a}hler metric is already diagonal
($\tilde{K}_{\bar{\alpha}\beta}=\tilde{K}_{\alpha}\delta_{\bar{\alpha}\beta}$),
the normalization matrix can be simplified :
$\mathcal{Q}_{\bar{\alpha}\beta}=(\tilde{K}_{\alpha})^{-1/2}\delta_{\bar{\alpha}\beta}$.
The normalized (physical) Yukawa couplings are
\cite{Nilles:1983ge}: \ba \label{ymu} Y_{\alpha\beta\gamma} &=&
e^{\hat{K}/2}\,\frac{\hat{W}^{\star}}{|\hat{W}|}Y'_{{\alpha}'{\beta}'{\gamma}'}\,
\mathcal{Q}_{{\alpha}'\alpha}\mathcal{Q}_{{\beta}'\beta}\mathcal{Q}_{{\gamma}'\gamma}\ea

It was shown in \cite{King:2004tx} that in the class of theories
with a hierarchical structure of the un-normalized Yukawa
couplings (in the superpotential), the K\"{a}hler corrections to
both masses and mixing angles of the SM particles are subdominant.
Since in these compactifications, it is very natural to obtain a
hierarchical structure of the un-normalized Yukawa couplings due
to their exponential dependence on the various moduli and also
because of some possible family symmetries, therefore one expects
the effects of the K\"{a}hler corrections which are less under
control, to be subdominant. The expressions for the
\emph{un-normalized} trilinears and scalar masses are given by
\cite{Nilles:1983ge}: \label{m'A'} \ba m'^2_{\bar{\alpha}\beta}
&=& (m_{3/2}^2 + V_0)\,\tilde{K}_{\bar{\alpha}\beta} -
e^{\hat K}F^{\bar{m}}(\partial_{\bar{m}}\partial_n\,\tilde{K}_{\bar{\alpha}\beta}-\partial_{\bar{m}}\,
\tilde{K}_{\bar{\alpha}\gamma}\,\tilde{K}^{\gamma\bar{\delta}}\partial_n\,\tilde{K}_{\bar{\delta}\beta})F^n
\\
A'_{\alpha\beta\gamma} &=& \frac{\hat{W}^{\star}}{|\hat{W}|}
e^{\hat{K}/2}\,F^m\,[\hat{K}_m\,Y'_{\alpha\beta\gamma}+\partial_m
Y'_{\alpha\beta\gamma}-(\tilde{K}^{\delta\bar{\rho}}\,\partial_n\tilde{K}_{\bar{\rho}\alpha}\,Y'_{\alpha\beta\gamma}+
\alpha \leftrightarrow \gamma + \alpha \leftrightarrow \beta)]
\nonumber \ea The \emph{normalized} scalar
masses and trilinears are thus given by:
\ba
m^2_{\bar{\alpha}\beta} &=& (\mathcal{Q}^{\dag}\cdot
m'^2 \cdot \mathcal{Q})_{\bar{\alpha}\beta} \\
\tilde{A}_{\alpha\beta\gamma} &=& A'_{{\alpha}'{\beta}'{\gamma}'}
\mathcal{Q}_{{\alpha}'\alpha}\mathcal{Q}_{{\beta}'\beta}\mathcal{Q}_{{\gamma}'\gamma}\nonumber
\ea

Let us discuss the implications for the soft terms, beginning with the
anomaly mediated corrections to gaugino masses.

\vspace{0.5cm} \hspace{3.5cm} {\it Anomaly mediated contributions
to Gaugino Masses} \vspace{0.5cm}

We saw in section \ref{treegaugino} that the gauginos are
generically suppressed relative to the gravitino. Since anomaly
mediated gaugino masses are also suppressed relative to the
gravitino (by a loop factor), they are non-negligible compared to
the tree level contributions and have to be taken into account.
Also, since anomaly mediated contributions for the three gauge
groups are \emph{non-universal}, these introduce non-universality
in the gaugino masses at the unification scale.

The general expression for the anomaly mediated contributions is
given by \cite{Gaillard:1999yb}: \be \label{anom123}(M)^{am}_a =
-\frac{g_a^2}{16\pi^2}[-(3C_a-\sum_{\alpha}C_a^{\alpha})e^{\hat K/2}W^{*}+(C_a-\sum_{\alpha}C_a^{\alpha})e^{\hat K/2}F^m\hat{K}_m+2\sum_{\alpha}(C_a^{\alpha}e^{\hat K/2}F^m\partial_m\ln(\tilde{K}_{\alpha}))]\ee
where $C_a$ and $C_a^{\alpha}$ are the casimir invariants of the
$a^{th}$ gauge group and $\alpha$ runs over the number of fields
charged under the $a^{th}$ gauge group. For a given spectrum such
as that of the MSSM, $C_a$ and $C_a^{\alpha}$ are known.

We first compute the F-term contributions
\ba
&&e^{\hat K/2}F^i\hat{K}_i=\frac{14}3e^{-i\gamma_W}\left(\tilde{L}_{1,+}+\frac 3 2\right)\times m_{3/2}\,\\
&&e^{\hat K/2}F^{\phi}\hat{K}_{\phi}=e^{-i\gamma_W}\left(\frac{a\tilde\alpha}{\tilde x}+\phi_0^2\right)\times m_{3/2}\,.
\nonumber
\ea
Then, equation (\ref{anom123}) gives \ba
\label{anomalyds} (M)^{am}_a &=&-e^{-i\gamma_W}\frac
{\alpha_{GUT}}{4\pi}
[-(3C_a-\sum_{\alpha}C_a^{\alpha})+\frac {14}3(C_a-\sum_{\alpha}C_a^{\alpha})\left(\tilde L_{1,+}+\frac 3 2\right) \\
&+&(C_a-\sum_{\alpha}C_a^{\alpha})\left(\frac{a\tilde\alpha}{\tilde
x}+\phi_0^2\right)-\frac 4 3\left(\tilde L_{1,+}+\frac 3
2\right)\sum_{\alpha}C_a^{\alpha}\,\sum_{i}\,\frac{1}{2\pi}\left(l\,\psi^{\alpha}_i\,\sin(2\pi{\theta}^{\alpha}_i)
\right)]\times m_{3/2}\,. \nonumber \ea where \be
\left(\alpha_{GUT}\right)^{-1}=\sum_{i=1}^{N}s_iN_i^{sm}\,,\ee and
we have defined the quantity: \ba
{\psi}^{\alpha}_i({\theta}^{\alpha}_i) \equiv \frac{d
\ln(\tilde{K}_{\alpha})}{d\,\theta^{\alpha}_i}\,,\ea where
${\theta}^{\alpha}_i$ implicitly depends on the moduli. However,
it is much simpler to keep the dependence as a function of
${\theta}^{\alpha}_i$, as is explained in the Appendix. Depending
on the values of the Casimir invariants $C_a$ and $C_a^i$ for the
three gauge groups, the anomaly mediated contribution can either
add to or cancel the tree level contributions. Here we also took
into account that $m_{3/2}=e^{K/2}\left|\tilde
x\right|A_2{e^{-b_2\vec\nu\cdot\vec a}}$  but $\tilde x < 0$,
since $Q-P\geq3$. Using the expression for $\tilde{L}_{1,+}$ in
(\ref{e61}) along with the definitions of $\tilde x$, $\tilde y$
and $\tilde z$ in (\ref{xyzw1}) in terms of $\tilde\alpha$ in
(\ref{e57}) and assuming that $Q-P\sim {\mathcal O}(1)$, in the
limit when $P$ is large we obtain \be\label{ltplus} \tilde
L_{1,+}=-\frac 3 2+\frac
3{2\,P\,\ln\left(\frac{A_1Q}{A_2P}\right)} \left(1+\frac
2{(Q-P)\,\phi_0^2}+\frac 7{\phi_0^2\,P\,\ln\left(\frac
{A_1Q}{A_2P}\right)}\right)\,. \ee Using (\ref{eqab}) and
substituting for $\nu$ from (\ref{app24}) into (\ref{eqab})
together with $a=-2/P$ we can express
\begin{equation}\label{e890}
\frac{a\tilde\alpha}{\tilde x}+\phi_0^2=\phi_0^2\left(1+\frac 2{(Q-P)\,\phi_0^2}+\frac{7}{\phi_0^2\,P\ln\left(\frac{A_1Q}{A_2P}\right)}\right)\,.
\end{equation}

Substituting (\ref{ltplus}) and (\ref{e890}) into
(\ref{anomalyds}) we obtain \ba \label{anomalyds1} (M)^{am}_a
&=&-e^{-i\gamma_W} \frac {\alpha_{GUT}}{4\pi}
[-(3C_a-\sum_{\alpha}C_a^{\alpha})+\left(1+\frac 2{(Q-P)\,\phi_0^2}+\frac
7{\phi_0^2\,P\,\ln\left(\frac
{A_1Q}{A_2P}\right)}\right)\\
&\times&\left((C_a-\sum_{\alpha}C_a^{\alpha})\left(\phi_0^2
+\frac 7{P\,\ln\left(\frac{A_1Q}{A_2P}\right)}\right)
-\frac{2\sum_{\alpha}C_a^{\alpha}\,\sum_{i}\,\frac{1}{2\pi}\left(l\,\psi^{\alpha}_i\,\sin(2\pi{\theta}^{\alpha}_i)
\right)}{P\,\ln\left(\frac{A_1Q}{A_2P}\right)}\right)]\times m_{3/2}\,.\nonumber \ea

Note that these $M$ theory vacua do not have a no-scale structure.
Therefore, the anomaly mediated gaugino masses are only suppressed
by loop effects, in contrast to the type IIB compactifications,
which exhibit a no-scale structure in the leading order
\cite{Conlon:2006us}, leading to an additional suppression of the
anomaly mediated gaugino masses.

As before, when one imposes the constraint (\ref{ior}), the anomaly mediated
gaugino mass contribution can be simplified further and is given
by:\begin{eqnarray}
\label{anomalyds13} (M)^{am}_a &=&-\frac
{\alpha_{GUT}{(e^{-i\gamma_W})}}{4\pi}
[-(3C_a-\sum_{\alpha}C_a^{\alpha})+\frac{29055+11374\sqrt{3}}{29448}{(C_a-\sum_{\alpha}C_a^{\alpha})}\nonumber \\
&-&\frac{139+396\sqrt{3}}{17178}{\sum_{\alpha}C_a^{\alpha}\,\sum_{i}\,\frac{1}{2\pi}\left(l\,\psi^{\alpha}_i\,\sin(2\pi{\theta}^{\alpha}_i)
\right)}]\times m_{3/2}\nonumber\\
(M)^{am}_a &\approx&-\frac {\alpha_{GUT}{(e^{-i\gamma_W})}}{4\pi}
[-(3C_a-\sum_{\alpha}C_a^{\alpha})+1.6556\,{(C_a-\sum_{\alpha}C_a^{\alpha})} \\
&-&0.048{\sum_{\alpha}C_a^{\alpha}\,\sum_{i}\,\frac{1}{2\pi}\left(l\,\psi^{\alpha}_i\,\sin(2\pi{\theta}^{\alpha}_i)\right)}]\times
m_{3/2}\,. \nonumber
\end{eqnarray}
From the left plot in Figure \ref{Am2} of the Appendix we note
that
$\left|\frac{1}{2\pi}\left(l\,\psi^{\alpha}_i\,\sin(2\pi{\theta}^{\alpha}_i)
\right)\right|<0.5$. In a generic case, we expect that parameters
$\theta^{\alpha}_i$ are all different and, as a result, the terms
appearing inside the corresponding sum over $i$ partially cancel
each other. Thus, in a typical case we expect that \be
\label{genericcond}
\left|{\sum_{\alpha}C_a^{\alpha}\,\sum_{i}\,\frac{1}{2\pi}\left(l\,\psi^{\alpha}_i\,\sin(2\pi{\theta}^{\alpha}_i)
\right)}\right|<1\,. \ee Neglecting the corresponding contribution
in (\ref{anomalyds13}), taking $\alpha_{GUT}=1/25$, and
substituting the Casimirs for an MSSM spectrum, we obtain the
following values in the leading order, up to an overall phase
$e^{-i\gamma_W}$: \be\label{ano456}
(M)^{am}_{U(1)}\approx\,\,0.01377\times m_{3/2},\,\,\,\,\,\,\,
(M)^{am}_{SU(2)}\approx\,\,0.02317\times m_{3/2},\,\,\,\,\,\,\,
(M)^{am}_{SU(3)}\approx\,\,0.02536\times m_{3/2}\,. \ee Finally,
combining the tree-level (\ref{gaugino36}) plus anomaly mediated
(\ref{ano456}) contributions, we obtain the following {\em
non-universal} gaugino masses at the unification scale: \be
M_1\approx\,\,-10.24\times 10^{-3}\,m_{3/2},\,\,\,\,\,\,\,
M_2\approx\,\,-0.84\times 10^{-3}\, m_{3/2},\,\,\,\,\,\,\,
M_3\approx\,\,+1.35\times 10^{-3}\, m_{3/2}\,. \ee We immediately
notice remarkable cancellations for $M_2$ and $M_3$ between the
tree-level and the anomaly mediated contributions. Recall that
since the distribution of $m_{3/2}$ peaked at $m_{3/2}\sim{\cal
O}(100)\,{\rm TeV}$, the possible range of gaugino masses is in
the desirable range $m_{1/2}\sim {\cal O}(0.1-1)\,{\rm TeV}$. One
of the consequences of these cancellations is a comparatively
lighter gluino. Furthermore, since $M_2$ is a lot smaller than
$M_1$, assuming R parity conservation, the neutralino LSP is
expected to be wino-like. This is confirmed by explicitly RG
evolving the gaugino masses to low scales, at least for the case
when the cosmological constant is tuned to be very small.

One should be extremely cautious however, since the predictive
expressions above are only true if (\ref{genericcond}) is
satisfied. The extra contribution neglected in the above estimates
is given by: \be \Delta_a=0.15\times
10^{-3}\,{\sum_{\alpha}C_a^{\alpha}\,\sum_{i}\,\frac{1}{2\pi}\left(l\,\psi^{\alpha}_i\,\sin(2\pi{\theta}^{\alpha}_i)\right)}\times
m_{3/2}\,. \ee Because of the large cancellations between the
tree-level and anomaly mediated contributions, it may happen that
these corrections become important in a relatively small region of
the overall parameter space, leading to a deviation from the above
result thereby altering the pattern of gaugino masses. Further
corrections may also come from varying $\alpha_{GUT}$ as well as
taking into account subleading corrections to the condition for
the cosmological constant to be very small. In section
\ref{subleading}, the effects of subleading corrections to the
very small cosmological constant condition on the gaugino masses
will be analyzed. A thorough study of these issues will be done in
\cite{ToAppear}.

\vspace{1.0cm} \hspace{7.0cm} {\it Trilinears} \vspace{0.5cm}

\noindent The normalized trilinear can be written as : \ba
\label{trilinears2}\tilde{A}_{\alpha\beta\gamma} &=&
\frac{\hat{W}^{\star}}{|\hat{W}|} e^{\hat{K}/2}\,
(\tilde{K}_{\alpha}\tilde{K}_{\beta}\tilde{K}_{\gamma})^{-1/2}
\,\left(\sum_{i} e^{\hat
K/2}F^m\,[\hat{K}_m\,Y'_{\alpha\beta\gamma}+\partial_m
Y'_{\alpha\beta\gamma}-\partial_m\ln(\tilde{K}_{\alpha}\tilde{K}_{\beta}\tilde{K}_{\gamma})]\right) \\
&=& Y_{\alpha\beta\gamma}\,\left(\sum_{i}
e^{\hat K/2}F^m\,[\hat{K}_m+\partial_m\,\ln
(Y'_{\alpha\beta\gamma})-\partial_m\ln(\tilde{K}_{\alpha}\tilde{K}_{\beta}\tilde{K}_{\gamma})]\right)
\nonumber \ea As stated earlier, the subscripts
$\{\alpha,\beta,\gamma\}$ stand for the visible chiral matter
fields. For example, $\alpha$ can be the left-handed up quark
doublet, $\beta$ can be the right-handed up quark singlet and
$\gamma$ can be the up-type higgs doublet. Our present
understanding of the microscopic details of these constructions
does not allow us to compute the three individual trilinear
parameters -- corresponding to the up-type Yukawa, the down-type
Yukawa and the lepton Yukawa matrices, explicitly. One can only
estimate the rough overall scale of the trilinears.

We see from (\ref{trilinears2}) that the normalized trilinears are
proportional to the Yukawas since the K\"{a}hler metric is
diagonal. If instead the off-diagonal entries in the k\"{a}hler
metric are small but non-zero, it would lead to a slight deviation
from the proportionality of the trilinears to the Yukawa
couplings. In most phenomenological analyses, the trilinears
$\tilde{A}$ are taken to be proportional to the Yukawas and the
\emph{reduced} trilinear couplings
$A_{\alpha\beta\gamma}\equiv\tilde{A}_{\alpha\beta\gamma}/Y_{\alpha\beta\gamma}$
are used. We expect this to be true in these compactifications
from above. If the Yukawa couplings are those of the Standard Model, then from (\ref{trilinears2}) the
normalized reduced trilinear coupling $A_{\alpha\beta\gamma}$ for
de Sitter vacua in general is given by \ba
\label{tri46}A_{\alpha\beta\gamma}&=&e^{-i\gamma_W}\left({1+\frac
2{(Q-P)\,\phi_0^2}+\frac 7{\phi_0^2\,P\,\ln\left(\frac
{A_1Q}{A_2P}\right)}}\right)(\phi_0^2+ \frac 1 {\,P\,\ln\left(\frac{A_1Q}{A_2P}\right)}[7
+2\ln\left|\frac{C_{\alpha\beta\gamma}}{Y'_{\alpha\beta\gamma}}\right|\\
&+&\sum_i\,\frac{1}{2\pi}(l\,\psi^{\alpha}_i\,\sin(2\pi{\theta}^{\alpha}_i)+{\alpha}\rightarrow
\beta + \alpha \rightarrow \gamma)])\times m_{3/2}\,.\nonumber \ea
If we then use (\ref{ymu}) together with (\ref{kmetric}) and
$\mathcal{Q}_{\bar{\alpha}\beta}=(\tilde{K}_{\alpha})^{-1/2}\delta_{\bar{\alpha}\beta}$,
we obtain the following expression for the trilinears \ba
\label{tri49}A_{\alpha\beta\gamma}&=&m_{3/2}\,e^{-i\gamma_W}\left({1+\frac
2{(Q-P)\,\phi_0^2}+\frac 7{\phi_0^2\,P\,\ln\left(\frac
{A_1Q}{A_2P}\right)}}\right)(\phi_0^2+ \frac 1 {\,P\,\ln\left(\frac{A_1Q}{A_2P}\right)}[7
+2\ln\left|\frac{C_{\alpha\beta\gamma}}{Y_{\alpha\beta\gamma}}\right|\\&-&3\ln(4\pi^{1/3}V_X)+\phi_0^2
-\sum_i\left(\left\{\frac 1 2\ln\left(\frac{\Gamma(1-\theta^{\alpha}_i)}{\Gamma(\theta^{\alpha}_i)}\right)
-\frac{1}{2\pi}l\,\psi^{\alpha}_i\,\sin(2\pi{\theta}^{\alpha}_i)\right\}+{\alpha}\rightarrow
\beta + \alpha \rightarrow \gamma\right)])\,.\nonumber \ea Imposing the constraint equation
(\ref{ior}) on the expression above, the reduced trilinears for a
dS vacuum with a tiny cosmological constant are simplified to: \ba
\label{trides}&A_{\alpha\beta\gamma}&=e^{-i\gamma_W}
(\frac{69+22\sqrt{3}}{72}+\frac{139+396\sqrt{3}}{34356}[7
+2\ln\left|\frac{C_{\alpha\beta\gamma}}{Y_{\alpha\beta\gamma}}\right|-7\,\ln\left(\frac{14(P+3)}{N}\right)+\frac
1{72}\left(15+22\sqrt{3}\right)\nonumber\\&&-6\,\ln\left(\frac
2\pi\right)-\sum_i\left(\left\{\frac 1
2\ln\left(\frac{\Gamma(1-\theta^{\alpha}_i)}{\Gamma(\theta^{\alpha}_i)}\right)-\frac{1}{2\pi}l\,\psi^{\alpha}_i\,\sin(2\pi{\theta}^{\alpha}_i)\right\}+{\alpha}\rightarrow
\beta + \alpha \rightarrow \gamma\right)])\times m_{3/2}\nonumber\\
&A_{\alpha\beta\gamma}&\approx e^{-i\gamma_W}(1.4876+0.024\,[10.45
+2\ln\left|\frac{C_{\alpha\beta\gamma}}{Y_{\alpha\beta\gamma}}\right|-7\,\ln\left(\frac{14(P+3)}
{N}\right)\\&&-\sum_i\left(\left\{\frac 1 2\ln\left(\frac{\Gamma(1-\theta^{\alpha}_i)}
{\Gamma(\theta^{\alpha}_i)}\right)-\frac{1}{2\pi}l\,\psi^{\alpha}_i\,\sin(2\pi{\theta}^{\alpha}_i)\right\}
+{\alpha}\rightarrow\beta + \alpha \rightarrow \gamma\right)])\times
m_{3/2}\,.\nonumber \ea

We see that compared to the gauginos, the trilinears depend on
more constants. The quantity $\left\{\frac 1
2\ln\left(\frac{\Gamma(1-\theta^{\alpha}_i)}{\Gamma(\theta^{\alpha}_i)}\right)-
\frac{1}{2\pi}(l\,\psi^{\alpha}_i\,\sin(2\pi{\theta}^{\alpha}_i)\right\}$
is of $\mathcal{O}(1)$. Therefore, in a generic situation, we
expect the terms inside the sum in $\sum_i \,\left\{\frac 1
2\ln\left(\frac{\Gamma(1-\theta^{\alpha}_i)}{\Gamma(\theta^{\alpha}_i)}\right)-
\frac{1}{2\pi}(l\,\psi^{\alpha}_i\,\sin(2\pi{\theta}^{\alpha}_i)\right\}$
to partially cancel each other and give an overall contribution
much smaller than the first three terms inside the square
brackets. Then, for known values of the physical Yukawa couplings
and reasonable values of $P$ and $N$, the trilinears generically
turn out to slightly larger than $m_{3/2}$.

\vspace{0.5cm} \hspace{7.0cm} {\it Scalar Masses} \vspace{0.5cm}

\noindent For an (almost) diagonal K\"{a}hler metric, the
normalized scalar masses reduce to : \be\label{sc567}
(m^2_{\bar{\alpha}\beta}) =
[m_{3/2}^2+V_0-e^{\hat K}F^{\bar{m}}F^n{\partial}_{\bar{m}}\partial_{n}\ln(\tilde{K}_{\alpha})]
\,{\delta}_{\bar{\alpha}\beta} \ee where we have used
(\ref{norm}). Using (\ref{ltplus}) in (\ref{sc567}), we obtain the
following expression for the scalar mass squared \ba
\label{scalarsds} (m_{\alpha}^2)&=& V_0 + (m^2_{3/2})\,[1-\frac
9{4\,P^2\left(\ln\left(\frac{A_1Q}{A_2P}\right)\right)^2}
\left(1+\frac 2{(Q-P)\,\phi_0^2}+\frac
7{\phi_0^2\,P\,\ln\left(\frac {A_1Q}{A_2P}\right)}\right)^2
 \nonumber \\ & &
\times\frac{1}{4\pi}\sum_i\,\{l^2\,{\psi}^{\alpha}_{\bar{i}i}\,\sin^2(2\pi{\theta}^{\alpha}_i)+l^2\,{\psi}^{\alpha}_i\,\sin(4\pi{\theta}^{\alpha}_i)
-2l\,{\psi}^{\alpha}_i\,\sin(2\pi{\theta}^{\alpha}_i)\}]\,. \ea
where we have defined another quantity:\ba
{\psi}^{\alpha}_{\bar{i}i}({\theta}^{\alpha}_i)\equiv
\frac{d{\psi}^{\alpha}_i}{d{\theta}^{\alpha}_i}\ea As in the case
of the trilinears, only the overall scale of the scalars can be
estimated, not the individual masses of different flavors of
squarks and sleptons. Once the cosmological constant is made small
by imposing the constraint (\ref{ior}), the scalars are given by
\begin{eqnarray}
\label{scalarsds2} (m_{\alpha}^2)&=& (m^2_{3/2})\,[1-
\frac{\left(139+396\sqrt{3}\right)^2}{524593216}\frac{1}{4\pi}\sum_i\,\{l^2\,{\psi}^{\alpha}_{\bar{i}i}\,
\sin^2(2\pi{\theta}^{\alpha}_i)+l^2\,{\psi}^{\alpha}_i\,
\sin(4\pi{\theta}^{\alpha}_i)
-2l\,{\psi}^{\alpha}_i\,\sin(2\pi{\theta}^{\alpha}_i)\}]\,\nonumber\\
&\approx&(m^2_{3/2})\,[1-\frac{0.0013}{4\pi}\sum_i\,\{l^2\,
{\psi}^{\alpha}_{\bar{i}i}\,\sin^2(2\pi{\theta}^{\alpha}_i)+l^2\,{\psi}^{\alpha}_i\,\sin(4\pi{\theta}^{\alpha}_i)
-2l\,{\psi}^{\alpha}_i\,\sin(2\pi{\theta}^{\alpha}_i)\}]\,\nonumber\\
&\approx&m^2_{3/2}\,.
\end{eqnarray}
Thus, to a high degree of accuracy, in the IIA limit, the scalar
masses for de Sitter vacua are flavor universal as well as flavor
diagonal and independent of the details of the matter K\"{a}hler
metric described by parameters $\theta_i$. Moreover, to a very
good approximation, they are equal to the gravitino mass. A
natural expectation away from the IIA limit is that the squark and
slepton masses are always of order $m_{3/2}$. Since $m_{3/2}$ is
of several TeV, the scalars are quite heavy, naturally suppressing
flavor changing neutral currents (FCNCs).


\subsection{Subleading Corrections to the Condition for small Cosmological Constant and Effects on Phenomenology}
\label{subleading}

In this subsection we would like to give a rough estimate of how
the amount of tuning of the cosmological constant might affect the
values of the soft parameters. In fact, the constraint (\ref{ior})
which sets the cosmological constant to zero in the leading order
still results in a very large value of the cosmological constant
$V_0\sim 0.01\times m_{3/2}^2m_p^2$, once the subleading terms are
taken into account. One can also do exact numerical computations
for manifolds with small number of moduli (say two).

Taking into account the terms in the subleading order as $\tilde
y\rightarrow 0$, the potential at the minimum with respect to the
moduli, as a function of the meson vev $\phi_0$ is given by
\be\label{eqn1} V_0=\frac{\left(A_2\tilde x\right)^2}{64\pi
V_X^3}\left[\phi_0^4+ \left(\frac{2a\tilde\alpha}{\tilde
x}-3+\frac 17\left(\frac{ab_1\tilde\alpha\tilde y}{\tilde x\tilde
z}\right)^2\left(\frac{a\tilde\alpha}{\phi_0^2\tilde
x}+1\right)^2\right)\phi_0^2+\left(\frac{a\tilde\alpha}{\tilde
x}\right)^2\right]\frac{e^{\phi_0^2}}{\phi_0^2}\left(\frac{A_1Q}{A_2P}\right)^{-\frac{2P}{Q-P}}\,.
\ee Hence, vanishing of the cosmological constant corresponds to
the vanishing of the combination \be\label{eqn2} \phi_0^4+
\left(\frac{2a\tilde\alpha}{\tilde x}-3+\frac
17\left(\frac{ab_1\tilde\alpha\tilde y}{\tilde x\tilde
z}\right)^2\left(\frac{a\tilde\alpha}{\phi_0^2\tilde
x}+1\right)^2\right)\phi_0^2+\left(\frac{a\tilde\alpha}{\tilde
x}\right)^2=0\,. \ee Recall that by imposing this condition in the
leading order as $\tilde y\rightarrow 0$ we had demonstrated that
the combination $P\ln\left(\frac{A_1Q}{A_2P}\right)$ is fixed by
\be\label{eqn3} 3-\frac
8{Q-P}-\frac{28}{P\ln\left(\frac{A_1Q}{A_2P}\right)}=0\,. \ee
Thus, the leading order condition on
$P\ln\left(\frac{A_1Q}{A_2P}\right)$ obtained from (\ref{eqn3}) is
given by \be\label{eqn9}
\left(P\ln\left(\frac{A_1Q}{A_2P}\right)\right)^{(0)}=\frac{28\left(Q-P\right)}{3\left(Q-P\right)-8}\,.
\ee In this case, the meson vev $\phi_0^2$ in the leading order is
fixed at the value \be\label{eqn4} (\phi_0^2)^{(0)}=-\frac 1
8+\frac 1{Q-P}+\frac 1 4\sqrt{1-\frac 2{Q-P}}+\frac 2{Q-P}
\sqrt{1-\frac 2{Q-P}}\,. \ee The subleading corrections to
$P\ln\left(\frac{A_1Q}{A_2P}\right)$ can be found iteratively if
we plug (\ref{eqn4}) into the subleading term $\sim\tilde y^2$ in
 (\ref{eqn2}) to obtain
\be\label{eqn5} \phi_0^4+ \left(\frac{2a\tilde\alpha}{\tilde
x}-3+\frac 17\left(\frac{ab_1\tilde\alpha\tilde y}{\tilde x\tilde
z}\right)^2\left(\frac{a\tilde\alpha}{\left(\phi_0^2\right)^{(0)}\tilde
x}+1\right)^2\right)\phi_0^2+\left(\frac{a\tilde\alpha}{\tilde
x}\right)^2=0\,. \ee Again, by setting the discriminant of the
biquadratic polynomial in (\ref{eqn5}) to zero and using
\be\label{eqn6} \tilde L_{1,\,+}=-\frac 3 2+\frac
3{2P\ln\left(\frac{A_1Q}{A_2P}\right)}\left(1+ \frac
2{\left(Q-P\right)\left(\phi_0^2\right)^{(0)}}\right)\,, \ee we
obtain the following condition on
$P\ln\left(\frac{A_1Q}{A_2P}\right)$ \be\label{eqn7} 3-\frac
8{Q-P}-\frac{28}{P\ln\left(\frac{A_1Q}{A_2P}\right)}+\frac{7}{\left(P\ln\left(\frac{A_1Q}{A_2P}\right)\right)^2}\left(3-\frac
4{\left(\phi_0^4\right)^{(0)}\left(Q-P\right)^2}+\frac
4{\left(\phi_0^2\right)^{(0)}\left(Q-P\right)}\right)=0\,. \ee To
compute the first subleading order correction to
$P\ln\left(\frac{A_1Q}{A_2P}\right)$ we express \be\label{eqn8}
P\ln\left(\frac{A_1Q}{A_2P}\right)=\frac{28\left(Q-P\right)}{3\left(Q-P\right)-8}+\delta^{(1)}\,,
\ee where the first term in (\ref{eqn8}) corresponds to the
leading order expression in (\ref{eqn9}) and $\delta^{(1)}$ is the
subleading order correction. Hence, plugging (\ref{eqn8}) into
(\ref{eqn7}) and keeping the terms linear in $\delta^{(1)}$, after
some algebra we obtain \be\label{eqn10}
\delta^{(1)}=\frac{\left(2+\left(\phi_0^2\right)^{(0)}\left(Q-P\right)\right)\left(2-3\left(\phi_0^2\right)^{(0)}\left
(Q-P\right)\right)}{4\left(\phi_0^4\right)^{(0)}\left(Q-P\right)^2}\,.
\ee Therefore, including the first subleading order, the condition
on the cosmological constant to vanish results in the following
constraint \be\label{eqn11}
P\ln\left(\frac{A_1Q}{A_2P}\right)=\frac{28\left(Q-P\right)}{3\left(Q-P\right)-8}+\frac{\left(2+\left(\phi_0^2\right)^{(0)}
\left(Q-P\right)\right)\left(2-3\left(\phi_0^2\right)^{(0)}\left(Q-P\right)\right)}{4\left(\phi_0^4\right)^{(0)}\left(Q-P\right)^2}\,.
\ee In particular, for the case when $Q-P=3$ we obtain
\be\label{eqn12}
P\ln\left(\frac{A_1Q}{A_2P}\right)=84-0.9977\approx 83.002\,. \ee
which yields a $\sim 1\%$ correction to the leading order. We have
confirmed that for a case when the compactification manifold has
two moduli, numerically imposing the constraint that the
cosmological constant has the observed value implies
\be\label{eqn14} P\ln\left(\frac{A_1Q}{A_2P}\right)\approx
82.9958\,, \ee which agrees with (\ref{eqn12}) to a very high
degree of accuracy.

One would now like to estimate the effects of tuning the
cosmological constant to its observed value on phenomenological
quantities. The quantity most sensitive to such corrections is the
gravitino mass, since it is proportional to
$\left(\frac{A_1Q}{A_2P}\right)^{-\frac P{Q-P}}$ and can therefore
change by a factor of order one. Of course, this hardly affects
the distributions of scales of $m_{3/2}$ and the emergence of the
TeV scale peak remains very robust. Moving on to gaugino masses,
recall that the tree-level gaugino mass is given by
\be\label{eqn13}
M_{1/2}\approx-\frac{e^{-i\gamma_W}}{P\ln\left(\frac{A_1Q}{A_2P}\right)}
\left(1+\frac 2{\phi_0^2\left(Q-P\right)}+\frac
7{\phi_0^2\,P\ln\left(\frac{A_1Q}{A_2P}\right)}\right)\times
m_{3/2}\,, \ee which for $Q-P=3$ and the leading order constraint
$\left(P\ln\left(\frac{A_1Q}{A_2P}\right)\right)^{(0)}=84$
resulted (up to an overall phase) in \be\label{eqn15}
\left(M_{1/2}\right)^{(0)}\approx -0.0240\times m_{3/2}\,. \ee
Including the first subleading correction in (\ref{eqn12}), the
tree level gaugino mass (up to an overall phase) is given by
\be\label{eqn16} M_{1/2}\approx-\frac{1}{83} \left(1+\frac
2{3\left(\phi_0^2\right)^{(0)}}+\frac
7{84\left(\phi_0^2\right)^{(0)}}\right) \times
m_{3/2}=-0.0243\times m_{3/2}\,, \ee resulting in a $\sim 1\%$
correction to the leading order expression in (\ref{eqn15}). For a
case when the compactification manifold has two moduli, the
numerically obtained result is given by \be\label{eqn17}
M_{1/2}\approx-0.0242\times m_{3/2}\,, \ee again confirming the
high accuracy of the approximate result in (\ref{eqn16}). Recall
that when the leading order constraint
$P\ln\left(\frac{A_1Q}{A_2P}\right)\approx 84$ is satisfied the
value of the cosmological constant is \be V_0\sim
0.01\,m_{3/2}^2\,m_{p}^2 \sim\left({\cal
O}\left(10^{10}-10^{11}\right){\rm GeV}\right)^4\,. \ee for
$m_{3/2}\sim{\cal O}\left(10-100\right)\,{\rm TeV}$. Thus, while
the cosmological constant changes by many orders of magnitute as
it is tuned to its observed value, the tuning has a very small
effect on the tree-level gaugino mass. However, due to the
cancellation between the tree-level and the anomaly mediated
contributions, the correction to the tree-level gaugino mass
computed above may be important and therefore, has been taken into
account.

Finally, the $\sim 1\%$ correction to the leading order
$\left(P\ln\left(\frac{A_1Q}{A_2P}\right)\right)^{(0)}$ due to the
tuning of the cosmological constant has almost no effect on the
anomaly mediated gaugino masses, trilinears and the scalars. This
is because the terms proportional to
$1/P\ln\left(\frac{A_1Q}{A_2P}\right)$ and
$1/\left(P\ln\left(\frac{A_1Q}{A_2P}\right)\right)^2$ are
subleading. These considerations indicate that a leading order
tuning of the cosmological constant is enough to allow reliable
particle physics phenomenology.


\subsection{Radiative Electroweak Symmetry Breaking (REWSB)}\label{rewsb}

It is very important to check whether the soft supersymmetry
breaking parameters in these vacua naturally give rise to
radiative electroweak symmetry breaking (REWSB) at low scales. In
order to check that, one has to first RG evolve the scalar higgs
mass parameters $m_{H_u}^2$ and $m_{H_d}^2$ from the high scale to
low scales. Then one has to check whether for a given $\tan\beta$,
there exists a value of $\mu$ which satisfies the EWSB conditions.
At the one-loop level, we find that EWSB occurs quite generically
in the parameter space. This can be understood as follows. The
gaugino mass contributions to the RGE equation for $m_{H_u}^2$
push the value of $m_{H_u}^2$ up while the top Yukawa coupling,
third generation squark masses and the top trilinear pull it down.
The suppression of the gaugino mass relative to the gravitino mass
causes it to have a negligible effect on the RGE evolution of
$m_{H_u}^2$. On the other hand, the masses of squarks and
$A$-terms are both of $\mathcal{O}(m_{3/2})$, which guarantees
that $m_{H_u}^2$ is negative at the low scale. Typically,
$m_{H_u}^2$ is proportional to $-m_{3/2}^2$, up to a factor less
than one depending on $\tan\beta$. Thus, the EWSB condition can be
easily satisfied with a $\mu$ parameter also of the order
$m_{3/2}$. Note that large $A$-terms (of $\mathcal{O}(m_{3/2})$)
are crucial for obtaining EWSB. Having large squark masses and
small $A$-terms cannot guarantee EWSB, as is known from the focus
point region in mSUGRA. Also, one has to ensure that the third
generation squarks have positive squared masses, which we have
checked. From the point of view of low scale effective theory,
EWSB appears to be fine-tuned, however a better understanding of
the underlying microscopic theory may help justify the choice of
parameters. We will report a detailed analysis of these issues in
\cite{ToAppear}.

\subsection{The $\mu$ and $B\mu$ problem}\label{muBmu}

We will not have much to say about the $\mu$ terms here, leaving a
detailed phenomenological study for our future work
\cite{ToAppear}. We will however, take this opportunity to
highlight the main theoretical issues. The normalized $\mu$ and
$B\mu$ parameters are : \ba \label{mubmu} \mu &=&
(\frac{\hat{W}^{\star}}{|\hat{W}|}e^{\hat{K}/2}{\mu}'+m_{3/2}Z-e^{\hat{K}/2}F^{\bar{m}}\partial_{\bar{m}}Z)\,
(\tilde{K}_{H_u}\tilde{K}_{H_d})^{-1/2} \nonumber \\
B\mu &=& (\tilde{K}_{H_u}\tilde{K}_{H_d})^{-1/2}\{
\frac{\hat{W}^{\star}}{|\hat{W}|} e^{\hat{K}/2}
{\mu}'(e^{\hat{K}/2}F^m\,[\hat{K}_m\,+\partial_m\,\ln{\mu}']-m_{3/2})+(2m_{3/2}^2+V_0)\,Z-m_{3/2}F^{\bar{m}}\partial_{\bar{m}}Z
\nonumber \\& &+m_{3/2}F^m[\partial_m Z -
Z\partial_m\log(\tilde{K}_{H_u}\tilde{K}_{H_d})]-F^{\bar{m}}F^n[\partial_{\bar{m}}\partial_n
Z - \partial_{\bar{m}} Z
\partial_n\log(\tilde{K}_{H_u}\tilde{K}_{H_d})]\}\ea

We see from above that the value of the physical $\mu$ and $B\mu$
parameters depend crucially on many of the microscopic details
eg. if the theory gives rise to a non-zero superpotential ${\mu}'$ parameter and/or
if a non-zero bilinear coefficient $Z$ is present in the
K\"{a}hler potential for the Higgs fields.  From section \ref{rewsb},
we see that one requires a $\mu$ term of $\mathcal{O}(m_{3/2})$ to get
consistent radiative EWSB. This is possible for eg. when one has a vanishing $\mu'$
parameter and an $\mathcal{O}(1)$ higgs bilinear coefficient $Z$, among other possibilities.

\subsection{Dark Matter}

For dS vacua with a small cosmological constant, $M_2 << M_1$ at
low scale. In addition, since $\mu$ should be of
$\mathcal{O}(m_{3/2})$ for consistent EWSB as seen in section
\ref{rewsb}, both $M_2$ and $M_1$ are much less than $\mu$. Hence,
the LSP is wino-like. As was discussed in section
\ref{subleading}, the tuning of the cosmological constant has
little effect on the gaugino masses, thereby preserving the
gaugino mass hierarchy. It is well known that winos coannihilate
quite efficiently as the universe cools down. Since the wino
masses in these vacua are $\mathcal{O}(100)$ GeV, the
corresponding thermal relic density after they freeze out is very
small. However there could be non-thermal contributions to the
dark matter as well, e.g. the decay of moduli fields into the LSP
after the LSP freezes out. In addition, one should remember that
the above result for a wino LSP is obtained after imposing the
requirement of a small cosmological constant. It would be
interesting to analyze the more general case where the results may
change. We leave a full analysis of these possibilities for the
future \cite{ToAppear}.

\subsection{Correlations}

As we have seen, the parameters of the MSSM
depend on the ``microscopic constants'' determined by a given $G_2$ manifold
and can be explicitly calculated in principle. Therefore, the parameters
obtained are correlated with each other in general. For instance we saw
that the gaugino and gravitino masses are related.
By scanning over the
allowed values of the ``microscopic constants'', by scanning the space of
$G_2$ manifolds,
one obtains a particular subspace of the parameter space of the MSSM at
the unification scale.
For a given spectrum
and gauge group, the RG evolution of these parameters to low
scales can also be determined unambiguously, leading to
correlations in soft parameters at the low scale
Finally, these
correlations in the soft parameters will lead to correlations in
the space of actual observables (for eg, the LHC signature space)
as well. In other words, the predictions of these vacua will only
occupy a \emph{finite} region of the observable signature space at
say the LHC. Since two different theoretical constructions will
have different correlations in general, this will in turn lead to
different patterns of signatures at the LHC, allowing us to
distinguish among different classes of string/$M$ theory vacua (at
least in principle). These issues, in particular the systematics
of the distinguishing procedure have been explained in detail in
\cite{Kane:2006yi}.

\subsection{Signatures at the LHC}

The subject of predicting signatures at the LHC for a given class
of string vacua requires considerable analysis. Here, we will make
some preliminary comments, with a detailed analysis to appear in
\cite{ToAppear}.

The scale of soft
parameters is determined by the gravitino mass. We saw in section
\ref{CClowsusy} that requiring the cosmological constant to be
very small by imposition of a constraint equation (eqn.
(\ref{er765})) fixes the overall scale of superpartner masses
to be of ${\mathcal O}(1-100)$TeV. Once the overall scale is fixed,
the pattern of soft parameters at $M_{unif}$ is crucial in
determining the signatures at the LHC. As explained earlier, these
$M$ theory vacua give rise to a specific pattern of soft
parameters at $M_{unif}$. We find non-universal gaugino masses
which are suppressed relative to the gravitino mass.
We furthermore expect that the
scalar masses and trilinears are of the same order as the gravitino.
The $\mu$ and $B\mu$ parameters are not yet understood. For phenomenological analysis however,
we may fix them by imposing consistent EWSB.

One can get a sense of the broad pattern of signatures at the LHC
from the pattern of soft parameters. Since gaugino masses are
suppressed and the fact that the anomaly contribution to the
gluino mass parameter approximately {\it cancels} the tree-level
contribution, one would generically get comparatively light
gluinos in these constructions, much lighter than the scalars,
which would give rise to a large number of events for many
signatures, in particular many events with same-sign dileptons and
trileptons in excess of the SM and many events with large missing
energy, even for a modest luminosity of $10\;fb^{-1}$. Since the
gauginos are lighter than the squarks and sleptons, gluino pair
production is likely to be the dominant production mechanism. The
LSP will be a neutralino for the same reason assuming R parity is
conserved.

It is also possible to distinguish the class of vacua obtained
above from those obtained in Type IIB compactifications, by the
pattern of signatures at the LHC. For the large volume type IIB
vacua, the scalars are lighter than the gluino
\cite{Conlon:2006wz} while for the KKLT type IIB vacua, the
scalars are comparable to the gluino \cite{Choi:2005ge}. This
implies that squark-gluino production and squark pair production
are the dominant production mechanisms at the LHC. Since the LHC
is a $pp$ collider, up-type squarks are preferentially produced
from $t$-channel valence u-quark annihilation if they are
reasonably light, leading to a charge asymmetry which is preserved
in cascade decays all the way to the final state with leptons. On
the other hand, for the class of $M$ theory vacua described here,
since gluino pair production is the dominant mechanism and the
decays of the gluino are charge symmetric (it is a Majorana
particle), the $M$ theory vacua predict a much smaller charge
asymmetry in the number of events with one or two leptons and
$\geq$ 2 jets compared to the Type IIB vacua.

\subsection{The Moduli and Gravitino Problems}

The cosmological moduli and gravitino problems can exist if moduli
and gravitino masses are too light in gravity mediated SUSY
breaking theories. Naively, after the end of inflation, the moduli
fields coherently oscillate dominating the energy density of the
universe. Since the interactions of the moduli are suppressed by
the Planck scale $(m_p)$, their decay rates are small perhaps
leading to the onset of a radiation dominated universe at very low
temperature ($T_R \sim {\mathcal O}(10^{-3})$ MeV for moduli of
${\mathcal O}$(100 GeV-5TeV)), compared to what is required for
successful BBN.

To check if the moduli and gravitino problem can be resolved in
these $M$ theory compactifications, one has to first compute the
masses of the moduli. After doing this, we will use the results to discuss
the moduli and gravitino problems.

The geometric moduli $s_i$ appear in the lagrangian with a kinetic
term given by \be\label{kinetic}
\sum_{i=1}^{N}{\frac{3\,a_i}{4\,s_i^2}\partial_{\mu}s_i\partial^{\mu}s_i}\,,
\ee which is non canonical. The canonically normalized moduli
$\chi_i$ are \be \chi_i\equiv\sqrt{\frac{3\,a_i}2}{\ln}\,s_i\,.
\ee The complete mass matrix including the mixed meson-moduli entries
is given by: \be
\left(m_{\chi}^2\right)_{i\,j}=\frac{2\,\nu^2(a_i\,a_j)^{\frac
1{\,2}}}{3\,N_iN_j}\frac{\partial^2\,V}{\partial s_i\partial
s_j}\,, \,\,\,\,\left(m^2\right)_{i\,\phi_0}=\sqrt{\frac{2 a_i}3}\frac\nu{N_i}\frac{\partial^2\,V}{\partial s_i\partial
\phi_0}\,,\,\,\,\,\left(m^2\right)_{\phi_0\,\phi_0}=\frac{\partial^2\,V}{\partial\phi_0^2}\,,\ee where we took into account the fact that
$\frac{\partial\,V}{\partial s_i}=0$, $\frac{\partial\,V}{\partial\phi_0}=0$
at the extremum and that $\nu_i=\nu$ for all $i=\overline{1,N}$.
A fairly straightforward
but rather tedious computation yields the following structure of
the mass matrix \ba\label{mmat}
&&\left(m_{\chi}^2\right)_{i\,j}=\left((a_i\,a_j)^{\frac
1{\,2}}\,K_1+\delta_{ij}\,K_2\right)\times m_{3/2}^2\,,\\
&&\left(m^2\right)_{i\,\phi_0}=(a_i)^{\frac
1{\,2}}\,K_3\times m_{3/2}^2\,,\nonumber\\
&&\left(m^2\right)_{\phi_0\,\phi_0}=K_4\times m_{3/2}^2\,,\nonumber
 \ea where,
in the large $\nu$ approximation, $K_1$, $K_2$, $K_3$, $K_4$ are given by \ba\label{ar45}
&&K_1\approx\frac{112}{27}\left(\frac{\tilde z}{\tilde
x}\right)^2\nu^4\approx\frac{48}{343}\left(\frac QP\right)^2\frac{
\left(P\,\ln\left(\frac{A_1Q}{A_2P}\right)\right)^4}{\left(Q-P\right)^4}\,,\\
\,\nonumber\\
&&K_2=-\frac{40}9\left(\tilde
L_{1,+}\right)^2-\frac{56}3L_{1,+}-8-2\phi_0^2\left(\frac{a\tilde\alpha}
{\tilde x\phi_0^2}+1\right)^2\nonumber \\\nonumber \\
&&\,\,\,\,\,\,\,\,\,\approx 10-\frac 8{Q-P}-\frac
8{(Q-P)^2\phi_0^2}-2\phi_0^2-\frac{36}{P\ln\left(\frac{A_1Q}{A_2P}\right)}\left(1+\frac 2{\left(Q-P\right)\phi_0^2}\right)\,,\nonumber \\\nonumber \\
&&K_3\approx-\sqrt{\frac 23}\frac{56\,ab_1\tilde\alpha\tilde z}{9\,\phi_0\tilde x^2}\nu^3
\approx\sqrt{\frac 23}\left(\frac{48}{49}\right)\left(\frac Q P\right)^2\frac
{\left(P\ln\left(\frac{A_1Q}{A_2P}\right)\right)^3}{\left(Q-P\right)^4\phi_0}\,,\nonumber\\
\,\nonumber\\
&&K_4\approx\frac{32}7\frac{\left(Q\ln\left(\frac{A_1Q}{A_2P}\right)\right)^2}
{\left(Q-P\right)^4\phi_0^2}\,.\nonumber
 \ea Using condition $P\,\ln\left(\frac{A_1Q}{A_2P}\right)=84$ to tune the cosmological constant
together with $Q-P=3$, we obtain from (\ref{ar45})
\be\label{leading}
K_1\approx 86016\left(\frac QP\right)^2\,,\,\,\,\,\,\,
K_2\approx 3.83\,,\,\,\,\,\,\,
K_3\approx 6815\left(\frac QP\right)^2\,,\,\,\,\,\,\,
K_4\approx 540\left(\frac QP\right)^2\,.\ee
We can diagonalize the matrix (\ref{mmat}) in two steps. We first construct a set
of orthogonal (unnormalized) $N+1$ vectors given by
\[\vec x_1=\left( \begin{array}{c}
   \sqrt{a_1} \\
   -\frac{a_1}{\sqrt{a_2}} \\
  0 \\
    . \\
   . \\
   . \\
    0 \\
    0
\end{array} \right),\,\vec x_2=\left( \begin{array}{c}
   \sqrt{a_1} \\
   \sqrt{a_2} \\
  -\frac{a_1+a_2}{\sqrt{a_3}} \\
    0 \\
   . \\
   . \\
    0 \\
    0
\end{array} \right),\,.\,.\,.\,,\vec x_{N-1}=\left( \begin{array}{c}
   \sqrt{a_1} \\
   \sqrt{a_2} \\
  \sqrt{a_3} \\
    . \\
   . \\
 . \\
   -\frac{\sum_{i=1}^{N-1}a_i}{\sqrt{a_N}} \\
    0
\end{array} \right),\vec x_{N}=\left( \begin{array}{c}
   \sqrt{a_1} \\
   \sqrt{a_2} \\
  \sqrt{a_3} \\
    . \\
   . \\
   . \\
    \sqrt{a_N} \\
    0
\end{array} \right),\vec x_{N+1}=\left( \begin{array}{c}
  0 \\
  0 \\
  0 \\
    . \\
   . \\
   . \\
    0 \\
    1
\end{array} \right)\]
Next, we normalize the above vectors and construct an orthogonal $(N+1)\times(N+1)$ matrix
\be\label{orthmatr}
{\cal R}=\left(\frac{\vec x_1}{|\vec x_1|},\,\frac{\vec x_2}{|\vec x_2|},
\,.\,.\,.\,\frac{\vec x_{N+1}}{|\vec x_{N+1}|}\right)
\ee
By applying the orthogonal transformation ${\cal R}$, the mass matrix in
(\ref{mmat}) is converted into a block-diagonal form given by:
\[ \left( \begin{array}{cccccc}
K_2  & 0     & \,.\,.\,.\,&0 &0                  & 0 \\
 0   & K_2  &  \,.\,.\,.\, &0&0                  & 0 \\
 .  &   .   &  \,.\,.\,.\, & . &.                  & . \\
 .  &   .   &  \,.\,.\,.\, & . & .                 & . \\
 .  &   .   &  \,.\,.\,.\, & . &  .                & . \\
 0   & 0&  \,.\,.\,.\,  & K_2 &  0                & 0\\
 0   & 0&  \,.\,.\,.\,  & 0 & \frac73K_1+K_2      &\sqrt{\frac 73}K_3\\
 0    & 0&  \,.\,.\,.\,  & 0 &  \sqrt{\frac 73}K_3 & K_4
\end{array} \right) \]
Hence, the first $N-1$ eigenvalues are given by
\be
\lambda_i=K_2\,\,\,\,{\rm for}\,\,\,i=1,...,N-1\,.
\ee
Diagonalizing the $2\times 2$ block is then trivial and boils down to solving a simple quadratic
equation.
Since for typical values of $Q$ and $P$ we have
\be
\frac{K_3}{K_1}\lesssim 0.07\,,\,\,\,{\rm and}\,\,\,\,\,\frac{K_2}{K_1}\ll 1\,,
\ee
the corresponding eigenvalues are given by
\be\label{twoother}
\lambda_N\approx \frac 7 3K_1\,,\,\,\,\,\,\,\,\,\lambda_{N+1}\approx K_4-\frac{K_3^2}{K_1}\,.
\ee
Because of the significant cancellation due to the minus sign in $\lambda_{N+1}$ we have to
beyond the leading order approximation in (\ref{leading}) when computing $K_1$, $K_3$ and $K_4$.
For dS vacua with a nearly zero cosmlogical constant we have verified numerically
that the corresponding eigenvalue is always positive, confirming that we have
obtained a stable minimum.
The $N-1$ degenerate light states have masses given by\be\label{modul1}
M_k\approx\sqrt{K_2}\,\times m_{3/2}\approx 1.96\times m_{3/2}\,,\,\,\,\,\,k=1,...,N-1\,, \ee
independent of any parameters of the model.
For dS vacua with a very small cosmological constant, numerical computations for the choice $P=27$, $Q=30$
give the following masses for the remaining two states
\be\label{modul2}
M_{N}\approx 600\times m_{3/2}\,,\,\,\,\,\,\,\,\,\,M_{N+1}\approx 2.82\times m_{3/2}\,.
\ee
Note that all the masses are independent of the number of moduli $N$ as well as the rationals $a_i$.

Since the gravitino mass distribution peaks at ${\mathcal O}(100)$
TeV, which is also in the phenomenologically relevant range, and
the light moduli are roughly twice as heavy compared to the
gravitino, the moduli masses are heavy enough to be consistent
with BBN constraints. A detailed study of this issue will appear
in \cite{ToAppear}.

\section{Summary and Conclusions}\label{summary}

A major goal of string/$M$ theory is to find solutions that incorporate the
Standard Model, important clues to physics beyond the
SM such as gauge coupling unification, and in addition explain phenomena the
SM cannot explain. The most important unsolved problem is explaining the
value of the weak scale (at or below about a TeV) purely in terms of the
Planck Scale of $\mathcal{O}(10^{19}$ GeV) or some other fundamental scale,
the Hierarchy Problem. One obstacle
to solving the hierarchy problem is that all the moduli that characterize
the string theory vacuum must be stabilized for a meaningful solution.

In fluxless $M$ theory vacua the entire effective potential is
generated by non-perturbative effects and depends upon all the
moduli. In this paper we have studied this potential in detail for
a particular form of the K\"{a}hler potential when the
non-perturbative effects are dominated by strong gauge dynamics in
the hidden sector and when such vacua are amenable to the
supergravity approximation. In the simplest case, we studied
$G_2$-manifolds giving rise to two hidden sectors. The resulting
scalar potential has $AdS$ vacua - most of them with broken
supersymmetry and one supersymmetric one. Then we studied the
cases in which there was also charged matter in the hidden sector
under the plausible assumption that the matter K\"{a}hler
potential has weak moduli dependence. In these cases the potential
receives positive contributions from non-vanishing $F$-term vevs
for the hidden sector matter leading to a unique de Sitter
minimum. The de Sitter minimum is obtained without adding any
``uplifting'' terms, such as those coming from antibranes, which
explicitly break supersymmetry. With the form of the K\"{a}hler
potential given by (\ref{kahler}), we have explicitly shown that
all moduli are stabilized by the potential generated by strong
dynamics in the hidden sector.

In the de Sitter  minimum we computed $m_{3/2}$ and found that a
significant fraction of solutions have $m_{3/2}$ in the TeV
region, even though the Planck scale is the only dimensionful
parameter in the theory. The suppression of $m_{3/2}$ is due to
the old idea of dimensional transmutation which has also been used
in heterotic string theory. The eleven dimensional $M$ theory
scale turns out to be slightly above the gauge unification scale
but below the Planck scale. The absence of fluxes is significant
for simultaneously having $m_{3/2} \sim$ TeV naturally and
$M_{11}$ not far below the Planck scale.

The problem of why the cosmological constant is not large is of
course not solved by this approach. We do however understand to a
certain extent  what properties of $G_2$-manifolds are required in
order to solve it and we suggest that one can set the value of the
potential at the minimum to zero at tree level and proceed to do
phenomenology with the superpartners whose masses are described by
the softly broken Lagrangian. One particularly nice feature is
that we are able to explicitly demonstrate that the soft-breaking
terms are not sensitive to the \emph{precise} value of the
potential at the minimum. A tree-level tuning of the cosmological
constant is enough to determine the phenomenology.

When we set the value of the potential at the minimum to zero at tree level
a surprising result occurs. Doing so gives a non-trivial condition on the solutions. When
this condition is imposed on $m_{3/2}$, for generic $G_{2}$ manifolds it
turns out that the resulting values of $m_{3/2}$ are all in the TeV region.
Thus we do not have to independently set $V_{0}$ to zero \emph{and} set  $m_{3/2}$
to the TeV region as has been required in previous approaches.

A more detailed study of the phenomenology of these vacua, particularly
for LHC and for dark matter, is underway and will be reported in a future
paper. In the present paper we presented the relevant soft-breaking
Lagrangian parameters and mentioned a few broad and generic features of the
phenomenology, for both our generic solutions and for the case where $V_{0}$
is set to zero at tree level. We presented a standard supergravity
calculation of the soft breaking Lagrangian parameters, and found that the
scalar masses $m_{\alpha },$ and also the trilinears, are approximately
equal to $m_{3/2}$, to the extent that our assumptions about the matter
Kahler potential are valid. Remarkably, the tree-level gaugino masses are
suppressed by a factor of $\mathcal{O}(10-100)$. This suppression
is present for all $G_2$-manifolds giving the de Sitter minimum. \ For calculating
the tree level gaugino masses the matter K\"{a}hler potential does not enter, so
the obtained values at tree level are reliable. Because the gaugino suppression is large,
the anomaly mediated mass contributions are comparable to the tree level ones,
and significant cancellations can occur. Gluinos are generically quite light, and
should be produced copiously at LHC and perhaps even at the Tevatron -- this is an
unavoidable prediction of our approach. We have also checked
that radiative electroweak symmetry breaking occurs
over a large part of the space of $G_2$ manifolds, and that the lightest neutralino
is a good dark matter candidate. It will be exciting
to pursue a number of additional phenomenological issues in our approach, including inflation,
baryogenesis, flavor and CP-violation physics, Yukawa couplings and neutrino masses.

The approach we describe here apparently offers a framework that
can simultaneously address many important questions, from formal
ones to cosmological ones to phenomenological ones (apart from the
cosmological constant problem,which might be solved in a different
way). Clearly, however much work remains to be done. In
particular, a much deeper understanding of $G_2$-manifolds is
required to understand better some of the assumptions we made
about the K\"{a}hler potential of these vacua.

\acknowledgments

The authors appreciate helpful conversations with and suggestions
from  Jacob Bourjaily, Lisa Everett, Sergei Gukov, Renata Kallosh,
James Liu, Joseph Lykken, David Morrissey, Brent Nelson, Aaron
Pierce, Fernando Quevedo, Diana Vaman, Lian-Tao Wang and James
Wells. GLK and PK thank the Kavli Insitute for Theoretical Physics
(KITP), UCSB for its hospitality where part of the research was
conducted. The research of GLK and PK is supported in part by the
US Department of Energy and in part by the National Science
Foundation under Grant No. PHY99-07949.The research of KB and JS
is supported in part by the Department of Energy.

\appendix
\section{K\"{a}hler metric for visible chiral matter in $M$ theory}

As stated in section \ref{others}, we will generalize the result
obtained for the K\"{a}hler metric for visible sector chiral matter
fields in toroidal orientifold constructions in IIA
\cite{Bertolini:2005qh} to that in $M$ theory. The result obtained
for the K\"{a}hler metric for the (twisted) chiral matter fields
($\phi_{\alpha}$) in the supergravity limit ($\alpha' \rightarrow
0$) in \cite{Bertolini:2005qh} is:
\begin{eqnarray}\label{metric1}
\tilde{K}^{0}_{\bar{\alpha}\beta} &=& \prod_{i=1}^3
\left(\frac{\Gamma(1-{\theta}^{\alpha}_i)}{\Gamma(\theta^{\alpha}_i)}
\right)^{1/2}
\,{\delta}_{\bar{\alpha}\beta} \nonumber \\
\tan(\pi{\theta}^{\alpha}_i)&=&
\frac{U_2^i}{U_1^i+q^{\alpha}_i/p^{\alpha}_i};\;\; U^i\equiv
U^{i}_1+iU^{i}_2
\end{eqnarray}
where $i=1,2,3$ denote the number of moduli in the specific IIA
example, $\{p,q\}$ are integers and $U^i$ are the complex
structure moduli in type IIA in the geometrical basis. As
mentioned before, we will restrict to factorized rectangular tori
with commuting magnetic fluxes, for which $U_1^i=0$. Then \ba
\tan(\pi\theta^{\alpha}_i)=\frac{p^{\alpha}_i}{q^{\alpha}_i}\,U_2^i\ea

The first step towards the generalization is to identify the $G_2$
moduli in terms of IIA toroidal moduli by imposing consistency
between results of IIA and $M$ theory. The consistency check is the
formula for ${\kappa}^2$ - the physically measured four
dimensional gravitational coupling. We have:
\begin{eqnarray}\label{kappa4}
{\kappa}^2 &=& \frac{{\kappa}^2_{11}}{\mathrm{Vol}(X_7)} ,\;\;\;\;M\;Theory\,[19].\nonumber\\
{\kappa}^2 &=&
\frac{{\kappa}^2_{10}\,g_s^2}{\mathrm{Vol}(X_6)},\;\;\;\;IIA\;String\;Theory\;\;(\mathrm{eqn.\,
18.2.2\,of\,}\, [36]).
\end{eqnarray}
where $X_7$ and $X_6$ are the volumes of the internal 7-manifold
and 6-manifold (in IIA) respectively. The $M$ theory gravitational
coupling ${\kappa}^2_{11}$ and the string theory gravitational
coupling ${\kappa}^2_{10}$ can be written in terms of the string
coupling in IIA ($g_s\equiv e^{{\phi}_{10}^A}$) and $\alpha'$:
\ba \label{kappa11}{\kappa}^2_{11}&=&\frac{1}{2}(2\pi)^8g_s^3(\alpha')^{9/2}\,[19] \nonumber \\
{\kappa}^2_{10}&=&\frac{1}{2}(2\pi)^7{\alpha'}^4\; (\mathrm{eqn.\,
13.3.24\,of\, [36]})\ea Also, the volumes of $X_7$ and $X_6$ (for
a IIA toroidal orientifold) can be written as: \ba
\label{volumes}\mathrm{Vol}(X_7) &=& V_X\,l_M^7\;\;\;\;
\mathrm{where}\;\; V_X=\prod^N_{i=1}(s_i)^{a_i};\;\;\;
\frac{l^9_M}{4\pi} = {\kappa}^2_{11}\;[18] \nonumber \\
\mathrm{Vol}(X_6) &=& (2\pi R^{(1)}_1)(2\pi R^{(1)}_2)(2\pi
R^{(2)}_1)(2\pi R^{(2)}_2)(2\pi R^{(3)}_1)(2\pi R^{(3)}_2)\ea
Using (\ref{kappa11}), we get \be \label{Mlength} l_M =
2\pi{\alpha'}^{1/2}g_s^{1/3}\ee which gives rise to the following
expression for $\kappa$ in $M$ theory: \be
\label{kappa4-M}{\kappa}^2 = \frac{\pi{\alpha'}g_s^{2/3}}{V_X}\ee

In IIA String theory, the definition of the IIA moduli $T_i$ and
$U_i$ in terms of the geometry of the torus is given below. We
will stick to the case of a factorized $T^6$, rectangular tori,
and commuting magnetic fluxes (in the type IIB dual) for
simplicity, in which case only the imaginary parts of the moduli
are important:
\ba \label{moduli} Im(T)^i &\equiv& T^{(i)}_2 = \frac{R^{(i)}_2}{R^{(i)}_1} \nonumber \\
Im(U)^i &\equiv& U^{(i)}_2 =
\frac{R^{(i)}_1\,R^{(i)}_2}{{\alpha}'};\;\;i=1,2,3.\ea where
$R^{(i)}_1,R^{(i)}_2$ are the radii of the $i^{th}$ IIA torus
along the $x$ and $y$ axes respectively.

\noindent Now, from (\ref{kappa4}), (\ref{kappa11}),
(\ref{volumes}) and (\ref{moduli})we get: \be \label{kappa4-II}
{\kappa}^2 =
\frac{\pi{\alpha'}g_s^{2}}{U^{(1)}_2U^{(2)}_2U^{(3)}_2} \ee
Combining (\ref{kappa4-M}) and (\ref{kappa4-II}) gives: \be
\label{Vx} \boxed{V_X =
\frac{U^{(1)}_2U^{(2)}_2U^{(3)}_2}{g_s^{4/3}}}\ee

The above formula (eqn.(\ref{Vx})) is quite general and should
always hold\footnote{within the limits of the IIA setup
considered.}, since it has been derived by requiring consistency
between formulas for the physically measured gravitational
coupling constant. To identify the individual $G_2$ moduli
however, is harder and model-dependent. In the next subsection, we
will do the mapping for the case of a simple toroidal $G_2$
orbifold ($T^7/Z_3$) considered in \cite{Lukas:2003dn}, where it
can be shown that eqn (\ref{Vx}) is satisfied.

\subsection{Particular Case}

In \cite{Lukas:2003dn}, the definitions of the $G_2$ moduli (eqn
2.4) and the K\"{a}hler potential (eqn 2.10) are not given in a
dimensionless form. Therefore, we will make them dimensionless as
is done in \cite{Acharya:2005ez}. So, we have: \ba \label{lukasG2}
s^i \equiv \frac{a^i}{l^3_M} = \frac{\int_{C^i}
\tilde{\phi}}{l^3_M};\;\;\; K = -3\ln
\left(\frac{\mathrm{Vol}(X_7)}{l^7_M}\right) + \mathrm{constant}
\ea The volume of the manifold and the moduli in this
compactification are explicitly given as \cite{Lukas:2003dn}: \ba
\label{explicit}
\mathrm{Vol}(X_7) = \prod_{i=1}^{7} R_i;\;a^1 = R_1R_2R_7;\;a^2 = R_1R_3R_6;\;a^3 = R_1R_4R_5;\nonumber\\
a^4 = R_2R_3R_5;\;a^5 = R_2R_4R_6;\;a^6 = R_3R_4R_7;\;a^7 =
R_5R_6R_7. \ea From (\ref{explicit}), we can write
$\mathrm{Vol}(X_7)=(a^1a^2a^3a^4a^5a^6a^7)^{1/3}$, which implies
that $V_X=\prod^7_{i=1}(s_i)^{1/3}$. Therefore, in the notation of
\cite{Acharya:2005ez}, $a_i=1/3;\,i=1,2,...,7$. We will identify
the $M$ theory circle radius as $R_7$, the remaining six radii can
just be identified as the $x$ and $y$ radii of the three tori in
IIA: \ba
R_1=(2\pi)R^{(1)}_1;\;R_2=(2\pi)R^{(1)}_2;\;R_3=(2\pi)R^{(2)}_1;\;
R_4=(2\pi)R^{(2)}_2 \nonumber \\
R_5=(2\pi)R^{(3)}_1;\;R_6=(2\pi)R^{(3)}_2;\;R_7=(2\pi)g_s{\alpha'}^{1/2}.\ea
With these identifications, and using (\ref{Mlength}), we can
write the individual $G_2$ moduli in terms of the IIA moduli: \ba
\label{G2moduli}
s^1&=&U^{(1)}_2;\;\;\;s^6=U^{(2)}_2;\;\;\;s^7=U^{(3)}_2; \nonumber \\
s_2&=&\frac 1
{g_s}\left(\frac{T^{(3)}_2U^{(1)}_2U^{(2)}_2U^{(3)}_2}{T^{(1)}_2T^{(2)}_2}\right)^{1/2};\;\;
s_3=\frac 1 {g_s}\left(\frac{T^{(2)}_2U^{(1)}_2U^{(2)}_2U^{(3)}_2}{T^{(1)}_2T^{(3)}_2}\right)^{1/2};\nonumber\\
s_4&=&\frac 1
{g_s}\left(\frac{T^{(1)}_2U^{(1)}_2U^{(2)}_2U^{(3)}_2}{T^{(2)}_2T^{(3)}_2}\right)^{1/2};\;\;
s_5=\frac 1
{g_s}\left({T^{(1)}_2T^{(2)}_2T^{(3)}_2U^{(1)}_2U^{(2)}_2U^{(3)}_2}\right)^{1/2}\,.\nonumber
\ea \noindent Therefore, we get: \ba V_X =
\prod^7_{i=1}(s^i)^{1/3}=\frac{U^{(1)}_2U^{(2)}_2U^{(3)}_2}{g_s^{4/3}}\ea
which is the same as (\ref{Vx}). So, for the particular case, this
suggests the following generalization: \ba \tilde{K}_{\alpha} &=&
\prod_{i=1,6,7}
\left(\frac{\Gamma(1-\theta^{\alpha}_i)}{\Gamma(\theta^{\alpha}_i)} \right)^{1/2}\nonumber \\
\tan(\pi{\theta}^{\alpha}_i)&=&
\frac{p^{\alpha}_i}{q^{\alpha}_i}\,s_i;\;\;i=1,6,7\ea

\subsection{General Case}
For more general $G_2$ manifolds with many moduli, the precise map
of the individual moduli is not completely clear. However, it
seems plausible that the complex structure moduli appearing in the
K\"{a}hler metric in IIA map to a subset of the $G_2$ moduli in a
similar way, as in the particular case.

Therefore, we use the following expression for the K\"{a}hler
metric for visible chiral matter fields in $M$ theory:\ba
\label{metric2} \tilde{K}_{\alpha} &=& \prod_{i=1}^p
\left(\frac{\Gamma(1-\theta^{\alpha}_i)}{\Gamma(\theta^{\alpha}_i)}
\right)^{1/2}; \;\;i=1,2,..,p \leq N (\equiv N)
\nonumber \\
\tan(\pi{\theta}^{\alpha}_i)&=&
c^{\alpha}_i\,(s_i)^l;\;\;c^{\alpha}_i=
\mathrm{constant};\;l=\mathrm{rational\;number\;of\;\mathcal{O}(1)}.\ea

The derivatives of the K\"{a}hler metric w.r.t the moduli are very
important as they appear in the soft scalar masses, the trilinears
as well as the anomaly mediated gaugino masses, as seen from
section \ref{others}. The first derivatives appear in the
trilinears and anomaly mediated gaugino masses while the second
derivatives appear in the scalar masses. For the metric in
(\ref{metric2}), these can be written as follows:
\begin{eqnarray}\label{derivatives} & &\partial_n
\ln(\tilde{K}_{\alpha}) = \psi^{\alpha}_n \left(\frac{\partial
{\theta}^{\alpha}_n}{\partial z^n}\right); \;\;\;\;\;
\psi^{\alpha}_n(\theta^{\alpha}_n)\equiv
\frac{1}{2}\frac{d}{d{\theta}^{\alpha}_n}\,\ln(\frac{\Gamma(1-{\theta^{\alpha}_n})}
{\Gamma(\theta^{\alpha}_n)})\nonumber\\
& &\partial_{\bar{m}}\partial_n \ln(\tilde{K}_{\alpha}) =
\delta_{\bar{m}n}\,\left[{\psi}^{\alpha}_{\bar{n}n}\left(\frac{\partial
{\theta}^{\alpha}_n}{\partial \bar{z}^n}\right)
\left(\frac{\partial{\theta}^{\alpha}_n}{\partial
z^n}\right)+{\psi}^{\alpha}_n\left(\frac{\partial^2{\theta}^{\alpha}_n}
{\partial\bar{z}^n\partial
z^n}\right)\right];\;\;\;{\psi}^{\alpha}_{\bar{m}n}(\theta^{\alpha}_n)\equiv\frac{d}
{d\theta^{\alpha}_m}{\psi}^{\alpha}_n
\end{eqnarray} The functions $\psi^{\alpha}_n$ and
$\psi^{\alpha}_{\bar{n}n}$ depend on the angular variable
$\theta^{\alpha}_n$, which in turn depend on the moduli. The first
and second derivatives of ${\theta}^{\alpha}_n$ with respect to
$z_n$ are given by: \ba \label{partial} \frac{\partial
{\theta}^{\alpha}_n}{\partial z^n} &=& \frac{1}{2i}\frac{\partial
{\theta}^{\alpha}_n}{\partial s^n} =
\frac{l}{4\pi i\,s^n}\,\sin(2\pi{\theta}^{\alpha}_n) \\
\frac{\partial^2{\theta}^{\alpha}_n} {\partial\bar{z}^n\partial
z^n} &=& \frac{1}{4}\frac{\partial^2{\theta}^{\alpha}_n} {\partial
({s}^n)^2} =
\frac{l}{16\pi\,(s^n)^2}\,\left[l\sin(4\pi{\theta}^{\alpha}_n)-2\sin(2\pi{\theta}^{\alpha}_n)\right]
\nonumber \ea

The dependence of the soft parameters in section \ref{others} on
$\theta^{\alpha}_n$ is extremely simple. Instead of reexpressing
the dependence on $\theta^{\alpha}_n$ in terms of the moduli, it
is much more convenient to retain the dependence on
$\theta^{\alpha}_n$, as $\theta^{\alpha}_n\in [0,1)$
\cite{Bertolini:2005qh} and the variation of relevant functions
with $\theta^{\alpha}_n$ in the allowed range can be plotted
easily. In particular, the function $F(\theta^{\alpha}_n)\equiv
\frac{1}{2\pi}\{{\psi}^{\alpha}_n\,\sin(2\pi{\theta}^{\alpha}_n)\}$
appears in the expression for the anomaly mediated gaugino masses
(eqn. (\ref{anomalyds13})) and the trilinears (eqn.
(\ref{trides})), the function $G(\theta^{\alpha}_n) \equiv
\ln\left(\frac{\Gamma(1-{\theta}^{\alpha}_n)}{\Gamma({\theta}^{\alpha}_n)}\right)$
appears in the expression for the trilinears (eqn. (\ref{trides}))
and the function $H(\theta^{\alpha}_n) \equiv
\frac{1}{4\pi}\{l^2\,{\psi}^{\alpha}_{\bar{n}n}\,\sin^2(2\pi{\theta}^{\alpha}_n)+l^2\,{\psi}^{\alpha}_n\,
\sin(4\pi{\theta}^{\alpha}_n)-2l\,{\psi}^{\alpha}_n\,\sin(2\pi{\theta}^{\alpha}_n)\}$
appears in the expression for the scalars (eqn.
(\ref{scalarsds2})). The variation of these functions with
$\theta^{\alpha}_n$ is quite mild as seen from Figure \ref{Am2}:

\begin{figure}[h!]\label{Am2}
\begin{tabular}{ccc}
      \leavevmode \epsfxsize 5.5 cm \epsfbox{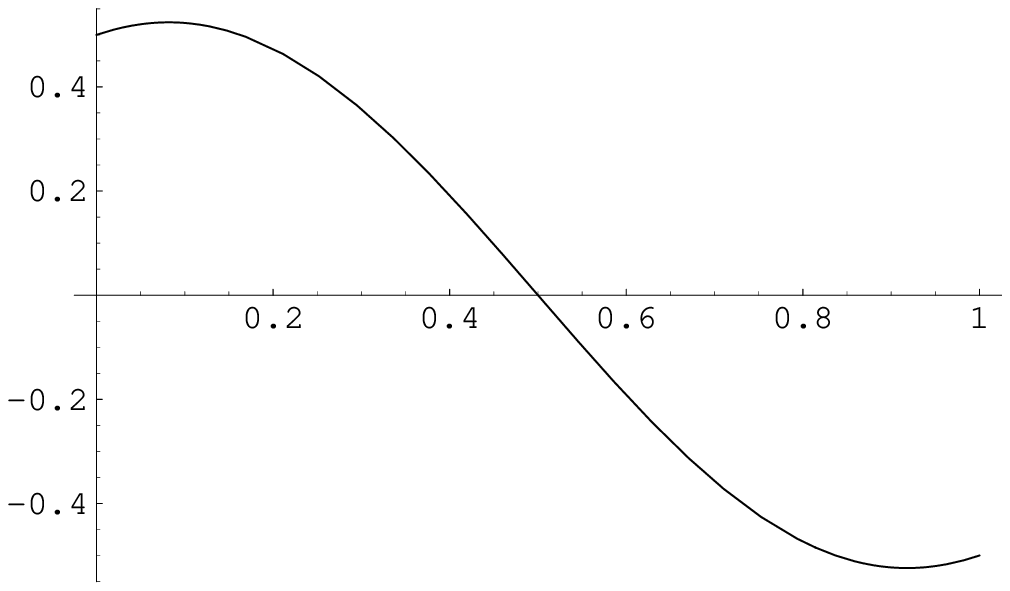}&
      \leavevmode \epsfxsize 5.5 cm \epsfbox{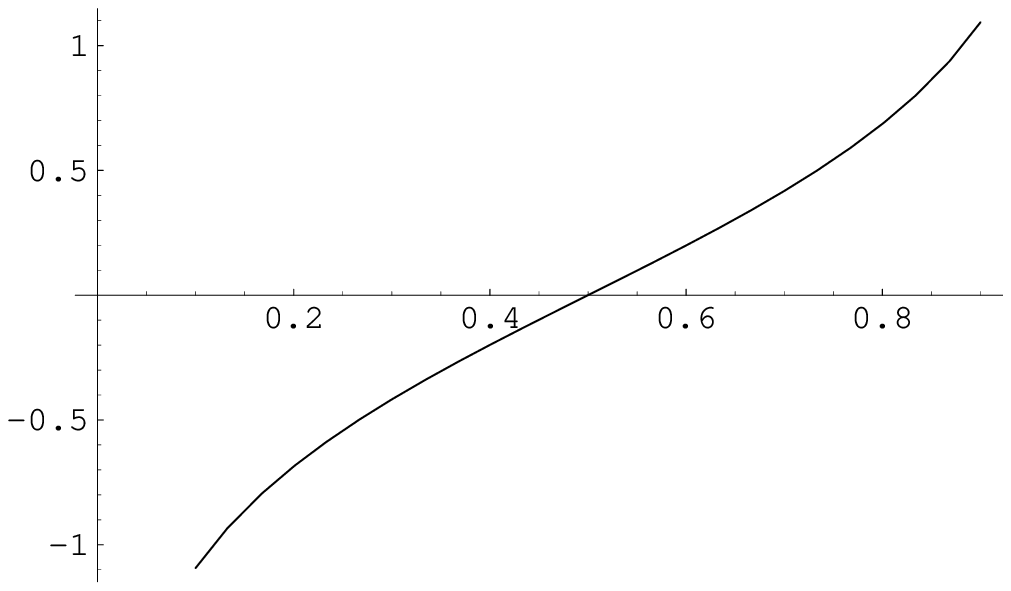}&
      \leavevmode \epsfxsize 5.5 cm \epsfbox{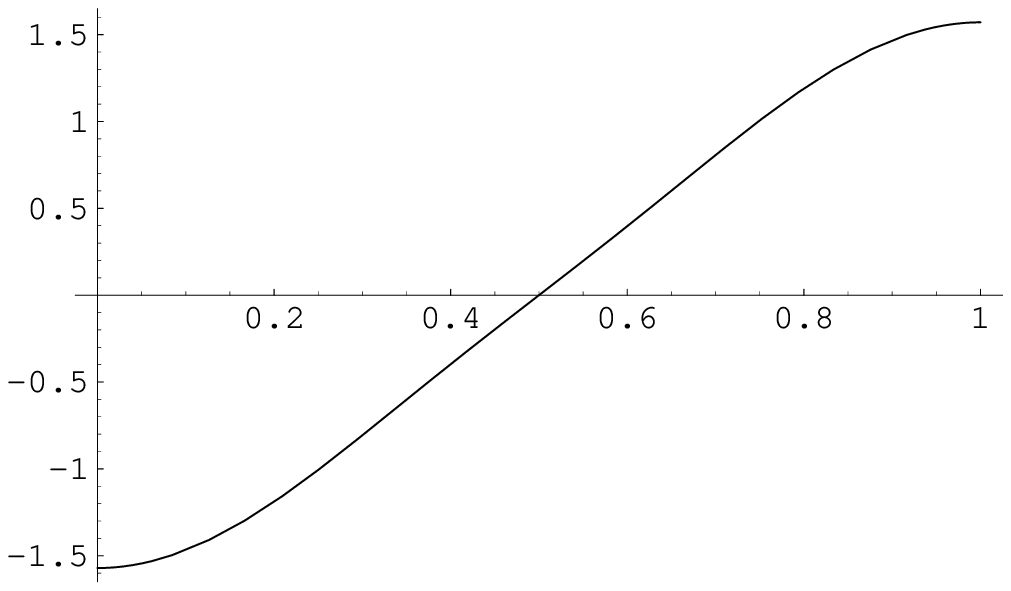}
\end{tabular}
\caption{\footnotesize{Left: $F$ as a function of
$\theta^{\alpha}_n$. Middle: $G$ as a function of
$\theta^{\alpha}_n$. Right: $H$ as a function of
$\theta^{\alpha}_n$ (for l=1). From \cite{Bertolini:2005qh},
$\theta_n \in [0,1)$.}}
\end{figure}
Since the functions $F$, $G$ and $H$ vary very mildly with
$\theta^{\alpha}_n$ and are $\mathcal{O}(1)$ in the whole range,
it is reasonable to replace the above functions by
$\mathcal{O}(1)$ numbers, as is done in section \ref{others}. This
is justified since we are only interested in estimating the rough
overall scales of various soft parameters.

\end{document}